# The impacts of incarceration on crime

David Roodman[1]

Open Philanthropy Project

September 2017

**Summary:** This paper reviews the research on the impacts of incarceration on crime. Where data availability permits, reviewed studies are replicated and reanalyzed. Among three dozen studies I reviewed, I obtained or reconstructed the data and code for eight. Replication and reanalysis revealed significant methodological concerns in seven and led to major reinterpretations of four. I estimate that, at typical policy margins in the United States today, decarceration has zero net impact on crime outside of prison. That estimate is uncertain, but at least as much evidence suggests that decarceration reduces crime as increases it. The crux of the matter is that tougher sentences hardly *deter* crime, and that while imprisoning people temporarily *stops* them from committing crime outside prison walls, it also tends to increase their criminality *after* release. As a result, "tough-on-crime" initiatives can reduce crime in the short run but cause offsetting harm in the long run. A cost-benefit analysis finds that even under a devil's advocate reading of this evidence, in which incarceration does reduce crime in U.S., it is unlikely to increase aggregate welfare.

[1] I thank Holden Karnofsky for guidance and support, Mark Schaffer for advice on applying the Anderson-Rubin test; Tim Carr for generous assistance with data from Georgia; Peter Ganong, Steven Levitt, and Thomas Marvell for sharing data; Ilyana Kuziemko for sharing code and arranging access to data; Donald Green and Alex Tabarrok for sharing or posting data and code; and David Abrams, John Berecochea, Paolo Buonanno, Chloe Cockburn, Gordon Dahl, Rafael Di Tella, Joseph Doyle, Peter Ganong, Donald Green, Randi Hjalmarsson, Ilyana Kuziemko, Gerry Gaes, Karalyn Lacey, Edward Miguel, Michael Mueller-Smith, Daniel Nagin, Emily Owens, Steven Raphael, Max Schanzenbach, Alex Tabarrok, Ben Vollaard, David Weisburd, and Crystal Yang for reviewing full or partial drafts. Thanks also to my GiveWell colleagues for valuable feedback. **All views expressed are attributable to me alone.**

## Contents

## 1. Introduction

When it comes to locking people up, the United States has become a world champion. In 1970, 196,000 people resided in American prisons, and another 161,000 in jails, which worked out to 174 inmates per 100,000 people (Census Bureau 1973, Tables 271, 273). In 2015, 1.53 million people languished in US prisons and 728,000 in jails, or 673 per 100,000 (BJS 2016a, Table 1).[2] Only North Korea, among major nations, may surpass the US in this regard.[3] Such statistics are almost always invoked and graphed when initiating discussions of criminal justice reform. Figure 1 and Figure 2 depict them afresh with photographs taken at the Eastern State Penitentiary in Philadelphia.[4] That fortress-like complex is now a museum, a window onto a criminal justice reform movement of some two centuries ago that sought to replace corporal punishment with solitary confinement, which was seen as humane and rehabilitative.[5]

The Open Philanthropy Project has joined a latter-day criminal justice reform movement. It too is motivated by the belief that something is wrong with the state's use of punishment to combat crime. Something is wrong, in other words, with those pictures. Higher incarceration rates and longer sentences, along with the "war on drugs," have imposed great costs on taxpayers, as well as on inmates, their families, and their communities (Alexander 2012). Yet even though the 59% per-capita rise in incarceration between 1990 and 2010 accompanied a 42% drop in FBI-tracked "index crimes," researchers agree that putting more people behind bars added modestly, at most, to the fall in crime (e.g., Levitt 2004; Tonry 2014; Roeder, Eisen, and Bowling 2015).[6]

**Figure 1. Prisoners per 100,000 residents by decade, US, 1900–2010 (Eastern State Penitentiary's Big Graph, southern view)**

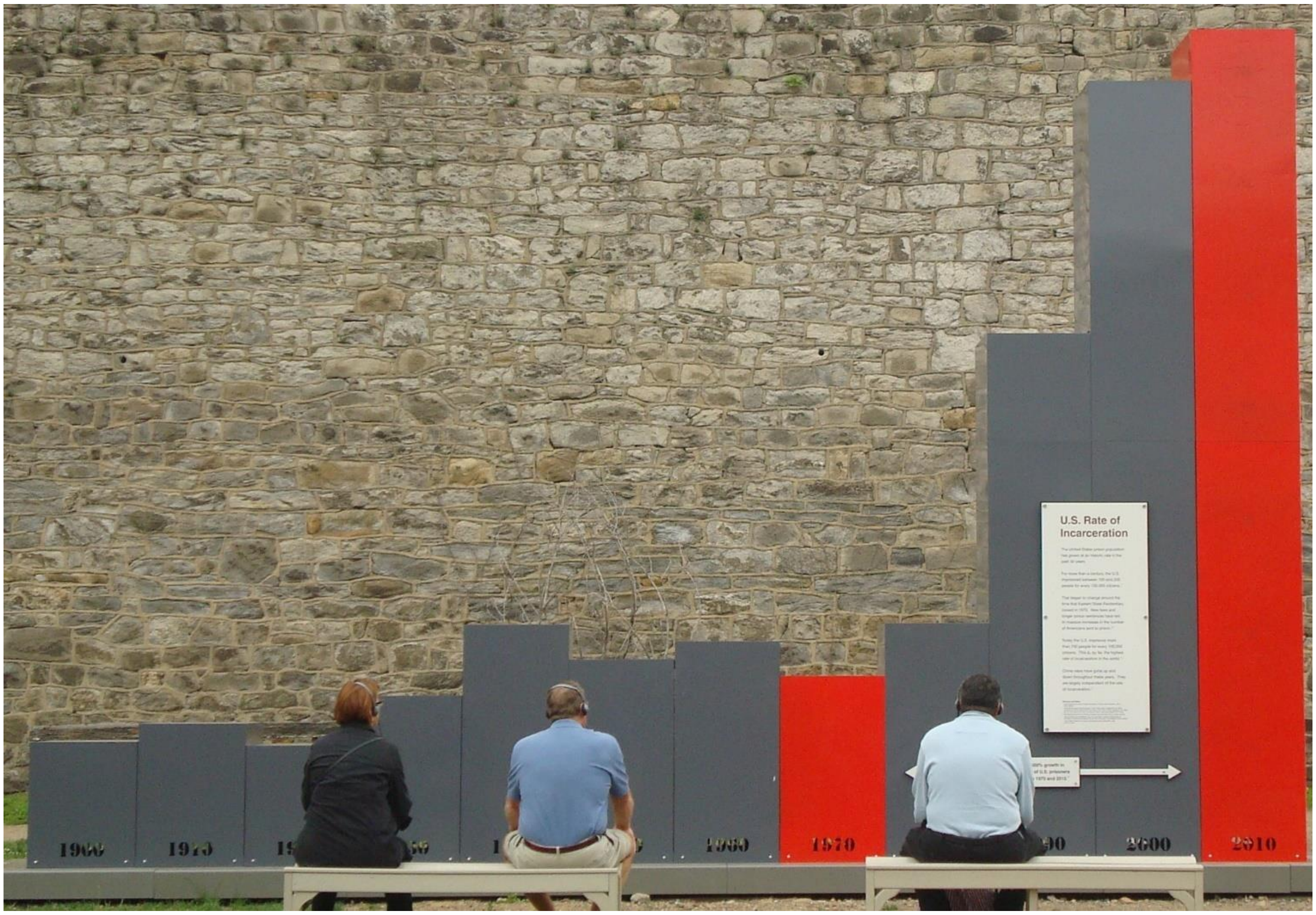

[2] For concision, I will sometimes use "prison" to mean jail and prison.
[3] prisonstudies.org/highest-to-lowest/prison_population_rate?field_region_taxonomy_tid=All; wikipedia.org/wiki/Prisons_in_North_Korea. Seychelles has a higher rate, apparently because it is home to a UN-funded prison housing Somali pirates (bbc.com/news/magazine-22556030).
[4] Images courtesy of Eastern State Penitentiary Historic Site. Figure 1 by Nicole Fox. Figure 2 by Rob Hashem.
[5] Eastern State Penitentiary website, j.mp/2bG0Izp.

Now, even if rising incarceration has not been a major factor behind falling crime, it might still have been a factor—and enough so that it ought to give pause to those pushing to reverse the incarceration boom. This report works to check that possibility, by reviewing empirical research on the impacts of incarceration on crime. It asks whether decarceration should be expected to increase or decrease crime. With the Open Philanthropy Project making grants for criminal justice reform, this review of the research is an act of due diligence.

**Figure 2. Prisoners per 100,000 residents by country, 2010 (Eastern State Penitentiary's Big Graph, eastern view)**

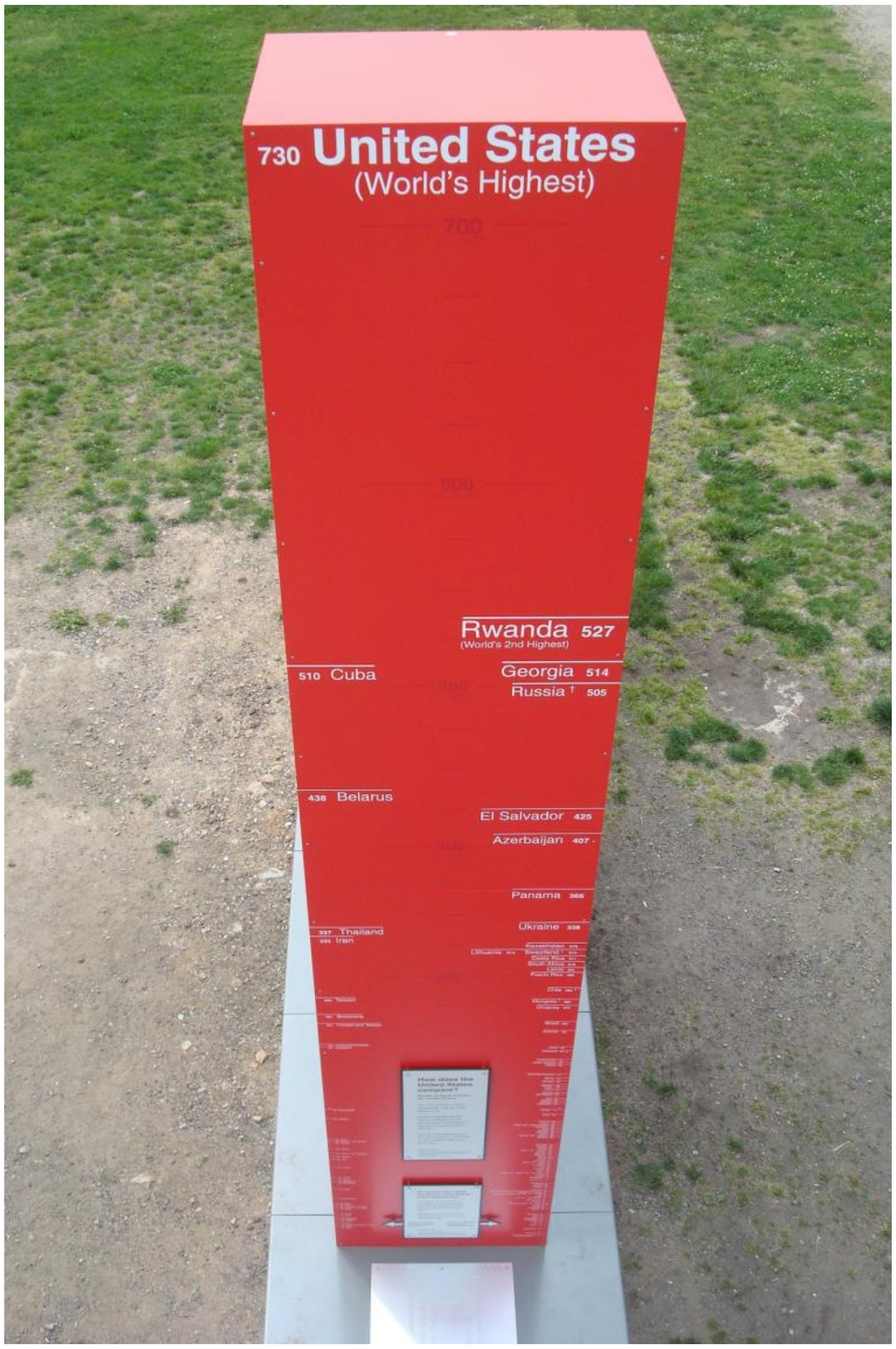

Any discussion of the impacts of incarceration should specify the alternative: incarceration as opposed to what? This review focuses mainly on studies that compare incarceration to ordinary freedom or traditional supervised released (probation and parole), as distinct from alternatives such as in-patient drug treatment and restorative justice conferences (Strang et al. 2013).[7] Those options may offer promise, and deserve more research and evidence reviews. Nevertheless, as a practical matter, if incarceration falls substantially in this country, ordinary and traditional supervised release will probably emerge as the main alternatives. That appears to have been the case in trend-setting California after decarceration reforms in 2011 and 2014.[8] Thus this review remains highly relevant to likely policy choices.

For manageability, this review restricts it to "high-credibility" studies: ones that exploit randomized experiments, or else "quasi-experiments" that arise incidentally from the machinations of the criminal justice system and ideally produce evidence nearly as compelling as experiments do (Angrist and Pischke 2010).

---

[6] 1,148,176 were incarcerated in the US at end-1990 (BJS 1992, Table 1.1), or 462 per 100,000. "Index crimes" are those long tracked by the FBI: homicide, rape, aggravated assault and robbery (violent crimes) and burglary, arson, motor vehicle, and larceny-theft (property crimes). They exclude drug crimes, fraud and identity theft, driving under the influence of alcohol, misdemeanors, and other crimes. Robbery is considered a violent crime because it is a crime against a person, as in a mugging. The FBI-reported violent and property crime rates were 729.6 and 5,073.1 per 100,000 in 1990 and 404.5 and 2,945.9 in 2010 (Sourcebook of Criminal Justice Statistics 2012, Table 3.106.2012).

[7] Probation is often thought of as being granted in lieu of incarceration while parole comes after. A more precise statement is that probation is granted by a judge while parole is granted by a parole board. In fact, a judge can sentence a person to incarceration followed by probation. A parole board might then split the incarceration sentence into two parts, one served behind bars, one served on parole.

[8] "At the state level, alternatives to custody are limited" in California. At the county level, "alternative custody placements for realigned offenders have increased but are being used for a low number of…inmates." (Martin and Grattet 2015, pp. 2, 3).

Further, in distilling generalizations and performing cost-benefit analysis, the review relies more heavily on the eight studies that I could replicate by accessing the underlying data and computer code.[9] Replication and subsequent reanalysis of these eight revealed significant econometric concerns in seven and led to major reinterpretations of four.

That experience led to an unexpected conclusion about the conduct of social science generally. For it raised doubts about the rest of the high-credibility studies included in this review, the ones that could *not* be so closely examined. It forced me to conclude that **even the best studies on incarceration and crime are less reliable than they appear**. And, like a car whose brakes fail once, **this raises questions about the reliability of published social science generally.** To put that more constructively, the scrutiny that research undergoes to appear in social science journals falls short of the optimum for policymaking. Perhaps the gap needs to be filled outside the normal academic research process, such as through reviews like this one.

As for the substance of this review, one can imagine that increasing incarceration either raises or lowers crime overall. Making incarceration likelier or longer may deter crime before it happens; prevent offenses by those behind bars; and make them more law-abiding afterward, by teaching job skills, treating drug addiction, or "scaring people straight." On the other hand, deterrence may be weak at the margin, especially for the most heinous crimes. Before attacking, does a potential rapist gather and weigh data on the local conviction rates and sentencing patterns? And putting more people in prison may cause more crime in prison—a possibility hardly studied. Finally, incarceration may be more *criminogenic* than rehabilitative. Having been imprisoned may make it harder for people to find legal employment, may psychologically alienate them from society, or may strengthen their social bonds with criminals, all of which could raise recidivism (Nagin, Cullen, and Jonson 2009, pp. 122–28).

Since plausible theories point in each direction, the question of the net impact of incarceration on crime must be brought to the data. Having reviewed and revisited published analyses in unprecedented depth, **my best estimate of the impact of additional incarceration on crime in the United States today is zero**. **And, while that estimate is not certain, there is as much reason overall to believe that incarceration increases crime as decreases it.**

Explaining the findings of this review more fully requires a conceptual preliminary hinted at above. Incarceration can be thought of as affecting crime before, during, and after: before incarceration, in that stiffer sentences may deter offending; during, in that people inside prison cannot physically commit crime outside; and after, in that having been incarcerated may shift one's chance of reoffending. The first is here called "deterrence," the second "incapacitation," and the third "aftereffects."

The reasoning that decarceration is unlikely to increase crime runs this way:

1. *Deterrence is de minimis*. Helland and Tabarrok's (2007) study of California's "Three Strikes and You're Out" law suggests that increasing sentences by 10% cut crime by 1%, for an "elasticity" of −0.1. Abrams (2012) looks at the impacts of two kinds of state laws—mandatory sentencing minimums and add-ons for crimes involving a gun—on two kinds of crime—gun-involved assault and gun-involved robberies. The study finds an impact in one of the four combinations, also with an elasticity of about –0.1. But reanalyses of the two studies calls even those mild estimates into question. Separately, a promising program in Hawaii that deploys swift sanctions to deter probation violations largely did not pan out in five replications on the mainland.
2. *Incapacitation is real, at least for acquisitive crime*: putting people in prison reduces crime outside of prison for the duration of their stays. (Of course, incarceration also creates new opportunities for crime *inside* of prison.) Credible estimates of incapacitation—defined here as the crime reduction outside of prisons—range widely by context. Particularly salient is the experience of California after the 2011 "realignment," which reduced confinement of people convicted of non-serious, non-sexual, nonviolent offenses. I tentatively estimate that each person-year of averted incarceration caused 6.7 more property crimes in the state—burglary, general theft, motor vehicle theft—among which the impact on motor vehicle thefts is clearest, at 1.2.[10]

[9] Exceptions are Iyengar (2008) and Roach and Schanzenbach (2015). See note 21.

3. *Most studies find that aftereffects are harmful:* more time in prison, more crime after prison. *In particular, all but one of the five studies that compare incapacitation and aftereffects in the same context find aftereffects to at least cancel out incapacitation.*[11] For example, Green and Winik (2010) calculate that drug defendants in Washington, DC, who happened to appear before longer-sentencing judges were at least as likely to be rearrested within four years as those appearing before shorter-sentencing judges—even though the first group spent more of those four years in prison, when they could not be rearrested. Evidently, while longer sentences temporarily suppressed criminality outside prison, they raised the odds of rearrest.

In short, incarceration's "before" effect is mild or zero while the "after" cancels out the "during."

Since this conclusion may be or look biased coming from an organization promoting decarceration, the review also develops a devil's-advocate position. From the evidence gathered here, how could one most persuasively contend that decarceration would endanger the public? I think the strongest argument would challenge as biased my critical reanalysis of the two studies finding mild deterrence (item 1). It would then invoke the minority of aftereffects studies that contradict item 3 above, concluding that longer sentences *do* reduce post-release criminality, notably Kuziemko (2013) and Ganong (2012)—setting aside my critical reanalyses of those as well. Then, incarceration would be seen as reducing crime before, during, and after.

Table 1 depicts the two interpretations considered here: the primary synthesis, and the devil's-advocate view. Impacts are expressed with respect to decarceration, so that a "+" means that decarceration would increase crime and a "–" means opposite.

Now, if the devil's advocate is right, the crime increase from decarceration might still be small enough that most people would view the tradeoff as worthwhile. After all, decarceration saves taxpayers money, increases the liberty of and economic productivity of citizens, and reduces disruption of their families and communities. To explore this possible trade-off more rigorously, the report closes with a cost-benefit analysis.

**Table 1. Thumbnail of primary and devil's-advocate estimates of the marginal impact of decarceration on crime in the US today based on replicable studies**

| | Primary synthesis of evidence | Devil's-advocate view |
|---|---|---|
| Deterrence | 0 | + (mild) |
| Incapacitation | + | + |
| Aftereffects | – | + |
| Total | 0 | + |

Overall, I estimate the societal benefit of decarceration at $92,000 per person-year of averted confinement. That figure is dominated by taxpayer savings and gained liberty. The crime increase perceived by the devil's advocate translates into $22,000–$92,000, depending on the method used to express crime's harm in dollars. I argue that the methodology behind the high figure is less reliable. It works from surveys that asked people how much they would pay for a 10% crime decrease, even though most Americans do not know how much crime occurs near them, thus what it would mean to cut it 10%. But if we accept the high figure, then in the worst-case valuation of the worst-case scenario plausibly rooted in the evidence, decarceration is about break-even. Given the great uncertainties in that calculation—about the crime impact of decarceration, the money value of crime victimization, the value of liberty—the precision in the worst-case assessment—$92,000 in costs, $92,000 in benefits—is an illusion. The worst case should be viewed as roughly break-even.

[10] This conclusion lines up well with that of the reviewed study, Lofstrom and Raphael (2016).

[11] Among the five studies—Green and Winik (2010), Loeffler (2013), Nagin and Snodgrass (2013), Mueller-Smith (2015), and Roach and Schanzenbach (2015)—only the last dissents. It is also the one where the quality of the quasi-experiment is least certain. See §9.8.

Meanwhile, this review's primary interpretation of the evidence puts the crime cost of decarceration at zero, which makes the benefit-cost ratio infinite.

In ending this review with a cost-benefit analysis, I implicitly invoke utilitarianism. But I do not mean to suggest that deontological moral frames, built more around notions of justice than cost, deserve no place in deliberations on criminal justice. Rather, I focus on cost-benefit analysis because I believe it has moral and political relevance, and because it is a place where I am especially suited to contribute.

Overall, it looks very hard to prove beyond a reasonable doubt that at typical margins in the US today, putting more people behind bars does society net good. More likely it is decarceration that passes the cost-benefit test.

## 2. Conceptual preliminaries

Before diving into individual studies, this section introduces some basic concepts and leitmotifs. Most have to do with the ways that incarceration can affect crime—or falsely appear to.

### 2.1. Swiftness, certainty, and severity

One could say the original criminal justice reformers were Enlightenment thinkers—Montesquieu, Voltaire, Beccaria, Bentham—who decried the cruelty of punishment in their day. They argued that government ought to impinge on liberty only to the extent necessary to secure its blessings. Torture and capital punishment, then routine, were therefore wrong.[12] Beccaria:

> *The degree of the punishment, and the consequences of a crime, ought to be so contrived as to have the greatest possible effect on others, with the least possible pain to the delinquent.* If there be any society in which this is not a fundamental principle, it is an unlawful society; for mankind, by their union, originally intended to subject themselves to the least evils possible. (Beccaria trans. 1819, p. 75; emphasis in original)

The utilitarian goal of minimizing total suffering led to an interest in the characteristics of punishment. Beccaria (trans. 1819, p. 76) urged that punishment be swift, "if we intend that, in the rude minds of the multitude, the seducing picture of the advantage arising from the crime should instantly awake the attendant idea of punishment." The swifter the punishment, the less severe it needed to be to achieve the same deterrence. That idea was enshrined in the US Constitution, in the Sixth Amendment's assertion of the right to a speedy trial.

In his more encyclopedic take on the principles of punishment, Bentham (pub. 1838, p. 401) completed what has become a standard triad of traits: the swiftness, *certainty*, and severity of punishment. "That the value of the punishment may outweigh the profit of the offence, it must be increased in point of magnitude, in proportion as it falls short in point of certainty."

Because of our concern about the rise of mass incarceration, this review dwells most on severity. Does a tougher sentencing regime—requiring more time in prison or incarceration in harsher, higher-security conditions—lead to more or less crime outside prison walls? However, the first studies reviewed will focus on swiftness and certainty, because of the hope that they can sometimes substitute for severity.

### 2.2. Causal channels from incarceration to crime

In another triad tracing back to Bentham, the effects of punishment are often categorized into general deterrence, incapacitation, and specific deterrence.[13] As noted in the introduction, the three terms map neatly onto the three timeframes: before, during, and after punishment. But the term "specific deterrence" fails to capture all the potential consequences of incarceration, so I have replaced it with "aftereffects." This

[12] Voltaire: j.mp/1QBL8Tz, j.mp/1QBLael. Montesquieu: j.mp/1QBMe1F. Beccaria: j.mp/1QBMsWL. Bentham: j.mp/1QBMHRI.

[13] "[P]unishment has three objectives: incapacitation, reformation, and intimidation." (Bentham pub. 1838, p. 396.)

then allows me to write "deterrence" in place of "general deterrence."

Organizing with respect to this triad of effects of punishment, this subsection lists theories about how incarceration affects crime. It does not try to judge the pervasiveness or strength of these channels. Reviewing the evidence comes later.

### 2.2.1. Before incarceration: deterrence

Deterrence is the prevention of crime by the threat of sanction. It almost certainly happens. How much is an empirical matter, and presumably varies by context.

### 2.2.2. During incarceration: incapacitation

Incapacitation is the prevention of crime outside prison by putting would-be criminals behind bars. Like deterrence, incapacitation almost certainly occurs, but how much is a question admitting many answers, depending on where, when, and who we are talking about.

This definition of incapacitation is subject to one important complication. More people doing time in prison means more people perpetrating crimes in prison. Prison crime is usually neglected in discussions of incapacitation, and is probably underrepresented in official statistics. What is normally meant by incapacitation is the prevention of crime outside prison walls. Put otherwise, incapacitation *could* be negative in some contexts, when it leads people to commit more crime in jail or prison than they would have outside. However, it would probably rarely show up that way in official statistics.

Because prison crime is so rarely studied, and because crime outside of prison has particular political salience, "incapacitation" in this review will refer to crime outside of prison.

### 2.2.3. Aftereffects

Because being locked up is a powerful experience, this channel from incarceration to crime is the most variegated. Incarceration can change one's life in many ways that in turn affect criminality after.

The traditionally favored term, "specific deterrence," captures the idea that doing time viscerally strengthens the fear of punishment, rather like swiftness and certainty, and thus deters people from reoffending. More succinctly, confronting criminals with consequences and removing them from society changes them for the better. The corrections system corrects. Penitentiaries elicit penitence.

No doubt, those things do often happen. And prisons do good in other ways. They may help (or force) people off of addictive substances, teach job and life skills, or improve literacy and self-control.

However, as Nagin, Cullen, and Jonson (2009, pp. 122–28) point out, the prison experience may also be criminogenic. It may alienate people from society, giving them less psychological stake in its rules. It may make people better criminals by giving them months together to learn from each other. It may strengthen their allegiances to gangs whose social reach extends into prisons. While some may get drug treatment, others may not, even as they suffer from withdrawal or preserve access to drugs. And incarceration can permanently mark people as felons, making it hard to find legal employment or obtain safety net social services, thus pushing them back to crime.

Thus whereas we can predict the sign of the measured effects of deterrence and incapacitation, if not the magnitude, net aftereffects could in principle go either way.

## 2.3. On measuring crime

It is hard to measure crime rates precisely. Many crimes never make it into official statistics because victims do not report them, out of shame, a sense of futility, or distrust of the police. Even for reported crimes, police may never identify perpetrators, which means that studies that track individuals over time may miss offenses that these individuals commit. The nature of available data also constrains how researchers define offenses and recidivism. Definitions appearing in studies reviewed here include being arrested, being charged, appearing in court, being convicted, and being sent to prison. A researcher working with court records, for example, may define recidivism as later reappearance in court, while one working with prison records may use later return to prison. *None* of these definitions fully captures criminality.

## 2.4. Five confounders

Even if criminal behavior were perfectly measured, the impacts of incarceration upon it would remain hard to assess. In this domain, as in so many others, causal arrows run every which way, with crime affecting incarceration, and third factors affecting both. If person A spends more time in prison than person B, and commits more crime after, that does not prove incarceration caused crime. Perhaps A served longer for committing a more serious crime, or for having more prior convictions, indicating a greater propensity to offend regardless of time behind bars.

By exploiting experiments or quasi-experiments, the studies reviewed here make strong claims to slice the Gordian knot of causality. And many do find effects on crime, however imperfectly measured, not easily ascribed to chance.

But even the results of the best studies can in principle be explained, or explained away, by other theories, which must be confronted in the course of reviewing them.

### 2.4.1. Aging

Of this bit of criminological wisdom, there is little doubt: young men commit the most crime. Adolphe Quetelet (1833, p. 66), a pioneer in the application of statistics to social science, documented the pattern in France in the late 1820s. The FBI does the same for the US today. In 2014, the peak ages for total arrests were 19–23, with about 345,000 arrests for each year of age in that range (FBI 2015, Table 38). Similar things can be said for other countries and decades (Farrington 1986). Hirschi and Gottfredson (1983) go so far as to call the pattern an "invariant" law of human nature.

Researchers do not fully understand how aging reduces crime. Possibly the brain matures and the criminal tendency wanes at the same rate whether or not a person is behind bars. That would support Hirschi and Gottfredson's thesis about the universality of the pattern. And in that case, releasing a young person from prison should on average increase crime more than releasing an older one. But if criminality is mainly tamed by growing bonds to workmates and lifemates, then incarceration could slow, even reverse, the normal processes by which aging reduces crime.

The universality of the crime peak among the youngest adults suggests that releasing relatively old people—say, in their 40s or later—is a promising tactic in the quest to cut incarceration while protecting public safety. However, the crime-age relationship contains some hidden complexities (NRC 1986; Farrington 1986; Piquero, Farrington, and Blumstein 2003; Farrington, Piquero, and Jennings 2013; Ulmer and Steffensmeier 2014). Whether a 35-year-old man drawn from the general population is more apt to break the law than a 20-year-old so drawn is one thing; whether a 35-year-old prison releasee is more apt to reoffend than a 20-year-old releasee is another.

The aging effect complicates the interpretation of some studies. Consider this example. In 1970, the government of California randomly paroled some prisoners six months early (Berecochea and Jaman 1981, reviewed below). 11.2% of the early parolees ended up back in prison within two years of release, more than the 8.0% of the control group ($p = 0.06$; see Table 12). So is the right conclusion that shortening prison spells caused more crime? Or is it just that people who got out six months sooner were six months younger, thus more crime-prone?

The problem is even more devilish than that example suggests. For to completely expunge the aging effect, an experiment would need to vary time served while holding constant both age of release from prison and age of entry—which is impossible. Or it could limit incarceration spells to a few weeks, which would reduce its relevance to understanding the impacts of longer spells.

Avoidance of the aging effect is one reason many studies reviewed here—those based on quasi-experiments in sentencing—start the follow-up clock not at release, but at (quasi-) randomization (Green and Winik 2010; Nagin and Snodgrass 2013; Loeffler 2013; Mueller-Smith 2015).

In fact, having borne the aging effect in mind during this review, I've concluded that it usually does not threaten study validity, for two reasons. First, age is observed. In the social sciences generally, the greatest worries about misleading results relate to unobserved factors, such as personality traits that lead both to longer sentences this year and more crime next year. Incarceration studies routinely control for age, which substantially lessens the concern.

Second, the aging effect appears too small or has the wrong sign to explain the results relating to aftereffects in most of the studies reviewed. In particular, the average subject in most is in his early thirties, when it appears that few people end their "criminal careers." In making this claim, and in judging the threat of the aging effect in this review, I lean heavily on Blumstein, Cohen, and Hsieh (1982), which analyzes arrest data from Washington, DC, in the 1970s. Although the study is old and purely descriptive—it does not claim quasi-experimental identification—it remains the best study of the aging effect I know of. Figure 3, adapted from the paper's Figure 13, shows the fraction of people that experienced their last arrest at each given age, which indicates the propensity to *exit* criminality at that age. The dotted lines connect actual data points while the solid ones show best fits within three age ranges. The exit rate is extrapolated to 50%/year at age 18 (marked as time zero in graph) and falls rapidly from there through about age 30, where it roughly plateaus at 8%. Then, in the early forties, the exit rate starts to climb, as people approach a kind of criminal retirement. In contrast, for example, the median subject in the California experiment was in his early 30's (Berecochea and Jaman 1981, Table 4), so the 29% relative drop in criminality (from 11.2% to 8.0%) in just six months looks far larger than an aging effect of 8%/year found in DC among people in their 30s.

**Figure 3. Probability of exit from criminal career as a function of years since 18th birthday, Washington, DC, mid-1970s, from Blumstein, Cohen, and Hsieh (1982)**

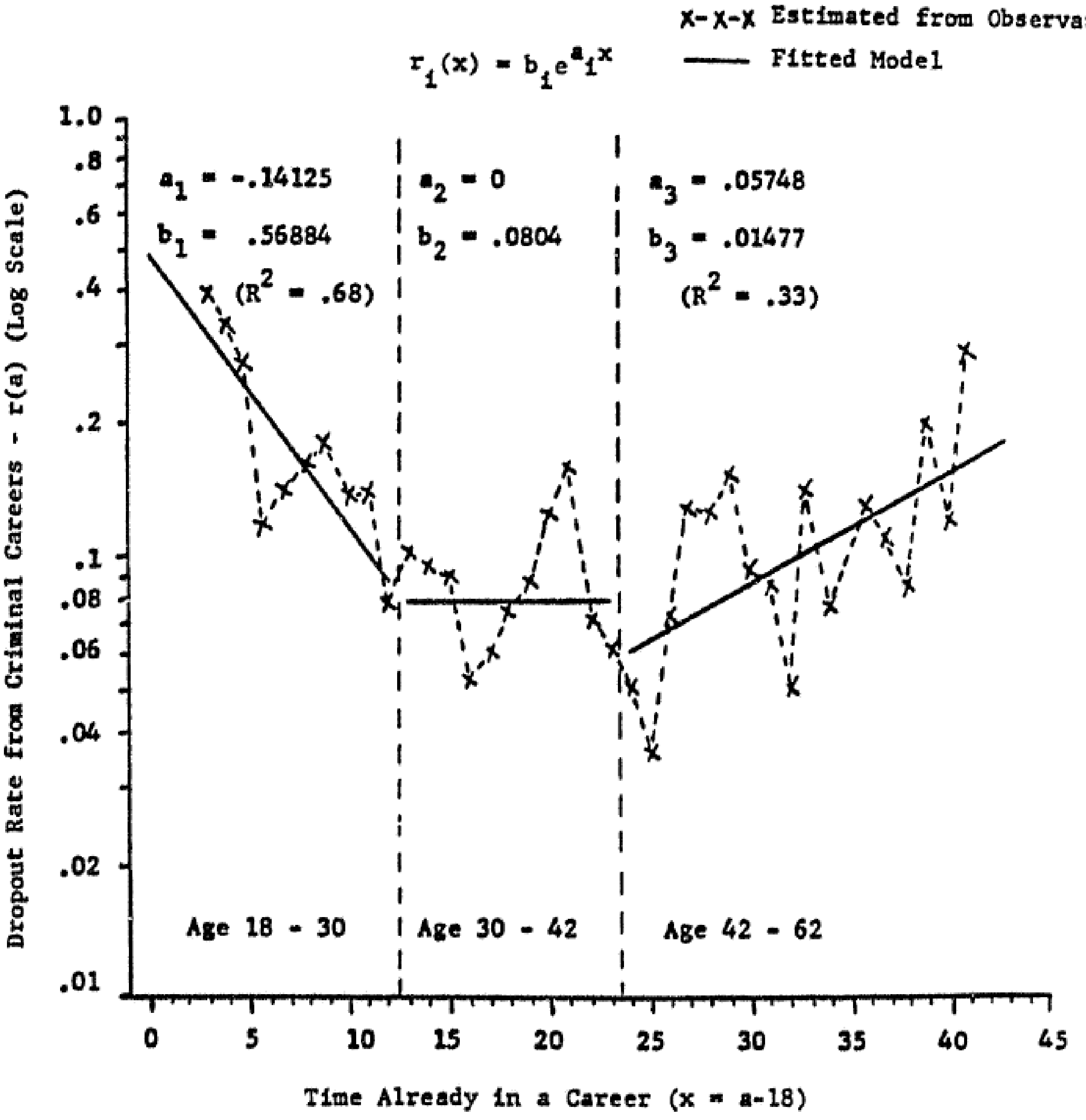

### 2.4.2. Replacement

Especially in crimes of commerce, jailing people for offenses tends not to cut offenses one-for-one (Ehrlich 1981, §II). Often when a street corner is vacated, other people spot a commercial opportunity and fill it. Certainly prostitution has survived prosecution. Nixon declared a "war on drugs" in 1971; despite millions of drug arrests (Snyder 2012, p. 13), that war seems not to have been won. Kuziemko and Levitt (2004, p. 2043) estimate that the 15-fold increase between 1980 and 2000 in people imprisoned for drug crimes lifted street prices for cocaine by 5–15%.

The upshot for consumers of research is an important reminder about the distinction between individual- and geography-level effects. A study that follows individuals might discover that incarceration raises or lowers crime among the individuals studied. But to the extent the crimes provoked or prevented are ones of commerce, offense totals will probably change less.

That said, the replacement effect probably does not operate to the same extent for drug buyers (as distinct from sellers), or for petty thieves, burglars, and perpetrators of more violent crimes. They hardly vie with each other, elbow-to-elbow, for limited opportunities to break the law.

### 2.4.3. Illicit industry destabilization

As shown in the failure of the war on drugs, an illicit industry can often dynamically adjust to the pressures placed upon it, much like the legal industries in textbook economic models. But the illegality of an industry can also make it fragile, so that enough external force can be highly disruptive and lead to more crime. Dills, Miron, and Summers (2008, p. 19) point out that in the 20th century, murder occurred most frequently in the US during major government efforts to suppress alcohol or drugs. Indeed, the alcohol business is often said to have been much more violent when it was illegal. Owens (2011) disputes that claim, citing lack of correlation between homicide rates and the timing of state-level enactments and repeals of temperance laws in the 1920s. I haven't investigated the question enough to opine. It nevertheless remains plausible that enforcement pressure sometimes perversely increase crime—and in a way that almost none of the studies here can control for.

Why might enforcement pressure increase crime? Miron (1999, pp. 83–84) argues that participants in illicit industries lack recourse to the conventional, non-violent ways of handling conflict, such as written contracts and courts. If the industry is stable enough, players in illicit industries can still reach informal agreements. Participants may view themselves in long-term relationships with other industry players, so that violating verbal promises or tacit rules today will make it harder to do business tomorrow. The propensity to knit such social arrangements is human, and is the ultimate historical basis of much formal commercial law (Benson 2011). But such informal arrangements can fall apart when the state is actively disrupting the business. Leave the illicit drug industry alone and it may tend toward internal peace; shorten the planning and trust horizons of participants by randomly arresting them and violence may surge as a means of conflict resolution.

### 2.4.4. Cognitive framing

The theory of specific deterrence is that having experienced incarceration in the past increases one's fear of incarceration in the future. Although the idea is old and straightforward, it is in a sense rather modern within economics. It recognizes that we are not perfectly informed, perfectly optimizing agents—else being in prison would not change one's views of that experience. Recently, two criminologists have enriched thinking about the aftereffects of incarceration by importing an idea from the modern subfield known as behavioral economics.

In July 2001, the Maryland State Commission on Criminal Sentencing Policy revised its voluntary guidelines on the length of sentences to be dispensed for various crimes. Like most such guidelines, these do not bind judges. In fact, a major goal of the 2001 revision was to conform the guidelines to practice rather than vice versa (MSCCSP 1999, p. 6). And that goal appears to have been attained. Although recommended sentences shortened for some crimes, actual average sentences for those crimes did not appear to change, relative to those for other crimes (Bushway and Owens 2013, p. 311).

Yet despite the stability in actual sentencing, people convicted of charges whose *recommended* sentences fell recidivated less, as compared to people convicted of other offenses. On average, a person whose ratio of actual to recommended sentence was 10% (not 10 percentage points) lower was 0.8 percentage points less likely to be rearrested within three years of release (se ≈ 0.4%; Bushway and Owens, Table 4, row 1, cols 2–5; Table 1, last row). For comparison, the pre-revision rearrest rate for this group was 55% (Table 1, last row, col. 3).

Why might convicts pay attention to changes in recommendations if judges did not? Bushway and Owens nominate a cognitive foible called "framing" (p. 302). Prosecutors may have cited the guidelines when negotiating plea bargains. So when recommended sentences were higher relative to actual sentences, freed convicts may have perceived the punishments as smaller, just as a bonus of $500 is disappointing when you expected $1,000. And, going forward, that may have reduced the fear of punishment. To bolster this theory, Bushway and Owens quote the late Melvin Williams ("Little Melvin"), who dominated the illegal drug business in West Baltimore for many years and is said to have inspired the character of Avon Barksdale in *The Wire*:

> They're not seeing that at a time when I spent twenty-six and a half years in [some] of the world's worst penitentiaries...I only did one-third of every sentence I ever had. (p. 301)

Like aging, framing can complicate interpretation of impact studies. A mass prisoner release, for example, creates quasi-experimental differences in actual time served, as distinct from time sentenced or time recommended by a parole board. But it also generates variation in the counterfactual punishment that may, in prisoners' minds, frame their actual punishment, thereby potentially influencing behavior through a competing channel. If two people are released after doing a year of time, but one had expected to serve a day more and the other a year more, then the latter may experience the year of incarceration as less punishing.

I find the theory of Bushway and Owens intriguing, but the evidence not completely persuasive. (See §9.12.2.) The key graph in their paper (Bushway and Owens, Figure 1) is reproduced here as Figure 4. It shows the three-year whether-rearrested rate for releasees convicted of crimes whose recommended sentences fell in 2001, in grey, and the rate for the rest, in black. The horizontal axis is the time of sentencing, expressed as months since January 1999. The vertical black line shows the moment of guideline revision, July 2001. The line for those convicted of crimes whose guidelines changed is noisy, probably because it is drawn from a tenth as many observations as the black. And to my eye, the fluctuations in the grey line do not change qualitatively in mid-2001. Bushway and Owens (Table 3, row 1) show that the grey line is lower after than before, on average and relative to the black one, with statistical significance. But such results are most convincing when they pop unambiguously out of graphical depictions.

I am unware of other tests of the framing theory, although I argue later in my review that it best explains one set of results in Kuziemko (2012).

**Figure 4. Fraction of releasees rearrested within three years, by whether crime's recommended sentence revised downward in 2001, four Maryland counties, 1999–2002, from Bushway and Owens (2013)**

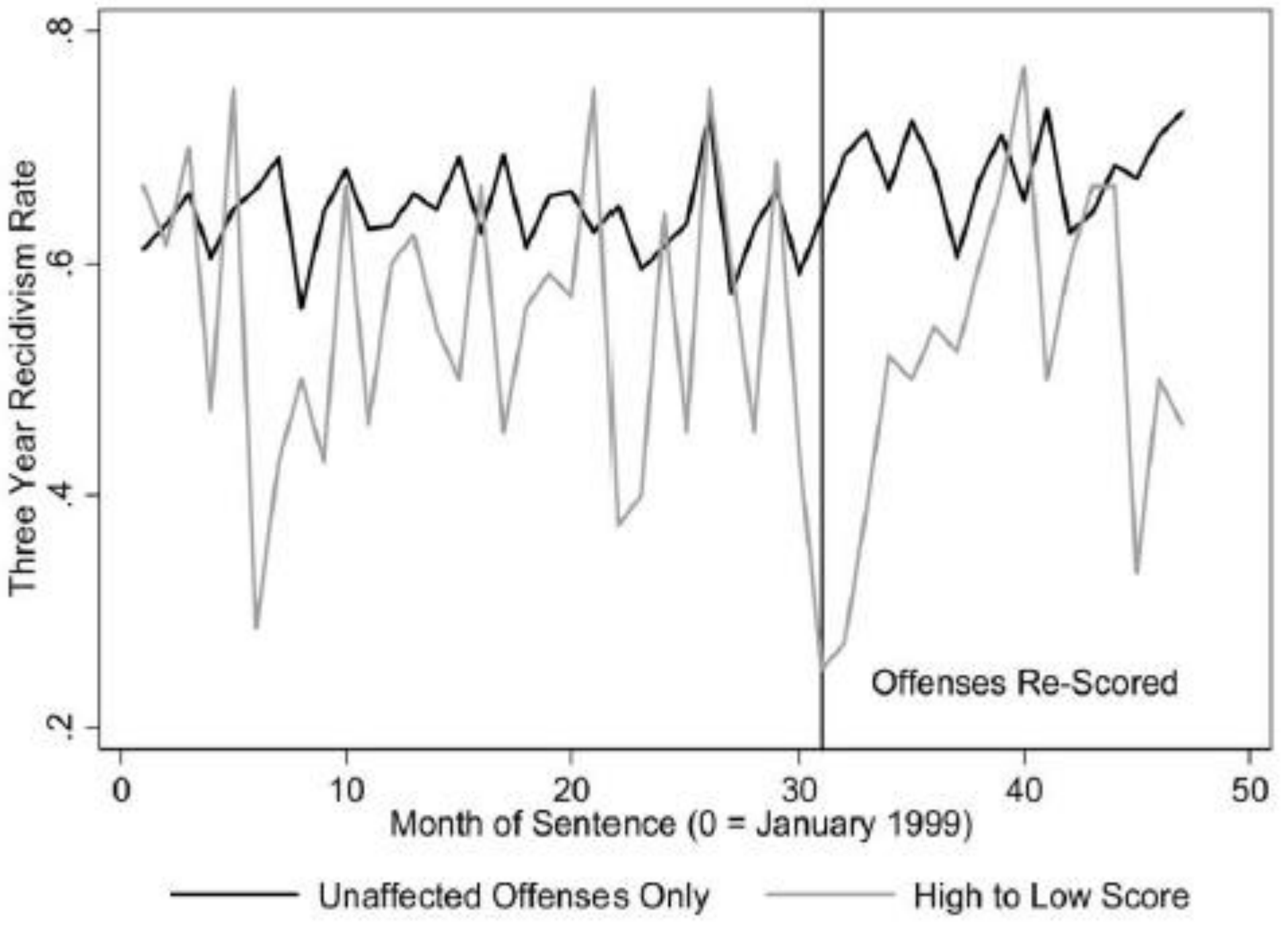

`

### 2.4.5. The parole effect

In reviewing the literature, I discovered another potential source of bias, which I have not seen discussed before. I call it "parole bias." It emerges from the split decision-making in many jurisdictions on how much time a person spends in prison. Judges sentence. Parole boards then decide how much of a sentence is served behind bars and how much in the "community" under the more or less watchful eye of a parole officer.[14]

While studies tend to implicitly characterize parole as the opposite of incarceration—as freedom—it is a transitional state on the way to freedom, during which the odds of reincarceration are elevated. Parole originated from the philosophy that a supervised transition to freedom helps rehabilitate convicts, by giving them a time in which they can reintegrate with society while knowing that the slightest infraction will yank them back to prison. The word "parole" derives from the French "*parole d'honneur,*" meaning "word of honor" (Alarid 2015, p. 44). Parolees may be reimprisoned for violating that word of honor. They may be returned for "technical violations" such as failing to keep regular appointments with a parole officer, or failing a drug test. And if arrested on a new charge while on parole, they may also be reimprisoned solely for that arrest, without the need for a conviction. And the parole board does not require the same quality of evidence as a court would in order to recommit, and is not subject to the same standards of due process.

Recognizing that parole is a distinct state between confinement and freedom complicates the interpretation of studies in which recidivism is measured as return to prison. For example, Kuziemko (2013), reviewed below, tracks inmates in Georgia who ended up on one side or the other of a threshold in the parole board's guidelines on how much time they should serve before release. On average, the people on the favorable side of the boundary spent less time in prison and a larger fraction of their three-year post-release follow-up periods on parole. That may have raised their odds of reimprisonment for technical violations, and, conditional on a new arrest, increased the swiftness and certainty of return within any given follow-up

[14] Exceptions include California and Illinois, which adopted determinate sentencing in the late 1970s (Washington University Law Review 1979).

period. The following causal chain may thus have operated:

> Less time in prison → more time on parole → more return to prison, for technical violations, misdemeanors, and felonies more swiftly and certainly punished

This could make it appear that early release increased crime, without any real impact on crime.

A cross-state comparison suggests that being on parole indeed causes a lot of return to prison. Working with data from the mid-1990s, Fischer (2005) observes that California easily surpassed six other large states on recidivism defined as return to prison, but not recidivism defined as rearrest or reconviction. He writes,

> Why is California so different with respect to its propensity to return offenders to prison for parole violations? A significant reason is that virtually all offenders released from California prisons go on parole supervision. Most large states do not have this policy. In Texas, for example, about 25% of prisoners are released without any parole supervision. In North Carolina, the figure is over 40%, and in Florida, more than 60% of all prisoners released have no parole supervision. (p. 2)

In other study contexts, parole bias can produce the appearance that more time leads to *less* crime. One potential example that arises in this report appears in the review of Ganong (2012), in which I modify the definition of return to prison to leave out returns triggered by parole revocations for technical violations, misdemeanors, and felonies not fully prosecuted. (See §9.13.) To the extent that parolees are charged with felonies more often, purely because they are easier to monitor and incarcerate, this exclusion compensates, reducing parole bias. But it may also go too far: to the extent that the government expeditiously revokes parole rather than prosecuting felony charges that *would have been led to conviction*, this modification undercounts felonies committed by parolees. In other words, this chain could operate too:

> Less time in prison → more time on parole → more felonies triggering revocation in lieu of prosecution → less return to prison for felony conviction

More generally, recidivism is defined and measured in many ways, and the likely sign and scope for parole bias vary in tandem. All else equal, being on parole might reduce the odds of reappearing in court if the path back to prison for parolees bypasses the courtroom. Depending on local practices, being on parole might also affect the odds of being arrested—conditional on being suspected of a felony—or of being booked in the local jail.

For example, compared to the examples above, which work from experimental variation in parole board decision-making, when the variation originates in the courts, parole bias can go also go the other way. To see how, imagine two statistically identical defendants, one who happens to be sentenced to a long term by one judge, the other sentenced to a shorter term by another judge. Faced with two similar inmates, a parole board might release both after the same amount of time, leaving the remainder of their sentences for parole. In effect, the first judge will have sentenced her defendant to more parole, not more prison. One causal chain at work would be:

> Longer sentence → more time on parole → more return to prison

This would make it look as if longer sentences cause more crime.

As it happens, the results from nearly all the studies of incarceration aftereffects reviewed here produce results predicted by the first and third pathways shown above. Studies exploiting variation in parole find that more time (in prison) leads to less crime. Those exploiting variation in sentencing tend to find the opposite.[15] *A priori*, nearly all their results could be a statistical mirage.

In practice, I think that is unlikely, as I will discuss in full as I review the studies. There are several reasons:

- In some contexts, parole boards have had more or less discretion than assumed here. Green and Winik (2010) and Loeffler (2013), take place in contexts where parole boards exercised little or no control over time served. Inmates had to serve (nearly) all their original sentences. In particular, Loeffler is set in Illinois, which adopted a completely *determinate* sentencing regime in the late 1970s (Washington University Law Review 1979), which all but eliminated parole. On the other hand, Berecochea and Jaman (1981) took place under California's fully *indeterminate* sentencing regime, where the parole board exercised such complete discretion over total time served, in prison and on parole as to substantially decouple the lengths of each. An extra month in prison did not lead mechanically to less time on parole.
- Some studies, such as Mueller-Smith (2015), confirm the robustness of their results by altering the definition of recidivism, and in ways that happen to invite opposite parole biases.

Still, it appears that parole bias has not been recognized before; and it is relevant to some important studies reviewed here, especially Kuziemko (2013) and Ganong (2012).

# 3. Process

This evidence review has been thoroughgoing, but not as systematic as a systematic review, in which best practice approximates the mechanical. Nevertheless, like any review, it entailed searching, filtering, and synthesis. Less typically, it entailed replicating and reanalyzing underlying statistical work where possible.

## 3.1. Searching

The search was an informal networking process. I found studies by:

- *Following citations in other papers*. For instance, the new Mueller-Smith (2015) working paper cites many papers using similar methods or touching upon similar substantive questions. I also relied on reviews such as Nagin, Cullen, and Jonson (2009), Nagin (2013), Chalfin and McCrary (2014).
- *Googling*. Google surfaced less-cited papers in response to keyword searches and listed all papers that cite a given one.
- *Spreading the word*. I sent preliminary drafts to the authors of nearly all the works[16]; four times, an author pointed me to a paper that was new to me.

## 3.2. Filtering

I filtered mainly on study design. I looked for studies that performed randomized experiments, or that strove to exploit the next-best thing, quasi-experiments. In truth, "quasi-experiment" admits no precise definition.[17] The studies I review exploit events or policies that sharply cleaved samples into groups that, it could reasonably be hoped, resembled each other statistically, aside from the incarceration regime. The studies employ these designs:

- *Randomized experiments*. Corrections agencies have cooperated with researchers to perform randomized trials, e.g., Berecochea and Jaman (1981), Descehenes, Turner, and Petersilia (1995), Killias, Aebi, and Ribeaud (2000), Gaes and Camp (2009), Weisburd, Einat, and Kowalski (2008), Hawken and Kleiman (2009), all of which are reviewed below.
- *Mass prisoner releases*. On August 1, 2006, Italy released 36% of its prisoners. Drago, Galbiati, and Vertova

[15] Hjalmarsson's (2009b) study of a sentencing discontinuity for juveniles is an exception. It is reviewed below.
[16] The exception was Ross (1982), a book whose author I could not find.
[17] Julian Simon—the man who invented airline flight overbooking—coined the term in Simon (1966, p. 195). He used it to refer to what would today be described as a US state panel study of the response of alcohol consumption to sudden, tax-induced price changes.

(2009) works at the individual level, asking whether releasees freed earlier on in their sentences—thus having more additional time hanging over their heads—recidivated less in the seven months after release. Buonanno and Raphael (2013) analyzes time series data for impacts on national and provincial rates of theft, assault, and other offenses. Kuziemko (2013) studies the release of 519 nonviolent offenders in Georgia on March 18, 1981, exploiting the fact that the governor's pressure for a mass release overrode the usual tight correlation between time served and time recommended by the parole board.

- *Discontinuities in sanctioning rules.* It has become common in the US for judges and parole boards to take guidance from point-based formulas in choosing a sanction—prison, probation, community service—and setting the length or security level of incarceration. Typically, the formulas influence but do not bind. And often they contain thresholds: one more point sends you to medium- instead of low-security prison, as in Chen and Shapiro (2007), or to prison instead of community service, as in Hjalmarsson (2009b). Kuziemko (2013) also takes advantage of a policy discontinuity, in Georgia's parole guidelines for time served. Since people on either side of such cutoffs are statistically similar, the split between them makes for a potentially compelling natural experiment. That said, the categorizations are often coarse, leaving the door open to bias. In Georgia, for example, one point can mark the difference between having no prior felony convictions and having one (j.mp/1MMXdSO), meaning that treatment and control groups may not be comparable enough to remove all bias. Exempt from this concern about coarseness is Lee and McCrary (2009), which focusses on the impact, by week of age, of turning 18 in Florida and thus becoming subject to harsher sentencing. Similarly, Hjalmarsson (2009a) tracks self-reported criminality among a national cohort of young people as they attain the age of criminal majority in their respective states.
- *Sharp policy changes.* In 2001 and 2004, Dutch cities implemented a national law increasing typical sentences for the most habitual of offenders from two months to two years. Vollaard (2013) looks for correlated breaks in the cities' crime trends. Ross (1982) draws together evidence from time series studies of the adoption of laws meant to deter drunk driving in various countries. In Georgia, on April 1, 1993, the parole board revised its length-of-stay guidelines, significantly increasing time served, a discontinuity that Ganong (2012) exploits. Abrams (2012) looks across states at the effects of the adoption of mandatory minimum sentences for some crimes, as well as of sentencing add-ons for crimes committed with guns.
- *(Quasi-)random judge/courtroom/prosecutor/public defender assignment.* Many court systems assign defendants to a judge, public defender, and/or prosecution team with a process that is substantially arbitrary, even random. Researchers can measure which judges or prosecutors, etc., are harsher, where harshness is defined by issuing longer or more frequent incarceration sentences. The situation can then be viewed as an experiment in which defendants are arbitrarily assigned to a harsher or milder sanctioning regime. For short, I will call all such studies "judge randomization studies." Martin, Annan, and Forst (1993) introduced this method, measuring the impacts of a couple nights in jail for first-time driving-under-the-influence (DUI) offenders. Kling (2006), not reviewed here, started the modern wave of such studies, looking at impacts of longer sentences on post-release earnings. Green and Winik (2010), Loeffler (2013), Mueller-Smith (2015), and Roach and Schanzenbach (2015) bring this method to impacts of time served on crime—in Washington, DC, Chicago, Houston, and Seattle—while Nagin and Snodgrass (2013) is set in several Pennsylvania counties. Aizer and Doyle (2015) applies it to juveniles in Chicago, and Di Tella and Schargrodsky (2013) to judges in Argentina choosing between prison and electronically monitored release.
- *Other.* Levitt (1996) exploits the somewhat arbitrarily timed progression of state-level lawsuits over prison crowding as an instrument for the incarcerated population: state prison growth tended to slow in the 1970s and 1980s when suits were filed and pick up years later after prisons were released from court control. Helland and Tabarrok (2007) gauge the deterrent effect of California's stringent "three strikes" law by comparing people tried and convicted twice for "strikeable" crimes—who thus live in the shadow of the law's twenty-five-to-life sentence for a third strike—to people tried twice but convicted one of those times of a lesser, non-strikeable offense.

### 3.3. Reanalyzing

In order to scrutinize studies more closely than is possible by reading them, I replicated and then reanalyzed them where data availability permitted. I was impressed by how often this process changed my interpretation, and in ways that matter for policy. This happened enough to convince me that we ought to rely less on opaque studies—ones that, for lack of access to data and/or code, cannot be replicated. More generally, it left me discouraged about the value of social science as practiced today, since most of the research is opaque.

For me, "reanalysis" embraces several activities.[18] It is more exploratory than mechanical, more like original research than systematic reviews. It includes: striving to replicate the published findings exactly, which is generally possible only when the authors share their analysis data sets and computer code; attempting to replicate results approximately otherwise; introducing new statistical tests or effect measurement approaches; and, in particular, complementing regression results with graphical depictions that often bring more insight. How I reanalyze is shaped by my analytical predilections, experience, biases, and skills. To partially compensate for my own limitations, I will post all code, as well as post data where possible, otherwise making it available upon request.[19]

I only reanalyzed eight of the 30 or so studies reviewed here. Reanalyzing a study typically takes weeks, and some studies, such as those examining swiftness and certainty in punishment, did not approach the question of the impact of mass incarceration on crime directly enough to earn the attention. For many other studies, data and code were not available.

I obtained data, and sometimes code too, in several ways. Some authors had published their data sets as required by a government funder (Deschenes, Turner, and Petersilia 1994) or posted their data and computer code pursuant to a transparency policy of the journal in which they published (Abrams 2012; Buonanno and Raphael 2013). Green and Winik (2010) posted their data and code outside any such policy. For Helland and Tabarrok (2007), Kuziemko (2013), and Ganong (2012), authors provided, or provided access to, data and code adequate to exactly match published results. Levitt sent a data sent enabling an approximate match. Finally, since Levitt (1996), Abrams (2012), and Lofstrom and Raphael (2015) draw mainly on public state-level data sets on prisoners, crime rates, and other variables, I could and did return to primary sources to replicate the construction, as well as the analysis of the studies' data sets.[20]

Bias may have crept into my search for data and code for replication and reanalysis. I may have disproportionately scrutinized the studies that concluded uncomfortably for the Open Philanthropy Project's criminal justice reform strategy. In truth, I believe my strongest bias is not for or against any proposition in criminology, but *contrarian skepticism of quantitative analysis*, which can be triggered by studies on all sides of an issue.

To guard against this bias, I wrote to the authors of all studies that a) were set in the modern US context and b) I had not yet reconstructed. This yielded no more data or code. Table 2 summarizes data and code availability of the US-based studies at the heart of this review.[21][22] All of the geography-level studies and four of the 11 individual-level ones could be approximately or exactly reconstructed. Researchers not providing data or code upon request cited various reasons: the code had been lost; confidentiality rules prohibited sharing of individual-level data; they had more pressing priorities.

[18] See Clemens (2015, Table 1) for a typology of "replication" analyses.

[19] Some providers of data prohibit redistribution.

[20] Helland and Tabarrok (2007) also works with public data (ICPSR data set 3355), but it is accessible only to people affiliated with academic institutions.

[21] Very late in the project I discovered that the data and code for Drago, Galbiati, and Vertova (2009) are posted online. Because the study is set in Argentina, and because it compares incarceration to intensively supervised release—both of which traits distance from the topic of mass incarceration in the US—I did not to invest time in reanalyzing it.

[22] Table 2 omits Iyengar (2008) and Roach and Schanzenbach (2015), which are included in this review because of serious claims to quasi-experimental identification, but are ultimately found to be less reliable than initially hoped. And it omits Buonanno and Raphael (2013), whose data and code are posted online and used in my review, but which is set in Italy.

**Table 2. Transparency of reviewed studies of the impact of the quantity of incarceration on crime in the US in recent decades, as compared to unsupervised or traditionally supervised release**

| Study | Context | Unit of observation | Main channel(s) | Data and code availability |
|---|---|---|---|---|
| Helland and Tabarrok (2009) | California, 1990s | Individual | Deterrence | Provided upon request |
| Abrams (2012) | US states, 1970–99 | Geography | Deterrence | Posted pursuant to journal policy; approximately reconstructed from primary sources too |
| Levitt (1996) | US states, 1972–93 | Geography | Incapacitation | Approximate data provided upon request |
| Owens (2009) | Maryland, 1999–2004 | Individual | Incapacitation | Not provided upon request; author no longer has access |
| Lofstrom and Raphael (2016) | California, 2011–13 | Geography | Incapacitation | Data and code not delivered upon request; data approximately reconstructed from primary sources |
| Green and Winik (2010) | DC, 2002–07 | Individual | Incapacitation, aftereffects | Data and code posted |
| Loeffler (2013) | Cook County (Chicago), 2000–08 | Individual | Incapacitation, aftereffects | Not provided upon request |
| Nagin and Snodgrass (2013) | Pennsylvania, circa 2000 | Individual | Incapacitation, aftereffects | Not provided upon request |
| Mueller-Smith (2015) | Harris County (Houston) | Individual | Incapacitation, aftereffects | Not provided upon request |
| Ganong (2012) | Georgia, 1993–95 | Individual | Incapacitation, aftereffects | Posted code, provided data upon request |
| Kuziemko (2013) | Georgia, 1981–2001 | Individual | Aftereffects | Code provided and access to original data arranged |
| Lee and McCrary (2009) | Florida, 1989–2002 | Individual | Deterrence and aftereffects (juvenile) | Not provided upon request |
| Hjalmarsson (2009a) | National youth sample, 1990s | Individual | Deterrence (juvenile) | Not provided upon request |
| Hjalmarsson (2009b) | Washington state, 1998–2000 | Individual | Deterrence (juvenile) | Not provided upon request |
| Aizer and Doyle (2015) | Chicago, 1990–2006 | Individual | Deterrence (juvenile) | Not provided upon request |

# 4. Summary of reviews

This multi-page table summarizes the individual study reviews that follow. Using shading from rich yellow to dark green, the final column articulates how each study supports or contradicts this report's dominant synthesis, which is that at typical policy margins driving mass incarceration in the US today, incarceration hardly deters; that it measurably reduces property crime, if not violent crime, through incapacitation; and that this short-term benefit is cancelled out over the long run by higher recidivism. Rows in blue indicate studies for which data was obtained. A fuller version of this table is available here.

| Study | (Quasi-)experiment | Setting, sample | Impact of stricter incarceration policy | Compatibility with report's synthesis |
|---|---|---|---|---|
| **Deterrence: Swiftness and certainty** | | | | |
| Weisburd, Einat, & Kowalski 2008 | Probationers delinquent on fees/fines/restitution randomly exposed to three enforcement regimes, two of which threatened jail time | 198 probationers apparently able to make court-ordered payments, 3 NJ counties, dates not clear | 35%→56% paid half of outstanding obligation within 6 months, 13%→39% paid all, after being served with Violation of Probation notice carrying threat of jail time. Additional threatened sanctions such as community service and employment training made no difference. | Mildly incompatible: shows deterrence from an immediate threat, in a setting removed from mass incarceration. In normal criminal proceedings, offenses may not lead to arrest, and arrests may not lead to imprisonment. |
| Hawken & Kleiman 2009; Hawken et al. 2016 | Under "HOPE" name, probation management overhauled to impose swift & certain sanctions, such as overnight jail time, for failure to get--or pass--drug test. | 330 probationers in treatment group, 163 in control, HI, 2009 | 12 months: 23%→9% appointments missed, 46%→13% drug tests failed, 15%→7%, probation revoked, 267→138 days incarcerated; 76 months: 7.1→6.3 appointments missed; 47%→42% multiple arrests for new crimes, 29%→21% return to prison | Mildly incompatible: same as above |
| Lattimore et al. 2016 | Randomized trials of HOPE replications | 1 county each in AR, MA, OR, TX. 1496 probationers total. | .83→.70 arrests; .37→.38 convictions | Mildly compatible: Impacts substantially smaller than in original HOPE. Shows modest deterrence in setting removed from mass incarceration |
| O'Connell, Brent, & Visher 2016 | Randomized trial of HOPE replication | City in DE, 384 probationers | 21%→30% probation revocation; 56%→53% any arrest for new crime ($p = 0.55$); 47%→49% any reincarceration ($p = .84$) | Mildly compatible: No clear deterrence found, in setting removed from mass incarceration |
| **Deterrence: Severity** | | | | |
| Ross 1982 | Institution of new sanctions for drinking & driving or heightened enforcement thereof | US, Canada, UK, Finland, Norway, Sweden, France, Australia, New Zealand, 1950s–70s | Well-publicized new laws or enforcement campaigns reduce traffic deaths, but effect mostly fades within 1 year, perhaps because true risk of getting caught stays low. | Mildly incompatible: shows deterrence, but mostly transient and in setting removed from mass incarceration |
| Drago, Galbiati, and Vertova 2009 | Because of overcrowding, sudden 3-year sentence reduction for most prisoners, causing 40% to be released immediately; commuted time added to *next* sentence if reconvicted within 5 years | 20950 releasees from Italian prisons, Aug. 2006 | −.16% points les return to prison within 7 months for each extra month in prospective punishment ( = each month *reduction* in time just served) | Moderately incompatible. Recidivism impact could be caused by having spent less time in prison (harmful aftereffects) or facing longer prospective sentences (deterrence). As for latter, threat had unusual cognitive salience. |
| Helland & Tabarrok 2007 | People with 2 trials and 2 "strike" convictions statistically similar to | 1447 people released from CA prison in 1994, | 15.1% rearrest reduction, per unit time; reanalysis shows effect essentially confined | Essentially non-contradictory: result can be explained by baseline imbalance, and if |

| Study | (Quasi-)experiment | Setting, sample | Impact of stricter incarceration policy | Compatibility with report's synthesis |
|---|---|---|---|---|
| | those with 2 convictions and 1 strike; yet the two groups differ in proximity to severe 3rd-strike punishment | year of adoption of "3 strikes" law, who had been tried twice for "strikeable" felons and convicted twice, except perhaps once for a non-strikeable offense | to drug crimes, at 31.1% | taken at face value implies mild deterrence |
| Iyengar 2008 (working paper) | Under CA "3 strikes" law, a serious or violent felony starts the strike count while any felony is asserted (incorrectly) to continue it. Thus two people convicted of same crimes in different order can differ in proximity to severe 3rd-strike punishment. | 17264 arrestees in Los Angeles, San Francisco, San Diego, 1990–99 | | Non-contradictory, the study being premised upon an incorrect reading of the law |
| Abrams 2012 | Many states imposed minimum sentences for crimes, especially by repeat offenders; or gun add-on laws to raise sentences for gun-involved crimes | 521 US police jurisdictions, 1970–99 | No clear impact of mandatory minimum laws on gun-involved assaults or robberies, nor impact of gun add-ons on gun assaults. Impact of gun add-ons on gun robberies: −6.4%, −13.8%, −16.4% after 1, 2, 3 years of implementation | Strongly compatible after reanalysis, which finds fragility in the perceived bend in gun-involved robbery trend after gun add-ons adopted. |
| **Incapacitation versus highly supervised release** | | | | |
| Deschenes, Turner, & Petersilia 1995 | Random assignment to early release from prison into intensive community supervision (ICS): frequent parole check-ins, drug tests, full-time drug treatment/training/job search. | 124 offenders who had been sentenced to ≤27 months or had violated probation, 7 MN counties, Oct. 1990–June 1992; average prison time 124 for ICS group, 228 days for prison group | No significant impacts on rearrest, reconviction | Neutral since compares incarceration to intensively supervised release |
| Di Tella & Schargrodsky 2013 | Random assignment of detainees to judges within a district's court, who vary in frequency of sentencing to prison vs. electronically monitored house arrest | 386 sentenced to electronic monitoring, 1140 to prison, Buenos Aires province, 1998–2007, all male, <40 | Return to prison by Oct. 2007 +15%, electronically monitored release *reduced* recidivism | Neutral since compares incarceration to intensively supervised release |
| **Incapacitation versus standard release** | | | | |
| Levitt 1996 | Sudden speed-ups and slow-downs in prison growth caused by progression of prison overcrowding lawsuits | US states, 1973–93 | Elasticities of violent & property crime with respect to prison population 1 year earlier: −0.379, −0.261 | Mildly supportive since imperfect fixes to discovered problems don't contradict the incapacitation finding |
| Owens 2009 | On Jul. 1, 2001, juvenile arrest records removed from consideration in sentencing offenders aged 23–25, cutting average time served 222 days | 133 convicted males who serve time, age 23–25, MD, 1999–2004 | −2.79 arrests/person/year, of which −1.65 drug arrests | Strongly supportive: evidence of modest incapacitation |
| Buonanno & | Same as Drago, Galbiati, & Vertova | Italy, 2004–08 | Initial release: −18 crimes/releasee/year, | Moderately supportive: evidence of |

| Study | (Quasi-)experiment | Setting, sample | Impact of stricter incarceration policy | Compatibility with report's synthesis |
|---|---|---|---|---|
| Raphael 2013 | 2009; also, refilling of prisons over following 2.5 years | | predominantly theft & robbery. Incapacitation from prison refilling: −25 – −47). | incapacitative effect on acquisitive crime in country with a much lower incarceration rate |
| Vollaard 2013 | In 2001 and 2004, two groups of cities implemented national law increasing sentences for habitual offenders from 2 months to 2 years; 1st-round cities chosen for having more crime | Netherlands, 1998–2008 | 0 for assault & sexual crimes; −3.66 acquisitive crimes/month/offender serving extended sentence; but lower impact in lower-crime cities | Moderately supportive: evidence of incapacitative effect on acquisitive crime from highly targeted incarceration in country with a much lower incarceration rate |
| Lofstrom and Raphael 2016 | Prison population reduction ("realignment") under pressure from overcrowding lawsuit, restricted to non-violent, non-sexual, non-serious offenders | CA, starting Oct. 1, 2011, total incarcerated fell from 232,000 to 213,000 in 6 months | Cross-state analysis: −2.8 reported property crimes/prisoner/year in 2012–13, with effect clearest for vehicle theft (–1.2); no change for violent crime | Mildly supportive since imperfect fixes to discovered problems don't contradict that incapacitation is real |
| **Aftereffects** | | | | |
| Berecochea & Jaman 1981 | Randomized 6-month release-to-parole acceleration | 1138 male felon inmates in CA who during Mar.–Aug. 1970 originally had parole date ≥6 months hence | 11.2%→8.0% sent back by court within 2 years | Moderately incompatible: other explanations possible but best explanation is beneficial aftereffects. Set ~50 years ago, in CA |
| Martin, Annan, & Forst 1993 | Random assignment to judges, some of whom did not comply with two-day jail requirement for first-time DWI offenders | 367 drunk driving cases, Hennepin County, MN, 1982 | No clear cross-judge differences in recidivism | Neutral: sentences only 2 days, impact roughly 0 |
| Chen & Shapiro 2007 | Assignment formula put statistically similar inmates on both sides of cutoff between minimum- and higher-security confinement (≤6 points vs. ≥7) | Inmates released from federal prison, 1st half of 1987, of which 91 scored 5–6 and 52 scored 7–8 | 19.8%→34.6%, 36.6%→55.8%, 48.4%→63.5% rearrested within 1, 2, 3 years | Moderately supportive, showing harmful aftereffects, only doubt being coarseness of point system within which discontinuity is exploited |
| Gaes & Camp 2009 | Randomized to Level I or III prison (Level I=lowest security, IV=highest) | 561 adult male inmates entering CA prison, Nov. 1998–Apr. 1999. | Level III confinement led to less time served & +31.1% prison re-entry per unit time post-release (relative change) | Neutral: finding of harmful aftereffects supports report's synthesis, but serious concern remains about baseline imbalance |
| Green & Winik 2010 | Quasi-random assignment among 9 judges, who varied in average sentence length given despite statistical similarity of their defendant pools | 1003 felony drug charges, DC, June 2002–May 2003 | +2.08% chance of rearrest within 4 years per month extra incarceration sentence ($p = .25$); but negative while incarcerated, implying positive after release | Moderately supportive, showing aftereffects offsetting incapacitation; best fixes to discovered problems don't change results |
| Loeffler 2013 | Random assignment among judges within a county's court | 20297 felony charges, Cook County (includes Chicago), 2000–03, apparently restricted by 67%-successful match to employment data | ~0 for felony rearrest ($p > .6$); but almost certainly − while incarcerated, suggesting + after release. ~0 for employment | Moderately supportive, showing aftereffects offsetting incapacitation; concern about post-treatment selection bias (1/3 of sample dropped for lack of employment information) could not be assessed for lack of data availability |
| Nagin & Snodgrass 2013 | Random assignment among judges within a county's court | 6127 offenders convicted in 6 PA counties, 1999 | ~0 for rearrest within 1, 2, 5, 10 years of sentencing; but almost certainly − while incarcerated, suggesting + after release | Moderately supportive, showing aftereffects offsetting incapacitation for net-zero effect; possibly some bias toward null finding |

| Study | (Quasi-)experiment | Setting, sample | Impact of stricter incarceration policy | Compatibility with report's synthesis |
|---|---|---|---|---|
| | | | | because of lower power from binary treatment and outcome indicators |
| Kuziemko 2013 | Assignment formula put statistically similar convicts entering prison on both sides of cutoff between medium- and high-risk in formula for recommended time served | 17373 GA prisoners entering after 1995 and leaving before 2006 | Each extra month served because of high-risk classification led to 1.3% points less return to prison within 3 years | Neutral after reanalysis. Quasi-experiment varies two treatments collinearly so impacts of incarceration unclear; parole bias could explain results |
| Kuziemko 2013 | Mass release of prisoners closest to release date, to make room for inmates in overcrowded *jails* | 519 non-violent-offender inmates in GA, Mar. 18, 1981, who on average served 13 of 17 months recommended | –3.2% points less return to prison within 3 years per extra month served | Perhaps neutral after reanalysis. Occam's razor favors cognitive framing explanation (surprise reduction in sentence reducing deterrence going forward) rather than rehabilitation |
| Kuziemko 2013 | After Jan. 1, 1997, inmates convicted of certain offenses required to serve 90% of original sentence, curtailing parole board discretion; time served for this group only rose 2 months relative to control group, but *incentive* for rehabilitation extinguished | 17437 GA prisoners sentenced in 1993–2001, some for 90%-rule offenses, some not, some released before rule, some not | Reduced incentive for good behavior in prison led to 6% points more return to prison within 3 years | Mildly contradictory: finding of benefit from longer prison terms based on lower-credibility, long-period difference-in-differences, not regression discontinuity; and are about incentives created by *opportunity* for early release, not early release per se |
| Ganong 2012 | GA updated parole board guidelines for time served on April 1, 1993. | 18,589 inmates who served 3 months–10 years and considered for parole Apr. 1, 1992–Mar. 30, 1994. | *Whether* returned within 1, 3, 10 years −3.9%, −5.9%, −3.7% points/year served; *number* of returns −3.9%, −6.1%, −7.8% points/year served | Neutral after reanalysis. Quasi-experiment varies two treatments collinearly so impacts of incarceration unclear; parole bias could explain results |
| Roach & Schanzenbach 2015 (working paper) | Quasi-random courtroom assignment to sentencing judges after conviction | 8780 convicts in King County near Seattle, 1999–2011 | −1.17%, −1.06%, −1.33% sentenced to new felony within 1, 2, 3 years of release, per month of previous judge's average sentence | Mildly contradictory: finds beneficial aftereffects, but marred by baseline imbalance |
| Mueller-Smith 2015 (working paper) | Random courtroom assignment | ~450,000 felony defendants, Harris County (includes Houston), 1980–2009 | −3.3%, −6.0%, −2.8% jailed, charged, convicted while serving time, mainly burglary & drugs; +6.7%, +5.6%, +3.6% after release, per year served. −$1632 quarterly earnings while serving time; −$247 after release, per year served | Moderately supportive, showing weak incapacitation and strongly harmful aftereffects; no problems evident, but study is complex, opaque, and an outlier in effect sizes |
| Dobbie, Goldin, & Yang 2016 | Random assignment among judges serving a given neighborhood district, some more likely to incarcerate *pretrial* defendants | ~420,000 bail hearing appearances, Philadelphia & Miami-Dade, 2007–14 | +4.1% rearrested within 2 years, including −13.4% before case disposition, +15.0% after | Strongly supportive in finding incapacitation and harmful aftereffects from pre-trial incarceration, with strong study design |
| Bhuller et al. 2016 | Random assignment among judges serving a given jurisdiction some more likely to incarcerate | 33,509 court appearances, Norway, 2005–09 | –24.8% rearrested within 1–24 months; −13.4% within 25–60 months; −27.4%overall | Moderately incompatible: strong study design shows beneficial aftereffects, but in Norway |
| **Juveniles** | | | | |
| Lee & McCrary 2009 (working | People just before or after 18th birthday are statistically similar but face different sentencing regimes | 64,073 juveniles in FL whose 1st arrest was before age 17, during | As 18th birthday passes, no statistically significant drop in odds of 1st serious arrest since 17, despite increased sentences. But | Strongly supportive: compellingly shows no deterrence, some incapacitation |

| Study | (Quasi-)experiment | Setting, sample | Impact of stricter incarceration policy | Compatibility with report's synthesis |
|---|---|---|---|---|
| paper) | | 1989–2002 | as date of 1st passes 18th birthday, chance of another within 30 days after falls 17.9%→8.4%, within 120 days 36.2%→24.0%, within 365 days 70.0%→24.0% | |
| Hjalmarsson 2009a | People just before or after age of criminal majority are statistically similar but face different sentencing regimes | National sample of 9000 males who were 12–16 on Dec. 31, 1996; interviewed yearly | −.171%, −1.267%, −.367%, −.428%, −.011% points/year for self-reported auto theft, theft of <$50, theft of >$50, drug sale, assault ($p$ = .55, .11, .48, .49, .22) | Strongly supportive: compellingly shows almost no deterrence |
| Hjalmarsson 2009b | Assignment formula put statistically similar juvenile convicts on both sides of cutoff between incarceration (≥15 weeks) and "local sanctions" (≤30 days detention/community service/community supervision/fine) | WA juvenile courts, Jul. 1998–Dec. 2000. 1147 incarcerated, 19395 not | Incarceration led to −36% returns to court per unit time post-release (relative change, $p < .01$) | Moderately incompatible, showing beneficial aftereffects, main doubt being coarseness of point system within which discontinuity is exploited |
| Aizer & Doyle 2015 | Random assignment among judges serving a given neighborhood district, some more likely to incarcerate juvenile defendants than others | 37692 juvenile offenders appearing in court 1990–2006 and turning 25 by 2008, Cook County (includes Chicago) | Those incarcerated (typically 1–2 months) graduated 12.5% points less and entered prison 23.4% points more before age 25 ($p < .005$) | Moderately supportive, showing harmful aftereffects, only doubt being about internally inconsistent definitions of neighborhoods within which judge assignment was randomized |

## 5. Deterrence: Swiftness and certainty

This review orders studies by the conceptual chronology of before, during, and after incarceration. The grouping is rough since some studies capture more than one channel. A final section focuses on studies of juveniles relating to any of the channels. The review of deterrence literature is split turn. It starts in this section with a pair of studies of reforms that made punishment more swift or certain. The next section moves to the deterrence of increased severity.

### 5.1. Weisburd, Einat, and Kowalski (2008), "The miracle of the cells: An experimental study of interventions to increase payment of court-ordered financial obligations," *Criminology & Public Policy*

That 2.2 million Americans are behind bars is often mentioned in discussions of criminal justice reform. Less noted is that incarcerated people constitute a minority of the *correctional population*. 870,000 people were on parole and 3.8 million on probation at the end of 2015 (BJS 2016a, Table 1). Probation and parole are circumscribed liberty. Those "enjoying" it are variously required to meet regularly with a probation or parole officer, submit to drug tests, avoid criminal activity, demonstrate that they have or are seeking work, stay away from certain people or neighborhoods, or pay certain fees (Blumstein and Beck 2005, p. 52). However reasonable or unreasonable, these rules do not always elicit compliance; for example, in Missouri in 2010, technical violations of terms of parole or probation accounted for an estimated 43% of all prison admissions (MWGSC 2011, p. 4).[23]

Community supervision is thus an important subdomain for the interplay of crime and punishment. It is distinctive in that officers can easily observe many violations. The government could never afford to breath-test every driver in order to deter drunk driving (see review of Ross 1982 below). But it can require the millions of people under community supervision to take regular drug tests and track compliance. In principle, this oversight can facilitate swift and certain punishment. In practice, many community supervision agencies receive less funding than prisons and are overwhelmed by caseloads (Blumstein and Beck 2005, p. 52). Inability to enforce the rules facilitates more violations, which further overwhelms caseworkers.

Wesiburd, Einat, and Kowalski cooperated with probation offices in three New Jersey counties to try to break that cycle of weak enforcement. In particular, they experimented with incentives for probationers to come current on court-ordered payments—fees, fines, restitution. The sample consists of 198 people deemed able to make the payments. The researchers split the subjects randomly into three groups. One control group received ordinary handling from the probation office. One treatment group received a notice of Violation of Probation (VOP), which threatened jail time. "At the time of the study, probationers were seldom served with VOPs…solely for nonpayment of court-ordered financial obligations" (p. 14). The second treatment group received VOPs and additional inducements: an obligation to perform community service 15 hours/week, and an offer of intensive support for job training and job search.

Despite the small sample, one finding emerged clearly. VOPs worked, while the additional inducements did not. 13% of the control group members eventually paid all the money they owed, while 39% of the pure-VOP group and 34% of the "VOP-plus" group did (Weisburd, Einat, and Kowalski, Tables 6 and 7). On this outcome, the treatment groups differed statistically from the control group (p = 0.0005, 0.004) but not from each other (p = 0.56).[24] Similarly, 56% and 61% of the treatment groups, respectively, paid at least half the money owed, compared to 35% for the control group (Tables 6 and 7).

It appears that the "miracle of the cells" does happen: when authorities credibly threaten jail time, many

[23] Possibly many of the reported technical violations actually occurred in response to arrests for new crimes.

[24] p values based on two-tailed tests of difference in proportions done in Stata with "prtesti 66 .39 69 .13", "prtesti 63 .34 69 .13", and "prtesti 63 .34 66 .39".

people who are able to respond do so.

## 5.2. HOPE

A man from Hawaii promoting HOPE—it is a great story in criminology, and Mark Kleiman tells it in his book *When Brute Force Fails*:

> Judge Alm had a problem. Probation officers were sending him reports of probationers—on probation for all manner of felonies from burglary to auto theft, from sexual assault to drug dealing—who were continuing to use methamphetamine, Hawaii's number-one problem drug. The files fairly bristled with violations: probationers accumulated multiple missed or positive drug tests, often as many as ten, before a report was made to the court. Those violations reflected very high levels of drug-taking, since tests were given only when probationers came in to meet their probation officers, and those meetings were scheduled weeks in advance: a probationer could avoid being caught simply by not using the drug for the three days before he was due to meet his probation officer. Nevertheless, one-fifth of all tests came back positive, and another one-tenth of the probationers called in for testing on any given day simply failed to show up…Either those probationers really could not quit, even for a few days, or they simply did not regard violating probation rules as anything to worry about. (Kleiman 2009, p. 34)

Steven Alm developed what came to be called Hawaii's Opportunity Probation with Enforcement (HOPE). The idea was to reengineer probation and court processes so that people failing to take or pass drug tests would experience swift and certain—but not severe—sanctions, such as an immediate night in jail.

> The HOPE intervention starts with a formal warning, delivered by the judge in open court, that any violation of probation conditions will not be tolerated: Each violation will result in an immediate, brief jail stay….Each probationer is assigned a color code at the warning hearing. The probationer is required to call the HOPE hotline each morning. The probationer must appear at the probation office before 2 pm that day for a drug test if his or her color has been selected. During their first two months in HOPE, probationers are randomly tested at least once a week (good behavior through compliance and negative drug tests is rewarded with an assignment of a new color associated with less-regular testing). A failure to appear for testing leads to the immediate issuance of a bench warrant, which the Honolulu Police Department serves. Probationers who test positive for drug use or fail to appear for probation appointments are brought before the judge. The hearing…is held promptly (most are held within 72 hours), with the probationer confined in the interim. A probationer found to have violated the terms of probation is immediately sentenced to a short jail stay….The probationer resumes participation in HOPE and reports to his/her probation officer on the day of release. (Hawken and Kleiman, p. 13)

In the short run, this activism would increase the burden on probation officers who were already struggling with caseloads of 180 people each (Kleiman 2009, p. 35). But once the program gained credibility, it might *save* probation officers time by deterring violations. Likewise, the net impact on how much time people spent in jail could go either way.

### 5.2.1. Hawken and Kleiman (2009), "Managing drug involved probationers with swift and certain sanctions: Evaluating Hawaii's HOPE"; Hawken et al. (2016), "HOPE II: A follow-up to Hawai'i's HOPE evaluation"

After a promising pilot, the Hawaii government worked with Hawken and Kleiman to put HOPE to a randomized test. Probation officers identified 507 people under supervision whom they deemed at high risk of violation, typically involving drugs. Of these, 493 met the study's inclusion criteria, and a random 330 were told to appear in court in order to enter HOPE, while the rest received probation as usual (Hawken and Kleiman, p. 35).

Twelve- and 76-month follow-ups show HOPE to be working. After one year, those assigned to HOPE missed fewer probation appointments, passed more drug tests, were arrested less, had probation revoked less often, and spent *less* time jail even though HOPE was quicker to send them there. (See Table 3.) Probably HOPE worked even better than these numbers suggest, because 30% of 330 people in the experiment's treatment group did not actually appear at the warning hearing at which they were to be enrolled in HOPE (Hawken et al., p. 30). (Rigor demands leaving the untreated 30% in the treatment group for statistical purposes, since *intent* to treat was randomized. Running the numbers this way also increases relevance to policymaking since in all HOPE-like programs not all intended participants will enter.)

Also striking were the views of HOPE probationers. Of the 28 surveyed probationers referred by a court to residential drug treatment, 19 reported positive views, while five were neutral and four negative. Among 16 interviewees *jailed* by HOPE, ten were positive on the program, four neutral, and two negative (Hawken and Kleiman, Figure 14). Perhaps those supporting HOPE appreciated how its swiftness and certainty helped them in their struggles with drug addiction.

Benefits persisted through long-term follow-up, even though circumstances further conspired against their detection. For not only did 30% of the people assigned to HOPE not participate, at least initially, but 35% of those *not* assigned to enter it eventually did (Hawken et al., p. 50). The de facto mixing of treatment and control after randomization presumably diminished measured impacts. Perhaps this is why the gap in whether someone was ever arrested for a new crime narrowed over the first 76 months, to 47% vs. 42%. At any rate, the gap was wider and more statistically significant when looking at the fraction of people arrested multiple times for new crimes, at 29% vs. 21%. And more people in the control group returned to prison: 27% instead of 13%. (See bottom half of Table 3.)

**Table 3. 3-. 12- and 76-month results in randomized trial of HOPE**

| Outcome | Assigned to control group | Assigned to HOPE | p value for difference (two-tailed) |
|---|---|---|---|
| **After 12 months** | | | |
| Missed probation officer appointments | 23% | 9% | 0.00 |
| Failed drug test | 46% | 13% | 0.00 |
| Arrested for new crime | 47% | 21% | 0.00 |
| Had probation revoked, returned to prison | 15% | 7% | 0.00 |
| Average days in jail or prison (sentenced) | 267 | 138 | 0.00 |
| **After 76 months** | | | |
| Missed probation appointments or drug tests | 7.1 | 6.3 | 0.09 |
| Arrested for new crime | 47% | 42% | 0.29 |
| Arrested multiple times for new crimes | 29% | 21% | 0.03 |
| Returned to prison | 27% | 13% | 0.00 |
| Number of subjects | 163 | 330 | |

Sources: Hawken and Kleiman (2009, p. 64), Hawken et al. (2016, pp. 49–53), except last and third-to-last p values computed by author.

Over 2,000 people are now enrolled in HOPE on the island of Oahu where it began (Hawken et al., p. 16). Thus there seems little doubt that HOPE works as intended, and at scale. In the logic of evidence-based policymaking, the next question is: can it be replicated elsewhere? According to an inventory in Hawken et al., "as of January 2015, HOPE or similar…programs are now employed in some twenty-eight states, one Indian nation, and one Canadian province, with even more jurisdictions considering doing so" (pp. 22–27).

### 5.2.2. Lattimore et al. (2016), "Outcome findings from the HOPE demonstration field experiment: Is swift, certain, and fair an effective supervision strategy?", Criminology & Public Policy; O'Connell, Brent, and Visher (2016), "Decide Your Time: A randomized trial of a drug testing and graduated sanctions program for probationers," Criminology & Public Policy

In November 2016, the journal *Criminology & Public Policy* published reports on several randomized trials of HOPE replications funded by the US Department of Justice. Lattimore et al. (2016) report on trials in counties of Arkansas, Massachusetts, Oregon, and Texas, while O'Connell, Brent, and Visher (2016) do so for a program in a Delaware city. The results feed cynicism more than hope. Applying swift and certain sanctions didn't work nearly so when transplanted to the mainland.

At the four sites where Lattimore et al. evaluated replications, they assessed fidelity to the original by tracking 11 indicators, such as the fraction of violations resulting in sanction and the fraction of those sanctions delivered within three days. The Arkansas implementer scored at least 80% on seven of the indicators; while Massachusetts scored that high on eight, Oregon on six, and Texas on ten. The researchers summarize:

> [T]he sites were largely successful at implementing HOPE programs that conformed to the established fidelity standards. In particular, the four programs certainly adhered to notions of "certainty" and "fairness" in that most violations were met with hearings and sanctions that were not overly onerous. "Swiftness" was also achieved in most sites, although all sites struggled to hold hearings within 3 days of violations, most hearings were held within a week—and most sanctions were imposed within 3 days of the hearings. (Lattimore et al. 2016, p. 1120)

While it is impossible to be certain whether any of the deviations greatly affected the impacts of HOPE—i.e., to know in advance which deviations might be innocuous and which not—it seems likely that all the replications are representative of what can happen when a jurisdiction works hard to increase swiftness and certainty in probation. While the participating agencies were importing a program invented elsewhere, they were chosen for the study through a competitive process, which they won by demonstrating commitment to the principles and capacity to implement.

Each of these four sites enrolled 300–400 people, making each local experiment nearly as large as the original HOPE evaluation. Follow-up began at randomization and lasted 650 days. For concision and statistical power, Table 4, below, reports the impacts by averaging across all sites. The HOPE-inspired programs appear to have *increased* the rate of probation revocation because of technical violations. This is unsurprising since the programs are meant to increase the certainty of revocation when it is formally warranted. Yet it cuts against the remarkable, contrary pattern that emerged in Hawaii (Table 3, above). As for another indicator of recidivism, subsequent arrests, Lattimore et al. determine that "there were no differences in the average number of recidivism arrests experienced during follow-up." In so doing, they seem to equate lack of statistical significance at $p = 0.05$ with no impact. Yet as Table 4 here shows, people in HOPE-style programs had 17 percentage points fewer arrests ($p = 0.06$). The fraction who got arrested at all was four points lower ($p = 0.11$); that reduction rises to 5 points when restricting to arrests for property crime ($p = 0.01$) and falls to 3 points for drug charges ($p = 0.05$).[25] In this sense, we can say that these four HOPE replications "worked" on average.

Yet the results from these four sites still disappoint. Statistical significance is not the same as practical significance. The fraction of people rearrested (as distinct from the average numbers of arrests cited above) only fell from 44% to 40%, as compared to a drop in Hawaii from 47% to 21% (after one year; see Table 3). And *convictions* for new crimes—arguably a better indicator of serious crime—were not clearly reduced (again, see Table 4). (The original HOPE evaluations do not report whether convictions fell in Hawaii, so it is unclear whether HOPE and its replications differed in this respect.)

[25] Lattimore et al. (p. 1125) also report that in survival modeling, HOPE increased time to arrest on a drug charge, with significance at $p = 0.06$. They do not report aggregates survival modeling results for total arrests or arrests for other categories of charges.

The statistical story is similar in Delaware. The evaluation reports that the Decide Your Time program achieved high fidelity to the principles of swiftness and certainty. Here too, the program appears to have increased revocations during the 18-month follow-up period, but it hardly reduced new arrests or convictions. (See Table 5.)[26]

Why didn't the replications live up to the original? The high impacts reported in Hawaii are unlikely to have occurred by chance, so the explanation probably lies in differences in execution and context. But in reading the commentaries in the same issue of *Criminology & Public Policy* and talking to the lead researchers on the replications, I have found no consensus on specifics. Kleiman (2016, p. 118), for instance, states that the four trials run by Lattimore et al., "by imposing a rigid program design without taking into consideration local conditions and opinions, and without building in flexibility even in the face of experience, violated the principles of good program design." But Alm (2016, pp. 1204–09), who visited the sites to advise local officials on implementation, fingers *deviations* from the HOPE model for compromising the replications.

I found a separate defense mounted by Alm, regardless of its merits as a defense, intriguing for what it says about deterrence. Alm argues that because HOPE *distinguished* itself with swift and certain sanctions, that researchers and replicators had incorrectly come to see those traits as *defining* HOPE:

> [S]ome researchers and academics…have misunderstood or mischaracterized HOPE as a sanctions-only program. HOPE has never been a sanctions-only program. It is, rather, a strategy to assist in the behavioral change process using EBPs by probation, treatment providers, and the judge. (2016, p. 1201)

That jargon, "EBPs," needs unpacking. In Alms's usage, it is shorthand for "Eight Evidence-Based Principles for Effective Interventions," an approach to probation supervision promoted by the National Institute of Corrections (NIC), part of the Department of Justice (Taxman, Shephardson, and Byrne 2004; DOJ and CJI 2004). As its name suggests, EBP is rooted in research on what works in corrections and, more generally, what helps people change their behavior. Among the key ideas (DOJ and CJI 2004, p. 6) are that supervision works best when:

- it produces incentives that are predominantly positive, such as reduced supervision after clean drug tests—carrots more than sticks;
- it applies the sticks (but not necessarily the carrots), with swiftness and certainty;
- along with incentives, it offers constructive paths forward, such as through cognitive-behavioral therapy.

This broader view of human motivation puts swift and certain sanctions in perspective. HOPE broke with convention in striving mightily to nail that second point. But it does not seem wise to rely on sticks alone. And contrary to popular perception, Alm argues, HOPE does not.

It is not clear that Alm is right to pin any misunderstanding so purely on the replicators. Alm too appears to emphasize swiftness and certainty in describing HOPE.[27] And the replications generally offered more than just sanctions. For example, 127 of the 371 Texas subjects received drug treatment, most of them at a residential facility (Lattimore et al., p. 1130).

[26] Hierarchical models with random effects by parole officer and demographic and other controls also find that the HOPE replication cut revocations, lowering the odds ratio by 22% (O'Connell, Brent, and Visher, Table 5, upper-right corner). But this result is only significant at p = 0.38 (email from John Brent, January 12, 2017).

[27] "I thought to myself, well, what would work to change behavior? And I thought of the way I was raised, the way my wife and I would—were trying to raise our son. You tell him what the family rules are, and then, if there's misbehavior, you do something immediately." (PBS NewsHour, February 2, 2014, j.mp/2kBi9XV).

Yet if there is something to this defense—if swift and certain punishment cannot do much good on its own—then that only compounds the impression from these replications that deterrence is a weak force against crime.

**Table 4. Impacts on recidivism of HOPE replications at sites in Arkansas, Massachusetts, Oregon, and Texas, from Lattimore et al. (2016)**

| Outcome | HOPE replication | Probation as usual | p value for difference |
|---|---|---|---|
| **Whether probation revoked** | 0.26 | 0.22 | 0.07 |
| | (0.44) | (0.41) | |
| **Whether arrested** | | | |
| For any charge | 0.40 | 0.44 | 0.11 |
| | (0.49) | (0.50) | |
| Violent crime charge | 0.10 | 0.11 | 0.28 |
| | (0.30) | (0.32) | |
| Property charge | 0.15 | 0.20 | 0.01 |
| | (0.36) | (0.40) | |
| Drug charge | 0.12 | 0.15 | 0.05 |
| | (0.32) | (0.36) | |
| Public order/other charge | 0.27 | 0.28 | 0.73 |
| | (0.44) | (0.45) | |
| Number of arrests | 0.70 | 0.83 | 0.06 |
| | (1.22) | (1.26) | |
| **Whether convicted** | | | |
| Of any charge | 0.28 | 0.26 | 0.44 |
| | (0.45) | (0.44) | |
| Violent crime charge | 0.05 | 0.05 | 0.48 |
| | (0.23) | (0.21) | |
| Property charge | 0.11 | 0.11 | 0.95 |
| | (0.31) | (0.31) | |
| Drug charge | 0.08 | 0.09 | 0.33 |
| | (0.26) | (0.29) | |
| Public order/other charge | 0.13 | 0.12 | 0.85 |
| | (0.33) | (0.33) | |
| Number of convictions | 0.38 | 0.37 | 0.70 |
| | (0.71) | (0.72) | |
| Observations | 743 | 761 | |

Standard deviations in parentheses. Follow-up lasted 650 days from the moment of randomization. For arrests and convictions, samples slightly smaller than shown.
Source: Lattimore et al. (2016), Table 3, with p values calculated by author's t tests.

**Table 5. Impacts on recidivism of HOPE replication ("Decide Your Time") in Delaware, from O'Connell, Brent, and Visher (2016)**

| | 6 months | | | 12 months | | | 18 months | | |
|---|---|---|---|---|---|---|---|---|---|
| | HOPE | PAU | p | HOPE | PAU | p | HOPE | PAU | p |
| Technical violation of probation | 0.22 (0.40) | 0.20 (0.41) | 0.63 | 0.29 (0.25) | 0.25 (0.44) | 0.28 | 0.30 (0.46) | 0.21 (0.44) | 0.05 |
| Arrest for new crime | 0.36 (0.48) | 0.38 (0.48) | 0.68 | 0.44 (0.43) | 0.46 (0.46) | 0.66 | 0.53 (0.50) | 0.56 (0.49) | 0.55 |
| Incarcerated | 0.43 (0.49) | 0.46 (0.49) | 0.55 | 0.57 (0.61) | 0.61 (0.49) | 0.48 | 0.67 (0.49) | 0.68 (0.47) | 0.84 |

*N* = 384. Standard deviations in parentheses. HOPE = Decide Your Time. PAU = probation as usual.
p = p value for difference.
Source: O'Connell, Brent, and Visher (2016), Table 3, with p values calculated by author's t tests.

### 5.3. Summary: Swiftness and certainty

In New Jersey and Hawaii, we see proof that swift and certain punishment *can* shape behavior as intended. But it comes from cases where violation can be quickly and easily detected and sanctions can be delivered promptly. Most crimes are not so easily monitored, nor perpetrators so easily identified and caught. And in the cases just examined, the probationers had temporarily lost certain civil rights: their right to freedom was contingent upon compliance. Often, due process—or just plain process—slows the criminal justice system.

The five locations that sought to replicate HOPE give us another caution: even when administrative impediments to swiftness and certainty are overcome, that may not change behavior very much. Probably most people who commit crime have few other options, because of poverty, addiction, and mental health problems. Sticks can change behavior, but carrots may work better, and people respond more to both when they see a clear path forward.

## 6. Deterrence: Severity

### 6.1. Ross (1982), Deterring the Drinking Driver

This short book reviews studies of policies intended to prevent driving under the influence of alcohol. Most relate to "Scandinavian-type laws" that make inebriated driving illegal even when no harm is done. These are also known as "per se" laws since they make driving under the influence a crime per se. The Scandinavian approach originated in Norway, in 1936, spread to Sweden in 1941, to much of Europe, Australia, New Zealand, and Canada in the 1960s and 1970s, and throughout the United States in the 1980s (Ross, ch. 4; NHTSA 2008, p. 19). The book reviews evidence on Scandinavian-style laws in all of these places. It also reports on a few studies of police crackdowns on drunk driving or suddenly increased punishments for existing offenses.

The book concludes rather discouragingly. Most efforts to use punishment to deter drinking and driving have not clearly succeeded. Some did for a few months or years, especially when launched amid great publicity, whether generated by high-profile political debate, as in France, or by government-funded outreach, as in the UK (pp. 28, 42).

The classic example of transient deterrence took place in the UK. After the Road Safety Act of 1967 went into force on October 9, making it a crime to drive with a blood alcohol level above 0.08%, the national rate of fatalities and serious injuries from *all* causes dropped noticeably, especially during weekends nights, 10pm–4am, when drunk driving was likely most common. (See Figure 5, which is copied from Ross 1973, Figure 10.)

Unfortunately, the same graph, along with other graphs and numerical analyses (Ross 1973, 1982), shows the gain largely fading within a few years. Ross hypothesizes that the publicity around the 1967 law initially led British drivers to overestimate the risk of getting caught under the law. Over time, they recalibrated to the true risk, which was low. In 1970, British police administered one breath test for every 2 million vehicle miles driven (Ross 1982, p. 33).

Though the evidence is thinner, the story for stepped-up *enforcement* of existing laws is similar, at least as of 1982, when this book was published. For example, in 1975, Chief Constable William Kelsall of Cheshire, in northwest England, was discouraged by the long-term failure of the 1967 Road Safety Act. So he used his position to conduct an "experiment…to go as far as we could within the law to breathalise all people driving between ten at night and two in the morning" (p. 72). The experiment did not go so far as to set up checkpoints at major roads, where, say, every tenth driver would be pulled over and tested. Rather, it had the police administer the breath test when investigating every accident and traffic-law infraction. "However," Ross writes, "as word of the experiment spread, the chief became the object of vehement protest by representatives of automobile clubs and local political figures who claimed that the effort was equivalent to random testing, which Parliament had specifically eliminated from the Road Safety Act" (p. 72).

Kelsall persisted for a month. The statistical results on the "Cheshire Blitz" resemble those for the country in 1967, but with more noise because of the smaller population. Some combination of the new policy and the surrounding publicity apparently cut the accident rate, especially at night—though this is less certain because of the volatility of the time series (Ross 1982, Figures 5–1, 5–2). At any rate, the effect soon faded.

Ross (ch. 6) looks last at instances in which governments increased the severity of penalties for already established offenses, and finds little impact. Finland doubled its maximum sentence for driving under the influence from two to four years in prison in 1950, and then raised it to eight in 1957 for fatal accidents. But in Finland, as well as in separate instances in Chicago and New South Wales, Australia, available time series on accidents and deaths do not obviously plunge in response to stiffer penalties.

Ross posits that two factors dampened the deterrence from increasing severity (p. 96). First, the lower the certainty of punishment, the less severity matters. If you are sure you will get caught, then you will probably be more sensitive to the difference between a day in jail and a year in prison. Second, the human beings who populate the criminal justice system may have subverted the harshest sanctions:

> Penalties considered unusually severe are unlikely to be forthrightly applied and that…may generate unexpected and undesired side effects when the attempt to apply them is made. Underlying the intellectual order of black-letter law is a social order of legal actors, and these can be expected to resist innovations that overturn established ways of doing things, especially when such innovations are considered extreme and unfair. This point has been noted elsewhere—for example, in discussion of the fact that the relatively common death sentences meted out in the United Kingdom in the late eighteenth and early nineteenth centuries were seldom carried out. (p. 96)

Overall, Ross shows that threat of punishment *can* deter. But when the threat it is weak, as it often is, so is the deterrence. In the case of drinking and driving, testing enough drivers to keep the threat live—with checkpoints and roadblocks—may be fiscally unsustainable (Kleiman 2009, pp. 45–46), or politically so.

**Figure 5. Night-time, weekend fatalities and serious injuries, by month, adjusted for seasonality and number of weekends in each month, UK**

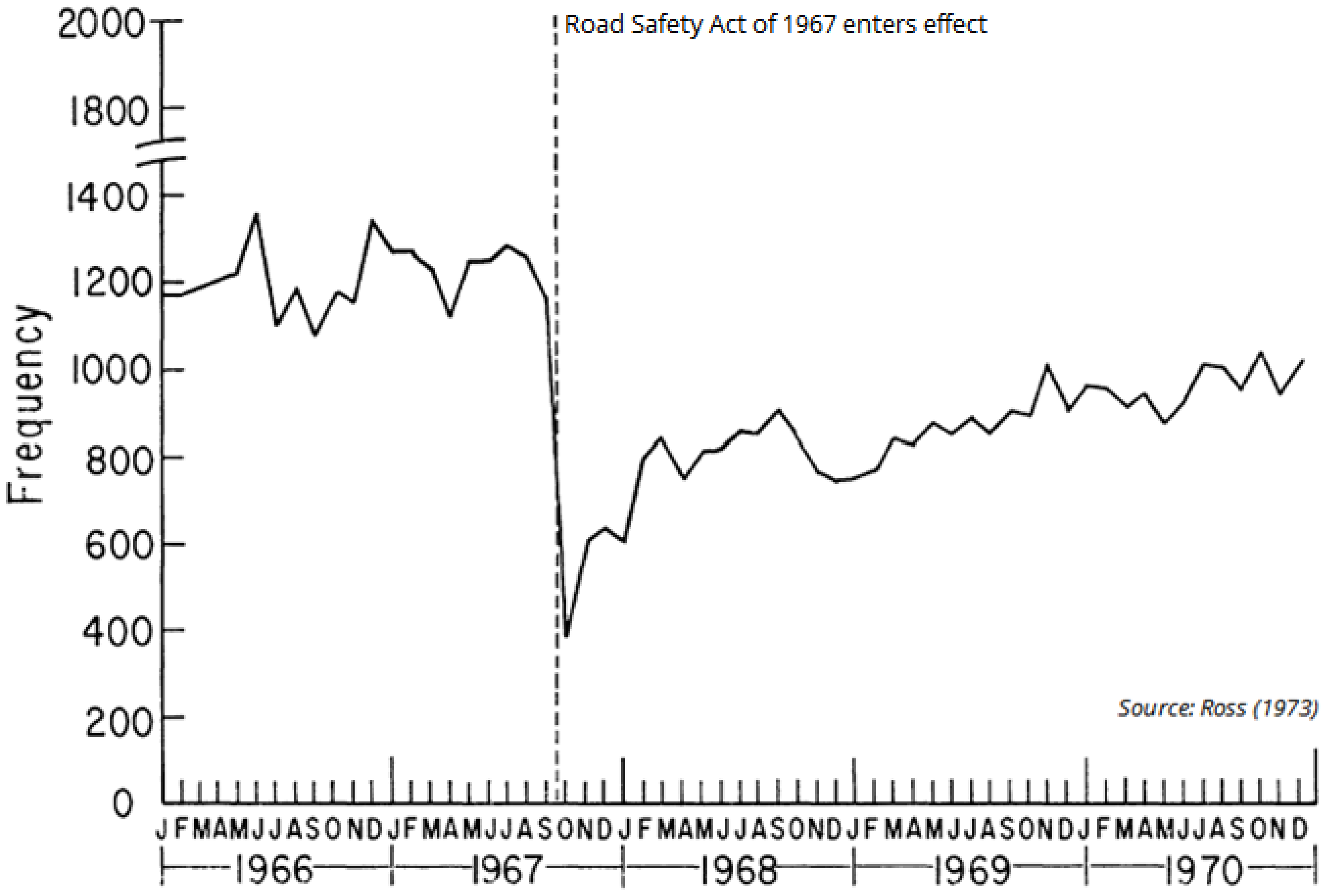

## 6.2. Drago, Galbiati, and Vertova (2009), "The Deterrent Effects of Prison: Evidence from a Natural Experiment," *Journal of Political Economy*

On August 1, 2006, the Italian government released a stunning 36% of all prisoners, some 22,000 people.[28] While distinctive in scale, the event fit with Italian history. Eight other times between 1962 and 1990, the government pardoned *en masse* (Barbarino and Mastrobuoni 2014). Pardons became less common after 1992, when the constitution was amended to require a two-thirds majority vote in parliament for approval (Buonanno and Raphael 2013, p. 2441). The great 2006 release came after years of debate spurred by Pope John Paul II, who was concerned about harsh prison conditions and overcrowding (Drago, Galbiati, and Vertova 2009, p. 264).

More precisely, the 2006 clemency law suspended the last three years of sentences for most crimes. Excluded were some of the most serious offenses, such as those relating to terrorism and the mafia, as well as "exploitation of prostitution." People with less than three years left to serve were freed immediately. Those with more than three years were brought that much closer to release. However, anyone receiving this clemency who recidivated within five years had the suspended portion of his old sentence tacked onto the new one, provided that the new sentence exceeded two years. Thus the suspensions converted to commutations only after five years. (Drago, Galbiati, and Vertova 2009, pp. 265–66.)

One way to study the impacts of the sudden fall in the national prisoner count is to study whether national crime rates jumped right after. Barbarino and Mastrobuoni (2014), reviewed below, take that approach. In contrast, Drago, Galbiati, and Vertova work at the level of the individual. They ask: If A received the same sentence as B, but was imprisoned closer to the clemency date and thus served less time, does A commit more or less crime after?

[28] The public Buonanno and Raphael (2013) data file "prisonertimeseries2.dta" records 60,710 prisoners just before and 38,847 a month later.

Interpreting their numerical answer, however, is tricky, for one can imagine two major stories at work. First, since A received more clemency, she would have a larger sentence add-on hanging over her head if she reoffended within 5 years, so she might be deterred more from criminality. But, second, she would have spent less time in prison, and so would have been subjected less to its rehabilitative or hardening influences. Both deterrence and aftereffects could operate.

With the data available to them (on prison re-entry), the authors are able to track the 22,000 releasees for only seven months, through February 28, 2007 (p. 267). They determine that each additional month of sentence suspension—one month less served for the last crime and potentially one more served for the next—cut the chance of prison reentry by 0.16 percentage points, from a base rate of about 11.5% (p. 268). If this reduction owes purely to deterrence, it is large: Drago, Galbiati, and Vertova (p. 273) estimate the elasticity of crime with respect to prospective sentence at −0.74. If the reduction owes purely to the (reduced) aftereffects of shorter incarcerations, then the elasticity is positive, meaning less time served led to less crime after. I estimate the elasticity under that interpretation at +0.48.[29]

It is impossible to determine the true roles of deterrence and incarceration aftereffects from the available data. Drago, Galbiati, and Vertova (p. 273) argue that deterrence dominates. They cite other studies that find predominantly *negative* (crime-reducing) incarceration aftereffects, which seem to rule out the +0.48 of my aftereffects interpretation. But they rely principally on Kuziemko (2013), about which I raise substantial doubts below. And as I explain in section 8.5, Kuziemko (2013) is in the minority among high-credibility aftereffects studies; most find more time leading to more crime. On the other hand, Italy's prisons refilled within a couple of years, which apparently reduced crime below pre-release levels (see review of Buonanno and Raphael below). This suggests that the released prisoners were less criminal than their replacements, despite probably having spent more time behind bars. Strong positive aftereffects (more time causing more crime) would make that outcome unlikely.

My best guess is that deterrence explains much of this study's key result. But because the pardon manipulated two variables at once—past time served and prospective time served—it is hard to be sure. Also, the dramatic and personal nature of the threat of punishment in this case may have given the prospect of enhanced punishment unusual cognitive salience and deterrent power. It is one thing for a legislature to increase sentences for a class of crimes—the sort of event discussed just below. It is another to be told that *your* sentence has been suspended but could be reinstated if you commit another serious crime.

## 6.3. Two studies of "Three Strikes"

California's "Three Strikes and You're Out" policy was proposed by a wedding photographer whose daughter had been murdered by a parolee, and was quickly adopted in the heat of the 1994 gubernatorial campaign (Zimring 1996, p. 245; Zimring, Hawkins, and Kamin 2001, pp. 4–6). The law was of a piece with the national "tough on crime" movement, yet singular in its severity. Zimring, Hawkins, and Kamin (2001, ch. 2) call it the "largest penal experiment in American history." If researchers are going to find deterrence from long sentences, they should find it in California.

---

[29] The average original sentence was 38.982 months, and the average suspended amount 14.511 (Drago, Galbiati, and Vertova, Table 1, col. 1). The average recidivism reduction was therefore about 14.511 months × .0016/month = .023218. Assuming that prospective sentences—before any add-on for revoked clemency—also averaged 38.982 months, add-ons threatened to increase sentences by 14.511. If we interpret the results as deterrence, this yields an elasticity of $\ln(1 - 14.511 \times .0016/.115)/\ln(1 + 14.511/38.982) = -.71$, which probably differs from the published $-.74$ because of rounding of the input values. (This neglects that only new sentences of at least two years are subject to the add-on.) If we instead interpret the results as aftereffects, then denominator is the proportional reduction in time served for the last conviction: $\ln(1 + 14.511 \times .0016/.115)/\ln(1 - 14.511/38.982) = .48$.

While avoiding the term, the law effectively defines a "strike" as a conviction for a serious or violent felony.[30] Having one strike doubles the sentence for a subsequent felony even if the latter is not serious enough to be another strike. Having two strikes makes the next sentence three times as long as normal—or extends it to twenty-five-years-to-life, whichever is *greater*. And after one or two strikes, you can only be paroled after serving 80% of the lengthened sentence, compared to a more standard 50% (Zimring, Hawkins, and Kamin 2001, pp. 8, 17–19; law text at j.mp/1NkK1CE). Notice that as enacted, and as in force during the periods of the studies discussed next, Three Strikes didn't quite match the baseball metaphor.[31] After serving sentences for two "strikeable" felonies, a person could then be put "out"— get twenty-five-to-life—even for a minor felony that itself would not count as a strike. But for convenience, I will sometimes refer to all people who received the law's maximum penalty as "three-strikers."

Partly because Three Strikes extends sentences for all felonies, many people doing twenty-five-to-life probably were and are serving much more time than similar people who were convicted of similar crimes but had not reached three strikes. Of the 6,900 three-strikers in prison in mid-2013, 500 were there for drug crimes, 1,450 for property crimes, 4,200 for crimes against persons (including 2,200 for robbery), and 750 for other crimes, mainly weapon possession (DCR 2013, Table 1). In 2004, the Justice Policy Institute found that only one state approached California's incarceration rate pursuant to a Three Strikes–style law: Georgia. It had 7,631 such inmates, or 0.9 per 1,000 residents. California had 42,322 two- and three-strikers doing time, or 1.2 per 1,000 residents. (Schiraldi, Colburn, and Lotke 2004, p.13.)[32]

As in rest of the country, crime fell in California starting in the early 1990s. California's re-elected governor was quick to credit Three Strikes (Butterfield 1996). I review two studies that attempt to examine the link more rigorously.

### 6.3.1. Helland and Tabarrok (2007), "Does Three Strikes deter? A nonparametric estimation," Journal of Human Resources

Three Strikes is severe, but like most criminal sanctions it is not certain, since its application is subject to the discretion of prosecutors, who decide which charges to bring, and to judges, who typically decide which stick. Two people might commit the same crimes yet not pay the same price under the law, thanks to having different strike counts. And arbitrariness, however unfortunate for justice, is gold for researchers wanting to study the impacts of severe sentences.

Helland and Tabarrok (2007) compare people who earned two strikes to contemporaries who *almost* did—who were charged twice with strikeable felonies but were convicted one of those times of a lesser, non-strikeable offense. The researchers ask: do the people who have served terms for two strikes, and now live in the shadow of twenty-five-to-life for a third, commit less crime? Does Three Strikes deter?

A thoughtful reader should doubt this comparison. If someone ended up in Helland and Tabarrok's two-trials-two-strikes group rather than the two-trials-one-strike group, perhaps a judge or prosecutor perceived something in the person's character or history that both argued for conviction for the more serious crime and pointed to greater risk of recidivism later on.[33] Then the weightier conviction would *predict* higher crime, not just *cause* lower crime by pushing people closer to the deterrent threat of Three Strikes. This mathematically opposite correlation would cause Helland and Tabarrok to underestimate the deterrence of Three Strikes.

Helland and Tabarrok respond to the concern in several ways. They perform checks that I discuss below.

---

[30] §667.5(c) of the California penal code defines violent felonies (j.mp/1lJ8FWs). §1192.7(c) defines serious felonies (j.mp/1lJ94rV).

[31] In 2012, after the follow-up periods of the studies reviewed here, California voters approved Proposition 36, which reduced the metaphorical mismatch by ending the twenty-five-to-life sentences for non-violent felonies and applying the change retroactively.

[32] As discussed below, Texas turns out to be a "three strikes" state in practice if not in name. It may rival California in the scale of application of its repeat-offender law; I lack data to check.

[33] Copying Helland and Tabarrok (note 9) I use "trial" as short hand for any process leading to conviction: a jury trial, a bench trial, or—by far the most common—a hearing in which the defendant pleads guilty.

And they make the asserted randomness of their treatment-control split look realistic by elaborating specifics of context:

> Our identification assumption is that these individuals are comparable because the outcome of the trial is to a considerable degree stochastic. How strong is the evidence? Is there a good eyewitness to the crime? How good is the defense lawyer relative to the prosecutor? How lenient or strict is the judge or jury? How eager is the prosecutor to cut a deal? How overcrowded are the jails? All of these factors will help to determine trial outcome but can be considered random with respect to other variables that might affect criminal disposition. (p. 312)

Furthermore, Helland and Tabarrok find that that their treatment and control groups hardly differ on observed traits. Out of 13 traits, such as age, race, and number of past arrests for various offenses, the groups' averages are shown to differ at $p < 0.05$ on only one, age at first arrest (Helland and Tabarrok, Table 1).

Helland and Tabarrok's recidivism variable is the *fraction of 1994 releasees who have not yet been rearrested, as a function of time.* The arrest database used is national.

Figure 6 shows their main results, plotting the share of subjects not yet rearrested, as a function of time since release. The dashed line declines more slowly, meaning that people living in the shadow of a third strike were arrested less—about 15% per year less (Helland and Tabarrok, Table 3, row for "CAStrike2"). I labelled Figure 6 to document the standard error of this estimate, about 6%, and the corresponding p value of 0.02.[34]

In elasticity terms, the effect looks modest. Helland and Tabarrok suggest 22 years as an average three-strike prison term. (A three-striker must serve at least 80% of the lower bound on those twenty-five-to-life sentences, meaning at least 20 years.) We don't know the counterfactual, how much time three-strikers would have served but for the law; Helland and Tabarrok use the average sentence for second strikes in this mid-1990s sample, which they cite as 43 months, and conservatively raise to 64 (pp. 327–28). On these figures, a quadrupling of time—from 64 months to 22 years—caused that 15% fall in crime among two-strikers. That's an elasticity of −0.12.[35,36] In other words, each 10% increase in sentencing reduced crime 1.2%.

To check the validity of their claimed quasi-experiment, Helland and Tabarrok rerun their analysis in another state with a Three Strikes–like law (Texas) and two major states without (Illinois and New York). Helland and Tabarrok classify releasees as having one or two strikes as if they had been convicted in California. If the resulting treatment and control groups are statistical twins, then they should recidivate at the same rate in Illinois and New York, but not in Texas. And that is how it turns out. Helland and Tabarrok find deterrence where they should and not where they shouldn't.

---

[34] Results come from a regression that is identical to that reported in Helland Tabarrok (Table 3) except that it is restricted to California. This restriction changes the two-strikes hazard ratio from 0.849 (se = 0.060) to 0.851 (se = 0.061).

[35] $\ln(1 - .151)\,/\ln(264/64) = -0.12$

[36] Helland and Tabarrok also compare the deterrence of Three Strikes to policing, on cost. After consulting, with them, I have reworked the comparison. The Vera Institute of Justice estimated that California spent $47,421/prisoner on its prison system in 2010 (Henrichson and Delaney 2012, p. 10). The 8,647 three-strikers in prison at mid-2010 (DCR 2010, p. 2) therefore cost some $400 million/year to incarcerate. If Helland and Tabarrok are right that three-strikers are spending four times as long in prison as they otherwise would have, then we can attribute 75% of this expense, $300 million, to Three Strikes. Redirecting $300 million to policing would have boosted state and local spending on police by 1.9%, from a base of $15.5 billion in 2010 (Census Bureau 2010). Surveying the literature, Klick and Tabarrok (2010, p. 134) estimate the elasticity of crime with respect to policing at −0.35, implying that the 1.9% spending increase would have reduced crime in California by 1.9% × 0.35 = 0.67%. The FBI tallied reports of 164,133 violent and 981,939 property crimes in California in 2010 (FBI 2011, Table 5). Following Helland and Tabarrok (p. 327), we divide these reported crime levels by the rates at which victims say they report crimes to the police—51.0% for violent and 39.3% for property crime (BJS 2011b, Table 7)—yielding a crime total of 2,800,000. Reducing that by 0.67% would avert 19,000 crimes/year.

For a final check, Helland and Tabarrok return to California and run their analysis on people with two trials and one or two *nonstrikeable* convictions (and evidently no strikeable convictions, though this is not clear). Again, since having one more nonstrikeable prior should invite no Three Strikes deterrence, if the Helland and Tabarrok strategy works, the two groups should have the same rearrest rates—and they do (Helland and Tabarrok, Figure 6).[37]

Helland and Tabarrok shared their data and code with me; reproducing key graphs and tables led me to revise my interpretation in two ways. First, a fuller presentation of the balance tests—of whether the treatment and control groups in California match each other in averages—reveals some imbalance, and in the direction that would naturally explain the headline results. For example, the (eventually less-recidivating) two-strikers were slightly younger at first arrest (21.12 versus 21.95) and had been arrested somewhat more before the study period (8.53 instead of 7.57), suggesting modestly greater criminal propensity at the outset of the quasi-experiments. Some of the associated p values exceed 0.05 and so escape mention in the Helland and Tabarrok presentation, and yet are fairly low. See the left half of Table 6, which perfectly matches Helland and Tabarrok, except for the row on total priors, where it reports a smaller difference but with much greater statistical significance.[38] The table (near the bottom) adds a test of whether the two sets of means differ overall, which returns a p value of 0.02, meaning that the differences are hard to ascribe to chance.

If the two-strikers came into the study with somewhat fewer priors, perhaps they tended to commit less crime (or get caught less) *during* the study not because of deterrence, but because they were different people to begin with.

Second, the impact found in study appears largely confined to *drug* crimes. To show this, Figure 7 progressively breaks Figure 6 out by crime type.[39] The top left shows that people with two strikes were arrested 0.3% less often for "index" crimes—the major ones in the FBI's standard crime rate statistics—with an insignificant p value of 0.98. Now, the 10.0% standard error on that estimate leaves a range of positive and negative values relatively compatible with the data. The true impact may not have been so close to zero. Nevertheless, the data are most consistent with the hypothesis that Three Strikes did not deter this major category of crime. The next two plots in the first row of Figure 7 show small, insignificant impacts on the violent and property subcategories. So do the next seven in reading order, after subdividing further.[40] The major exception occurs outside the index crime super-category: in drug crimes. People with two strikes were arrested 31.1% less often for drug offenses.

Especially since the two-strikers had fewer drug priors before their second-strike convictions, it is hard to confidently attribute their lower drug arrest rate after to the deterrence of Three Strikes.

What about Texas, where Helland and Tabarrok also find that escalating sentences deter? Analogous graphs for that state (not shown) do reveal more apparent impact across crime types. But the quasi-experiment in Texas also looks less clean and compelling, with the less-recidivating two-strikers being 3.63 years younger and having 5.64 instead of 9.34 prior arrests. (See right half of Table 6.)

And even if California's two-strikers were deterred from drug crime, the net impact on public safety was probably minimal. To the extent they were deterred from selling, probably others took their place (see discussion of replacement effect in section 2.4.1). To the extent that they were deterred from using illegal drugs, this may well have benefited them while making little difference for the rest of society. These sorts of doubts explain why most cost-benefit analyses of crime exclude drug crimes and other crimes of commerce, implicitly treating them as having little societal net cost next to offenses such as robbery and assault. The cost-benefit analysis at the end of this report does the same.

[37] These falsification tests also tend rule out parole bias. In principle, those convicted of a second strike could serve longer times on parole as well as prison, lengthening the period of vulnerability to parole revocation-in-lieu-of-conviction, which could depress recorded convictions in the two-strikes group.

[38] Helland and Tabarrok (Table 1) reports 12.2 prior arrests for one-strikers and 10.3 for two-strikers, and a standard error of 2.92 for the difference.

[39] Impacts and p values displayed in Figure 7 come from Cox proportional hazard regressions like that in Helland and Tabarrok (Table 3) but restricting the definition of failure to a given crime category.

[40] Surprisingly, two-strikers were arrested 453.6% *more* for murder, a difference with notable statistical significance, at p = 0.1; but probably not too much should be read into this since murder is rare, statistical flukes happen, and no such impact appears for the cousin crime of assault.

**Figure 6. Rearrest rate as function of time since release after second strikeable conviction, Helland and Tabarrok (2007) sample**

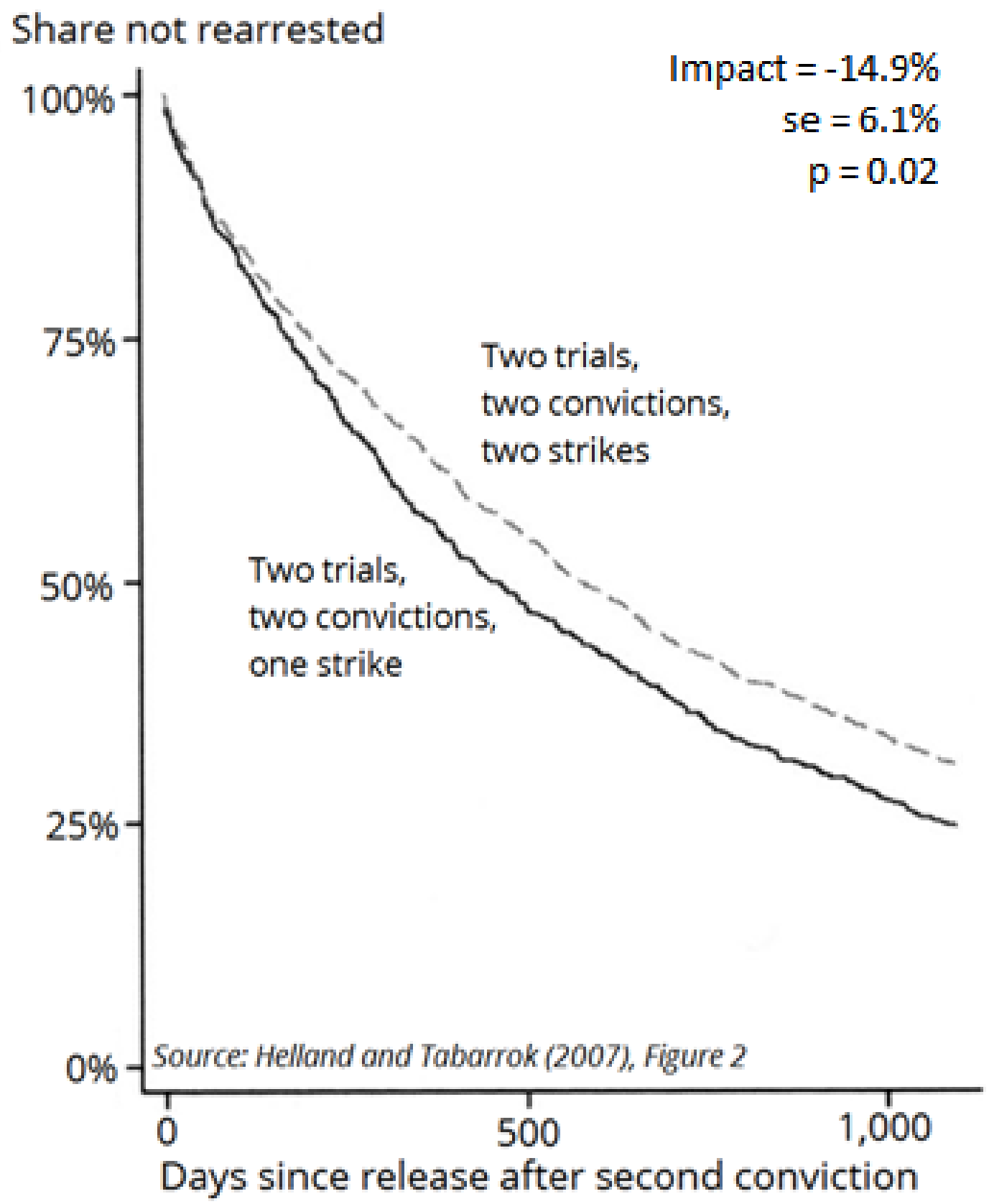

**Table 6. Characteristics of treatment and control groups, California and Texas, Helland and Tabarrok (2007)**

| | California | | | Texas | | |
|---|---|---|---|---|---|---|
| Variable | Two convictions, one strike | Two convictions, two strikes | p value of difference | Two convictions, one strike | Two convictions, two strikes | p value of difference |
| Age | 33.39 | 32.94 | 0.34 | 37.68 | 34.05 | 0.02 |
| Age at first arrest | 21.12 | 21.95 | 0.02 | 21.40 | 20.92 | 0.68 |
| Black | 0.28 | 0.26 | 0.46 | 0.41 | 0.43 | 0.81 |
| Hispanic | 0.29 | 0.32 | 0.37 | 0.23 | 0.23 | 0.96 |
| Prior arrests | 8.53 | 7.57 | 0.01 | 9.34 | 5.64 | 0.00 |
| Murder | 0.04 | 0.03 | 0.27 | 0.05 | 0.05 | 0.98 |
| Robbery | 0.26 | 0.23 | 0.31 | 0.33 | 0.22 | 0.20 |
| Assault | 0.73 | 0.64 | 0.14 | 0.33 | 0.36 | 0.86 |
| Other violent | 0.11 | 0.09 | 0.12 | 0.08 | 0.03 | 0.10 |
| Burglary | 0.58 | 0.50 | 0.08 | 0.59 | 0.71 | 0.36 |
| Larceny | 1.08 | 0.76 | 0.00 | 0.95 | 0.50 | 0.01 |
| Arson | 0.02 | 0.01 | 0.48 | 0.03 | 0.03 | 0.85 |
| Weapon possession | 0.34 | 0.30 | 0.34 | 0.18 | 0.11 | 0.32 |
| Drug | 2.10 | 1.75 | 0.01 | 1.10 | 0.80 | 0.12 |
| Months in prison | 85.57 | 82.00 | 0.84 | 243.08 | 351.12 | 0.55 |
| All at once | | | 0.02 | | | 0.00 |

| | | | | | |
|---|---|---|---|---|---|
| Observations | 473 | 974 | | 39 | 519 |

Note: All subjects released from California prison in 1994 and then tried twice for "strikeable" offenses. Source: Helland and Tabarrok (2007), Table 1; author's estimates, adapting Helland and Tabarrok's code and data.

**Figure 7. Rearrest rate as function of time since release after second Three Strikes–era conviction, by type of charge upon rearrest, Helland and Tabarrok (2007) sample**

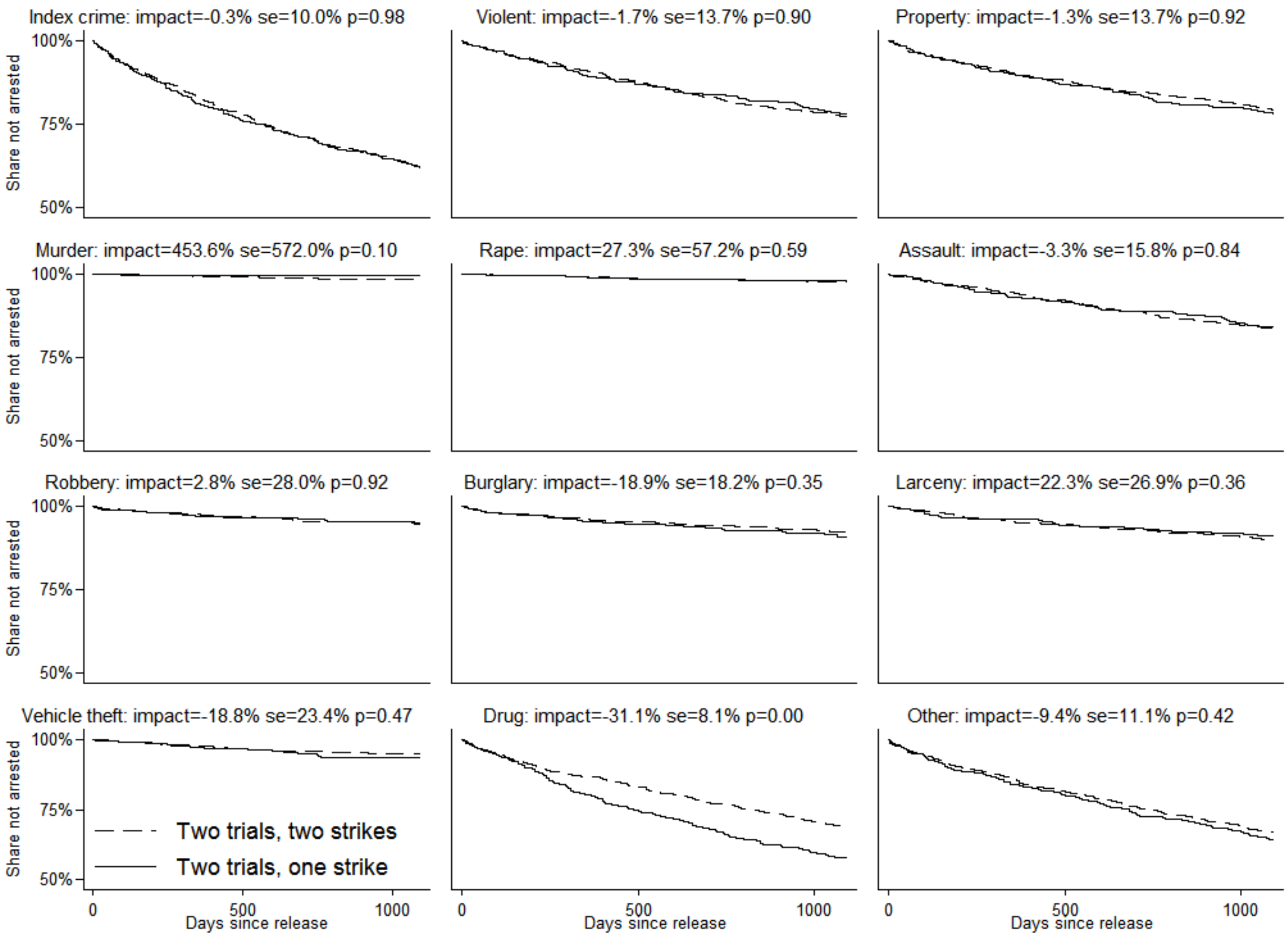

### 6.3.2. Iyengar (2008), "I'd rather be hanged for a sheep than a lamb: The unintended consequences of 'three-strikes' laws," NBER working paper

Iyengar (2008) comes at the same question, whether California's Three Strikes deterred, from another angle. The idea here is that only serious or violent felonies can count as strikes, but any strike lengthens prison terms for *all* later felonies, so the same crimes in a different order can put one at risk for different sentences going forward. This looked to me like the basis for a clever econometric strategy. People with similar criminal histories but at risk for different sentences could make good comparators.

But the paper appears premised on an incorrect reading of the law. Iyengar describes the Three Strikes sentencing rule this way:

> This mismatch between strikes and felonies arises because while *all felony convictions count as strikes after the first strike*, only certain felonies are covered as record aggravating or "triggering" offenses.
>
> …
>
> [C]onsider the following example with two criminals both of whom have previously committed a theft and a burglary. Criminal A first committed a theft and then committed

> burglary. Criminal B first committed a burglary and then committed a theft. Under sentencing guideline prior to Three-Strikes, both these individuals would face similar sentencing eligibility if they committed a third offense. However, after the Three-Strikes law change, the ordering of the crimes committed matters. Because burglary is a triggering offense, it activates Three-Strikes sentencing. *All felonies committed after the activation of Three-Strikes then count as strikes*. Thus, if individual A commits a new offense, that offense will count as a second strike since he has committed no offenses after the burglary. In contrast, a new offense committed by individual B will count as a third strike because he committed a theft after committing a burglary. [emphasis added] (p. 14)

The picture of the law created here is that only a serious or violent felony can *start* the strike count but any felony can continue it. If this were true, it would indeed generate a good quasi-experiment, since a burglary-theft sequence would put you at two strikes while theft-burglary would put you at one.

However, the facts appear otherwise. Section 1(e) of the law (j.mp/1NkK1CE) defines the key three-strike sentence enhancements, with phrases including "If a defendant has one prior felony conviction…" and "If a defendant has two or more prior felony convictions as defined in subdivision (d)…" As the latter quote makes explicit, section (d) defines "prior felony convictions" for purposes of the law; and it defines them as serious or violent felony convictions only:[41] Less-serious felonies never count as strikes. Thus, the burglary-theft sequence in Iyengar's example still leaves a person at one strike.[42] Perhaps Iyengar's apparent confusion arises from the imperfect correspondence I noted earlier between the law and the "Three Strikes" metaphor.

## 6.4. Abrams (2012), "Estimating the deterrent effect of incarceration using sentencing enhancements," *American Economic Journal: Applied Economics*

US states have adopted several kinds of laws to deter crime through punishment. Some impose or increase minimum sentences for certain crimes, curtailing the discretion of judges and parole boards over time served. As in California, these often focus on people convicted of repeat offenses. Another popular step has been to increase sentences specifically for crimes involving guns. Such "gun add-on" laws appeal across the political spectrum because they take aim at gun violence without limiting gun rights.

Abrams studies the impacts of the both types of law, and from a much wider-angle view than Helland and Tabarrok. The unit of observation is the city-year combination, where "cities" are the approximately 500 most populous law enforcement agency jurisdictions (LEAs) in the US and years range from 1965 to 2002 (p. 36). The New York Police Department, for example, is the biggest LEA. With this two-dimensional "panel" data set, Abrams analyzes whether robberies and assaults involving guns fell after states enacted mandatory minimums or gun add-ons.

Zooming out has pros and cons. The disadvantage is that the implicit quasi-experiment—before-after comparisons in hundreds of cities—is less valid than, say, Ross's tightly focused UK study. Possibly, raising sentences for gun crimes deters no robbery, and yet appears to because the add-ons are enacted more often when gun robberies happen to be falling—whether because of a long-run trend, or because such laws are more often passed during transient crime spikes. On the other hand, more data bestows more statistical power. And Abrams's regressions, unlike Helland and Tabarrok's are immune to the replacement effect critique. Abrams's outcome variable is total reported crime in an area, not rearrest among a small subgroup of interest, so it factors in replacement.

Since gun add-ons mainly *extend* sentences for people who would be incarcerated anyway, they can especially help shed light on deterrence, as distinct from incapacitation and aftereffects. In the first year or so after adoption, most of the people subject to such laws would be in prison anyway. So in the short-term, gun add-ons should not increase incapacitation or aftereffects. But their deterrent effects could kick in immediately. Citing estimates that people convicted of representative crimes typically serve three years, Abrams (p. 40) therefore focusses on impacts in the first three years.

---

[41] "[A] prior conviction of a felony shall be defined as…[a]ny offense defined in subdivision (c) of Section 667.5 as a violent felony or any offense defined in subdivision (c) of Section 1192.7 as a serious felony."

[42] Selena Teji, Research Manager at Californians for Safety and Justice, confirmed my reading (email, June 23, 2016).

In regressions, Abrams finds that mandatory minimum laws do not affect crime to a statistically definitive degree.[43] In contrast, gun add-on enactment was followed by statistically significant drops in gun-involved robberies. According to a representative regression, gun robbery fell an average 6.61%, 14.8%, and 17.9% one, two, and three years after add-ons went into effect (standard errors = 4.60%, 4.85%, 6.12%; Abrams Table 3, col. 6; these regressions exclude pre-1974 data because of data problems and control for state-specific linear time trends; standard errors clustered by state). Like Helland and Tabarrok, Abrams (p. 54) infers an elasticity of deterrence of about −0.1. The apparent impacts on gun-involved assaults are also consistently negative, but much smaller and hard to distinguish from zero, at 1.81% and 0.82% after two and three years (se = 1.33%, 2.13%; Abrams Table 4, cols. 4 & 8).

Abrams also performs a graphical check, by recasting the analysis as an event study (Abrams, Figures 4 & 5). The graph shows how the rate of a crime such as gun robbery evolved on average in the years before and after adoption of a mandatory minimum or gun add-on law, after controlling for year and state effects. Abrams (p. 45) discovers that the gun robbery rate starts declining about a year *before* an add-on goes into effect, and suggests that the cause is the publicity around a law when it is passed, which can take place as much as a year earlier (pp. 50–51). Abrams's Figure 5 recalibrates with respect to date of passage rather than implementation. It shows the gun robbery rate stable before passage and dropping after. This strengthens the case that gun add-ons deterred gun robberies. (See Figure 9, pane A.)

Nevertheless, the conclusions appear questionable to me.

One caution is that the study tests two policy changes—mandatory minimums and gun add-ons—against two outcomes—assaults and robberies involving guns—and finds clear impacts in only one of the four combinations, and that only after switching from effective date to enactment date. The mind is drawn to those most significant, persuasive results, but it is a mistake to focus on the significantly non-zero results and leave aside the rest. Overall, deterrence does not emerge strongly.

Replication surfaces one issue in the first set of regressions that is noteworthy, even if not central to our interest in gun add-on laws. Contrary to the description in the paper, the study does not symmetrically test for impacts of gun add-on and mandatory minimum laws. As just stated, the regressions include a dummy for whether an add-on was enacted within the last 1, 2, or 3 years and a dummy for whether a mandatory minimum law was in effect in a given year, which constitutes asymmetric treatment. E.g., the results labelled as "Three year impact…After MM law date" are coefficients on a dummy for whether a mandatory minimum law was in effect in a given year, not whether one was passed within the last three years.

The larger concern is fragility, i.e., that arguably modest changes to the analytical approach cause large changes in the conclusions.

The data set constructed for the paper is posted on the journal's website. In addition to working with this data set, I went back to primary sources to reconstruct the data set—or a version of it. The largest challenge was handling what is a well-recognized problem: the quality of the FBI crime data (Abrams, pp. 33–34; Maltz and Targonski 2002; Lott and Whitley 2003; Maltz 2006; Lynch and Jarvis 2008). The FBI began collecting crime totals from LEAs in 1929; participation was and is voluntary (Maltz 2006, p. 2). The computerized files begin in 1960 (j.mp/1WPlQoR), with monthly resolution. Over time, more LEAs have supplied data, which means that some trends in state totals could reflect expansion in coverage rather than evolution in criminality. In addition, many LEAs' series contain gaps, which are not reliably flagged: sometimes zero means zero and sometimes it means "unknown." The meaning is obvious in many cases, as when the NYPD's gun robbery count drops from 251 to zero between May and June of 2002 and remains there. But because the data set has 10.6 million LEA-year-month rows for 1960–2014, a person cannot review it all. Algorithmic treatments will inevitably commit errors, marking some real zeroes as missing and vice versa.

[43] In fact, the panel regressions appear to contain an error, but not one that changes interpretations. The apparent intention is to focus on short-term (deterrence) effects by regressing crime on whether a law was implemented in the last one, two, or three years only. The dummy for whether an add-on was passed is restricted to these time frames with the line, "replace yaddon = 0 if relyr > `loop'", in the public "AbramsDeterrence.do" code file. But the dummy for mandatory minimums is not, so that, contrary to the labelling in Abrams (Table 3), it always only indicates whether mandatory minimums were ever enacted. I find that restricting the mandatory minimum dummy as apparently intended does not affect results significantly.

Abrams's (pp. 32–38) strategy for confronting these problems is to focus on the approximately 500 largest LEAs with (nearly) complete data for 1965–2002, which cover about 40% of the US population; and then to hand-clean the data, filling missing values by a process that is not formally documented. The cleaning evidently required interpolation and extrapolation to fill some gaps, as some crime counts in the Abrams data set are fractional, unlike primary data. Inevitably, the process is imperfect: Abrams, for example, appears not to have detected the mid-2002 cessation of New York City's gun robbery series and treats the subsequent zeroes that year as real.

I take a different approach, which while also inevitably imperfect, tests the robustness of the Abrams findings. Instead of restricting to LEAs with nearly complete data over a long period, I retain more LEAs: the 3,522 with at least 17,581 people in 2000, which were home to exactly 80% of the population that year (NACJD 2005).[44] Missing data are identified algorithmically. For example, a gun-involved robbery count is marked missing for a given month and LEA if it is zero and if all robberies totaled at least 100.[45] For the half century 1965–2014, I apply Multiple Imputation (MI) to fill identified data gaps. MI is generally considered a rigorous response to missingness because it works to maximize the use of available information while taking account of the uncertainty arising from the gaps (e.g., Allison 2002, Honaker and King 2010). MI generates several copies of the data set—I make five, as is typical (Honaker and King 2010, p. 561)—each inserting different values for the missing observations. The imputations are generated in a way that is random but conforms to overall patterns in the data.[46] MI-based regression then analyzes each copy, producing somewhat different results each time. The results are combined to generate final estimates, along with confidence intervals designed to incorporate the added uncertainty from missingness.

Scatter plots in Figure 8 show that the two data sets broadly agree on per-capita rates of crimes of interest in 1970–99, by state and year. Each data point is marked with a two-letter state code and two-digit year. Correlations across the data sets in total robberies, total gun robberies, total assaults, and total gun assault cluster 0.98 when weighting by state population, except for gun assaults at 0.95. (For this graph, the five imputed data sets were averaged together. All values are taken in logarithms.) Unfortunately, because both data sets are large and embody numerous idiosyncratic judgments about which zeroes are data gaps and how gaps should be filled, it is hard to pin down which differences in data most account for any differences in results.

---

44 The imputation model failed to converge when I expanded to 90% of the population.

45 I mark an assault or robbery entry as missing if it is negative; if the count for the gun subcategory exceeds that for the larger category; if it equals zero or the count for the larger category while the latter exceeds 100; if total crime in a given year does not follow a perfect monthly, quarterly, semiannual, or annual pattern of non-zeroness; or if the total crime count is identified as missing by Targonski (2004, covering 1977–2000; codes −99, −98, −93, −94, −90, −85, −80 in a "CI" field).

46 I impute missing values of log population, robberies, assaults, gun robberies, and gun assaults. The pattern of missingness is nearly monotone, meaning that each variable in that list is rarely missing unless the preceding one is. To reduce the computational burden of multivariate imputation, I force perfect monotonicity by changing some entries to missing, then apply sequential conditional imputation. The model for log population is OLS. Those for the rest are Poisson, taking population as the exposure variable. All models include LEA, year and calendar month fixed effects. Previously imputed count variables are included in models in logs, zero-inflated (with additional dummies for zero values). In order to avoid imposing statistical continuity across event boundaries, all of these controls are interacted with a dummy for whether an LEA is in a state with an add-on law passed by the given time. The process is run separately for the data set limited to the 521 Abrams LEAs and for the larger data set covering 80% of the population.

Aside from the use of Multiple Imputation, the new event study regressions incorporate two specification changes:

- Some of the regressions work with monthly data in order to extract sharper information about timing of crime changes.
- Like Abrams's panel regressions, the samples of the new event study regression include states where no add-on was ever passed. This step is meant to improve the precision of the estimates of the coefficients on the demographic controls and year effects, and thereby of the impact estimates of interest. A dummy is added to the model that equals 1 for the added states.

Figure 9 shows how these changes affect the Abrams graphical event study, the one examining whether, on average, states' gun robbery trends broke distinctly downward upon passage of an add-on law. Pane A of the figure corresponds exactly to Abrams Figure 5. To formally check for a trend break, the presentation here adds best-fit lines extending one, two, or three years forward and backward from the moment of passage, in red, and shows the p values for tests for equality of slope for corresponding lines. For example, working from the original Abrams analysis, the hypothesis that the gun robbery trend was the same in the year before and year after adopt cannot be easily rejected ($p = 0.52$). But extending to two or three years exposes a significant break ($p = 0.03, 0.01$).[47] However, moving to the new data set—still restricting to the Abrams LEAs but handling missing data differently—greatly weakens these rejections (panel B). Now it is hard to confident of a trend break. Doubling population coverage to 80% further weakens the results (pane C). Moving to monthly data in order to measure timing effects more precisely does the same (panes D and E).

Finally, I apply Abrams's panel regressions—the ones that generate the impact estimates cited above—to the new 80%-population data set. (See Table 7.) The results are surprisingly unstable now, with positive, strongly significant coefficients outnumbering negative ones; taken at face value, these results suggest that add-on laws *increased* gun robberies.

On the evidence displayed in Abrams and this reanalysis thereof, the strongest thing one can say for deterrence is that one kind of law—gun add-ons—might deter one kind of crime—gun robberies—temporarily. And even that does not clearly hold.

[47] To perform these tests, the event study regressions are rerun with all dummies corresponding to lags within 1, 2, or 3 years of enactment replaced with a pair of linear spline terms for those spans. A Wald test is then performed of the hypotheses that the slopes of corresponding before- and after- splines are the same.

**Figure 8. Log per-capita crime rates by state and year, 1965–2002, new vs. Abrams data set**

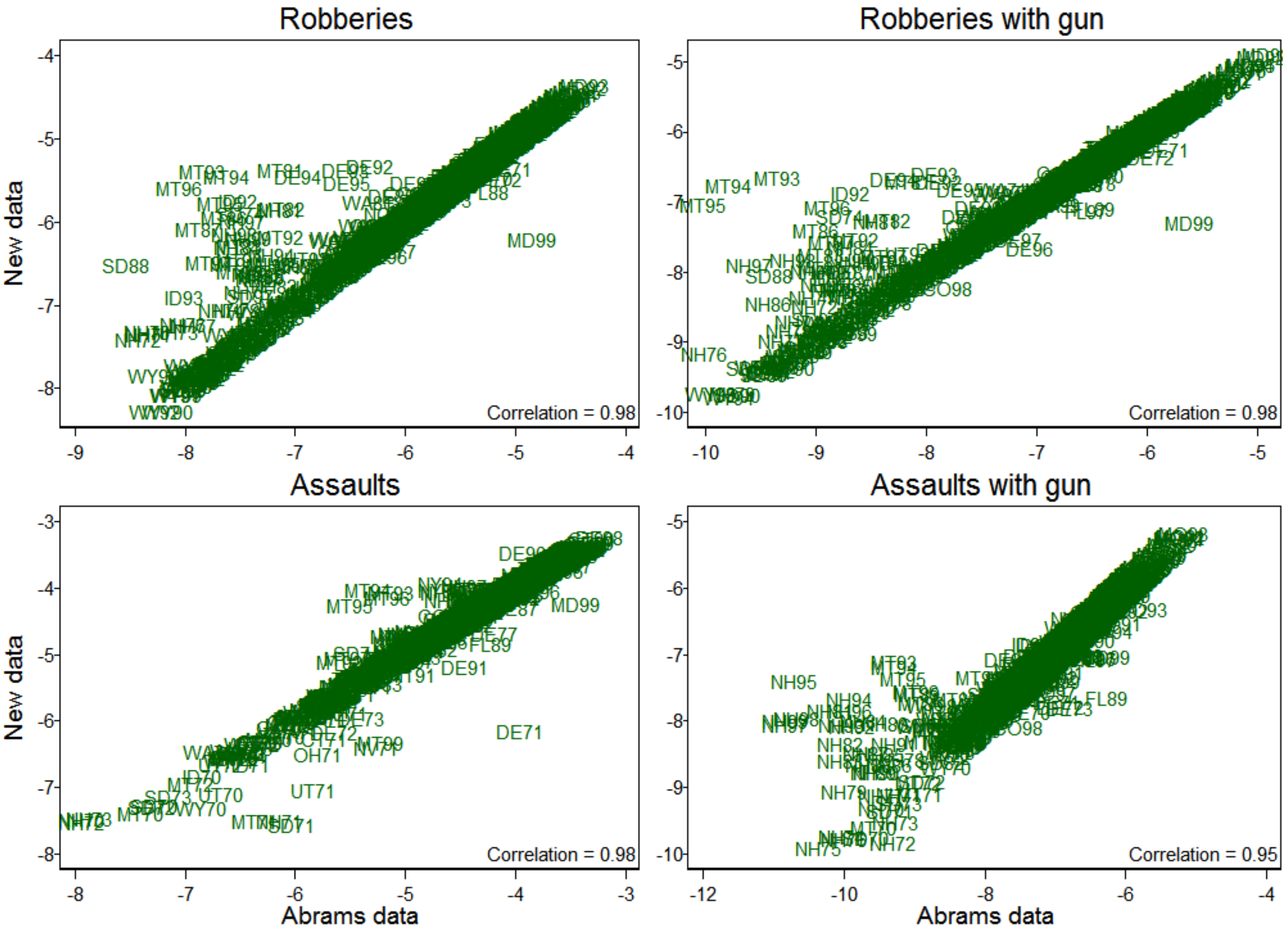

**Figure 9. Gun robbery rate versus time until/since gun sentence add-on passage, controlling for year fixed effects**

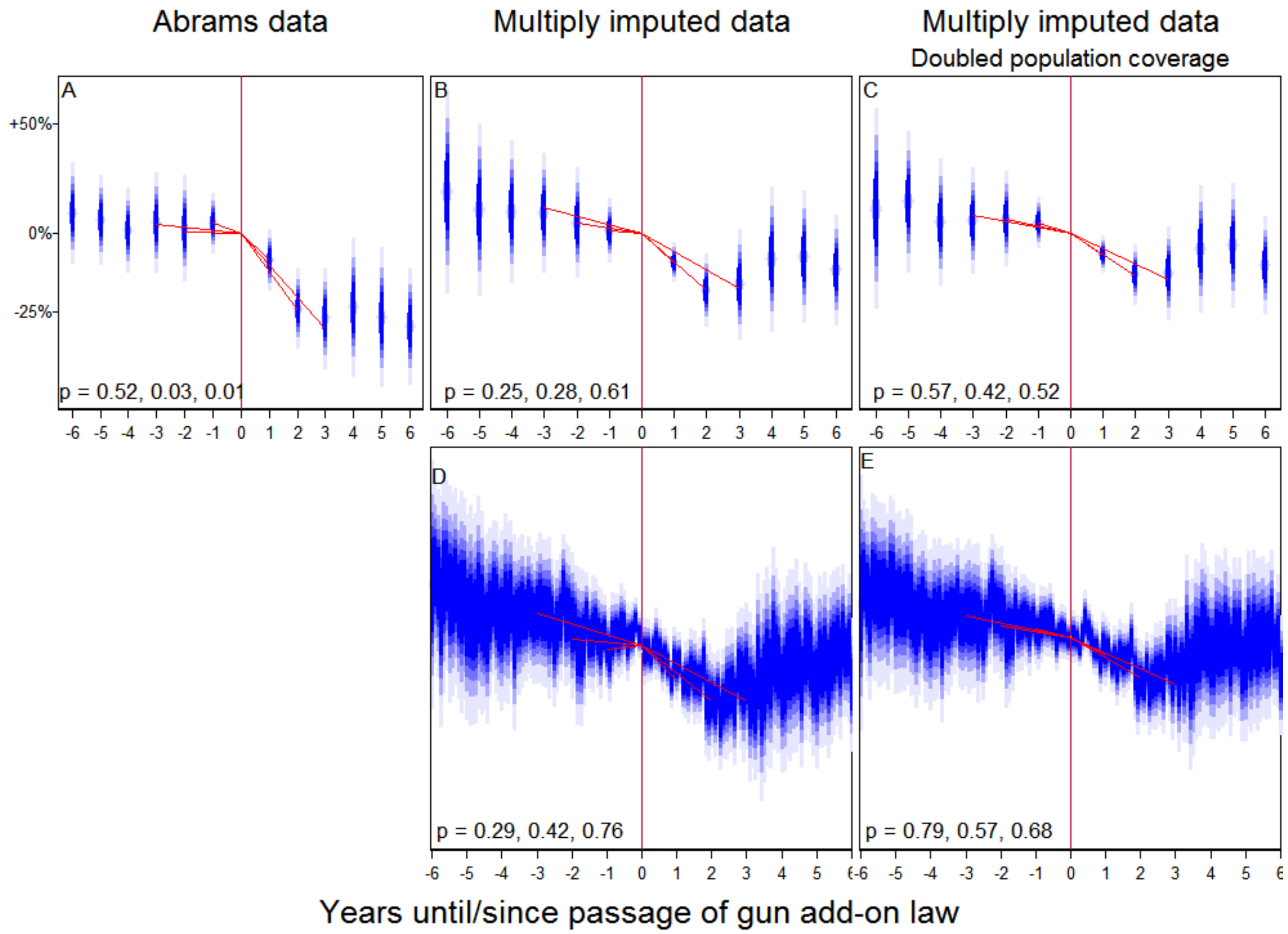

**Table 7. Impact of gun add-ons and mandatory minimum laws on state gun robbery rates, 1970–2006, following Abrams (2012, Table 3) panel regressions**

| | | | | | | | | |
|---|---|---|---|---|---|---|---|---|
| 1 year after add–on law | 0.0344 | 0.0242 | 0.1137** | 0.1499*** | 0.0627 | 0.0248 | 0.1183** | 0.1208** |
| | (0.0535) | (0.0404) | (0.0407) | (0.0521) | (0.0578) | (0.0443) | (0.0485) | (0.0464) |
| 2 years after add-on law | –0.0105 | –0.0413 | 0.0930** | 0.1299** | 0.0459 | –0.0127 | 0.1239** | 0.1369*** |
| | (0.0519) | (0.0389) | (0.0368) | (0.0524) | (0.0602) | (0.0454) | (0.0457) | (0.0443) |
| 3 years after add-on law | –0.0456 | –0.0880** | 0.0218 | 0.0207 | 0.0024 | –0.0750 | 0.0399 | 0.0276 |
| | (0.0525) | (0.0374) | (0.0327) | (0.0342) | (0.0648) | (0.0505) | (0.0388) | (0.0354) |
| State-specific time trends? | No | Yes | No | Yes | No | Yes | No | Yes |
| Data within 6 years of passage? | No | No | Yes | Yes | No | No | Yes | Yes |
| Post-1974 only? | No | No | No | No | Yes | Yes | Yes | Yes |
| Observations | 1,845 | 1,845 | 306 | 306 | 1,598 | 1,598 | 222 | 222 |

Following Abrams (2012, Table 3, "add-on" rows), dependent variable is log per-capita gun-involved robberies/year; regressions include state and year fixed effects and demographic and poverty controls; and standard errors (in parentheses) clustered by state. Unlike in Abrams, data are pre-aggregated to state level; and multiple imputation is used. **Significant at $p<0.05$.

## 6.5. Summary: Severity

We are left with little convincing evidence that at today's margins in the US, increasing the frequency or length of sentences deters aggregate crime.

Ross's inventory of drunk driving laws and their consequences points mainly to transient deterrence.

Drago, Galbiati, and Vertova find strong deterrence in Italy. Conceivably, much or all what they find is not deterrence, but an aftereffect of incarceration, since releasees with longer suspended sentences hanging over them were also the ones who had served less time. But the deterrence interpretation looks reasonable. The special circumstance of that release made the threat of added punishment especially salient in the minds of releasees.

Of the three studies set in the higher-incarceration US context, the two that pass initial muster—Helland and Tabarrok in California and Abrams nationwide—agree on a mild impact elasticity of −0.1, which itself looks open to challenge. Upon reexamination, their data favor the hypothesis of an aggregate impact even closer to zero. Even if Three Strikes deterred drug crime among two-strikers—a conclusion shadowed by the doubt that treatment and control were not quite comparable—replacement probably offset much of the effect. And the Abrams finding that one kind of severity enhancement deterred one kind of gun crime appears fragile; my reanalysis favors the hypothesis that such laws did not perturb crime trends.

# 7. Incarceration versus highly supervised release: incapacitation and aftereffects

## 7.1. Deschenes, Turner, and Petersilia (1995), "A dual experiment in intensive community supervision: Minnesota's prison diversion and enhanced supervised release programs," *Prison Journal*

As the US prison population boomed in the 1980s, states grappled with the fiscal consequences: prisons are expensive. A program in Georgia that explored a middle way between incarceration and traditional probation and parole gained national attention after an encouraging evaluation was published in 1987 (Erwin and Bennett 1987). An Intensive Supervision Programs (ISP) allowed people convicted of crimes to live in the "community" while monitoring them much more closely than traditional probation and parole, with lower caseloads per officer, more frequent check-ins and drug tests, and sometimes electronic monitoring. Many states soon copied Georgia. The hope was that ISP would incapacitate more than traditional supervision while imposing lower fiscal and human costs than prison. (Petersilia and Turner 1993, pp. 281–85.)

In 1986, the US Department of Justice initiated attempts to credibly evaluate ISP. Under the supervision of RAND researchers, 12 coordinated, randomized experiments took place across the country. According to two of the researchers, "The ISP demonstration programs constitute the largest randomized experiment in corrections undertaken in the United States" (Petersilia and Turner 1993, p. 292).

Unfortunately for our interest, only two of the experiments were in "prison diversion," that is, in assigning to people to ISP instead of incarceration. The rest compared ISP to traditional probation or parole. And these two essentially failed as experiments. In Marion County, Oregon, only 28 people cleared all the hurdles to enter ISP, such as having no history of violent crime. In Milwaukee, 72 made it, but judges sent nearly all to prison anyway. (Petersilia and Turner 1993, pp. 303–306; Table 1, upper-right corner.) Of course, while these failures prevent us from learning about the impact of ISP relative to incarceration, they tell us something about practical barriers to ISP.

Possibly because of this disappointment, the Department of Justice hired RAND to experimentally evaluate another instance of diversion to ISP. This one produced statistically meaningful results. In 1990, the Minnesota legislature enacted a form of ISP, which it called intensive community supervision (ICS). Deschenes, Turner, and Petersilia evaluates it in comparison to both prison and traditional community supervision (parole or probation), but I focus on the comparison to prison.

The authors describe Minnesota's program this way:

> During the first phase (about 6 months, or one half the presumptive sentence or time to sentence expiration), offenders are under house arrest and must remain in their approved residence during all hours except those where specific permission to leave (e.g., for work) has been granted. Offenders have four face-to-face meetings per week with an ICS agent. They also must submit to random, weekly, unannounced drug and alcohol tests.
>
> During the second phase (about 4 months), offenders have at least two face-to-face meetings with the ICS agent, are subject to twice-monthly drug tests, and are under a modified house arrest.
>
> The third phase lasts for at least 2 months and subjects offenders to one face-to-face meeting with their ICS agent per week. Drug tests may be done at the discretion of the ICS agent, and offenders must live under a modified house arrest arrangement.
>
> The fourth phase, which lasts until the supervised release for ICS cases and until sentence expiration for ISR cases, is the least onerous and requires two face-to-face meetings with the ICS agent per month, discretionary drug testing, and a curfew instead of house arrest.
>
> If offenders in the ICS program violate one of the rules of the program, for example, fail a drug test, leave their house for other than an approved activity, and so on, they may be sent back to prison….If sent back, ICS offenders serve the original term of imprisonment or until the expiration of sentence, whichever is shorter. (p. 333)

Intake again happened slowly, but not fatally so for the study. From October 1990 to June 1992, 124 offenders in seven counties entered the experiment; 48 were randomly assigned to prison and 76 to ICS. Adherence to treatment was imperfect, however: 4 of the 48 went into ICS instead while 21 of the 76 went to prison, for such reasons as not meeting program criteria and judge refusal. Properly, the authors analyze the groups as originally randomized (pp. 336–37). The ICS group, as randomized, spent an average 124 days in prison, while the prison group did 228 days (pp. 342).

No statistically significant impacts on recidivism appeared at the two-year follow up regardless of whether recidivism was defined as ever being arrested, ever being jailed, or ever being imprisoned within one or two years of random assignment (Deschenes, Turner, and Petersilia, Figure 2, bottom half).

However, those findings are only presented in low-resolution graphs and the paper's definition of "significant" is not entirely clear. ($p<0.05$ appears meant. That is the usual default and is explicit in one place, Table 6.) It appears that ICS-assigned subjects were arrested more in the first year, at ~32% vs. ~20%, which could be significant at $p<.2$ or .25 (Deschenes, Turner, and Petersilia, Figure 2, graph 3). By two years, the ever-arrested rates are identical for the two groups, at ~45% (Deschenes, Turner, and Petersilia, Figure 2, graph 4).

The finding of no difference in recidivism after prison and recidivism after intensively supervised release is positive in the sense that, according to the authors' calculations, ICS costs $17,631/person/year, as opposed to $23,040/person/year for prison. This suggests that at least among nonviolent offenders, intensively supervised parole or probation can substitute for prison with little increase in crime and some fiscal savings. In addition, most convicts probably would prefer ISP—or at least prefer to have the option.

## 7.2. Di Tella and Schargrodsky (2013), "Criminal recidivism after prison and electronic monitoring," *Journal of Political Economy*

Not unlike the US, the Province of Buenos Aires experienced a prison boom starting a few decades ago. "The inmate population held in prisons and jails experienced a large increase (from 12,223 in 1994 to a peak of 30,721 in 2005) without a corresponding increase in infrastructure investment" (Di Tella and Schargrodsky, p. 35). The predictable result: crowded prisons.

The Buenos Aires judicial system moves slowly enough that most people in jail or prison are *awaiting* trial (p. 41). Many are released before trial once they serve the time they would be sentenced to if convicted. Punishment before conviction seems perverse but, for what it is worth, the majority of the cases involve "flagrancy"—being caught in the act—which to the judiciary makes guilt look likely (p. 37).

In 1997, Buenos Aires introduced an alternative to incarceration, which judges could prescribe at their discretion:

> Under the program, offenders stay at home wearing a bracelet on their ankle. The bracelet transmits a signal to a receptor installed in the offender's house. If the signal is interrupted, manipulation is detected, or vital signs of the individual are not received, the receptor sends a signal to the service provider through a telephone line. The private provider investigates the reason for the signal and, whenever necessary, reports to the [electronic monitoring] office of the Buenos Aires Penitentiary Service, which sends a patrol unit to the inmate's house. (p. 36)

Scarcity of equipment limited scale. Between 1997 and 2007, some 900 offenders cycled through 300 ankle bracelets (p. 37).

But, fortunately for social science, the scarce resource was allocated in an apparently quasi-experimental way, which allowed the authors to track the during and after effects of electronically monitored supervision versus prison. "Whenever a person is detained by the police, she or he is assigned to the judge who was on duty on that day, and duty turns are assigned by a lottery" (p. 31). And, crucially, these judges differed in their use of electronic monitoring. Two-thirds never assigned it while the rest used it in 2.68% of cases (p. 46). Di Tella and Schargrodsky describe an ideological split—or spectrum—between the *garantista* judges, who might be called liberal in the US context, and *mano dura* ("tough hand") judges, who slanted oppositely (p. 31). This ideological split is also fortunate for social science, for it makes the judge assignment instrument look strong as well as valid. When instrumenting actual electronic monitoring assignment with a judge's average rate, the instrument earns a t statistic of at least 9 (Di Tella and Schargrodsky, Table 5).

Di Tella and Schargrodsky define recidivism as reentry into the Buenos Aires prison system. One peculiarity in their definition is that they do not set the follow-up period to a given amount of time after randomization or release, such as three years, but end it for all subjects on October 2007. This produces an average post-release period of 2.85 years (p. 54). Since the follow-up period shrinks as one moves forward through the sample in time, if use of electronic monitoring also exhibited a time trend, this could create spurious results.

In fact, that risk is a special case of a larger risk that could threaten this study more than most in this review because of the long time frame and low number of subjects per unit of time. Even if electronic monitoring is quasi-randomly assigned at any given time, if its use, say, rose while crime in Buenos Aires fell, spurious correlations could emerge. The authors combat this risk in two ways. First they include dummies for each year. Second, they construct a matched sample, so that the treatment and control groups have similar follow-up period profiles. For each of the 386 subjects entering electronic monitoring, they select three control subjects close in age, imprisonment date and length, crime type, judicial status, and number of previous incarceration spells (p. 45; all subjects are male and under 40).[48]

Di Tella and Schargrodsky (Table 5) run their main impact regression a few ways, getting similar results. Despite the greater liberty, "outmates," imprisoned at home rather than in a cell, recidivated 11–16 percentage points less, despite greater freedom (se = 4.9–7.0 percentage points). To the extent that electronically monitored liberty reverse-incapacitated—freeing people to commit crime when they otherwise would have been in prison—the reduction in harmful aftereffects of imprisonment evidently more than compensated.

[48] The Buenos Aires government required researchers to copy their data by hand, so they had to limit themselves to a small subset of control subjects.

Di Tella and Schargrodsky may somewhat overestimate the impact in one respect. It turns out that 66 of the 386 treatment group members *escaped* their electronic monitoring (p. 61). To the extent that they left Buenos Aires and were convicted of crimes elsewhere, the study would miss that and underestimate the treatment group's recidivism. But for this to completely explain the study's impact estimate of 11–16 percentage points, it would need to be the case that 11–16% × 386 = 42–62 escapees recidivated elsewhere. Among the 66 escapees, that would require a combined out-migration and recidivism rate of 64–94%. That rather improbably exceeds the recorded recidivism in the raw data—22.5% in the control group and 51 / (386 − 66) = 15.9% in the non-escapee treatment group (p. 54). Most likely then, the bulk of the apparent impact did occur through a reduction in harmful aftereffects.

### 7.3. Summary: Incarceration versus highly supervised release

Coming from US data, the Deschenes, Turner, and Petersilia finding that highly supervised release led to about the same amount of recidivism as incarceration deserves more weight in the US policy context than the Di Tella and Schargrodsky finding that it led to less. Either way, highly supervised release looks no worse than incarceration, and cheaper too. Unfortunately, another consistent theme is that scaling up and sustaining highly supervised release is hard. The challenges in the US context were described above. In Buenos Aires, the program was suspended after an outmate escaped monitoring and killed a family of four (Di Tella and Schargrodsky, p. 63). Whatever the average effect, the program came to be seen as unsupportable after that tragedy.

## 8. Incapacitation

### 8.1. Levitt (1996), "The effect of prison population size on crime rates: Evidence from prison overcrowding litigation," *Quarterly Journal of Economics*

This paper is an early and characteristic work of Steven Levitt of "Freakonomics" fame. It uses a clever insight to construct a quasi-experiment out of otherwise non-experimental data. The setting resembles that of Abrams (2012), a panel data set with one observation for each US state and each year in 1973–93. The research question is whether, within a state, higher or lower growth in prisoners/capita causes higher or lower growth in crime in the following few years. The clever insight is that prison overcrowding lawsuits, brought against states by groups such as the ACLU on the argument that overcrowding constituted cruel and unusual punishment, caused rather arbitrarily timed changes in prison growth. As the cases proceeded over years, courts took control of state prison systems in order to restrain or reverse prison growth, and later released them from court control.[49]

To turn these complex legal battles into numerical variables that can be viewed as spawning quasi-experiments, Levitt identifies five milestones in the progression of overcrowding litigation: filing the suit; receiving a preliminary court decision (requiring a reduction in overcrowding by some date); obtaining a final decision from a judge; further court action such as appointing a special monitor; and release from court supervision. Restricting himself to the dozen states, mostly southern, whose entire prison systems fell under court control, Levitt develops the chronologies reproduced in Table 8.

[49] The Prison Litigation Reform Act of 1996 all but eliminated such suits. The biggest exception has been Plata v. Brown, which led to the realignment reform in California in 2011. Lofstrom and Raphael, reviewed below, investigate its consequences.

**Table 8. Prison overcrowding litigation events, states in which courts took statewide control of prison system, 1971–93**

| State | Suit filing | Preliminary decision | Final decision | Further court action | Release by court |
|---|---|---|---|---|---|
| Alabama | 1974 | 1976 | 1978 | 1979 | 1984 |
| Alaska | 1986 | | 1990 | | |
| Arkansas | | | 1971 | 1974 | 1982 |
| Delaware | | | 1988 | 1992 | |
| Florida | 1972 | 1975 | 1977 | 1980 | |
| Mississippi | 1971 | | 1974 | | |
| New Mexico | 1977 | 1980 | 1990 | 1991 | |
| Oklahoma | 1972 | | 1977 | | 1986 |
| Rhode Island | 1974 | | 1977 | 1986 | |
| South Carolina | 1982 | 1985 | 1991 | | |
| Tennessee | 1980 | | 1982 | 1985 | |
| Texas | 1978 | 1980 | 1985 | 1992 | |

Source: Levitt (1996), Table I.

Levitt calculates that in these states, per-capita prison populations grew:

- 2.3%/year above the national average in the years leading up to a lawsuit filing;
- 5.1%/year below in the first and second calendar years after;
- about 5% below in the year of a final court decision and the two years after that;
- and 5.2% faster again in the year of release from court control (Levitt, Table IV, col. 2).

Meanwhile, across each of the stages, violent and property crime moved oppositely with prison population (Levitt, Table IV, cols. 3 & 4).

These patterns fit some natural hypotheses: lawsuits were more common where prison populations were rising fast; initiating an overcrowding suit and obtaining a court decision counteracted expansion; release of the prison system from court control enabled new growth spurts; and, most important, prison growth reduced crime. Since the apparent impacts come quickly, and in response to one-time events rather than permanent changes in sentencing, incapacitation probably explains them more naturally than does deterrence or aftereffects.

Levitt's formal regressions reinforce these hypotheses. To make instruments, Levitt uses the five milestones to break a state's progression through the lawsuit experience into six stages. Except for the first, pre-filing, each is further split into three substages: the year a milestone is reached, the 1–2 calendar years following, and even later. Within each of the five triplets, the first two are retained, yielding the basis for 11 one-zero indicator variables, each for a different stage and substage.[50] OLS regressions of per-capita growth in prison population, violent crime, and property crime directly on the instruments reveal the two *post–final decision* dummies as especially predictive of lower prison growth and higher crime growth (Levitt 1996, Table V). Instrumented regressions put the elasticities of violent and property crime with respect to the incarcerated population at −0.379 and −0.261 (se = 0.180 and 0.117; Levitt 1996, Table VI, columns 3 and 6). These say that a 10% increase in prisoners per capita cut the violent crime rate 3.79% and the property crime rate 2.61%. More concretely, they suggest that at the margin of release or imprisonment an average prisoner would have committed about 1.2 reported violent crimes and 6.7 reported property crimes in that first year

[50] I believe the exclusion of the third dummy in each triplet is unexplained.

of freedom.[51]

Hoping to replicate the Levitt analysis, I reconstructed the data set. This surfaced one noteworthy issue. First, I built the variables from original official sources, except that I took police counts from a data set provided by Thomas Marvell (whom Levitt also thanks for data).[52] After, Steven Levitt sent me a data set, while emphasizing that it might not be exactly the one used in the paper. The two sets largely match. Correlations exceed 0.99 for all variables except for growth in police officers and violent crimes per capita (both at 0.96) and growth in per-capita prisoners (at just 0.769). That last discrepancy, in the treatment variable, matters most. Between 1926 and 1976, the federal government defined a state's prison population for statistical purposes as those people who were sentenced to at least a year and were housed in facilities run by the state government. In 1977, the definition switched from *custody* to *jurisdiction* in order to recognize that someone sent to prison by the state of Connecticut, say, might be housed in a local jail or a privately run prison. The old custody-based time series have been carried forward but the new jurisdiction-based ones only begin in 1977 and 1978. (ICPSR 2015, p. 6.) The data set provided to me by Levitt switches from the old to the new definition in 1977 or 1978 in a period of much lawsuit activity, creating significant discontinuities for states such as Alabama and Alaska. These discontinuities can interfere with proper measurement of prison growth. The new replication data set therefore keeps to the old definition throughout.

With the two data sets in hand, I worked to reproduce results in Levitt. I found what appear to be minor problems in the initial tabulations and regressions, and am unable to exactly match the key results. But the overall findings are the same.[53] My best replication of Levitt's estimates of the elasticities of violent and property crime per capita with respect to prisoners per capita are −0.456 and −0.250 (se = 0.177 and 0.16). See the upper left of Table 9.

Overall, the replication and reanalysis suggest that Levitt (1996) is right that in response to sudden changes in prison growth, crime probably did move oppositely in 1973–93. However, modern methods produce wider confidence intervals, leaving the magnitude of the effect more uncertain than the paper suggests. The primary concern is instrument weakness; that is, that the ripples that the litigation sent into the prison population time series were overshadowed by other influences. As is better appreciated today, weak instruments can bias quasi-experiments, quietly eroding their validity (e.g., Murray 2006).

Technical issues, and my adjustments for them, run as follows (and see results in Table 9):

- *Heteroskedasticity*. While Levitt's impact estimation approach is asymptotically robust to it, heteroskedasticity can still reduce precision, which is a particular concern if regressions are rendered delicate by instrument weakness. A primary form of heteroskedasticity appears to be greater crime rate volatility in smaller states. To compensate, like Katz, Levitt, and Shustorovich (2003), I prefer to weight by state population, as reported in the second and fourth sections of Table 9. This tends to cut the apparent impact on violent crime by a third to a half while increasing it for property crime. When weighting by population, the two data sets converge in the results they generate, suggesting that this change indeed adds stability.
- *Serial correlation.* The Levitt standard errors are theoretically robust only to heteroskedasticity, implicitly assuming that successive data points from a given state are statistically independent after conditioning on controls. But Arellano-Bond tests for serial correlation over various lag distances, done while weighting by population, suggest some correlation out to 15–20 years. I therefore cluster standard errors by state, which I believe is more the norm now.[54] As shown in column 2 of Table 9, this added conservatism actually does not systematically widen the standard errors.

---

[51] Weighting by state population, the sample averages for prisoners, property crimes, and violent crimes are 186.78, 582.96, and 4,802.45 per 100,000. $1/186.78 \times -.379 \times 582.96 =- 1.18$ and $1/186.78 \times -.261 \times 4802.25 =- 6.71$.
[52] The major differences appear in Illinois crime totals for 1975–82, with the values in the new data set derived from a current public source (ucrdatatool.gov/Search/Crime/State/StatebyState.cfm).
[53] Levitt Table II states that the maximum share of people who are black in any year and state is 92.2%, whereas in both data sets it is about 70.9%—in DC in 1972. Table IV reports 9 observations for "Year of status change"/"Further court action" even though according to Table I, only 8 states experienced further court action. Similarly, Table IV reports 17 observations for "Two or three years following status change"/"Further court action" even though those 8 states could have spent at most 2 years each in that stage, for a maximum of 16.

- *Many and weak instruments*. Levitt (Table V, cols 1 & 2) shows that most of the 11 litigation stage indicators used as instruments have little predictive power for prison population growth or shrinkage. Yet they are retained in the paper's instrumented regressions. Perhaps this is the reason that once serial correlation is accounted for by clustering, the regressions perform poorly on the Kleibergen-Paap test for underidentification; p values on this test are ideally close to 0, but most in Table 9, col. 2, are around 0.4–0.5.[55] Taken at face value, these test results say that we cannot be sure that the instruments have *any* strength. In search of more reliable inference, I apply three strategies:
    - *LIML*. In the face of many potentially weak instruments, many sources (e.g., Angrist and Pishke 2009, p. 213) advise checking 2SLS results with Limited-Information Maximum Likelihood.[56] In this case, the results are compatible with Levitt's 2SLS ones, but less certain because of the larger standard errors (Table 9, col. 3).
    - *IJIVE*. I applied the Improved Jackknife Instrumental Variables Estimator (IJIVE) procedure of Ackerberg and Devereux (2009), which they argue is more reliable than LIML under heteroskedasticity. Since errors are correlated within states, I jackknife by state rather than observation.[57] Overall, the results cohere with LIML and 2SLS for violent crime but swing more extreme (yet insignificantly) for property crime. Yet, except for property crime in the new data set, they appear more reliable, because they perform well on the underidentification test. Perhaps this improvement owes to the reduction in identification noise from pruning extremely weak instruments.
    - *Graphical Anderson-Rubin*. The Anderson-Rubin test checks up to two hypotheses at once: whether the impact of treatment equals some given value, such as zero; and whether the instruments are valid, relating to the outcome only via the treatment. And the test works in such a way that if instruments are weak, it will not produce unrealistically narrow confidence intervals. A modern, computationally intensive approach to inference with weak instruments is to run this test hundreds of times, each for a different hypothesized value of the impact rate. One then graphs the results in order to synopsize which values look most compatible with the data.[58]

      Figure 10 demonstrates the approach with a simulated data set. The data set has a million observations, a treatment variable that has an impact of 1.0 on the outcome, and an instrument for the treatment that is both valid and strong.[59] The p value peaks at an impact rate of 0.99926, which is very close to the true value. As one moves left or right from there, the p value falls. It crosses below 0.05, which is marked by the dotted line, once one reaches 0.9978 or 1.0016. Those two numbers therefore bound the 95% confidence interval, the range of potential values for the true impact rate that the data do *not* allow us to rule out with 95% confidence. The true value of 1.0 fits inside this confidence interval, as it should.

      To bring this method to the Levitt study while minimizing my own discretion, I applied it using Levitt's 11 instruments one at a time, and then using all at once. See Figure 11, which pertains to violent crime, and Figure 12, for property crime; both figures follow my preferences

---

54 Katz, Levitt, and Shustorovich (2003) clusters by state-decade. And in turning the Levitt paper into a textbook example, Wooldridge (2010, p. 364) clusters by state.

55 The pattern is similar if Newey-West standard errors with a bandwidth of just 5 years are used.

56 I also tried the Continuous Updating Estimator (CUE; Hansen, Heaton, and Yaron 1996), which can be viewed as a generalization of LIML that drops the estimating assumption of homoskedasticity (Baum, Schaffer, and Stillman 2007, pp. 477–78). However, results appeared less stable, sometimes being far more negative that 2SLS, LIML, and IJIVE, sometimes much closer to zero. Perhaps this instability properly indicates uncertainty wrought by weak instruments.

57 JIVE is designed to assure that even in finite samples, each observation's value for a constructed instrument is independent of that observation's realizations of the endogenous variables. If errors are correlated within groups, that design principle requires jackknifing by group.

58 I thank Mark Schaffer for suggesting this approach and providing guidance. The distributions of the Anderson-Rubin statistics are bootstrapped using the Wild Restricted Efficient bootstrap of Davidson and MacKinnon (2010), as implemented in my "boottest" package for Stata, with 1,000 replications per data point.

59 The data-generating process is $y = x + e_2$, $x = z + e_1 + e_2$, and $z, e_1, e_2$ are standard normal.

in running on the new data set, weighting by state population, and clustering errors by state. Overall, the individual instruments produce a mixed bag. Some reject negative impacts more confidently, and some reject positive impacts more confidently. In many cases, the instrument looks weak, meaning that the prison population appears not to have been affected by the type of litigation event in focus. For example, plot 4 in both figures never comes close to 0.05 within the graphed range. Evidently, a judge issuing a preliminary decision did affect prison growth the next year, so an analysis using only this is instrument is unable to rule out with much certainty *any* possible value for the impact of incarceration on crime. This failure to opine with confidence is not a vote against Levitt's thesis, only an abstention.

Finally, plots 12 in both figures run all instruments at once, as in Levitt's regressions. For violent crime (Figure 11), the test strongly rejects positive impact, and is skeptical of most negative impact levels too, which might indicate instrument invalidity. For property crime (Figure 12), the test again strongly rejects positive impacts, but now views values near –0.66 as highly plausible.

In conclusion, Levitt's suggestion that increased incarceration reduced crime in the short run is plausible and is generally corroborated by the regressions and tests run here. However, in light of plot 12 of Figure 11, the conclusion looks more credible for property crime than violent crime. Interestingly, Lofstrom and Raphael, reviewed below, find an impact on property crime, but not violent crime, after California's "realignment" reform in 2011, which itself was precipitated by prison overcrowding litigation.

**Table 9. Impact of prisoners per capita on crime per capita, following Levitt (1996)**

| | 2SLS | 2SLS-cluster | LIML-cluster | IJIVE-cluster |
|---|---|---|---|---|
| **Levitt data, unweighted** | | | | |
| Impact on violent crime | –0.456 | –0.456 | –0.552 | –0.610 |
| | (0.177)** | (0.232)* | (0.302)* | (0.348)* |
| Kleibergen-Paap underid. p | 0.01 | 0.47 | 0.47 | 0.01 |
| Kleibergen-Paap F | 3.32 | 5.27 | 5.27 | 14.18 |
| Hansen overid. p | 0.27 | 0.29 | 0.31 | |
| Impact on property crime | –0.250 | –0.250 | –0.282 | –0.931 |
| | (0.160) | (0.133)* | (0.174) | (0.583) |
| Kleibergen-Paap underid. p | 0.04 | 0.45 | 0.45 | 0.08 |
| Kleibergen-Paap F | 2.59 | 16.14 | 16.14 | 3.85 |
| Hansen overid. p | 0.56 | 0.57 | 0.57 | |
| **Levitt data, weighted by state population** | | | | |
| Impact on violent crime | –0.243 | –0.243 | –0.360 | –0.272 |
| | (0.110)** | (0.098)** | (0.187)* | (0.150)* |
| Kleibergen-Paap underid. p | 0.01 | 0.47 | 0.47 | 0.01 |
| Kleibergen-Paap F | 3.32 | 5.27 | 5.27 | 14.18 |
| Hansen overid. p | 0.12 | 0.45 | 0.43 | |
| Impact on property crime | –0.369 | –0.369 | –0.534 | –0.561 |
| | (0.112)*** | (0.074)*** | (0.169)*** | (0.264)** |
| Kleibergen-Paap underid. p | 0.04 | 0.45 | 0.45 | 0.08 |
| Kleibergen-Paap F | 2.59 | 16.14 | 16.14 | 3.85 |
| Hansen overid. p | 0.19 | 0.57 | 0.55 | |
| **New data, unweighted** | | | | |
| Impact on violent crime | –0.284 | –0.284 | –0.328 | –0.420 |
| | (0.105)*** | (0.113)** | (0.154)** | (0.210)* |
| Kleibergen-Paap underid. p | 0.01 | 0.39 | 0.39 | 0.04 |
| Kleibergen-Paap F | 3.48 | 15.30 | 15.30 | 3.56 |
| Hansen overid. p | 0.15 | 0.37 | 0.35 | |
| Impact on property crime | –0.231 | –0.231 | –0.333 | –1.092 |
| | (0.169) | (0.186) | (0.399) | (1.111) |
| Kleibergen-Paap underid. p | 0.05 | 0.58 | 0.58 | 0.39 |
| Kleibergen-Paap F | 3.02 | 21.48 | 21.48 | 0.70 |
| Hansen overid. p | 0.54 | 0.58 | 0.57 | |
| **New data, weighted by state population** | | | | |
| Impact on violent crime | –0.155 | –0.155 | –0.200 | –0.214 |
| | (0.080)* | (0.069)** | (0.112)* | (0.130) |
| Kleibergen-Paap underid. p | 0.01 | 0.39 | 0.39 | 0.04 |
| Kleibergen-Paap F | 3.48 | 15.30 | 15.30 | 3.56 |
| Hansen overid. p | 0.12 | 0.39 | 0.48 | |
| Impact on property crime | –0.357 | –0.357 | –0.665 | –0.995 |
| | (0.106)*** | (0.088)*** | (0.383)* | (1.121) |
| Kleibergen-Paap underid. p | 0.05 | 0.58 | 0.58 | 0.39 |
| Kleibergen-Paap F | 3.02 | 21.48 | 21.48 | 0.70 |
| Hansen overid. p | 0.25 | 0.79 | 0.82 | |

$N$ = 1,063 in Levitt data, 1,029 in new. Dependent variable is change in log per-capita violent or property crime since previous year. Independent variable reported is change in log per-capita custodial prison population since previous year. All regressions include state and year dummies and other economic and demographic controls. Levitt data provided by Steven Levitt in January 2016. New data set constructed from primary sources, except police counts from Thomas Marvell. IJIVE regressions jackknifed by state. Standard errors in parentheses, heteroskedasticity-robust in first column, clustered by state in rest. *Significant at p<0.1. **Significant at p<0.05. ***Significant at p<0.01.

**Figure 10. Anderson-Rubin p values for various hypothesized impact rates, artificial data set (true value = 1.0)**

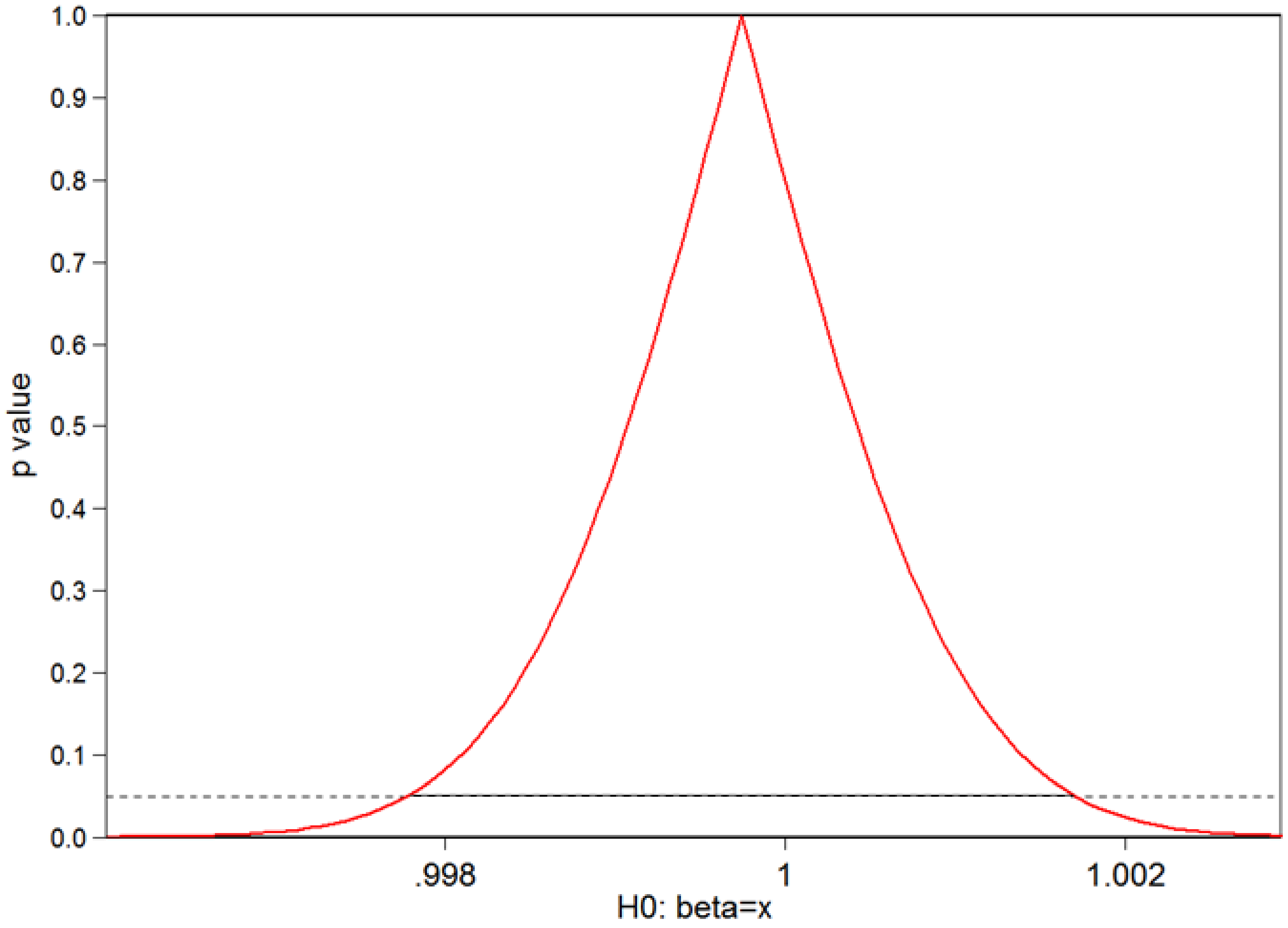

**Figure 11. Bootstrapped p values for various hypothesized impact rates of prison growth on violent crime, by choice of instrument, new data set**

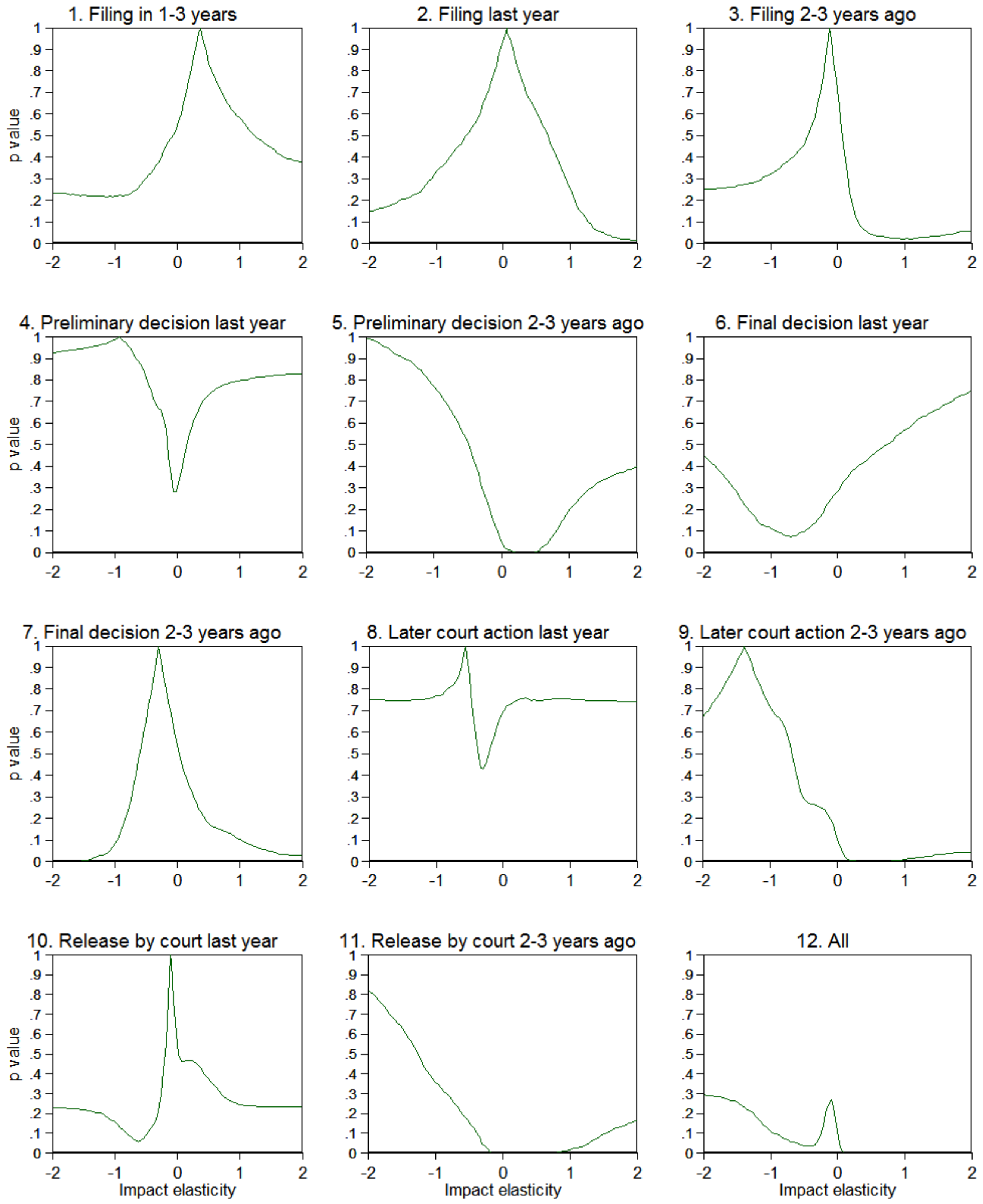

**Figure 12. Bootstrapped p values for various hypothesized impact rates of prison growth on property crime, by choice of instrument, new data set**

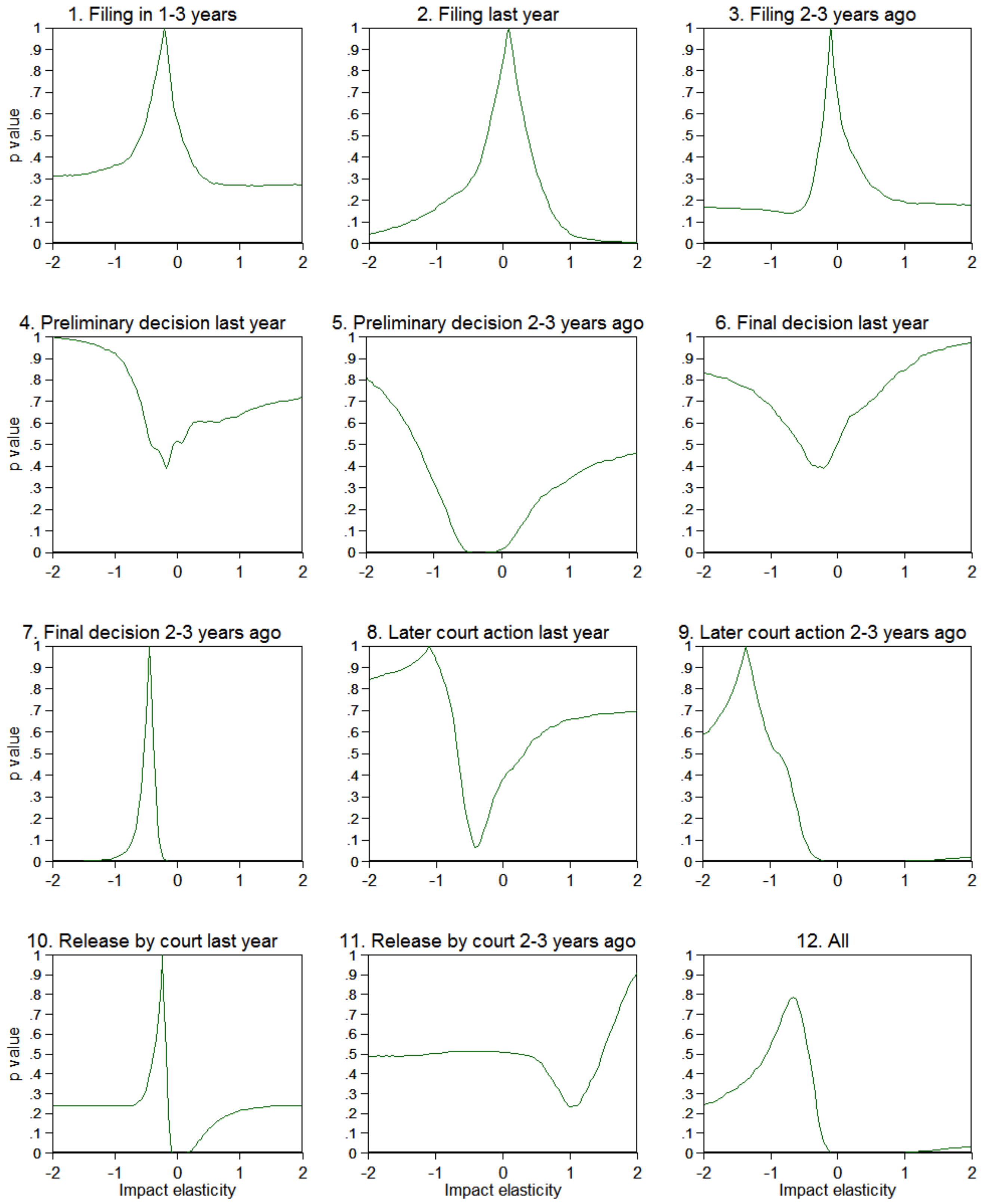

## 8.2. Owens (2009), "More time, less crime? Estimating the incapacitative effect of sentence enhancements," *Journal of Law and Economics*

The main challenge to estimating incapacitation—how much crime prisoners would commit were they free—is that they are not free, and they may differ in criminal propensity from whatever comparison group of unincarcerated people a researcher may construct. Perhaps those behind bars are more apt to offend; or perhaps less so, if those outside prison walls are more experienced and effective criminals, more often breaking the law and avoiding the arm of the law too. One effective research strategy would to be randomly release some prisoners early and track how much crime they commit, or at least how often they are arrested or convicted (see Berecochea and Jaman, below). But to avoid the framing bias hypothesized by Bushway and Owens (§2.4.4), the earlier-than-usual release should not come as a late surprise, and rather arise from normal criminal justice processing. Owens (2009) finds a case that fits meets these criteria.

Like many states, Maryland has a public body that is responsible for reviewing judges' sentencing practices and formulating guidelines. With effect July 1, 2001, the Maryland State Commission on Criminal Sentencing Policy lowered the age after which it recommended ignoring a defendant's juvenile record when setting sentences, from 26 to 23 (Weber 1996, p. 12; MSCCSP 2001, Table 5-2). Though nonbinding, the revision cut prison time for those affected, meaning people aged 23–25 with juvenile records. Before, in this age group, having been a juvenile offender approximately doubled time served for non-serious offenses. After, the differential disappeared. (See Figure 13, which is copied from Owens, Figure 1.)

In a preliminary regression controlling for factors such as age and seriousness of most recent crime, Owens estimates that time served for the affected group fell an average 222 days (Owens Table 3, col. 2).

To estimate the reverse-incapacitation from this drop—the additional crime committed by the "lucky" former juvenile offenders released earlier than they would once have been—Owens focuses on the 73 such people in the data who were sentenced during 2002–04. Crucially, the paper estimates the dates when each *would* have been released using a model calibrated to data for unlucky offenders sentenced before the policy change. It then tallies the number of arrests in these individualized counterfactual windows. Owens calculates that the lucky offenders were arrested at an annualized rate of 2.79 when they might otherwise have been in prison (se = 0.72), of which 1.65 were for drug offenses (se = 0.52; Owens Table 5, col. 1).

However, to count arrests is to undercount crimes, since offense does not always lead to arrest. So Owens scales up the numbers in two steps, albeit only for non-drug crimes because of data limitations. She first multiplies by the ratio of crimes reported to the police to crimes "cleared" by arrest, as reported by local law enforcement agencies to the FBI. That gives an impact on reported crimes. She then adjusts for unreported crimes by multiplying by the ratio of *crime victimizations*, as inferred from the National Crime Victimization Survey, to crimes reported to the FBI (pp. 565–66). In sum, the 1.14 non-drug arrests/year indicate commission of 1.44 reported index crimes/year (se = 0.66, Owens Table 5, col. 1) and 2.9 committed index crimes/year (se ≈ 1.33; p. 566).[60]

Overall, I think it is reasonable to focus and rely on the estimate that incapacitation of 23–25-year-olds in Maryland prevented 2.9 violent or property crimes/person/year, and leave aside the impact on drug arrests. For lack of data, Owens does not translate drug arrests into drug crimes. And to the extent that the arrests are for selling drugs, they may be subject to a substantial replacement effect (see section 2.4.1).

As Owens (p. 567) notes, this incapacitation estimate is smaller than most others. Recall that Levitt's study implied 1.2 *reported* violent crimes and 6.7 reported property crimes per prisoner-year, against Owens's 1.44. Perhaps the criminality of the average releasee fell between Levitt's study period, 1973–93, and Owens's study period, circa 2001—because crime was dropping generally, or because the massive prison growth of the 1980s and 1990s interned people of less and less criminal propensity at the margin.

[60] Owens (Table 5, col. 1) reports 2.79 arrests/subject, including 1.65 drug arrests, leaving 1.14 non-drug arrests. Scaling the standard error for the estimated 1.44 reported crimes, 0.66, to the 2.99 committed crimes yields ~1.33 (and does not factor in the uncertainty in victimization rates).

**Figure 13. Average time served for non-serious offenses, 23–25-year-old males, by juvenile delinquency status, 1999–2004, from Owens (2009)**

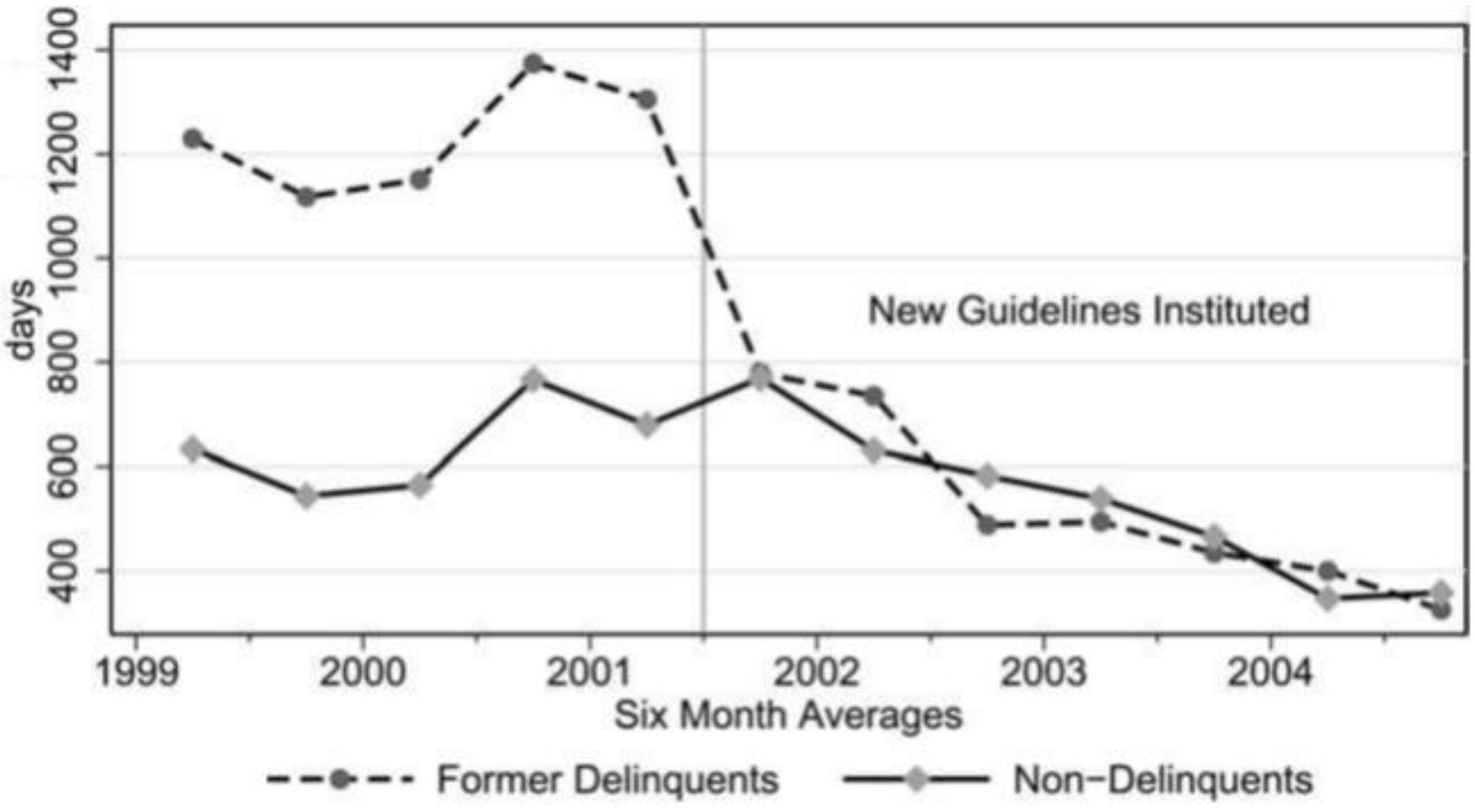

## 8.3. Buonanno and Raphael (2013), "Incarceration and incapacitation: Evidence from the 2006 Italian collective pardon," *American Economic Review*

This paper exploits the same natural experiment in Italy as Drago, Galbiati, and Vertova, but in a different way. Where the earlier paper compares individual releasees on recidivism as a function of remaining sentence, this one studies whether total crime reported to police jumped after the mass amnesty. And it adds an interesting twist, by contending that the clemency created two natural experiments: the sudden release on August 1, 2006, and then a historically rapid *rise* in the incarcerated population as the prisons refilled over the next 2–3 years. (See Figure 14, which is made from the study's publicly posted data.)

Figure 15 shows Italian crime trends over the years surrounding the mass release. It tells a clear story. Theft and robbery—robbery being theft off a person—tracked inversely with the prison population, jumping after the release, then descending steadily over the following years. Because theft accounted for roughly half of reported crimes, total crime exhibited the same pattern, as shown in the bottom right of the figure. The other crime categories—those involving violence, sex, or drugs—do not exhibit breaks and reversions nearly so sharp.

Buonanno and Raphael's (Table 4, col. 4) most conservative regressions estimate that the mass release quickly increased reported crime by 57.0 per 100,000 residents per month (se = 11.6), 41.5 of which were theft or receiving stolen property (se = 6.9). Put otherwise, upon release, each ex-prisoner committed an average of 1.5 crimes per month reported to the police, or 18.0 per year (se = 4.0; Buonanno and Raphael, Table 4, col. 1). This surpasses by an order of magnitude what Owens found in Maryland.

Shifting to the post-release prison population run-up as the basis for impact analysis, Buonanno and Raphael find incapacitation averaging 46.8 reported crimes/person-year at the six-month mark (se = 16.2; Table 5, row 1). This exceeds by a factor of roughly 2.5 the estimated impact rate of the initial release (18.0, as just mentioned). That difference shows up in Figure 15 too: while the prison population returns to its original level, theft and robbery drop *below* their pre-release levels. Taken at face value, this implies that the people initially released from prison had less propensity to commit crime than the people who replaced them, so crime fell more, proportionally, when the prisons were refilled than it rose when they were partially emptied.

The impact estimates derived from the post-release run-up should, at least initially, be greeted with more skepticism. As one correlates crime and prison population over longer time spans, the usual concerns about alternative explanations (endogeneity) creep in. Perhaps unrelated forces sent crime in Italy downward in 2006–08.

However, I tend to believe the results in this case. The crime and prison population and contours (Figure 14 and bottom right of Figure 15) match well once the first is flipped. The apparently larger effect of putting new people *in* prison—especially of the first ones to be arrested and imprisoned—makes sense if the newly incarcerated are more criminal than those released. Buonanno and Raphael (pp. 2452–53) offer several reasons this may be so. At 39, the releasees are older on average than the newly incarcerated. The two groups do overlap, with some of those receiving clemency soon ending up back in prison; but probably these are the releasees with above-average criminal propensity.

Conceivably, the new prisoners had more propensity to offend than the old simply because the old carried with them the crime-depressing effects of the release itself, as studied in Drago, Galbiati, and Vertova (reviewed above). But probably that does not suffice to explain the net drop in crime. Buonanno and Raphael (pp. 2452–53) extrapolate from Drago, Galbiati, and Vertova's numbers that the releasees' criminality was depressed by 16%. That falls far short of the factor-of-2.5 difference found here, which equates to a 60% crime reduction for mass-releasees relative to their replacements.

Buonanno and Raphael close by breaking the data down by province in order to estimate incapacitation another way, testing whether provinces with more releasees returning home saw bigger crime jumps. I do not put the same stock in this analysis because the regional variation in prison returnees could easily be correlated with third factors that could also drive crime. It does not look like a clean experiment. Nevertheless, the results from this second approach match those from the first, which somewhat strengthens my confidence in the overall findings.

The major caveat for an American reader is that Italy incarcerates much less of its population than the US does: one per thousand just before the release, compared to seven per thousand in the US.[61] America's returns to incapacitation may have diminished in recent decades if people with less and less criminal propensity were placed behind bars.

---

[61] Italy had 58.7 million people in 2006 and 60,710 prisoners just before the mass release (Buonanno and Raphael public data). The US had 322.7 million people at the end of 2015, 2.2 million of them in prison (BJS 2016a, Table 1).

**Figure 14. Prisoners in Italy, circa August 2006 mass release, from Buonanno and Raphael (2013) data**

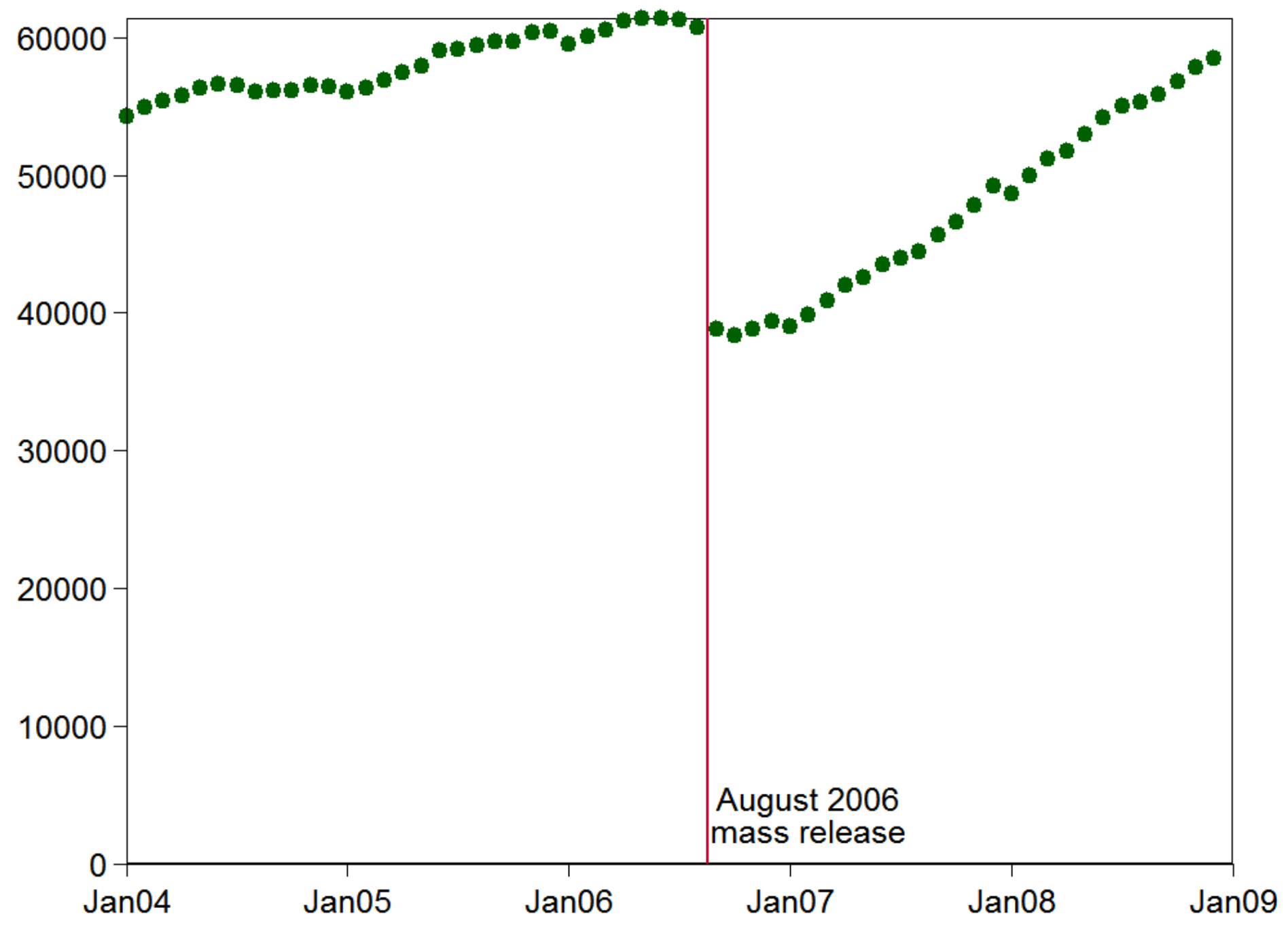

**Figure 15. Reported crimes by type and month, Italy, circa August 2006 mass release, Buonanno and Raphael (2013) data, seasonally adjusted, with local first-order polynomial fits and 95% confidence intervals**

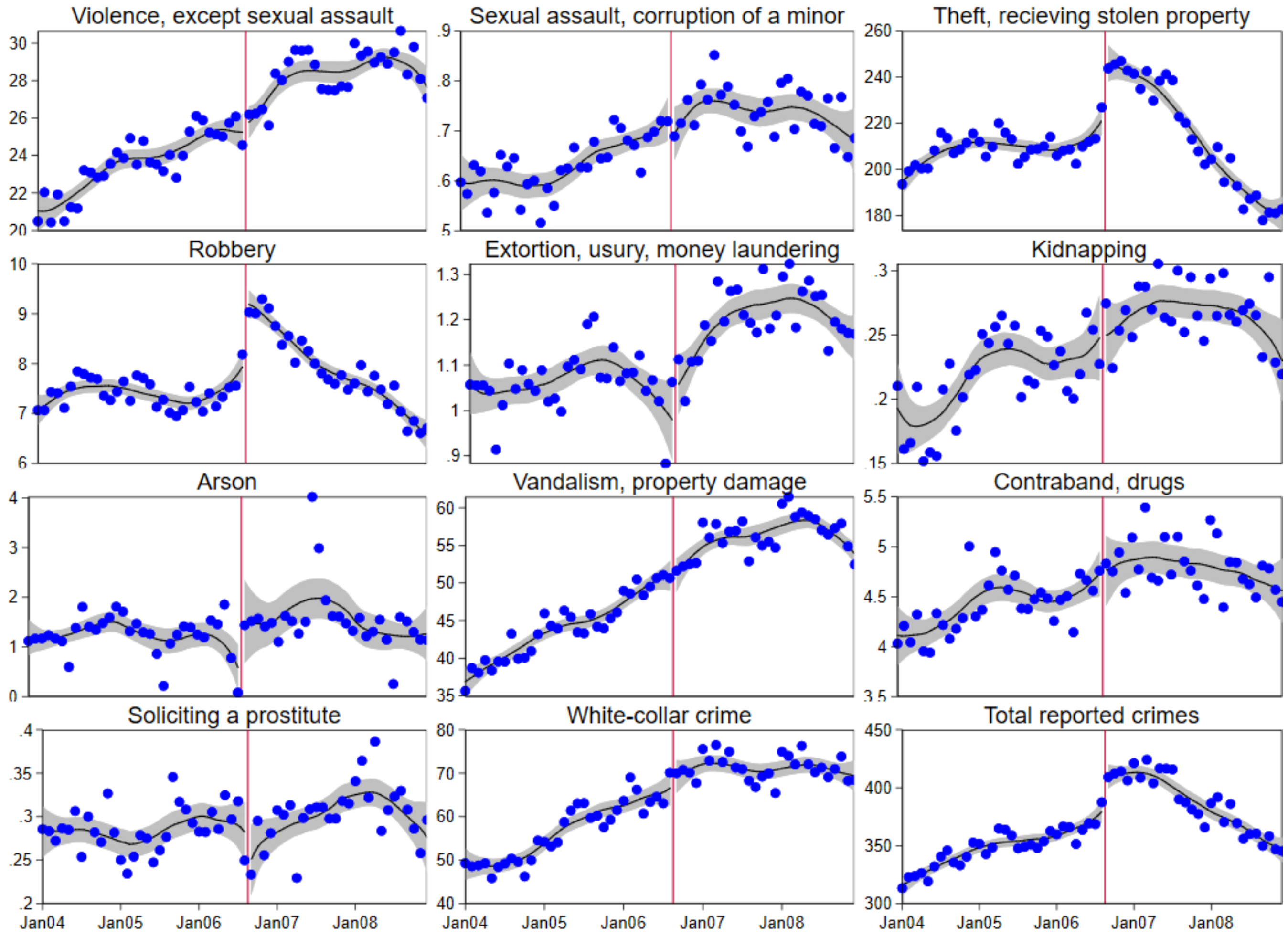

## 8.4. Vollaard (2013), "Preventing crime through selective incapacitation," *Economic Journal*

In April 2001, the Netherlands adopted a policy allowing judges to sentence the most prolific criminals to much more time than before. Somewhat like Helland and Tabarrok, Vollaard assesses the crime impact of this sentence enhancement for repeat offenders. Yet the study is best seen as one of incapacitation, not deterrence. Since the typical sentence lengthening was from 2 months to 2 years, rather than from several years to many more as in California, incapacitation quickly enters the results. And—in contrast with the sweeping application envisioned in California's Three Strikes law—the Dutch judges selectively targeted truly prolific, drug-addicted offenders who were no strangers to jail: evidently they were not easily deterred.

The statistics of this select group of prolific offenders are remarkable. According to one survey, people receiving enhanced sentences under the new law admitted committing an average of 256 crimes per year, in order to earn 50–100 euros/day to support a drug habit. They had been convicted 31 times before on average, and a few had passed 300 convictions. They spent an average of four months in jail each year. Eighty percent reported theft as their main source of income in the last month. "By 2001, many of these highly prolific offenders were aged 40 or over: they had fallen victim to the heroin-epidemic that swept Europe back in the 1980s" (Vollaard, p. 266, citing Koeter and Bakker 2007).

Dutch authorities implemented the new regime in 2001 on a pilot basis, in 10 cities with many prolific offenders. They deferred the national rollout until after 2004. Vollaard collected monthly crime data for the 10 cities in the first wave, plus an additional 21 second-wave cities. His core results are captured in two pairs of graphs. The first pair plots 1) domestic burglaries and car thefts and 2) assault and sexual crimes, breaking out both crime groups by the implementation wave of the city in which the crimes were perpetrated. See Figure 16 (copied from Vollaard, Figure 2). The top half of Figure 16 shows that before the sentence enhancement law went into effect in the 10 first-wave cities in 2001, they had about 50% more crime than the 21 second-wave cities. After, it appears, crime fell in both groups, but more so in the first-wave cities, even in percentage terms. The data hint that the gap stopped narrowing after the law went into effect in the second-wave cities. As for violent crimes (bottom half of the figure), no differences appear anywhere along the time span.

Vollaard then processes the data to remove fixed city and time effects, and centers each city's times series around its date of sentence enhancement adoption. Taking averages produces the second pair of graphs (with 95% confidence intervals rendered as vertical lines; see Figure 17, which is copied from Vollaard, Figure 3). We see that, so processed, the acquisitive crime trend held level in the months leading up to adoption of the law and declined afterward, while the violent crime trend rose slowly both before and after, making no suggestive break with the past. This more-rigorous pair of graphs confirms the impression created by the first pair: incarcerating prolific offenders a couple of years instead of a couple of months substantially cut the overall theft rate.

I see two potential explanations for these graphs. First is incapacitation: confining prolific thieves may indeed have cut property crime, but not violent crime for which the known thieves were less responsible. Second is regression to the mean: the 10 first-wave cities may have earned their place in the first wave by virtue of random, transient crime waves circa 2000. That would have set them up for declines anyway, on average, in the early 2000s.[62]

However, I doubt regression to the mean can fully explain the results. For it to hold, the top halves of Figure 16 and Figure 17 should be roughly symmetric, with the property crime gap between first- and second-wave cities widening before 2001 and narrowing after. While Figure 16 allows for a modest long-term widening before 2001, the gap narrows much more afterward.

[62] I thank Donald Green for pointing out this possibility.

Turning from graphs to regressions, Vollaard (Table 3, col. 5) finds that 3.66 acquisitive crimes per month were prevented for each person imprisoned under the law (se = 1.18). Violent crime was not affected. Overall, the law appears to have reduced crime 25% by 2007 (p. 274).

Vollaard checks whether the strategy of imprisoning prolific offenders hit diminishing returns. Using police estimates of the number of offenders in each city meeting the law's definitions, he forms a variable that is the fraction of such people in prison at a given time. The product of this variable and the number of such prisoners should, when added to regressions, receive a positive coefficient if there are diminishing returns. That would work against the negative coefficient on the prisoner count alone (which in itself indicates that more prisoners lead to less crime). This is in fact what happens (Vollaard, Table 3, col. 9). Vollaard calculates that moving from the 25th percentile to 75th percentile among cities in the rate of incarceration of prolific criminals—from 9% to 20%—cuts the crime reduction from any further incarceration by 25% (p. 279).

Interestingly, Vollaard cites another study that concludes that "some 60% of the convicted offenders state that they feel substantially better after the enhanced prison sentence. Only a small group considers the sentence as unfair" (p. 282, citing Koeter and Bakker, 2007). As in the Hawaii HOPE program, the people sent longer to Dutch jails probably recognized that they have self-control problems. And they may have benefited—or hoped to benefit—from the mandatory drug treatment during the long spells behind bars.

In my view, Vollaard demonstrates strong incapacitation. Selective incarceration as practiced in the Netherlands may well work well in the US too. As Vollaard suggested to me in e-mail, sentences for first-time offenders could be shortened, but second-, third-, etc., offenders could serve gradually lengthening terms. Past failure to participate in required drug treatment or other rehabilitation programs could count toward prior offense totals.

But as with the Buonanno and Raphael study of Italy, this one of the Netherlands should not be naively extrapolated to the margin of mass incarceration in the US. The US incarceration rate is far higher than the Dutch one, making generalization vulnerable to the caveat of diminishing returns demonstrated here.[63] The US, with its vastly expanded prison system, appears to have incarcerated much less selectively in recent decades. Presumably the majority of US prisoners would offend at lower rates than the small group targeted in the Netherlands.

[63] Vollaard states that the program incarcerated 1400 people, representing 5% of all prisoners, which implies a Dutch national incarceration rate of about 0.2%.

**Figure 16. Monthly rates of domestic burglaries and car thefts and of assault and sexual crimes, selected cities, Netherlands, 1998–2008, from Vollaard (2013)**

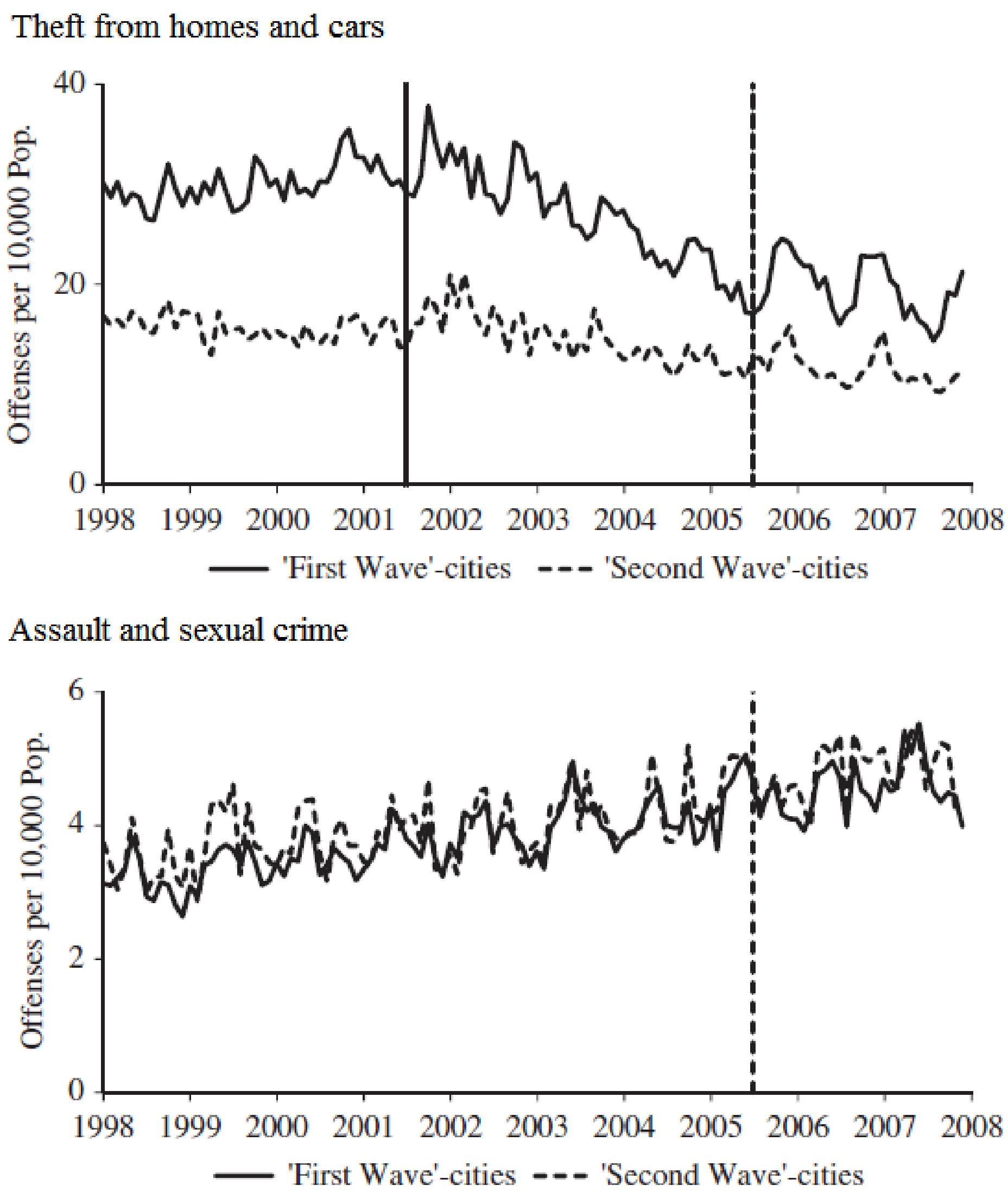

**Figure 17. Monthly rates of domestic burglaries and car thefts and of assault and sexual crimes, controlling for month and city effects, selected cities, Netherlands, relative to time of each city's sentence enhancement implementation, from Vollaard (2013)**

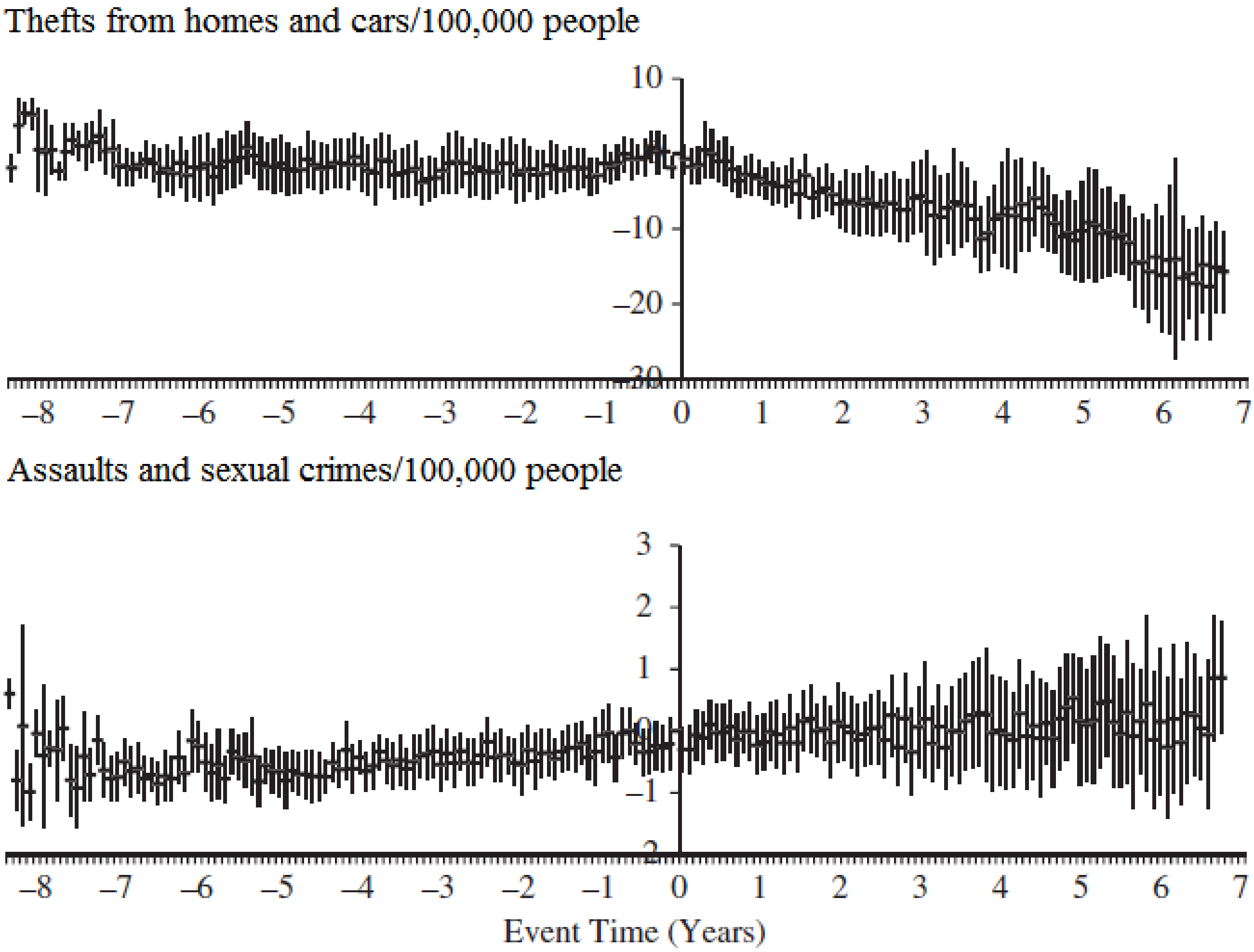

## 8.5. Lofstrom and Raphael (2016), "Incarceration and crime: Evidence from California's public safety realignment reform," *Annals of the American Academy of Political and Social Science*

California's criminal justice policy pendulum has reversed. Not only did voters approve Proposition 36 in 2012 to scale back Three Strikes. The year before, successful prison overcrowding lawsuits drove the legislature to enact "realignment" reforms in order to shrink the prison population. The new law grants judges more discretion to apply sanctions other than incarceration, and moves inmates convicted of non-serious, non-sexual, nonviolent offenses from prisons to jails, where they can more quickly earn "good-time credits" toward early release. (Lofstrom and Raphael, p. 200.) Minimum parole time and maximum reincarceration time for technical violations of supervision were cut from 12 to six months (Sundt, Salisbury, and Harmon 2016, p. 318).

California's prison population began falling as soon as the law went into effect, from 160,295 on October 1, 2011, to 138,903 six months later. However, because the law diverted some parole violators to jail instead of prison (Lofstrom, Raphael, and Grattet 2014, p. 6), the jail population rose from 71,848 to 74,049. (See Figure 18, which presents data from California's Department of Corrections and Rehabilitation,

j.mp/1rvkPWi, and Board of State and Community Corrections, j.mp/1s5MTQz.)[64]

A glance at monthly FBI data suggests that crime began climbing in California in 2011, though possibly before the reforms went into effect.[65] The rise reversed in 2013–14, but property crime, if perhaps not violent crime, remained above the trend of the 2000's. (See Figure 19. The violent crime line excludes rape because the FBI broadened its definition of rape in 2013 and the monthly data do not contain consistent series under either the old or new definition.)

To more rigorously assess this surface association, Lofstrom and Raphael analyze the data in two ways. The first parallels an approach in Buonanno and Raphael. Instead of looking across provinces of Italy, it looks across counties in California, and examines whether changes in prison and jail populations correlated with crime rate changes. In reviewing the Italy paper, I passed over its cross-geography results as less compelling than its pure time series results. The general issue with the cross-geography analysis in both places is that the natural experiment in incarceration—the relatively sudden, arbitrary departure from context—operated far more cleanly in the time dimension. In principle, Los Angeles the day before made an excellent comparator for Los Angeles the data after—far more so than San Francisco the day after. Realignment did indeed affect counties differently, in that some reduced their prison populations more than others. But as Lofstrom and Raphael (Table 2) show, poorer counties with more prisoners saw the biggest drops. The worry is that factors such as poverty also perturbed the evolution of crime rates or correlated with other factors that did, spoiling the quasi-experiment.

Lofstrom and Raphael's most conservative cross-county regressions include month and county dummies, so they should eliminate spurious associations generated by third factors to the extent their effects on crime rates are fixed over time or space. These show no impact on violent crime. As for property crime, each additional person-month of incarceration was associated with 0.1–0.15 fewer reported crimes, depending on how the regressions are run. The apparent impact is lower and statistically weak in the toughest regressions, which include both the month and county effects (impact = 0.089, se = 0.087). (Lofstrom and Raphael, Table 4, final column.)

Looking at finer crime categories, Lofstrom and Raphael (Table 4) find no definite impact on burglary or robbery, which is somewhat surprising since these crimes were the ones whose rates changed most in Italy and the Netherlands. Rather, the locus of the impact appears to be vehicle theft, at the rate of 0.103 thefts per person-month of incarceration avoided (se = 0.036), or 1.2/year. (Lofstrom and Raphael, Table 4, final column.)

Lofstrom and Raphael's second analytical tack operates across states instead of counties. The authors check whether California's overall crime trends deviated after the reform from those of other states. To make the comparison, they form a "synthetic control." This is mathematical blend of other states, with the weights chosen by an algorithm that seeks a close match with California on pre-reform crime trends. For example, their synthetic control for violent crime is 34% Florida, 16% Maryland, 7% Montana, 21% New York, 19% Rhode Island, and 3% South Carolina (Lofstrom and Raphael working paper, Appendix Table A1, col. 1).

As shown in Figure 20, which is made from Lofstrom and Raphael's Figures 6 and 7, by design the synthetic controls almost perfectly match California through 2010, the last full year before the reform. After, the violent crime lines still match, but the property crime ones diverge, as California's rises relative to the control. Again the suggestion that incarcerating fewer perpetrators of non-serious, non-sexual, non-violent

[64] In a sense, the jail population expanded by design, since the law also shrank the ranks of people on probation or parole—from 398,000 at end-2010 to 380,900 at end-2011, about where the total has hovered since (BJS 2013, Table 6; 2014, 2015a, Appendix Tables 1). The shortening of minimum parole time was presumably the primary cause, given that the paroled population dropped from 111,100 to 89,300 in 2011 (BJS 2013, Table 6). As a result, Sundt, Salisbury, and Harmon (2016, p. 318) point out, we cannot view realignment as a pure decarceration policy. Possibly, the mild contraction in supervision also affected crime.

[65] Figures seasonally adjusted by partialling out calendar month dummies from the per-resident crime rates in logs. All data extracted from FBI "Offenses Known and Clearances by Arrest" files at icpsr.umich.edu/icpsrweb/ICPSR/series/57.

crimes increased property crime only.

And again, formal number crunching finds the clearest effect on property crime, at 3.8 more per prisoner-year in 2012–13 than in 2010.[66] That includes 1.2 motor vehicle thefts/year (p. 216), which itself matches the cross-county regressions (one-tailed $p<0.04$, 0.02; working paper, Table 7, col. 6; final paper, Table 5, second panel).

Having closely replicated the cross-state regressions, I have once more developed significant doubts about methodology. Yet I have also struggled to construct a compelling alternative, which has forced me to conclude that the available data contain irreducible uncertainty about the impact of realignment. My best estimates align well with those just cited.

One problem in the Lofstrom and Raphael synthetic control analysis looks solvable. The paper appears to use an inappropriate method to gauge the statistical significance of the impacts found. Following Abadie, Diamond, and Hainmueller (2010), which develops the synthetic control method, Lofstrom and Raphael (p. 214) apply the same measurement to every state in turn—as if Alabama had passed realignment, then Arkansas, and so on. For example, Figure 21, modeled on graphs in Abadie, Diamond, and Hainmueller (2010, p. 502), plots treatment–synthetic control differences for all states, in the case of burglaries. The central black line shows the difference between treatment and control for California. Before 2011, when treatment and control match almost perfectly, the plot of their difference hugs zero. After realignment begins, when the California burglary rate rises above the control's rate, the black line in this figure that shows their difference rises too. The light grey lines do the same for every other state, as if they too had enacted realignment. Amid the tangle of other states' grey lines, California's post-2011 rise does not look unusually far from zero. This is why Lofstrom and Raphael's approach puts a weak p value of 0.4 on the burglary impact (one-tailed p value 0.204; Lofstrom and Raphael 2015, Table 5, panel 2, col. 3).

This technique implicitly assumes that states are statistically homogenous (at least on unobserved traits); that is, that aside from the lack of realignment, Alabama, Arkansas, and the rest can stand in statistically for California. Yet for many states, the grey lines in Figure 21 do not look mathematically comparable. They do not hug nearly so close to zero as California's line despite the synthetic control algorithm's hunt for a perfect match for them too. And we can expect the crime series of states poorly matched to their controls before 2011 to stay poorly matched after, causing them to deviate more from the controls, if randomly. This makes other states a misleading basis on which to judge the significance of California's post-2011 deviation.

Several factors might explain other states' poorer pre-treatment matches. Smaller states' crime rates may be more volatile because they are smaller, and so less noise is averaged out of the yearly numbers. With more idiosyncratic variation, we expect them to be less well matched by their synthetic controls. Or some states may experience more or less unique time trends for less random reasons. South Dakota might find a great match in North Dakota, while Hawaii might stand alone.[67] In technical language, states are heterogeneous.

In fact, in their inaugural application of synthetic controls, Abadie, Diamond, and Hainmueller also study the impact of a policy change in California, a tobacco control program passed in 1989. Encountering the problem just described, they propose judging statistical significance by comparing California to other states not based on difference-in-differences impacts, as above, but on ratios of their post- to pre-treatment mean squared predictions errors (p. 503). This view makes California's post-realignment burglary rise, exhibited in Figure 21, look extremely improbable if by chance, because dividing its magnitude by the near-zero pre-2011

[66] Lofstrom and Raphael (working paper, Table 7, col. 6) estimate that property crime increased 227.02 per 100,000 relative to 2010. Dividing that by their denominator, an incarceration reduction of 60 per 100,000 (working paper, p. 31) gives 3.8.

[67] If other states were fully comparable, there would be no need to synthesize a special control through matching. The theoretical parts of Abadie, Diamond, and Hainmueller (2010) seem to implicitly assume heterogeneity in observables—making synthetic controls potentially superior—but homogeneity in unobservables, meaning homoskedasticity. I am aware of no theory or simulation evidence supporting the synthetic control method in the presence of heteroskedasticity.

prediction errors produces a huge value.

I closely replicate the Lofstrom and Raphael synthetic control regressions and revise them to judge significance in the way proposed by Abadie, Diamond, and Hainmueller.[68] Before running the new tests, I make one other change. To sharpen the focus on the policy break on October 1, 2011, I take advantage of monthly FBI data to shift the statistical year to begin October 1.[69] In contrast, because calendar year 2011 includes both pre- and post-treatment months, Lofstrom and Raphael drop it from both the pre- and post-treatment periods, thus removing the data closest to the break. Alone, this modification largely preserves the results: compare the bottom graph of Figure 20 to the top one of Figure 22.

But deriving p values in the alternative way alters the texture of the results. The first two rows of Table 10 compare Lofstrom and Raphael's original estimates of impacts on 2012–13 crime rates to revised estimates that shift the statistical year to October 1, incorporate 2011 data, and compute p values as Abadie, Diamond, and Hainmueller suggest. The revision largely preserves magnitudes of impacts while raising the statistical significance for burglary and lowering it for motor vehicle theft. I view these estimates as more reliable.

Yet they, like the original estimates, harbor another issue: wholes are not sums of parts. Lofstrom and Raphael estimate the impact of total property crime at 227 per 100,000 Californians, including 45 burglaries, 21 larcenies, and 72 motor vehicle thefts…which add to only 138 (first row of Table 10 below).[70] The revised results (second row) are even more out of kilter, the impact on the whole of property crime being twice the impacts on the parts. The contradiction arises because the benchmark is re-synthesized for each crime category and subcategory. For example, for property crime, Lofstrom and Raphael's benchmark is 52% Wyoming and 16% Nevada, etc. (working paper, Appendix Table A1, col. 6). Yet for larceny, the control is 1% Wyoming and 35% Nevada, etc.—rather different, even though larceny constitutes 60% of reported property crime in California. As a result, the synthetic control method appears to produce unstable results.

To produce more consistent results, I develop one synthetic control for all violent crime subcategories and another for all property crime subcategories. For each, two natural bases for matching beckon. Taking property crime as an example, one can task the computer with finding the mathematical mix of states that best matches California on total property crime for each year in 2000–11, as in the top graph of Figure 22. Or one can have it search for the best match on burglary, larceny, and motor vehicle theft all at once. The latter seems superior because it brings more information to bear.

To my surprise, doing the latter slashes the apparent impact of realignment (see bottom graph of Figure 22 and bottom row of Table 10). The apparent property crime change in California (in addition to now equaling the sum of the impacts on subcategories, as desired), collapses from +175 to –7 per 100,000 Californians per year (p = 0.03 to p = 0.57).

Why the change? As the subtitles in Figure 22 document, the weights forming the synthetic control when matching on total property crime favor an almost completely different set of states than when matching on crime subcategories. The biggest difference is in the weight put on the one state of overlap, Nevada. Where Lofstrom and Raphael's synthetic control for property crime is 16% Nevada, and the replication's is 10%, matching on burglary, larceny, and motor vehicle separately lifts Nevada's weight to 50%. And it turns out that property crime rose after realignment in Nevada too. (See Figure 23.) So when Nevada figures heavily

---

68 All synthetic control regressions run here are performed with the "synth" module for Stata by Hainmueller, Abadie, and Diamond. Although use of the programs "nested" option to obtain the "V matrix" appears to be the norm (Abadie, Diamond, and Hainmueller 2010, p. 496), here, it caused estimation to crash or run indefinitely. So it was dropped, leading to reliance on a regression-based V matrix provided automatically by the module.

69 Figure 21 uses the same convention.

70 As is common, the fourth property crime category, arson, is omitted here because it is much smaller.

in the control, crime in California rises much less by comparison.

The discovery that Nevada's crime trends mimicked California's complicates interpretation of the California trends. The 2011 crime uptick may have reflected a regional pattern; indeed, graphs not shown for Oregon and Arizona suggest increases, or at least cessation of decreases there too. Why would crime rise in the West? Perhaps a third factor was at work, such as a spreading illegal drug epidemic. In that case, California's neighbors would make excellent controls for studying the impact of realignment, which California alone implemented. On the other hand, perhaps realignment *caused* the regional inflection, as some people who would have been incarcerated in California travelled to nearby states and committed crimes there.[71] In that case, California's neighbors would constitute tainted controls, for they would be partially treated too. Benchmarking against them would cause us to underestimate California's post-realignment changes.

Either way, this dilemma teaches two lessons, one new, one old. The new lesson is that the synthetic control method can produce unstable results. The underlying problem again appears to be that states are heterogeneous, making the "best match" as evanescent as it is effervescent. The old lesson is more general, about the black box problem. Almost every econometric method, by distilling large data sets down to a few numbers, obscures as much as it reveals. Complexity can obscure more effectively.

In the face of the instability of the synthetic controls, I run regressions in a more traditional mode, with one data point for each US state and year. The hypothesized data generating process is:

$$y_{it} = \beta T + \mu_i + \nu_t + \epsilon_{it}$$
$$\mathrm{E}[\epsilon_{it} | T, \mu_i, \nu_t] = 0$$

where $y_{it}$ is the log per-capita rate of crime of a given type in state $i$ in year $t$, $T = 1$ only for California post-realignment, and $\mu_i, \nu_t$ are state and year fixed effects. As in Lofstrom and Raphael, the data start in 2000. As in the revised synthetic control regressions, the statistical year starts on October 1 and 2011 data are retained. The OLS regressions are run in differences, eliminating the $\mu_i$. Two steps are taken to combat the heterogeneity that evidently destabilizes the synthetic control results. Standard errors are clustered by state, which improves robustness to arbitrary patterns of heteroskedasticity and within-state serial correlation. And states are weighted by total crime so that smaller states with more volatile crime series do not receive undue influence.[72] Since the classical clustered standard errors for $\beta$ turn out implausibly small—evidently because of something akin to the singleton dummy problem—wild-bootstrapped 90% confidence intervals are reported instead.[73]

For presentation—and for use in the cost-benefit analysis at the end of this review—impacts on log crime rates are re-expressed with respect to unlogged crime rates, both per 100,000 Californians, and per prisoner-year of averted incarceration. They are further scaled to adjust for the under-reporting of most crimes.[74] (The statistics used to this point are only for *reported* crimes.) Nevada, it bears emphasizing, is treated as just one control among many, down-weighted because of its modest population.[75]

---

[71] Reviewer Steven Raphael points out that this would not have happened, to the extent that otherwise-incarcerated people were placed under community supervision that prevented them from leaving their counties of residence.

[72] When the dependent variable is the log of a group average and underlying populations vary by size, the optimal weight for removing heteroskedasticity and inefficiency caused by differing group sizes is $n_i p_i / (1 - p_i)$ where $n_i$ is population and $p_i$ is the proportion, here crimes per capita of some type (Maddala 1983, p. 29). In this case $1 - p_i \approx 1$, so the optimal weight is essentially $n_i p_i$, the number of crimes. But since the actual number of crimes is endogenous, Greene (2003, pp. 687–88) suggests a two-step procedure to use *predicted* rather than actual values. First the regression is run weighting by state population. Then it is rerun weighting by the number of crimes predicted by the first regression.

[73] These are derived by inverting the bootstrapped Wald test for the null $\beta = \beta_0$ and using my "boottest" module for Stata.

[74] Conversion from crimes reported to crimes committed is based on national-level estimates of reporting rates from BJS (2015, Table 6). The reporting rate for murder is assumed to be 100%. Uncertainty in the BJS estimates of reporting rates is not factored into any of the confidence intervals in Table 11.

[75] Sundt, Salisbury, and Harmon (2016) run regressions somewhat akin to these, except the samples are cross-sections for 2012, 2013, and 2014

The results appear in Table 11, and in Figure 24, which depicts the table's first row in graphical form, much as was done for Levitt in Figure 11. The new estimates broadly cohere with those from the synthetic control regressions in the middle row of Table 10, the ones that do not heavily weight Nevada. A post-realignment change is again clearest for motor vehicle theft, at a reported rate of 49 per 100,000 Californians. Increases in burglary and larceny also look likely; they are not significant at p = 0.1 but the second row of Figure 24 makes clear that a positive impact rate is much more likely than negative. Meanwhile, the regressions hardly suggest an increase in violent crime.

Even if these estimates surpass Lofstrom and Raphael's in building a stable, statistically comparable control set—which is by no means certain—they do not slay the deeper threats to causal interpretation here. The regressions suggest only that property crime rose in California after realignment more than in other states. They do not speak directly to whether the climb began before realignment, nor whether realignment's effects spilled over to California's neighbors, compromising them as controls.

On reflection, the largest barrier I see to a confident judgment on realignment's impacts on crime is the uncertainty about timing. Crime in California may or may not have started rising in mid-2011, before realignment. Possibly the rise then is a mix of statistical noise and optical illusion, owing in large part to anomalously low numbers for February 2011 (look again at Figure 19). The annual trend using October 1 years hardly decelerates in the 12 months before realignment (solid lines in Figure 22). Or possibly the seeming early rise is a true effect of realignment: reduced *deterrence*. Governor Brown had signed the realignment law in April 2011. The next month, the US Supreme Court rejected California's final appeal in the protracted legal battle that had prodded passage of the law. Both events received press attention, which may have reached the eyes and ears of people contemplating whether to commit crime.[76]

I often invoke Occam's Razor, which is the principle that the simplest explanation is mostly likely right, all else equal. The simplest explanation for the evidence before us is that realignment caused *all* the crime increases in California and its neighbors; in particular, any increase seen before October 2011 is a statistical illusion.

In this review, the realignment evidence, ambiguous as it is, plays a special role. Because of California's national significance, the recency of realignment, and the potentially pessimistic conclusion from the point of view of decarceration proponents, I will take these incapacitation estimates as a key input to the devil's advocate cost-benefit scenario in section 11.2. In that skeptical spirit, I will select the results from my alternative regressions (Table 11) which conclude fairly pessimistically, largely by deemphasizing Nevada as a comparator: realignment caused 1.5 burglaries, 4.0 larcenies, and 1.2 motor vehicle thefts per person-year of incarceration prevented.

Those estimates agree with the ones from Italy and the Netherlands in that the main impact is on acquisitive crime, not violent crime. But they are smaller, and smaller too than the impacts found in Levitt, a study set in the US a few decades earlier. Where the California reform apparently reverse-incapacitated roughly 2.9 *reported* crimes/year (summing the entries in the third row of Table 11), the numbers were 7.9/year in Levitt (1.2 violent and 6.7 property; see review of Levitt) and 1.5 and 3.66 per *month* in Italy and the Netherlands (see reviews of Buonanno and Raphael and Vollaard). Since California imprisoned more of its population, these results are consistent with diminishing incapacitation returns to incarceration. (California incarcerated 0.6% of its residents versus 0.2% in the Levitt 1973–93 national sample, 0.2% in the Netherlands, and 0.1% in Italy.[77]) That said, the focus of California's admission reform on non-serious offenders might also explain and instead of taking differences, they control for 2010 crime rates. No steps are taken to reduce heteroskedasticity, such as weighting by population; just as argued for synthetic controls, treating states as homogeneous in this way tends to bias downward the estimated significance of the impact estimate.

[76] Google News search results for January–August 2011 at j.mp/2aviTr4.

[77] On the Netherlands, see note 63. On Italy, see note 61. Levitt (Table II, upper left) reports 0.1681%. Pre-realignment California incarcerated 231,000 out of 37.7 million: see Figure 18.

the fall in incapacitation since Levitt's study period, just as it explains the lack of impact on violent crime.

**Figure 18. Thousands of inmates in prison and jail, by month, California, January 2011–January 2013**

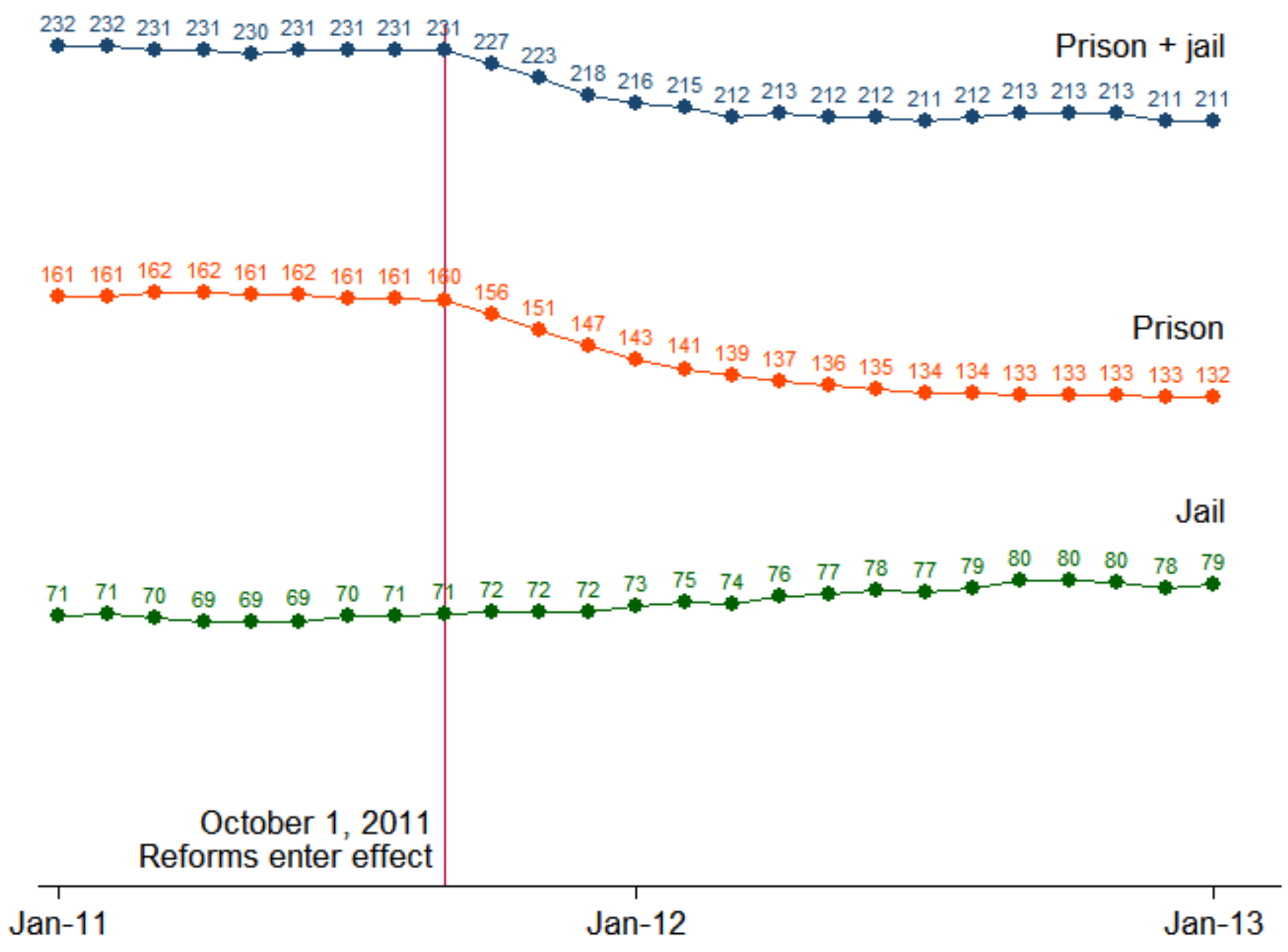

**Figure 19. Property and violent crimes (excluding rape) by month, California, seasonally adjusted, 2000–14**

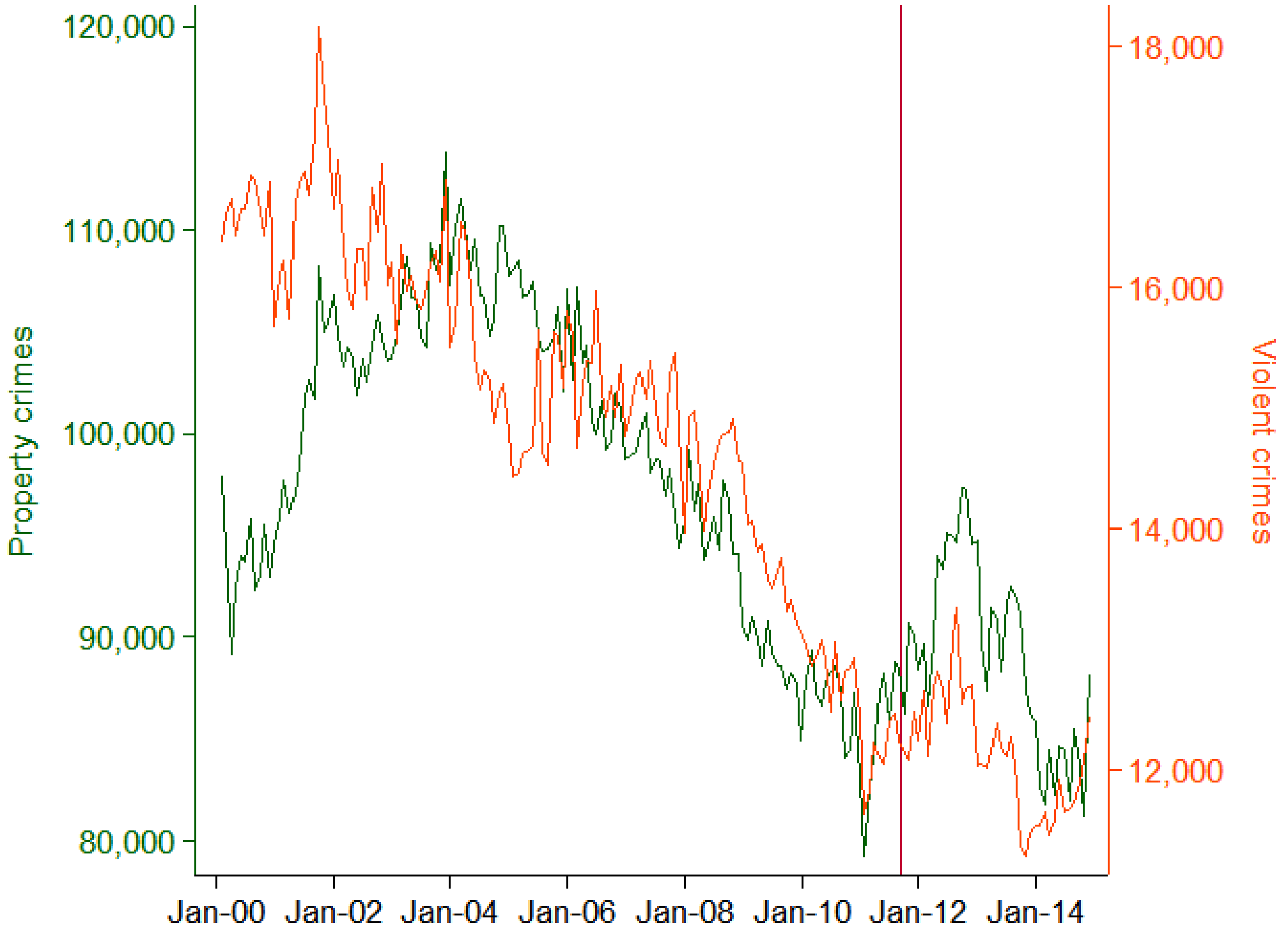

**Figure 20. Violent and property crimes, per 100,000 people in California and comparison states (weighted average), 2000–13, per 100,000 people, from Lofstrom and Raphael (2016)**

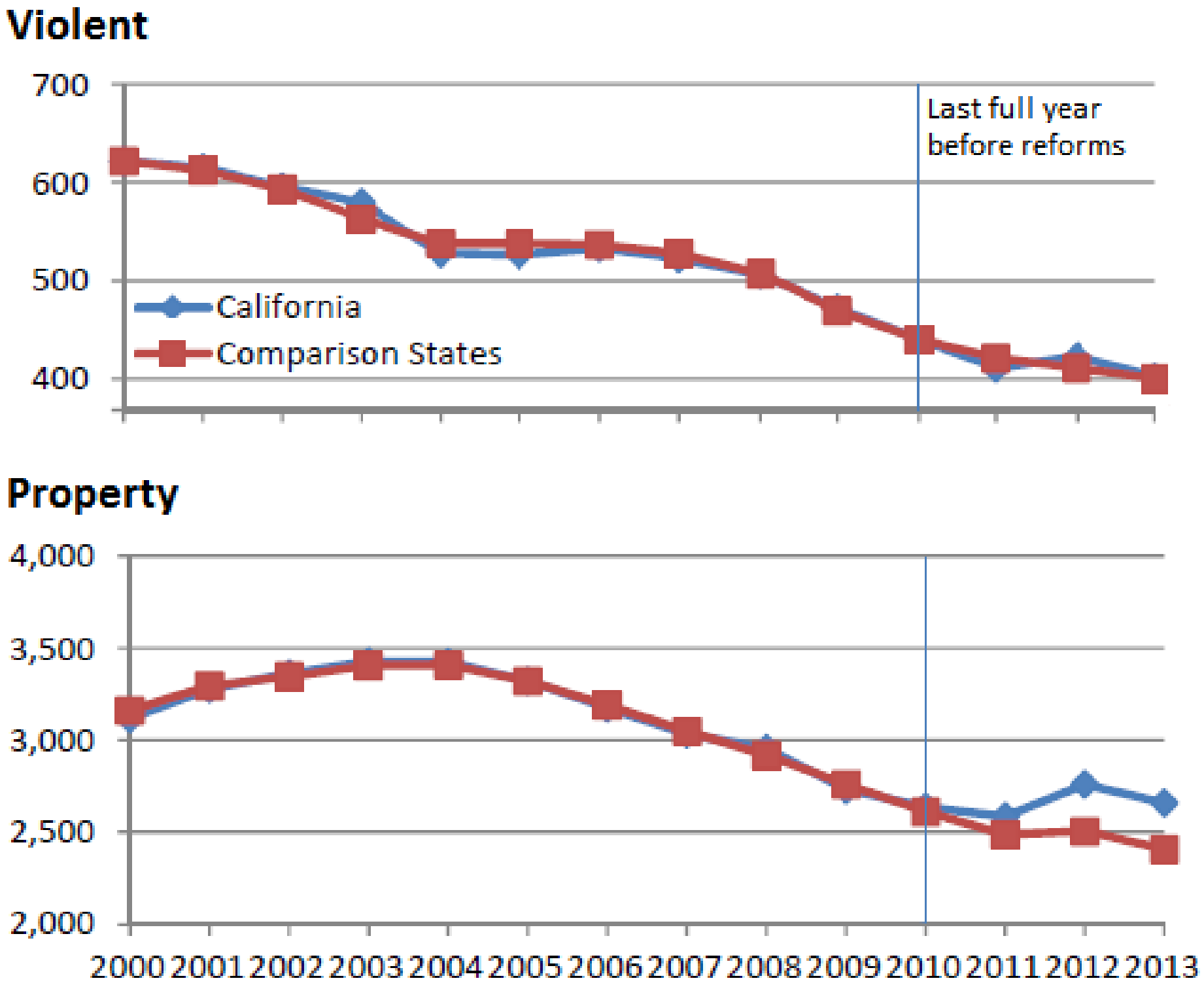

**Figure 21. Treatment-control difference in burglaries per 100,000 residents, by state, using synthetic control method with 2000–11 pre-treatment period and October 1 years**

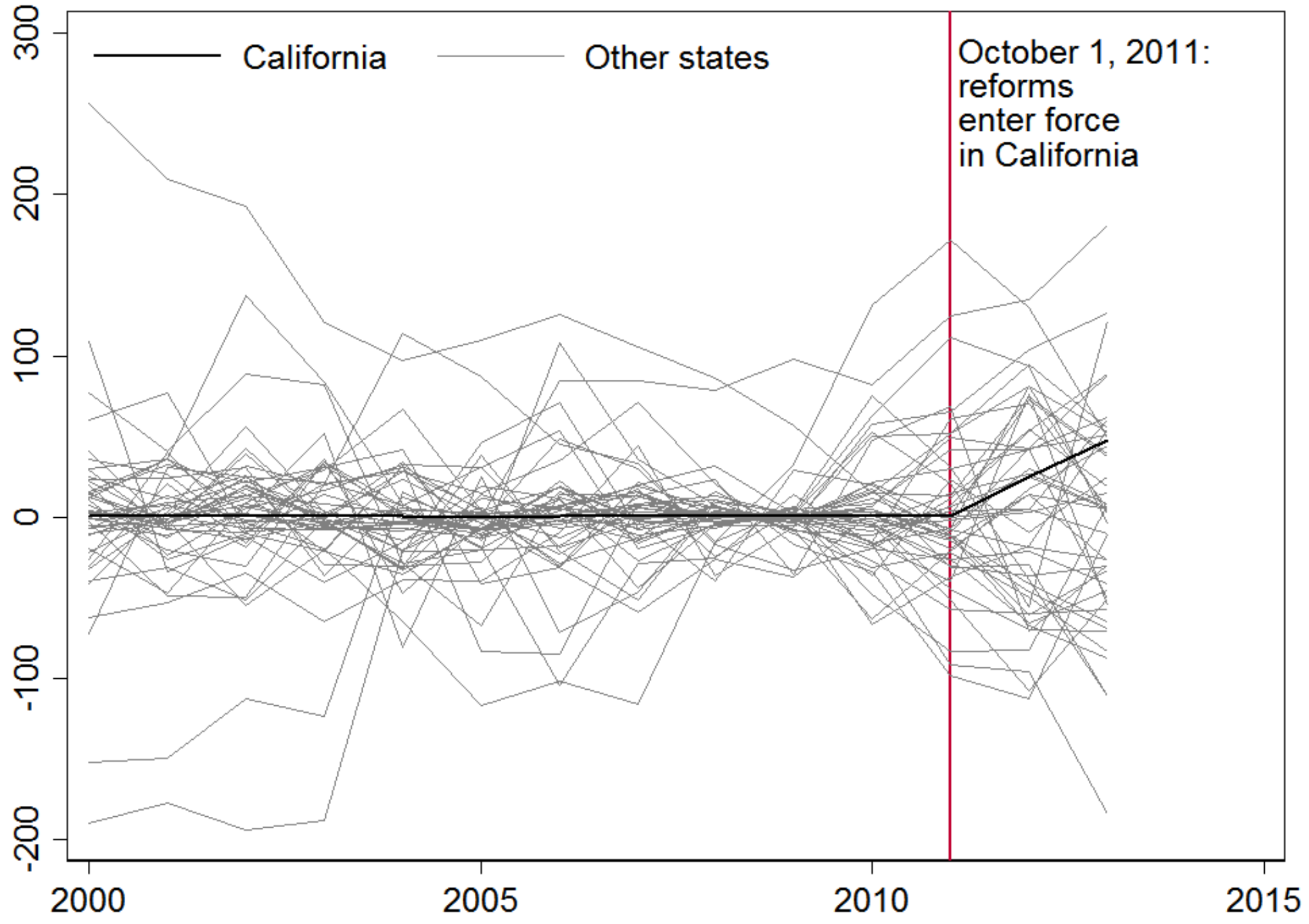

**Figure 22. Property crimes per 100,000 people in California and comparison states (two weighted averages), 2000–13, following Lofstrom and Raphael (2016)**

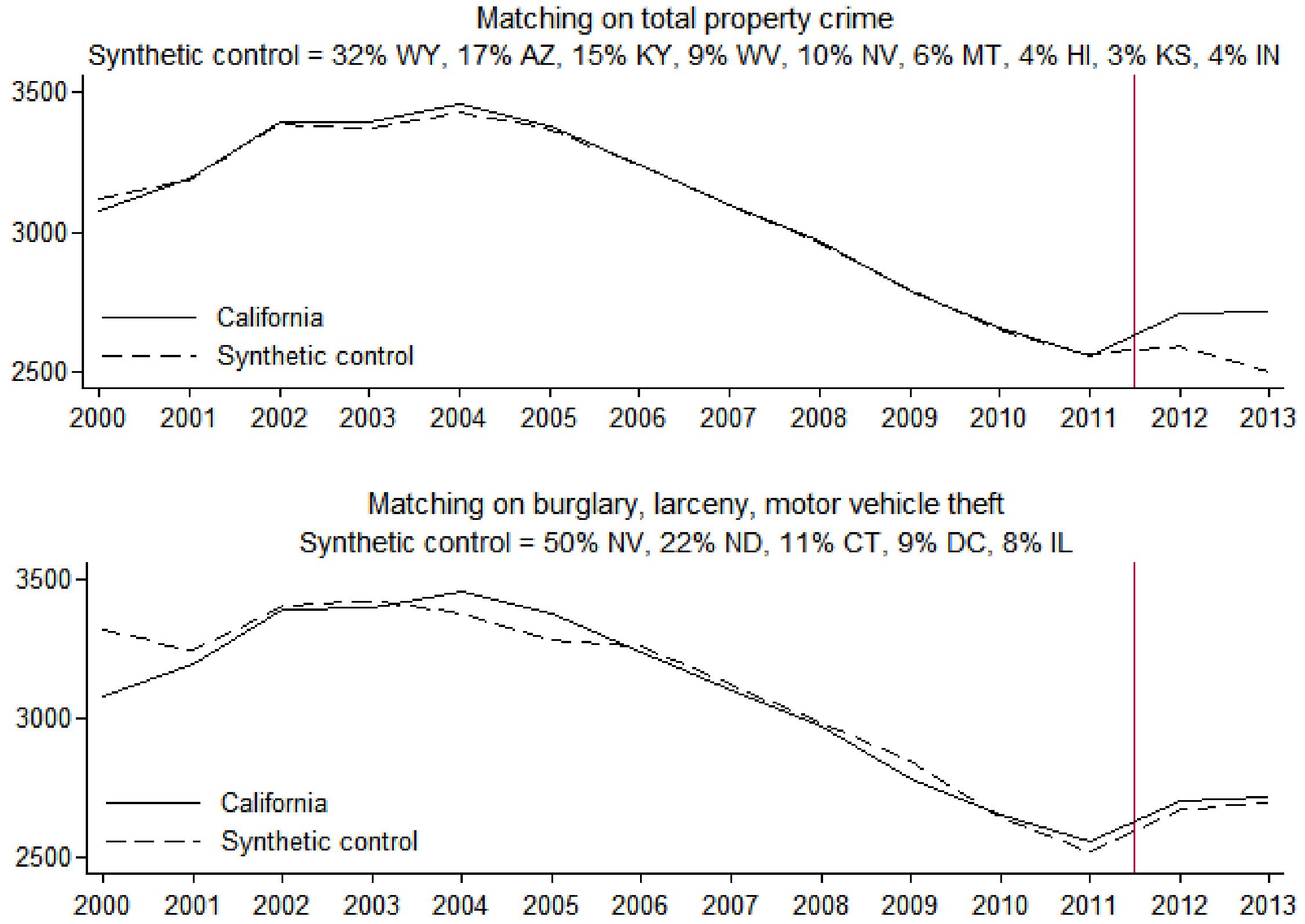

**Table 10. Estimated impacts of California's "realignment" on reported crimes per 100,000 residents, original and revised synthetic control results following Lofstrom and Raphael (2016)**

| | **Total violent** | Murder | Rape | Aggravated assault | Robbery | **Total property** | Burglary | Larceny | Motor vehicle theft |
|---|---|---|---|---|---|---|---|---|---|
| Original | 6.28 | 0.07 | –0.71 | 10.76 | 4.05 | 227.02 | 44.64 | 20.53 | 71.89** |
| | (0.65) | (0.65) | 0.82) | (0.41) | (0.73) | (0.08)* | (0.41) | (0.73) | (0.04) |
| Revised | 6.26 | 0.14 | –1.38* | 6.63 | –6.18* | 174.98** | 35.04** | –21.62 | 56.34 |
| | (0.74) | (0.68) | (0.09) | (0.95) | (0.09) | (0.03) | (0.01) | (0.80) | (0.20) |
| Revised, | 17.70 | 0.72 | –1.16* | 16.89 | –0.01 | –7.22 | 4.08 | –52.11 | 40.81** |
| common control | (0.89) | (0.80) | (0.05) | (0.97) | (0.91) | (0.57) | (0.34) | (0.47) | (0.03) |

Original results are impacts on crime rates in calendar years 2012–13 relative to 2010, from Lofstrom and Raphael (2016, Table 5; working paper, Table 7); p values are two-tailed, based on the empirical "placebo" distributions of impact estimates for all states. Revised results start with monthly FBI crime data; include Washington, DC, but exclude Alabama, Florida, and Minnesota for lack of monthly or quarterly data; shift the beginning of the statistical year to October 1; add 2011 to the pre-treatment period; extend the baseline period from 2010 to 2010–11; shorten the treatment period for rape from 2012–13 to 2012 because of a definitional change in 2013; and base p values on the empirical "placebo" distributions of ratios of post- to pre-treatment mean squared prediction errors. "Common control" regressions use the same synthetic control for violent crime and its subcategories, and separately for property crime and its subcategories, matching on 2000–11 histories for all the given subcategories at once. *Significant at p<0.1. **Significant at p<0.05.

**Figure 23. Property and violent crimes (excluding rape) by month, Nevada, seasonally adjusted, 2000–14**

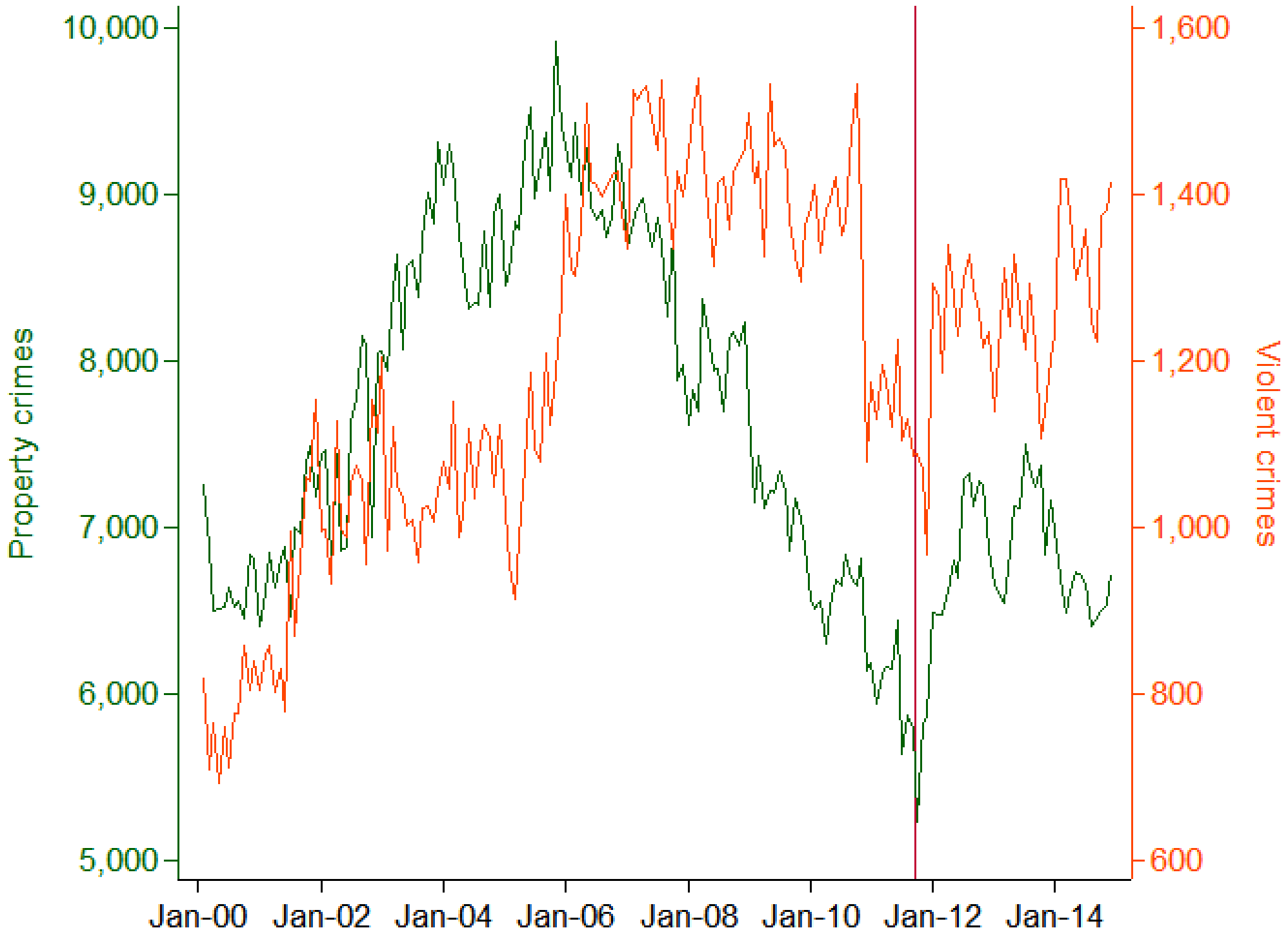

**Table 11. Static panel estimates of the impact of the 2011–12 California prisoner reduction on various categories of crime, two years following realignment**

| Mathematical form of outcome | Murder | Rape | Aggravated assault | Robbery | Burglary | Larceny | Motor vehicle theft |
|---|---|---|---|---|---|---|---|
| **Log crimes** | | | | | | | |
| Per 100,000 | 0.025 | –0.0095 | –0.0030 | 0.0047 | 0.073 | 0.036 | 0.12* |
| residents | [–0.17, 0.22] | [–0.11, 0.08] | [–0.056, 0.055] | [–0.12, 0.15] | [–0.018, 0.204] | [–0.045, 0.194] | [0.015, 0.273] |
| **Crimes reported** | | | | | | | |
| Per 100,000 | 0.12 | –0.20 | –0.44 | 1.1 | 46 | 58 | 49* |
| residents per year | [–0.75, 1.15] | [–2.2, 1.7] | [–8.0, 8.3] | [–27, 39] | [–10, 136] | [–70, 336] | [5.8, 121.3] |
| Per prisoner-year | 0.0023 | –0.0037 | –0.0083 | 0.022 | 0.86 | 1.10 | 0.92* |
| | [–0.014, 0.022] | [–0.041, 0.033] | [–0.15, 0.16] | [–0.51, 0.74] | [–0.20, 2.57] | [–1.3, 6.4] | [0.11, 2.29] |
| **Crimes committed** | | | | | | | |
| Per 100,000 | 0.12 | –0.58 | –0.71 | 1.8 | 81 | 210 | 63* |
| residents per year | [–0.75, 1.15] | [–6.4, 5.1] | [–13, 13] | [–43, 62] | [–19, 242] | [–254, 1223] | [7.6, 157.5] |
| Per prisoner-year | 0.0023 | –0.011 | –0.013 | 0.034 | 1.5 | 4.0 | 1.2* |
| | [–0.014, 0.022] | [–0.12, 0.10] | [–0.24, 0.25] | [–0.80, 1.17] | [–0.35, 4.58] | [–4.8, 23.1] | [0.14, 2.97] |

$N$ = 2,208, except $N$ = 2,093 for rape. Rape sample ends in 2012 because of definitional change in 2013. Statistical years begin October 1. DC included; Alabama, Florida, and Minnesota excluded for lack of monthly or quarterly data. Only first row presents primary regressions results; remaining rows re-express first row as $(\exp(\beta) - 1)\mu$ where $\beta$ is the first row's point estimate and $\mu$ is the 2011 California mean for each cell's crime type. All regressions weighted by total state-level crimes as predicted by initial population-weighted regressions (Greene 2003, pp. 687–88). Conversion to per-prisoner-year figures based on total prison and jail population reduction from 614 to 562 per 100,000 between September 30, 2011, and March 31, 2012. Conversion from crimes reported to crimes committed based on national-level estimates of reporting rates from BJS (2015). Wild-bootstrapped, state-clustered 90% confidence intervals in brackets. *Significant at $p<0.1$.

**Figure 24. Bootstrapped p values for various hypothesized impact rates of realignment on log crime rates in California, 2012–14**

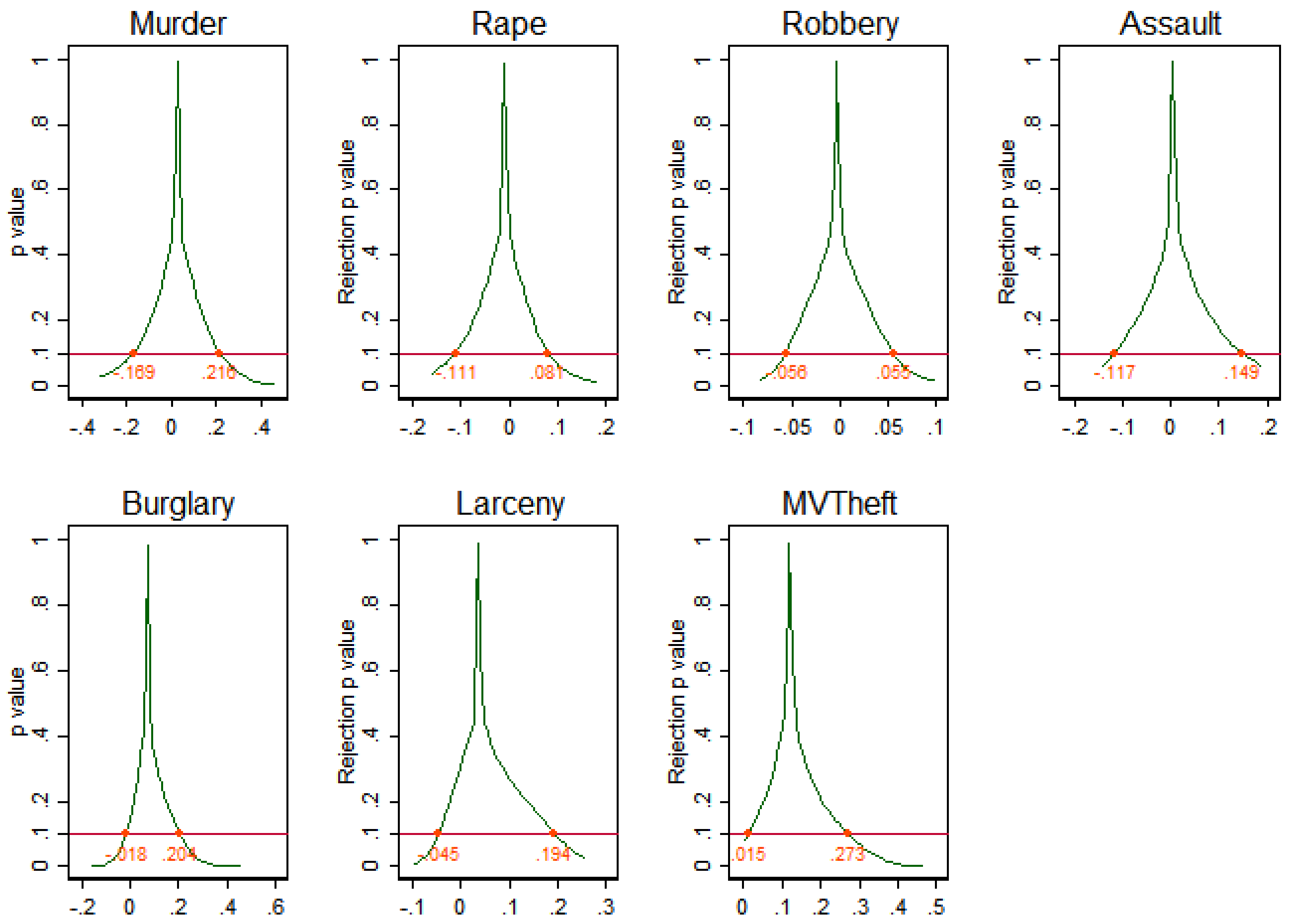

## 8.6. Summary: Incapacitation versus standard release

Surveying these studies of incapacitation relative to standard release reveals a few patterns:

- All find incapacitation.
- Incapacitation emerges more clearly for property crime than violent crime. Possibly this is merely because some of the policies studied (in the Netherlands and California) focused on people convicted of property crimes. But it may also be that the propensity for violence is more evenly distributed in the population, so that incarcerating some people does less to contain it.
- Incapacitation looks lower at margins where incarceration is higher, which suggests diminishing returns to incarceration. The US studies find less incapacitation per prisoner-year than the European ones. Owens finds the least, at 2.9 violent or property crimes/prisoner-year, of which 1.44 are reported; and that appears in a relatively crime-prone group, 23-25–year-olds with juvenile priors. Levitt's estimates imply about 1.2 reported violent crimes and 6.7 reported property crimes/year in 1973–93. And in contemporary California, realignment arguably increased violent crimes not all and reported property crimes by 2.9 per prisoner-year (third row of Table 11).

# 9. Aftereffects

## 9.1. Berecochea, Jaman, and Jones (1973), "Time served in prison and parole outcome: An experimental study: Report number 1"; Berecochea and Jaman (1981), "Time served in

prison and parole outcome: An experimental study: Report number 2"

In 1970, the California's Department of Corrections and parole board ran a pioneering randomized experiment in early release from prison. The goal was to study whether the length of incarceration affected how well people did on parole. At the time, American prisons operated much more on the theory of rehabilitation than punishment. In California, this meant that the legislature gave broad sentencing leeway to judges; judges in turn set sentences that were meant as capacious upper limits, leaving it to the parole board to decide when to move people from incarceration to parole, and when to release them from parole (Bushway and Paternoster 2009, p. 120).

In California, each inmate's case underwent annual review, at which time the parole board might set a release date, typically less than 12 months off. The experiment applied to men whose release dates had been set at least six months hence. Of the 2,282 such men during the study's intake period, March–August 1970, 972 were excluded from the experiment for reasons such as having committed first-degree murder or needing to complete an in-prison rehabilitative program. Among the remaining 1,310, about half had their release accelerated by an experimental six months. After randomization, 99 control and 73 treatment subjects were deleted from the study because of death in prison or other reasons (Berecochea and Jaman, pp. 4–6).

Table 12 displays results with reference to the originally randomized sample. The follow-up periods are two years, starting at release from prison. The table's first column shows the results for attrition (just mentioned), and the rest for various possible outcomes, ordered from best to worst. The best was a favorable parole outcome, which could mean discharge from parole without incident, or serving less than three months in jail, among other possibilities. Next come an unfavorable parole outcome not resulting in return to prison (but possibly return to jail); parole outcome pending; return to prison by a parole board, such as for a violation of the terms of parole; and return to prison by a court, presumably for conviction for a new crime.

The greatest treatment-control differences are at the extremes: those incarcerated longer had more favorable parole outcomes (51.6% vs. 46.5%, $p = 0.07$) and fewer returns to prison on the order of a court (8.0% vs. 11.2%, $p = 0.06$).[78] More time led to less crime, or at least better parole outcomes.

The apparent impact probably does not owe to aging. True, those released earlier were six months younger and thus perhaps more likely to offend and get caught. But at a median age of about 31, a six-month, ~25% relative decline in recidivism looks steep as an aging effect (going by the last column of Table 12). It compounds to 48% annually.

Cognitive framing bias (see section 2.4.4) also looks like a weak candidate to explain the results. Recall that Bushway and Owens estimate that in Maryland around 2001, each 10% fall in actual relative to recommended sentence cut the three-year ever-rearrested rate by about 0.8 percentage points, or roughly 1.6% of the baseline rearrest rate of ~50%. We might assume that the Berecochea and Jaman subjects on average expected to serve 37.9 months, as the control group actually did, and thus that the treatment group experienced a 17.5% cut relative to expectation (to 31.3 months). If we extrapolate from Bushway and Owens assuming the same elasticity of impact, we expect in California a framing effect on the probability of being ordered back to prison by a court of $17.5\% / 10\% \times 1.6\% \times 11.2\% = 0.31$ percentage points, where 11.2% is the baseline rate in the control group (Table 12, last column). This is only a tenth of the treatment-control gap (11.2% – 8.0% = 3.2%).

Conceivably a flavor of parole bias infects the numbers. This research is, as the report titles state, about *parole outcomes*. If someone completed parole without incident and then robbed a bank, all within the two-year follow-up window, that would still have been counted as a favorable parole outcome (Berecochea and Jaman, p. 9).[79] And if parole officers on average discharged sooner those who had done more time in prison, then this oddity in measurement would have affected treatment and control asymmetrically. It would have mechanically generated more favorable outcomes for the longer-serving inmates. In effect, they would have disproportionately exited the study before getting into trouble. Arguably bolstering this speculation is the report that corrections staff knew about the experiment (Berecochea, Jones, and Jaman, pp. 5–6). Possibly some sought to compensate for leniency in incarceration with prolonged parole supervision. On the other hand, this theory assumes that some control subjects *did* earn favorable discharge from parole and then *did* behave in ways that *would* have triggered unfavorable outcomes had they still been on parole. And that apparently did not happen much. "Discharges in less than two years were uncommon; those which did occur were typically the result of an arrest-free first year on parole, which is highly predictive of no serious difficulties thereafter" (p. 9). Whether these uncommon events were too uncommon to explain a 7.9% differential in favorable outcomes (Table 12, col. 2) is not clear.

[78] Unlike in Table 12, the papers do not compute the significance of the treatment-control difference in the individual kinds of favorable parole outcome, but rather in the favorable/unfavorable/pending splits taken together. As a result, all of the tests performed return p values above 0.05, which the papers interpret as "no impact."

A critique relating to attrition looks more trenchant, if still speculative. In general, randomizing trials does not solve does not solve the problem of potential attrition bias. In the case at hand, more of the control subjects, the ones imprisoned six months extra, attrited from the study because of "death in prison, loss of parole date, erroneous inclusion in the project, escape from prison, and other reasons" (Table 12, col. 3). The study provides no more information on which of these factors mattered most often, nor what might have caused the differential. Possibly, people who were to stay in prison longer were more likely to cause or experience events, such as escape, that would have predicted higher recidivism had they stayed in the study. If so, keeping them in prison longer gradually filtered out of the control group people more likely to have recidivated. The attrition differential—99 versus 73—is comparable to the new-conviction differential, at 54 vs. 71 (Berecochea and Jaman, Table 7, upper-right corner).[80]

Unlike Berecochea and Jaman, who adhere to a 0.05 significance threshold, I see the treatment-control difference in new convictions, significant at p = 0.06, as a real statistical pattern. And it may well mean what it seems to mean, that more time in prison, in California in the early 1970s, caused less crime after. The potential for attrition bias, though speculative, causes me the most doubt, for I also see the attrition differential, significant at p = 0.08, as unlikely to be caused by chance.

**Table 12. Parole outcomes over two years following release, from Berecochea and Jaman (1981)**

| | | | | Not returned to prison | | | Returned to prison | |
|---|---|---|---|---|---|---|---|---|
| | Sample size | Average months served | Attrited[1] | Favorable parole outcome[2] | Unfavorable parole outcome[3] | Parole outcome pending[4] | By parole board | By court (new conviction) |
| Control group | 671 | 37.9 | 14.8% | 51.6% | 13.6% | 0.9% | 11.2% | 8.0% |
| Treatment group | 636 | 31.3 | 11.5% | 46.5% | 17.5% | 0.8% | 12.6% | 11.2% |
| p value for difference | | | 0.08 | 0.07 | 0.05 | 0.83 | 0.43 | 0.06 |

One control and two treatment subjects excluded because they were released less than two years before follow-up. [1]Causes of attrition are "death in prison, loss of parole date, erroneous inclusion in the project, escape from prison, and other reasons." [2]Includes those serving <90 days in jail, or receiving suspended jail time, or misdemeanor probation, or completing parole without problems. [3]Includes jail ≥90 days, felony probation ≥5 years, suspended prison time, and more. [4]Parole violation occurred but disposition pending at two-year point. Sources: Berecochea and Jaman (1981, p. 6, Tables 5 & 7, Appendix); author's tests for significance of differences in proportions.

## 9.2. Martin, Annan, and Forst (1993), "The special deterrent effects of a jail sanction on first-time drunk drivers: A quasi-experimental study," *Accident Analysis & Prevention*

This paper pioneered the now-popular "judge randomization" strategy (see §3.2), albeit in a context distant from current debates over criminal justice reform. It takes place in 1982, in Hennepin County, which includes Minneapolis. There, judges in the municipal court agreed to impose two-day jail sentences on all first-time driving-under-the-influence (DUI) offenders (p. 562). Evidently the agreement was nonbinding, for the judges did not equally adhere to the policy:

> Data were collected on all first-offender drunk driving cases adjudicated by two judges of the Hennepin County Municipal Court during an 11-month study period. One of the judges was known to sentence virtually no first offenders to jail ("no jail" judge); the other was known to sentence virtually all first offenders to two days in jail ("jail" judge).…Hennepin County Municipal Court cases are assigned approximately at random through normal assignment procedures… (p. 562)

[79] Presumably it would have taken a major effort, or have been impossible, to link the parole board's outcome data to a separate agency's individual-level data on crimes not impinging on the supervision of parole.

[80] The cited source reports 71 but this looks like an arithmetic error because displayed subtotals to not sum to displayed totals. 72 looks correct.

When instrumented with an indicator for the "jail judge," receipt of a jail sentence had little predictive value for whether a person was reconvicted for DUI within two years.

One can lob some criticisms at the study. Martin, Annan, and Forst seem not to detail the "approximately at random" assignment process, so it is hard to know how close to random it was. They partially reassure by showing that the treatment and control groups resembled each other in gender and in frequency of having prior non-alcohol-related driving convictions—as they would if randomly assigned (Martin, Annan, and Forst, Table 1). However, the "jail judge" got more under-26 defendants, 50% instead of 41%, $p = .03$[81], and across the both judges this subgroup recidivated more (Martin, Annan, and Forst, Table 3). Conceivably the true effect of jail time was negative (lower reoffense), but was offset by the wayward youthfulness of the jail judge's defendant pool.

Another concern, Martin, Annan, and Forst raise, is that "treatment" by a judge is multidimensional (p. 566). And while the "no jail judge" was softer when it came to jail time, the judge was tougher in meting out fines. Of the no-jail judge's 185 defendants, 110 had to pay fines greater than $200, about $500 in today's dollars. Only seven of the jail judge's did (p. 564). Thus the quasi-experiment is not purely in added jail time, but in the combination of that and lower fines. And the effects of the two cannot be disentangled. Possibly the negligible impact found results from the offsetting effects of each.

Since current concerns about incarceration have to do with people serving sentences measured in months and years, not nights, we need not belabor the interpretation of Martin, Annan, and Forst. Its greatest significance contribution is initiating an important new approach and illustrating one of its pitfalls.

## 9.3. Chen and Shapiro (2007), "Do harsher prison conditions reduce recidivism? A discontinuity-based approach," *American Law and Economics Review*

Chen and Shapiro broke new ground by constructing a quasi-experiment in the *quality* of incarceration instead of the quantity. "Quality" in this case means the security level of a prison, maximum security being the harshest. If toughness in quality and quantity similarly affect post-release criminality, then studying one can give insight into the other. And a pure experiment in quality would sidestep aging bias, since if people entered prison at the same average age they would exit at the same average age too.

The quasi-experiment arises because federal prison managers, rather like the judges in Maryland (see review of Owens), used a non-binding point schedule in assigning inmates to security levels. The schedule factored in the severity of the current and past crimes, past history of violence and escape attempts, and other information. A score of 0–6 points led to a recommendation of minimum security, 7–9 to low, 10–13 to low/medium, and 14 and above to medium (Chen and Shapiro, Table 2). The discontinuities between 6 and 7, 9 and 10, and 13 and 14 created quasi-experimental variation in prison security level if people on either side of these divides were statistically similar upon entering the prison system.

[81] p value is from my test for difference in proportions.

The study's sample is a set of inmates released from prison in the first half of 1987 (p. 8). Although the guidance rule contains at least three thresholds, Chen and Shapiro focus on that between minimum and low security—between 6 and 7 points—probably because that's where the bulk of the data lies (p.10). In general when exploiting a threshold, the closer one requires subjects to be it, the better those on either side should match, making for a stronger study—but the smaller the sample. Facing this tradeoff, Chen and Shapiro run their analysis with several bandwidths.

Table 13 shows some of their results. The table is split into thirds, by bandwidth. The top third has the widest: it compares those who scored 4–6 to those who scored 7–9. It shows that 44.71% of the 170 people in the 4–6 band, which nominally corresponded to minimum security, were actually assigned to a higher security level. But a higher share in the 7–9 band, 92.94%, were so assigned, leaving a difference big enough to offer hope for a useful quasi-experiment. The first two rows of the top third also show that more of those in the 7–9 band were rearrested within over one, two, or three years, and with statistical significance. Those differences can be viewed as impacts of assignment *recommendation* since recommendation depends on points; the fourth row infers impacts of *actual* assignment, by taking ratios. For example, being above the 6–7 cutoff, in addition to increasing the placement rate in higher security by 48.24 points, appeared to raise the one-year whether-rearrested rate by 10 percentage points, for an average impact of 10 / 48.24 = 20.73 percentage points.

The major concern about that last result is that the wide bandwidth makes the samples somewhat incomparable, with the higher-scoring group perhaps more likely to recidivate even before assignment to higher security. The rest of the table therefore narrows the bandwidth. The impact estimates remain stable, although they lose significance at standard p value thresholds in the narrowest sample. Given the similarity in the central impact estimates—compare the last line for the first and last thirds—the impact is probably still there, but hard to detect because of the tiny sample.

Chen and Shapiro buttress their findings with two "falsification tests." The first applies the same procedure to 56 inmates who were exempted from the scoring system and were instead placed in special facilities because of medical needs. Within this group, happening to score just above 6 did not affect their incarceration experience, because of the medical exception. And it also did not appear to affect their recidivism (Chen and Shapiro, Table 4, panel A). This is reassuring—if only moderately so, given the small sample. Second, the authors use the same method to check for "impacts" where there could be none, such as on traits such as race and sex. Here, the results are mostly good, but not perfect. E.g., the 5–6 group was 89% male while the 7–8 group was 100% male, a statistically significant difference. (Chen and Shapiro, Table 4, panel B.) Addressing this evident treatment-control difference, Chen and Shapiro (note 17) report getting similar results if they restrict just to men.

Chen and Shapiro do not mention that one input to the scoring determining incarceration quality is incarceration quantity. So this may not be a pure experiment in quality after all. An inmate gets an extra point if expected to serve more than a year, three if more than five years, and five extra points for more than seven years (Chen and Shapiro, Figure 1). Chen and Shapiro appear not to provide statistics on time served, which would help in checking this possibility. They do report that the comparison groups differ little in age at release—36.34 for the 5–6 group versus 35.77 for the 7–8 group—which is backward compared to what one would expect if the 7–8 group waited significantly longer to get out (Chen and Shapiro, Table 4, panel B).

On balance, I think the Chen and Shapiro results are more likely right than wrong. The quasi-experiment is not as compelling as it would be if the scoring system were more fine-grained and if a larger sample could be constructed close to the boundary. One can always wonder if the difference of at least a point between the two groups indicates pre-existing difference in future recidivism risk, biasing the results. But the falsification tests reduce this concern. The hypothesis that time in higher-security prison raises recidivism looks most compatible with the data.

**Table 13. Impact of recommendation for and assignment to higher- vs. minimum-security prison at scoring cutoff between 6 and 7 points, from Chen and Shapiro (2007)**

| Score range | Observ-ations | Share in above-minimum security | Share rearrested within 1 year | 2 years | 3 years |
|---|---|---|---|---|---|
| **Sample: within 3 points on either side** | | | | | |
| 4–6 points | 170 | 0.4471 | 0.2176 | 0.3529 | 0.4647 |
| 7–9 points | 85 | 0.9294 | 0.3176 | 0.4824 | 0.5765 |
| Impact of *recommendation* to higher vs. minimum | | 0.4824*** | 0.1000* | 0.1294** | 01118* |
| Impact of *assignment* to higher vs. minimum (IV) | | | 0.2073* | 0.2683* | 0.2317 |
| **Sample: within 2 points on either side** | | | | | |
| 5–6 points | 91 | 0.4725 | 0.1978 | 0.3626 | 0.4835 |
| 7–8 points | 52 | 0.9423 | 0.3462 | 0.5577 | 0.6346 |
| Impact of *recommendation* to higher vs. minimum | | 0.4698*** | 0.1484** | 0.1951** | 0.1511* |
| Impact of *assignment* to higher vs. minimum (IV) | | | 0.3158* | 0.4152** | 0.3216* |
| **Sample: within 1 point on either side** | | | | | |
| 6 points | 44 | 0.5227 | 0.2273 | 0.4091 | 0.5227 |
| 7 points | 32 | 0.9688 | 0.3438 | 0.5625 | 0.6250 |
| Impact of *recommendation* to higher vs. minimum | | 0.4460*** | 0.1165 | 0.1534 | 0.1023 |
| Impact of *assignment* to higher vs. minimum (IV) | | | 0.2611 | 0.3439 | 0.2293 |

*significant at p<.10; *significant at p<.05; *significant at p<.01.

Source: Chen and Shapiro (2007), Table 3.

### 9.4. Gaes and Camp (2009), "Unintended consequences: Experimental evidence for the criminogenic effect of prison security level placement on post-release recidivism," *Journal of Experimental Criminology*

Like Chen and Shapiro, Gaes and Camp examine how the security level of a prison experience affects outcomes after release. Unlike Chen and Shapiro, this paper exploits a randomized experiment, which the California prison system executed between November 1998 and April 1999. The experiment was originally run to test whether a new algorithm for assigning inmates to higher-security confinement in California better predicted in-prison misconduct. But Gaes and Camp recognized the opportunity to recast it as testing whether security level affected subsequent recidivism.

For most inmates, the old and new assignment systems agreed. Where they disagreed, the new system usually mapped inmates to level III while the old system mapped to level I. (The lowest security level was I and the highest, IV.) Since inmates were randomly sent through the old or new assignment systems, a de facto sub-experiment put 264 into level I and 297 into level III (pp. 148–49).

Rather than defining the outcome as *whether* a releasee was rearrested or reincarcerated within some time horizon, Gaes and Camp, like Helland and Tabarrok, use information about the timing of each subject's reentry into prison, if any, to estimate how security level affects the probability per unit time that a still-free releasee would return (p. 49). Gaes and Camp (Table 2) find that assignment to higher-security (level III) raised this probability by 31.1% (not 31.1 percentage points; se ≈ 10%). A harsher prison experience led to more crime after release.

Or maybe not. As is standard in writing up randomized trials, the first table of the paper checks whether the two groups are statistically indistinguishable, as they should be after randomization. On predetermined traits such as race and type of crime, the test is easily passed. However, the groups differ with statistical significance on two closely related "post-determined" traits: total time served and age at release. Those put in higher security got out sooner and younger. In particular, those assigned to level III did 16.3 instead of 23.0 months on average (p = 0.000) and were 23.8 years old at release instead of 24.8 (p = 0.029) (Gaes and Camp, Table 1).

These differences may have arisen by chance: unlikely is not impossible. Or perhaps the experiment only effectively involved a small handful of prisons, and one of them happened to work more with higher-security inmates and favor earlier releases. Or possibly doing time in higher-security *caused* earlier release, if the parole board saw tougher conditions as substituting for longer time.

Whatever the cause, the difference arose, and it muddies interpretation. One group did less time under tougher conditions, while the other group did opposite. So it becomes unclear which group experienced more punishment, however defined.

In defense of Gaes and Camp, Chen and Shapiro conclude similarly (and I think more credibly). But for me it remains the case that the Gaes and Camp results do not shift priors as much.

### 9.5. Green and Winik (2010), "Using random judge assignments to estimate the effects of incarceration and probation on recidivism among drug offenders," *Criminology*

Green and Winik brings the judge randomization strategy of Martin, Annan, and Forst to the heart of our inquiry: not DUI cases with maximum incarceration spells of two days but drug cases that put months or years of liberty at stake. The sample is 1,003 defendants, all of whom entered the DC Superior Court between June 1, 2002, and May 9, 2003 (p.357). (Why these dates is not clear.)

As is common in judge randomization studies, the assignment of subjects to the "treatment" of appearing in a particular courtroom was fairly arbitrary, but not actually random:

> During 2002 and 2003…the Court used a mechanical wheel to rotate the assignment of new cases among the calendars—assigning one case to calendar 1, the next case to calendar 2, and so on. The arraignment court coordinator…explained that she deviated from the cycle when the caseload of a calendar was out of balance with the rest, generally because the judge in question had processed cases faster or slower than the norm. When such imbalances occur, she explained, the coordinator can skip an overloaded calendar in the cycle or assign additional cases to an underloaded one…. Cases remain on the same calendar through the final disposition, but the judges assigned to each calendar sometimes rotate at the beginning of each year. We, therefore, considered calendar assignment rather than specific judge assignment to be the randomly assigned treatment. (pp. 365–66)

And as in other judge randomization studies, Green and Winik define the follow-up periods to include incarceration time—in this case beginning when a judge "disposes" of a case, deciding guilt and potentially assigning a sentence. Starting the follow-up period at release, as in Berecochea and Jaman, would allow confounding from age effects, as explained in section 2.4.1. This choice mixes incapacitation and aftereffects together in the measured results. Of course, the distinction between the two may be secondary in policymaking, where the bottom-line question is typically how incarceration affects crime, regardless of channel. And for that question, starting the clock at case disposition gives the cleanest answer.

In the Green and Winik data, the nine courtrooms, or "calendars" range rather widely in average prison sentence (5.1–11.9 months), in share of defendants given probation instead (29.4–60.2%), and in average probation sentence (6.4–14.9 months). These differences help assure that courtroom assignment dummies will be strong instruments (at least for sentencing and probation length taken one at a time). In contrast, if the courtrooms had all sentenced at the same rates, there would have been no quasi-experiment.

My preferred Green and Winik regressions are a pair that include incarceration as well as probation sentence: both are channels by which calendar assignment might affect recidivism. And they use limited-information maximum likelihood (LIML) instead of two-stage least squares (2SLS): LIML is known to be more reliable when instruments are weak (e.g., Anderson, Kunitomo, and Sawa 1982, p. 1025). Table 14 below ("original" column) reproduces the results of interest from the one of these regressions that controls for demographic and other traits. An extra month of incarceration sentence is found to increase the four-year rearrest rate by 2.08 percentage points. The standard error of 1.79% makes the estimate statistically different from 0 at p = 0.25 according to the usual two-tailed test. Equivalently, the hypothesis of negative impact, based on a one-tailed test, rejects at 0.125. So even if we interpret that mildly significant positive effect as zero, it implies that longer sentences increased post-release criminality, and enough so to cancel out incapacitation. Meanwhile, probation's effect is essentially nil. And since Open Philanthropy Project is funding organizations working to reduce incarceration, it is conservative for us to interpret this positive but only weakly significant coefficient the way Green and Winik do, as zero. That is: considering incapacitation and aftereffects together, longer sentences for drug offenders in DC did not increase crime, only failed to decrease it.

In a side-exercise, Green and Winik take a stab at isolating criminogenic aftereffects from incapacitation, essentially by shifting forward the follow-up window for incarcerated defendants to begin upon release.[82] They recognize that this reduces rigor by allowing in aging effects. Taken with the proper salt, this exercise puts the impact of an extra month in prison on the rearrest rate over the four years after release at 3.02 percentage points (p. 380). Compare that to the more-rigorous 2.08 that is depressed by incapacitation. Green and Winik's standard error for the 3.02% estimate is 1.86%, making it significantly different from 0 at p = 0.1. However, their posted code computes this as the average standard error from individual simulations, making its meaning questionable. I bootstrapped the whole process and obtained a median impact estimate of 2.08%, and 10th and 90th percentiles at –0.05% and 7%.[83] Thus it still seems likely that an additional month in prison raises the four-year rearrest rate by a couple of percentage points.

Rerunning the preferred regression—the one in the first column of Table 14—with Green and Winik's posted data and code, I performed several specification tests not done in the paper. As shown in Table 14, the Kleibergen-Paap underidentification test reassures as to instrument weakness, while the Kleibergen-Paap rk statistic of 2.74 suggests roughly only 10% of the weak-identification bias toward OLS. But the regression does poorly on the Hansen *J* test of joint instrument validity (p = 0.09; if using 2SLS, 0.04). To pinpoint why, I experimentally added each calendar dummy alone into the second stage. Including the dummies for calendars 5 and 9 improved the Hansen test most, in this regression and in the one for felony reconviction discussed later. Possibly the apparent statistical problem here is pure randomness. Otherwise, either some inmates were assigned to calendars 5 and 9 in a non-arbitrary, non-quasi-experimental way; or the judges in calendars 5 and 9 "treated" their defendants in some way not captured by the incarceration and probation variables, which in turn affected recidivism.

Addressing this issue by adding the dummies for calendars 5 and 9 to the structural equation cuts the apparent impact of incarceration by a third while greatly raising that of probation (Table 14, col. 2). But the continuing lack of significance at conventional levels mainly reinforces Green and Winik's finding of "no detectable effect on rearrest" (p. 358), and continues to suggest that harmful aftereffects cancelled out incapacitation.

I supplemented those regressions with analogous ones for whether a person was reconvicted of a felony within four years, since that is more relevant to how incarceration affects serious crime. As the right half of

[82] They first run a Weibull time-to-failure regression of actual arrest times on a set of controls. For the incarcerated, time to failure (arrest) is counted from release date. From these results, they compute the probability of rearrest within four years for each subject. For subjects who were incarcerated and not rearrested within the first four years from case disposition, they use the probabilities to generate 1,000 Monte Carlo data sets. They run their preferred 2SLS and LIML regressions, excluding probation as a treatment, on each (p. 380).

[83] My nonparametric "pairs" bootstrap stratifies by calendar and, as in the original, clusters by codefendant group. 100 replications were run. In each replication the Weibull regression is first run, then 100 simulations are performed, yielding a mean criminogenic effect. Because of a couple of outliers, the standard deviation of the bootstrapped coefficients is 25%.

Table 14 shows, the net impact of time served remains near zero, and with similar standard errors.

Out of continuing concern about instrument weakness—incarceration and probation sentences might be well instrumented individually but too correlated across courtrooms to be well distinguished from each other after instrumentation—I next perform a graphical Anderson-Rubin test. I focus on the modified LIML regression for rearrest (Table 14, col. 2). Unlike in the reviews of Levitt and Lofstrom and Raphael, this regression instruments two variables instead of one—incarceration and probation—which literally adds of dimension of complication. Figure 25 depicts p values from 0.05 to 0.95 with shades of grey (light to dark).[84] Possible values for the impact of an additional month of incarceration sentence on the four-year rearrest rate fall on the X axis while those for probation go on the Y axis. Consistent with the regression's point estimates (Table 14, col. 2), the darkest area in the middle has X and Y coordinates of about +1%; those are the values hardest to reject as incompatible with the data. The widest blob is the 95% confidence domain; although it includes zero for both variables, the Anderson-Rubin test if anything rejects negative impact rates more confidently that the LIML regressions in Table 14.[85]

Parole bias is probably not substantial in these regressions. During the period of study, the law required inmates to serve at least 85% of their sentences (Green and Winik, note 3), depriving the parole board of most discretion. Moreover, in the Green and Winik data set, *probation* revocations do not show up as rearrests, so probably parole revocations do not either. Possibly, the 15% parole time, though small, increased parole revocations in lieu of reconvictions, which would slightly downward-bias the estimated impacts on reconvictions and help explain why those impacts appear a bit smaller in Table 14.

My final concern is about Green and Winik's focus on the four-year rearrest rate. Whether a person is arrested for any reason over an arguably long period might not be very sensitive to the amount of crime committed. Perhaps a pre-determined and uniform fraction of releasees will be arrested again sooner or later, yet having been incarcerated causes them to commit more or less crime. To the extent that incarceration affects crime more than the long-term arrest odds, this study would miss the full impact.

I explored this concern by rerunning my preferred Green and Winik regression repeatedly while varying the follow-up period from two days to four years. In Figure 26, the dashed brown line shows the share of people sentenced to prison who, according to their original sentence, were free within the given number of days of their case disposition. The black line shows the estimated impact on the rearrest rate through that period, with 95% confidence intervals in green. In the first year, the impact tends negative, presumably from incapacitation. But by two years, it swings positive, and stays there out to four. Something that would have validated my concern—yet which I do not find—would be a *larger* positive swing that then decays in the lead-up to the four-year mark. That would have suggested that the choice of four years was hiding the full impact, by giving more of the control group time to "catch up" in the sense of experiencing at least one arrest. Indeed, Figure 27 repeats the exercise for the indicator of more serious crime—convictions of felonies instead of arrests for misdemeanors or felonies—and its impact estimates hover even closer to zero.

The conclusion of Green and Winik largely withstands scrutiny: at least at one current margin, the criminal justice system appears to lose in criminogenic aftereffects what it gains from incapacitation. That said, the need to include dummies for two courtroom calendars raises a general concern that quasi-randomization failed.

---

[84] Produced with the "weakiv" program for Stata (Finlay, Magnusson, and Schaffer 2013).

[85] However, since the Anderson-Rubin test is a joint test of a null hypothesis about the impact of the instrumented variables and the hypothesis that instruments are valid, the rejection when impact parameters are hypothesized to be negative could indicate implied instrument invalidity in that range as well as violation of the null of negative impact.

**Table 14. Impact of incarceration and probation for drug offenders in Washington, DC, on four-year recidivism rate, from Green and Winik (2010)**

| | Rearrest | | Felony reconviction | |
|---|---|---|---|---|
| | Original | Modified | Original | Modified |
| Impact of incarceration | 2.08 | 1.32 | 0.99 | –0.04 |
| (% points per month sentenced) | (1.79) | (1.01) | (1.42) | (0.81) |
| Impact of probation | 0.14 | 0.75 | 0.28 | 1.35 |
| (% points per month sentenced) | (0.73) | (0.72) | (0.56) | (0.59) |
| Hansen J overidentification p | 0.09 | 0.73 | 0.08 | 0.85 |
| Kleibergen-Paap underidentification p | 0.004 | 0.0008 | 0.004 | 0.0008 |
| Kleibergen-Paap rk Wald F | 2.74 | 3.60 | 2.74 | 3.6 |

Notes: Standard errors clustered by defendant in parentheses. Original regressions use LIML and the full control set. Modified regressions remove exclusion restriction on dummies for court calendars 5 and 9. Original rearrest regression from Green and Winik (2010), Table 7, col. 6. Original felony reconviction reportedly in online appendix, which is not now online, and produced by original code, which is.

**Figure 25. Anderson-Rubin p values for various hypothesized impact rates of incarceration or probation sentence on subsequent four-year rearrest rate**

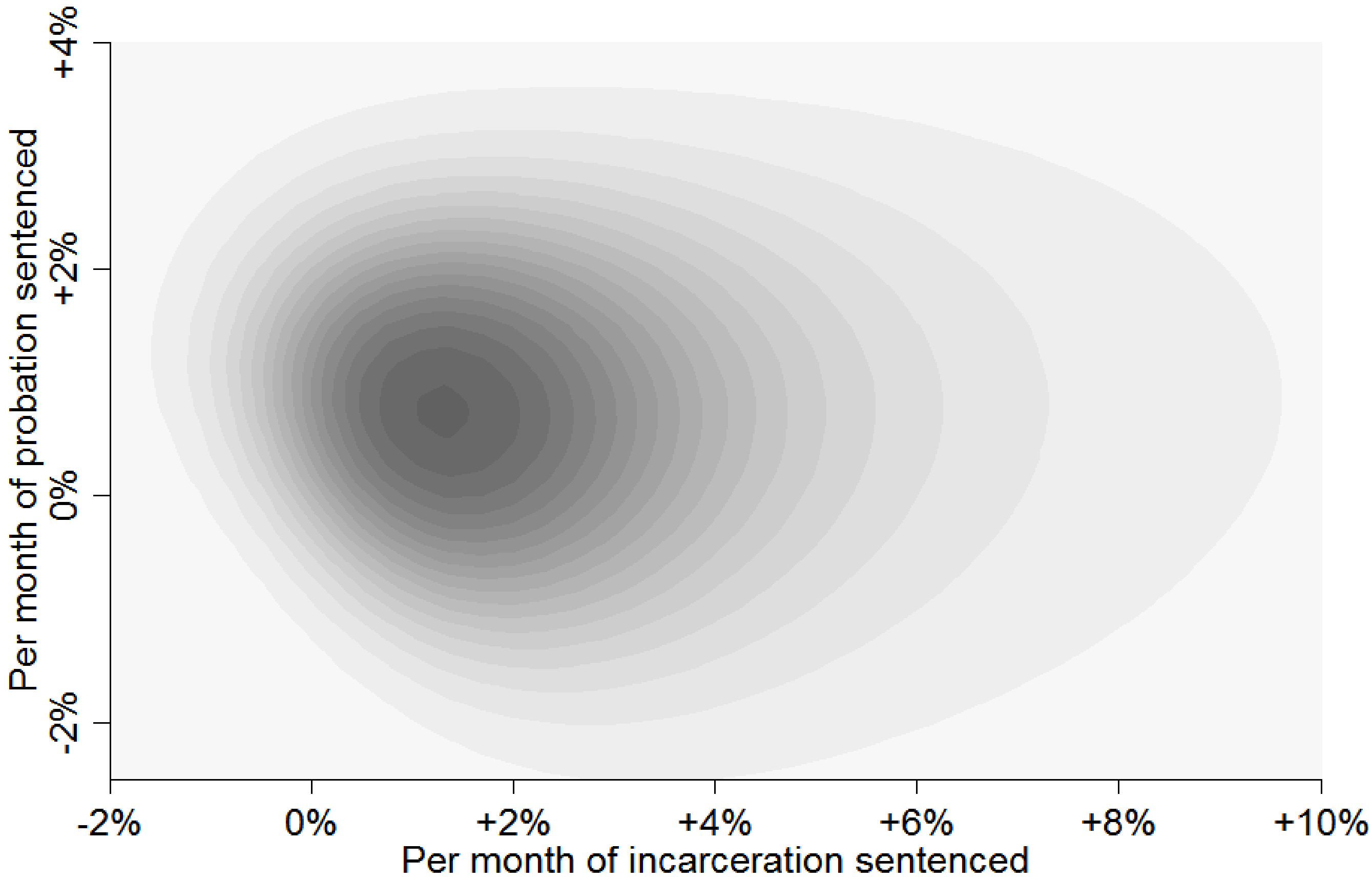

**Figure 26. Estimated impact of a month of incarceration on cumulative any-arrest probability as function of follow-up length, based on Green and Winik (2010)**

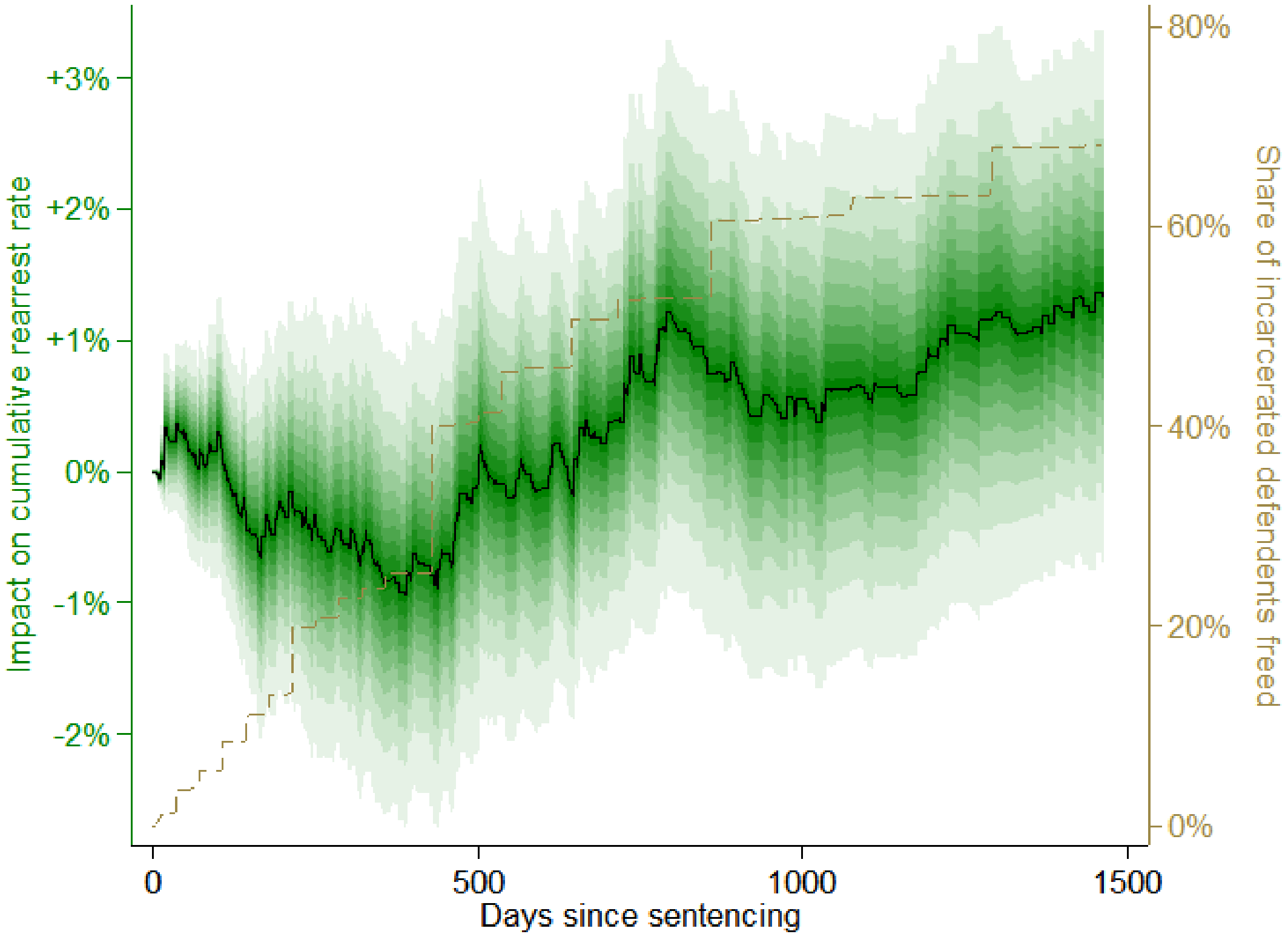

**Figure 27. Estimated impact of a month of incarceration on cumulative any–felony conviction probability as function of follow-up length, based on Green and Winik (2010)**

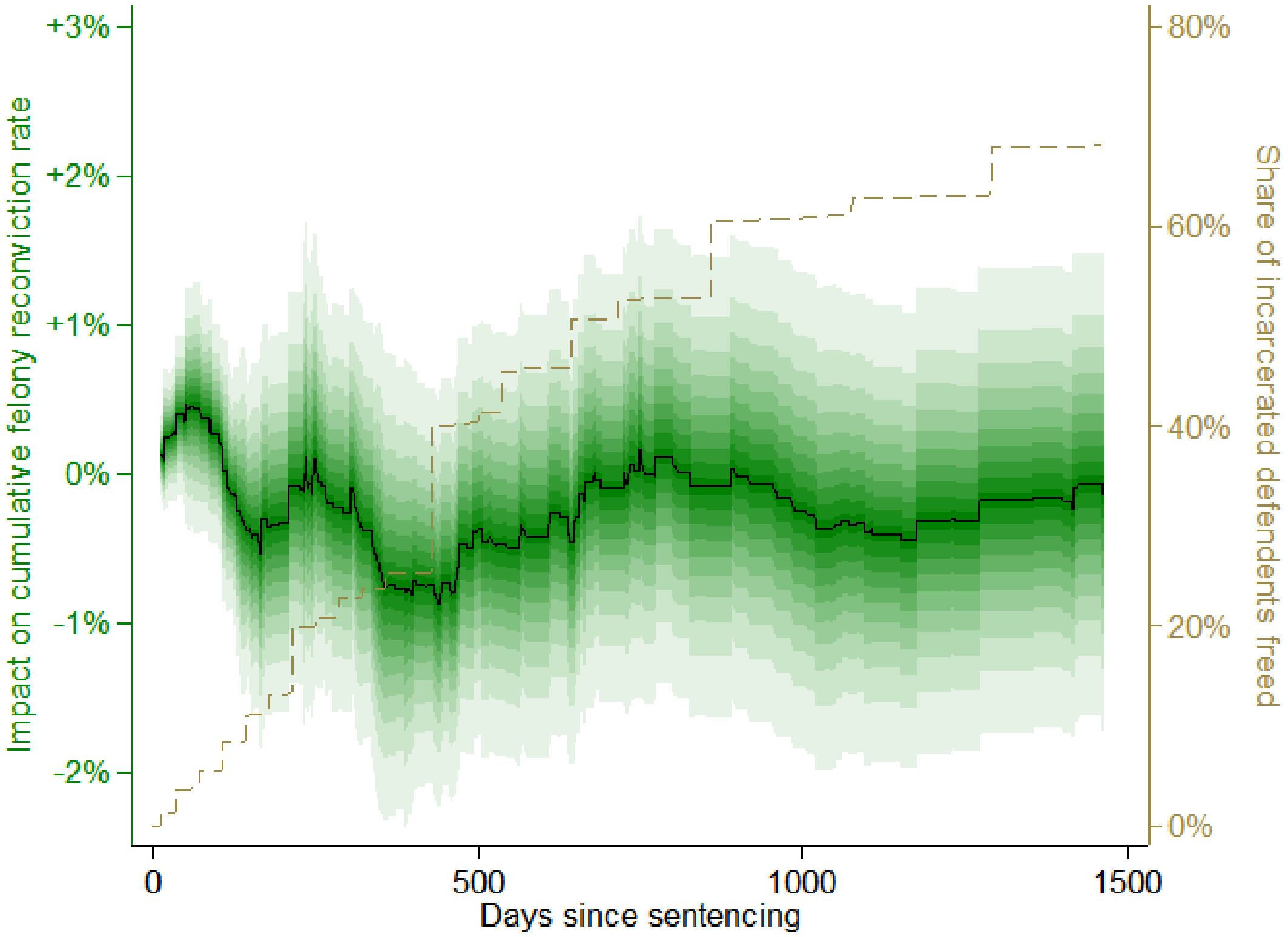

## 9.6. Loeffler (2013), "Does imprisonment alter the life course? Evidence on crime and

employment from a natural experiment," *Criminology*

What Green and Winik do in Washington, DC, this paper does in Chicago. The study design differs mostly in small ways. The follow-up period is five years instead of four. The outcome is whether arrested for a felony rather than for any infraction. The sample, from 2000–03, is much larger, with 20,297 defendants instead of 1,003. And the Cook County Circuit Court had more judges: 25 instead of nine.

In order to limit incapacitation's contribution to the measured impact, the sample is restricted to people in the three lowest charge classes. "In practice, this produced a sample in which relatively less serious offenses—including drug possession, larceny, and weapons charges—were overrepresented relative to the Circuit Court as a whole" (p. 10).

Perhaps the largest design difference from Green and Winik is that the treatment variable indicates only whether a defendant was sentenced to prison, not how long. Which is better? Two issues arise. First, *whether* and *how much* a person is incarcerated might have a low or even negative correlation: perhaps judges who sentenced to prison more often sentenced to prison for fewer months, and enough so that their average prison spells, factoring in the zeroes for those put on probation, were lower. Then, switching from incarceration frequency to incarceration amount would flip the impact results. As far as I saw, the paper does not provide information to check this possibility, which admittedly seems unlikely. On the other hand is the second issue, the possibility of nonlinear impacts. Perhaps the first month of prison is what leaves lasting effects, with additional time making little difference to post-release behavior. If so, then Loeffler's treatment dummy would adequately capture the relevant aspect of each sentence. If the marginal impact of doing time declines more slowly, then the quantity of time served looks more relevant.

On balance, the "how much" looks more meaningful because it contains more information, and information bestows statistical power.

Because Loeffler evaluates impacts on employment as well as rearrest, it appears that a third of the sample was dropped for lack of employment information (p. 11)—even in the rearrest regressions, which do not need this information. (The recidivism and employment regressions contain the same number of observations.) If this attrition was significantly correlated with potential outcomes, then it could bias the results. Perhaps the risk is small; but it also appears to have been avoidable.

Figure 28 distills the paper's main results on the impact of incarceration on crime by plotting the fraction of each judge's defendants who are arrested within a given year against the fraction sentenced to prison. Scanning horizontally, we see that the average of the binary treatment indicator ranges across judges from 26% to 47% (Loeffler, Figure 2a), which suggests cross-judge variation adequate for a quasi-experiment. More formally an F test of the treatment indicators as predictors of assignment to prison returns 6.49 ($p <$ .0001; p. 14). But the treatment does not correlate with the outcome, which is confined to the 60–70% range across all judges. Regressions (Loeffler, Table 2, cols. 3–4) confirm the lack of impact.

Parole bias is again unlikely to figure because since the late 1970s the prison system of Illinois has operated under *determinate sentencing*, which all but eliminated parole (Washington University Law Review 1979).

Unlike Green and Winik, Loeffler also assesses the impact on employment, specifically whether someone had a job five years after indictment. The story is the same: a graph and regressions find no impact. This again suggests that aftereffects cancel out incapacitation.

The findings are plausible. My one doubt is that the study's bottom line reflects not just lack of effect, but lack of power. A binary indicator of whether arrest happens at all over a long time—arrest over five years—contains less information on total crime impact than what is available in the public Green and Winik dataset, with its information on timing of first arrests and convictions after release. An unfairly extreme example that illustrates the concern would be a study of the impacts of smoking on 100-year survival rates. And as for the outcome variable, so for the treatment variable. The indicator of *whether* someone is incarcerated also reduces power. Since a crucial and important finding from studies of this ilk is that incarceration aftereffects offset incapacitation, it is valuable to minimize the risk that this impression arises from lack of power.

**Figure 28. Five-year rearrest rate vs. percent sentenced to prison, by judge, Cook County, 2000–03, from Loeffler (2013)**

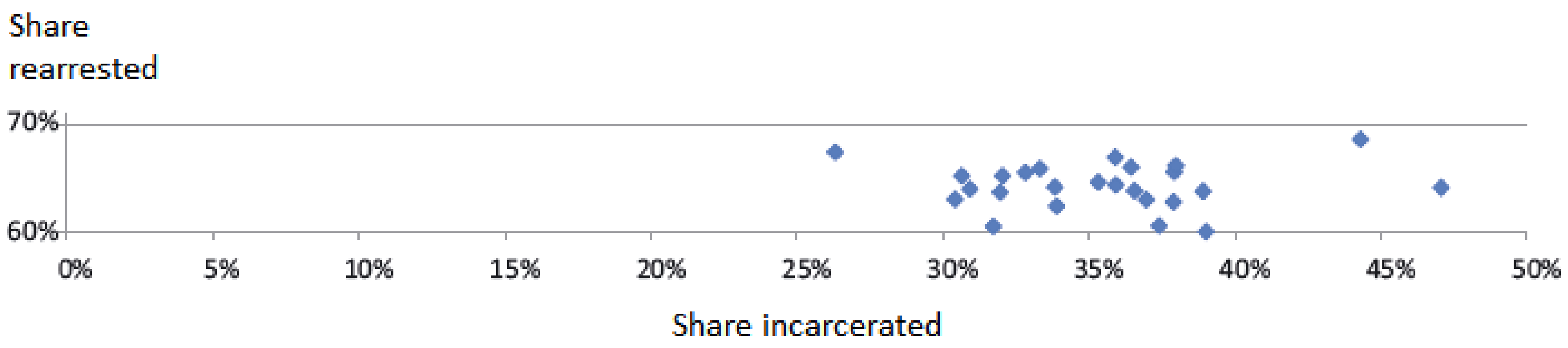

## 9.7. Nagin and Snodgrass (2013), "The effect of incarceration on re-offending: Evidence from a natural experiment in Pennsylvania," *Journal of Quantitative Criminology*

This study too resembles its predecessor within this review. The treatment and outcome variables are again binary. The research question is whether being incarcerated affects the chance of any rearrest over one, two, five, or ten years. And again no information is included on average sentence length, which could facilitate checking its relationship with incarceration frequency. Unlike the earlier studies, however, the authors state that assignment to judges was truly random (p. 612).

The authors sample 6,515 defendants who entered one of five Pennsylvania county court systems in 1999, each of which practiced random courtroom assignment. The five counties were selected from the state's 67 on several criteria, including cross-judge similarity on observable defendant traits such as age and race—meaning that randomization succeeded in creating statistical balance—and cross-judge diversity in sentencing severity (p. 605, 608). For a good quasi-experiment, defendants needed to look similar going into the courtrooms, yet carry different burdens coming out.

The paper formally departs from a clean quasi-experimental design in filtering the sample *after* randomization, and on a trait that might be correlated with both treatment and outcome. Possibly for lack of data, only convicted people are studied (p. 608). In principle, this filtering could bias results. For example, suppose incarceration does not affect criminality. And suppose that, of two judges receiving identical defendant pools, one convicts more and sentences more the convicted to prison. The borderline defendants—the ones convicted only by the tough judge—could be ones with lower criminal propensity. In the Nagin and Snodgrass set-up, they would enter only the tough judge's sample, reducing the average criminal propensity of that sample. Thus, higher incarceration frequency would appear to go with lower recidivism. This bias would make the impact estimates conservative from the point of view of the Open Philanthropy Project as a criminal justice reform funder, and for Daniel Nagin, who argues that incarceration probably increases recidivism (Nagin, Cullen, and Jonson 2009; Cullen, Jonson, and Nagin 2011).

However, Daniel Nagin states in e-mail that while Pennsylvania "does not publish statistics on conviction rates, they are extremely high—certainly over 90%." Approaching 100% leaves little room for such bias.

The study also resembles its predecessors in this report in discerning no overall impact. As an example, Figure 29 shows rearrest versus incarceration rates for the seven judges in the Dauphin County court system (Nagin and Snodgrass, Figure 4). The grey bars show incarceration rates, and are the same in all four panes; within each pane they vary substantially, with two judges looking much more lenient than the rest. The black bars are rearrest rates, and are effectively flat across all four panes. Regressions for each county and for all at once confirm the lack of correlation (Nagin and Snodgrass, Tables 6 & 7). To deal with the possibility of weak instruments, Nagin and Snodgrass (pp. 610–11) construct confidence intervals using a method equivalent to graphical Anderson-Rubin, without calling it that.

Despite the same concern about power as in Loeffler, the similarity of these results to those in Green and Winik and Loeffler, and their basis in true randomization makes these results credible.

**Figure 29. Rearrest rate over one, two, five, and ten years vs. incarceration rate, by judge, Dauphin County, PA, from Nagin and Snodgrass (2013)**

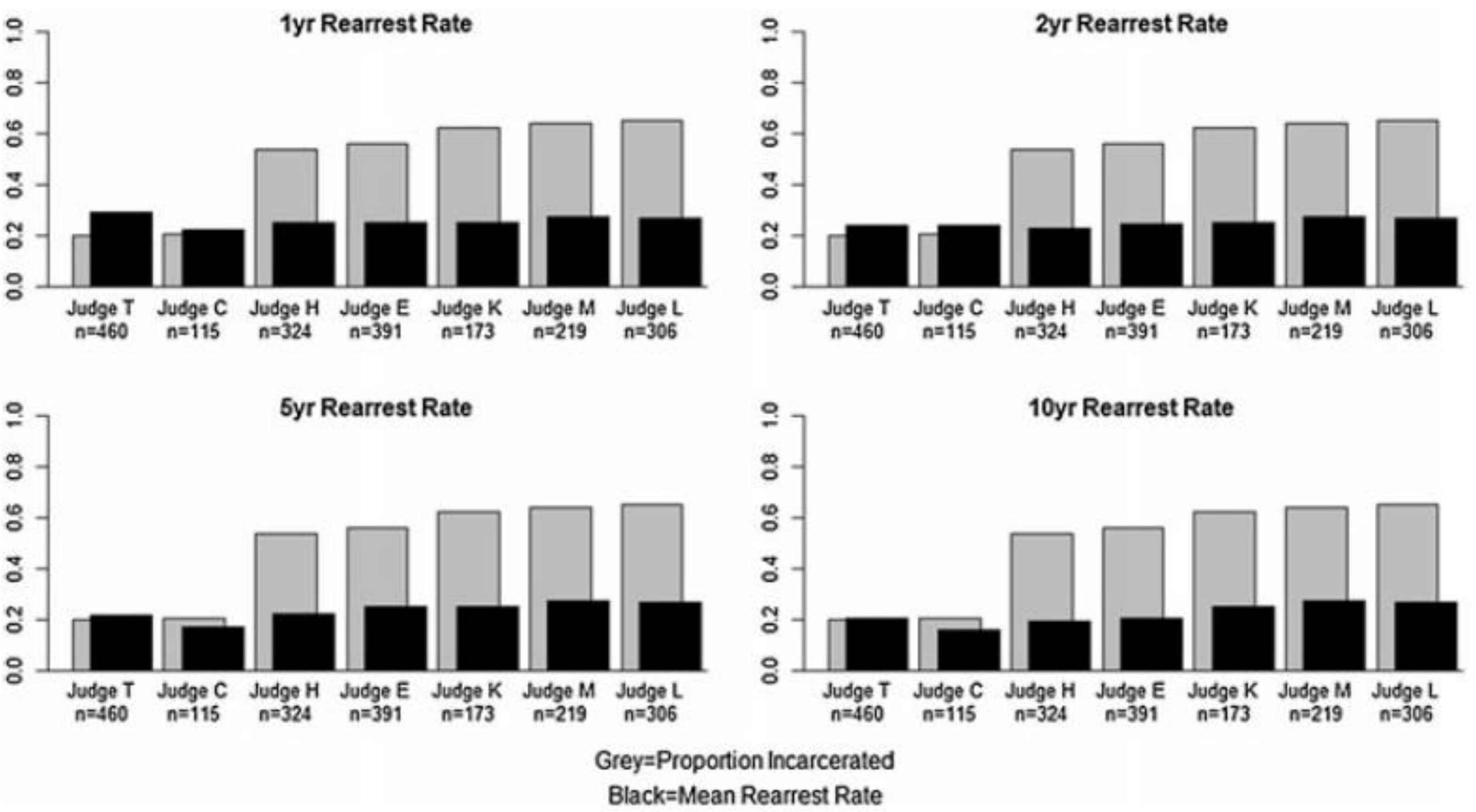

## 9.8. Roach and Schanzenbach (2015), "The effect of prison sentence length on recidivism: Evidence from random judge assignment," working paper

This judge randomization study takes place in King County, which includes Seattle. There, defendants who are convicted are assigned new judges for sentencing (p. 5).

It is not certain that the assignment process produced a good experiment. As usual, the process was not literally random. Roach and Schanzenbach's describe it most fully in quoting a government manual:

> If a defendant pleads guilty in the plea court, at omnibus or at case scheduling, the case shall be assigned a sentencing judge by the Criminal Department Sentencing Coordinator(s)… Sentencing hearings are set by the Sentencing Coordinators for Friday afternoons at three times: 1:00, 1:45, and 2:45 p.m. An average of four sentencing hearings are set in each time slot….The Sentencing Coordinators shall endeavor to assign sentencing hearings equally among all criminal and civil department judges and will assign each judge no more than twelve defendants for mainstream sentencing hearings…. The Sentencing Coordinator assigns a sentencing judge and a sentencing date immediately after the defendant enters a guilty plea or is found guilty. (p. 6)

This requires immediate assignment and balance in quantity across judges, but does not, in effect, specify how arbitrarily assignment is to be achieved. In a bid to maximize the arbitrariness, Roach and Schanzenbach only include people who were convicted and sentenced on the same day, Friday, on the idea that this allowed less time for non-arbitrary factors to influence a person's assignment to a sentencing judge.

This filter yields a sample of about 8,000 defendants convicted between 1999 and 2011, who were sentenced to 9 months on average (Roach and Schanzenbach, Table 1, row 1).

Despite this step, of the two King County court locations, Roach and Schanzenbach (p. 6) discard one as a data source because its samples differ statistically across judges. In the data from the retained location, the Kent Regional Justice Center, F tests do not point to cross-judge differences on age, race, or gender; but they do suggest differences in offense severity, prior nonviolent convictions, and, possibly, prior serious violent convictions ($p = 0.051, 0.066, 0.203$; Roach and Schanzenbach, Table 2). It is not clear whether those deviations from balance correlate with judges' sentencing severity in a way that could explain the study's results.

Roach and Schanzenbach graph their estimated judge effects, i.e., the average sentence handed down by each judge, expressed for convenience relative the most lenient judge. They also plot 95% confidence intervals. (See Figure 30, based on the paper's Figure 2.) Most of the judges do not differ from each other in average sentence—at least not with great statistical significance—but the most lenient and most severe differ clearly. (Whether this diversity suffices to prevent problematic instrument weakness is also unclear because the authors do not report weak identification diagnostics.)

In addition to instrumenting sentencing with the judge dummies, Roach and Schanzenbach run separate regressions that instrument with a single "leave-out" treatment variable. It is, for each week and judge, the judge's deviation from the cross-judge average sentence in all *other* weeks. Leaving out the current week prevents an outlier observation, such as a bunch of 5-year sentences from one judge in one week, from throwing the average against which it is benchmarked.

Perhaps the most straightforward and reliable regressions are those of recidivism—defined as reappearance in the King County Superior Courts—directly on that judge severity variable. These "reduced form" regressions do not instrument actual sentence length with each judge's average, and so avoid any concern about instrument weakness. Instead they directly correlate recidivism with average judge severity.

This is the one judge randomization study finding negative aftereffects, meaning more time leading to lower crime. The regressions suggest that each month of additional sentence reduced one-, two-, and three-year recidivism by 1.17, 1.06, and 1.33 percentage points (se = 0.403%, 0.487%, 0.547%; Roach and Schanzenbach, Table 4, row 2). Compare those impacts to the sample averages of 12%, 20%, and 25% (Roach and Schanzenbach, Table 1). Since the reductions look the same over the three timeframes—one percentage point per month sentenced—it seems that any impact from incarceration is short-lived, playing out in the first year.

Both the magnitude and the transience of the impact make it unlikely to be caused by aging. Nor does parole bias appear to be a factor as in there is no parole in Washington (p. 7).

The possibility of parole bias and the apparent imbalance between the treatment and control groups prevent me from fully relying on the paper.

**Figure 30. Average incarceration sentence by judge, relative to judge 25, 1999–2011, Kent Regional Justice Center, King County, WA, from Roach and Schanzenbach (2015)**

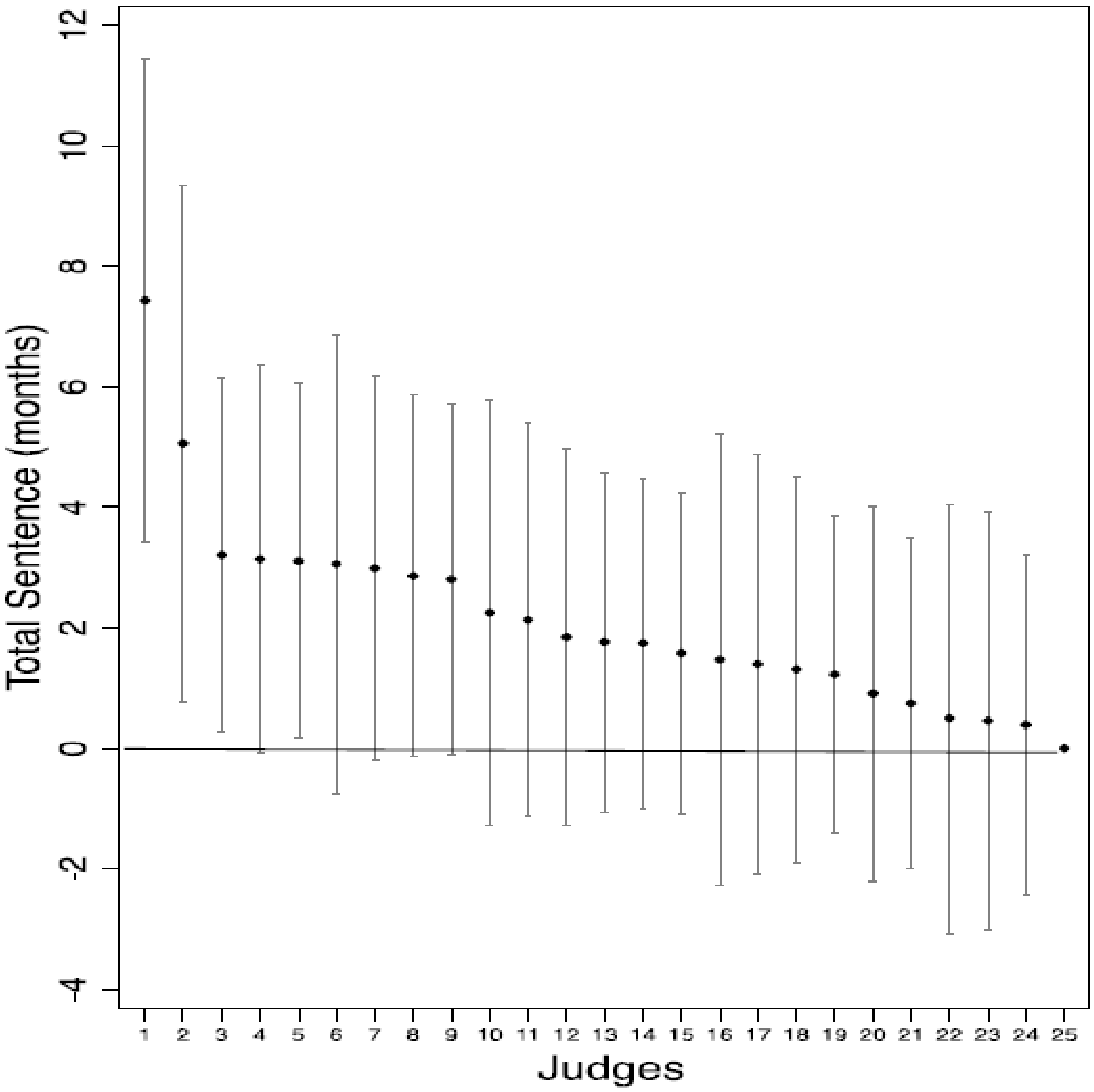

## 9.9. Mueller-Smith (2015), "The criminal and labor market impacts of incarceration," working paper

This is a formidable new judge-randomization study. With data on some 450,000 felony defendants—tried between 1980 and 2009 in the courts of Harris County, which includes Houston—the sample exceeds those of all the other individual-level studies reviewed here, combined. The study also links in other government data sets in order to check impacts on employment, earnings, and recidivism (p. 11, 25). Recidivism is defined several ways: whether booked in county jail, whether charged in Harris County court, whether convicted anywhere in Texas (p.23). And while I have classed the study under "aftereffects," it estimates incapacitation too (p. 3).

The study beings by making the case that two econometric issues may distort other judge randomization studies. Both go to the idea that the severity of a judge is multi-dimensional. The first is what Mueller-Smith calls "omitted treatment bias." The "treatment" of a criminal trial has many facets: whether a person is incarcerated, and how long; whether put on probation, and how long; whether assessed a fine, and how much; and so on. Attributing an increment of recidivism to one dimension while ignoring the others can produce misleading results (pp. 13–14). (This issue arose in the discussion of Martin, Annan, and Forst, where judges applied both fines and jail time, and in Green and Winik, which considered probation as well as incarceration.) Mueller-Smith (Table B.1) shows an example using real data from Harris County. A stripped-down regression instrumenting with courtroom dummies shows that being sentenced to prison raised the chance of being charged with a new crime within one year by six percentage points. However, if the *length* of incarceration is then controlled for, the impact jumps to 15 points. Why? The judges who sentenced defendants to prison more often also did so for longer, on average. And, over the first year in prison, longer sentences led to lower recidivism, probably because of incapacitation. Until controlled for, this correlated but countervailing effect masked much of the impact of the incarceration decision per se on recidivism. In this simplified example, Mueller-Smith finds that when additional treatment dimensions such as fines and probation are factored out, incarceration per se leads to 26 percentage points more in court (re)appearances.

To prevent omitted treatment bias, the regressions include treatments of secondary interest, such as whether the sentence included a probation or a fine.[86] These multiple treatment variables demand multiple instruments, which Mueller-Smith provides in the hundreds. All are products of three factors: a dummy for some courtroom trait such as which judge presided or which chief or assistant prosecutor participated; a dummy for the type of crime the defendant is charged with, a distinction that can increase monotonicity, as we saw; and a defendant trait such as age, race, or total prior felony convictions. All are exogenous, provided the instruments—the judge and prosecutor dummies—are valid, and provided that the variables for crime type and individual traits are all controlled for in the main regression equation. In e-mail, Mueller-Smith confirmed that this is the case, and reported that 464 instruments are generated altogether.

Mueller-Smith's second econometric concern is "non-monotonicity." Most instrumental variables studies implicitly assume that the instrument (such as the average severity of a judge) does not simultaneously raise the actual treatment amount for some subjects while reducing it for others (Angrist and Imbens 1995, p. 435). Violations of this assumption can blow up impact estimates. To see why, consider an example adapted from Mueller-Smith. Suppose two judges receive identical caseloads: 100 DUI offenders and 100 drug possession offenders. Judge A imprisons 50 of the DUI convicts and 25 of the drug offenders. Judge B does the reverse, so that they both incarcerate 75 people, and appear equally severe overall. Now suppose that incarceration raises recidivism by 10% on average, but not exactly, because there is an element of randomness. Then recidivism will differ slightly across the two judges' caseloads, say, by 1%. The mathematics of instrumental variables will estimate the treatment effect as the difference in recidivism per unit of difference in incarceration: 1% divided by 0%, which is infinity. But if one were to separately study the DUI and drug samples, one would find little difference in recidivism, and in both samples estimate an impact of about zero, not infinity. The root of the problem is that appearing before Judge A instead of B can raise or lower one's odds of conviction, depending on the crime. That is non-monotonicity.

In addition to highlighting and confronting these two econometric complications, the study introduces methodological innovations aimed at disentangling incapacitation from incarceration aftereffects. It constructs a "panel" data set with one data point per defendant and quarter-year since initial criminal charge.[87] Where previous judge randomization studies are cross-sections, relating whether or how long a person is incarcerated to whether or how much the person recidivates over some time frame, the panel set-up allows Mueller-Smith to relate whether people recidivate in a given quarter to whether they are incarcerated then; as well as to *whether* they were incarcerated in the last five years; and, if now free, *how long* they were incarcerated. The first of these captures incapacitation, the second dependence of aftereffects on incarceration per se, the third their dependence on the length of incarceration.

With more than 400,000 defendants and 13 million defendant-quarters in the regressions of primary interest here, 464 instruments is arguably too few to cause maladies associated with instrument proliferation (Roodman 2009). Nevertheless, Mueller-Smith applies a modern method called "Post-LASSO" (Belloni and Chernozhukov 2013) to prune the instruments, selecting the those that best predict each treatment (pp. 50–

[86] To reduce computation time, Mueller-Smith (p. 14) partials out these "non-focal" treatment variables in an initial stage. This preserves point estimates for the focal treatments, but, contrary to the August 2015 draft I read, does change the standard errors.

[87] Standard errors are clustered by defendant and dummies are included for time of being charged and number of quarters since.

52). However, Mueller-Smith (Table D.7, col. 1) also shows that skipping this step preserves the texture of the results.

Before estimating impacts, Mueller-Smith performs the standard step of testing whether the study's quasi-randomization is successful. The results seem to put a question mark over the study's conclusions. If randomization is successful, then the groups of people appearing before various judges in various years should look demographically similar. In fact, they don't. Mueller-Smith (Table 2) reports F tests for similarity in the fraction that is female, the fraction that is Caucasian, and so on, but does not provide the associated degrees of freedom needed for a strict interpretation of these tests. The text (p. 10) states that "The test statistics for defendant characteristics generally range between 1 and 1.4. These indicate a technical rejection of the null hypothesis" of no difference. But it goes on to argue that the groups differ much more on sentencing outcomes; judges really do differ. "Together these results indicate that assigned caseloads look very similar ex-ante but quite different ex-post." But it seems to me that if the sample is large enough that small ex-ante differences in observable traits are detectable, then it is large enough that small differences in unobservable traits can create the appearance of statistically significant impacts.

Mueller-Smith separately studies people charged with misdemeanors or felonies.[88] I focus on the latter because felony convictions are more serious and contribute more to prison growth and crime worries in the US. The felony defendants were 81% male, 30% Caucasian, 46% African American, 23% Hispanic, and 30.26 years old on average at time of charge (Mueller-Smith, Table 1). Average time sentenced or served appears not to be reported, but Mueller-Smith (Figure 4) shows the paper's results come mainly from differences among people who served less than one year, meaning that the results should be read as pertaining most certainly to that range.

Table 15 contains some of those results. They strongly suggest that incarceration in Houston increased crime through the incapacitation and aftereffects channels, at the margin. Each of the table's columns corresponds to a different definition of recidivism. The first row shows, unsurprisingly, that being in prison reduces crime outside of prison. This incapacitation is estimated about 3 percentage points each quarter, depending on the definition, meaning that about 3 percent of those incarcerated would booked, charged, or convicted per quarter. The second row checks the post-release impact of incarceration per se, and finds none. This can be interpreted to mean that being incarcerated for a tiny spell, such as one day, does not affect recidivism. The last row suggests that the aftereffects as a function of time served, however, are substantially negative: for each additional year behind bars, post-release criminality is 3.6–6.7 percentage higher in each quarter. So, for example, going by the penultimate column, a person who spends a year in prison is 2.8 points less likely to be convicted of a new crime in each following quarter, for a one-year crime reduction of some 4 × 2.8% = 11.2%. But, once free, the person commits 3.6 points more crime in each successive quarter, on average, so after 11.2% / 3.6% = 3.1 quarters of freedom, barely nine months, the harmful criminality aftereffects surpass the incapacitative benefits.

By the standards of the literature, Mueller-Smith finds little incapacitation. For example, the 3.4% per-quarter impact found on being charged in court for a new felony (top row, second column of Table 15) implies that former defendants return to court for felonies about 4 × 3.4% = 0.136 times per year. Contrast that with the Owens (Table 5) incapacitation estimate for Baltimore, 2.79 arrests per year, which itself comes in much lower than other incapacitation studies reviewed above. Why the difference? Mueller-Smith (note 23) points out that these numbers are dominated by the experiences of people who were on the margin of any incarceration, i.e., who would have been incarcerated by some judges and not others. They may tend to commit crime than the people in Owens's study, 79% of whom had been incarcerated for some period (Owens, Table 1). The Mueller-Smith paper does not report the share of its felony defendant sample that was incarcerated.

---

[88] Separate court systems handle misdemeanors and felonies, so this splitting occurs before courtroom assignment, and partitioning the econometrics along the same line does not jeopardize the quasi-experiment.

Meanwhile, possibly for the same reason, the aftereffects more harmful than in most other studies. Where Green and Winik tentatively estimated that an extra year of prison raised the four-year any-rearrest rate by 3.02 percentage points (p. 380), Mueller-Smith finds 6.7% more rearrests every *quarter* (see Table 15, row 3, col. 1).

In comparative context, the low incapacitation and strong aftereffect estimates are what lead Mueller-Smith to conclude that incarceration has almost certainly increases crime at the margin in Houston.

It is hard to see how parole bias could cause the apparent aftereffects. Those sentenced to longer terms probably spent more time on parole. In turn, the added risk of parole revocation *may* have elevated the risk of being booked in county jail, Mueller-Smith's first definition of recidivism. But extra time on parole would probably not inflate recidivism defined the other ways: charged with a new offense in Harris County or convicted of a new offense anywhere in Texas.

The sheer size and sophistication of the Mueller-Smith study earn my respect. Many robustness checks are carried out. And the results are consistent across definitions of recidivism. No other study has so assessed the "during" and "after" effects of incarceration on crime in such a high-powered way.

Yet other traits make me cautious. There is a hint of randomization failure. The conclusions are extreme within the literature. The methods are complex. The study is opaque in that the data and code are inaccessible, and the working paper omits many details.[89] It is hard to see exactly what is going on with the large number of triple-interaction instruments, the modern method of synthesizing them into a single instrument for each treatment, and the many implementation details such as quarterly or biannual recalculation of instruments. Perhaps after the study is published in a journal, the data and code will become more accessible and these concerns can be probed.

**Table 15. Quarterly impact of current incarceration status and past incarceration history on recidivism defined three ways, felony defendants, Harris County, TX, from Mueller-Smith (2015)**

| | Booked in Harris County jail | Charged in county criminal court with new felony | Convicted anywhere in Texas | Earnings ($/quarter) |
|---|---|---|---|---|
| (Still) in jail or prison | –0.033*** | –0.034*** | –0.028*** | –1632.1*** |
| | (0.0080) | (0.0047) | (0.0074) | (293.0) |
| If free, whether incarcerated before | 0.0038 | –0.0022 | –0.00071 | –683.5** |
| | (0.0074) | (0.0046) | (0.0058) | (345.3) |
| If free, years incarcerated before | 0.067*** | 0.047*** | 0.036*** | –246.5 |
| | (0.0058) | (0.0041) | (0.0047) | (150.3) |

Standard errors clustered by defendant in parentheses. Regressions include crime type and defendant trait controls, and dummies for time of charge and quarters since. ***significant at p<.01 ** significant at p<.05.
Source: Mueller-Smith (Tables 4, 5, 7).

## 9.10. Dobbie, Golding, and Yang (2016), "The effects of pre-trial detention on conviction, future crime, and employment: Evidence from randomly assigned judges," working paper; Gupta, Hansman, and Frenchman (2016), "The heavy costs of high bail: Evidence from judge randomization," *Journal of Legal Studies*

Between May and November of 2016 there appeared a spate of remarkably similar papers on a stage of the criminal justice process that is otherwise neglected in this review: that between arrest and trial. Nearly two-thirds of the 745,000 people in US jails as of mid-2014—468,000—were not serving time for a conviction, but awaiting trial (Minton and Zheng 2015, Table 3). Some had been detained unconditionally by a judge while others remained in jail because they could not make bail.

[89] I did not find (precise) statements of the average incarceration spell, the incarceration rate, the reference year of the presumably inflation-adjusted income variables, the control set, the list of crime groups, the lists of focal and non-focal treatments, the instrument sets, the numbers of instruments fed into Post-LASSO and the numbers retained, the degrees of freedom in the F tests of statistical balance, and the partial correlations across the many LASSO-constructed instruments.

Five new papers examine the consequences of bail and pre-trial detention for such outcomes as conviction, sentence length, employment, and recidivism. All are set in large cities and most exploit judge randomization. Two are passed over here because they do not examine crime impacts (Stevenson 2016 in Philadelphia; Leslie and Pope 2017 in New York) while another does not exploit what I consider a strong natural experiment (Heaton, Mayson, and Stevenson 2016 in Harris County).[90] Of the remaining two, Gupta, Hansman, and Frenchman (2016) analyzes data from Philadelphia, whose system supports a judge randomization design (Table 3), and Pittsburgh, whose system does not (Table A2); while Dobbie, Golding, and Yang (2016) study Philadelphia and Miami-Dade County, both of which support a judge randomization approach. If we set aside Pittsburgh for lack of a strong experiment, then the second study effectively subsumes the first. (The second also uses more years of data, 2007–14 instead of 2010–15.)

Dobbie, Goldin, and Yang (p. 14) construct their "leave-out" judge randomization instrument much as do Roach and Schanzenbach. For each person arrested, it is the fraction of *other* people appearing before the same judge in the same year who obtained release by either direct judicial fiat or making bail.[91,92] A randomization check—examining whether leniency so measured is correlated with race, sex, priors, and other variables—returns a reassuring p value of 0.72 (Dobbie, Goldin, and Yang, January 2017 version, Table 3, bottom right).[93]

With the instrument in hand, Dobbie, Goldin, and Yang then estimate the impacts of attaining pre-trial release on downstream events, from conviction weeks later to employment years later. Like all the papers in this set, they find that obtaining quick release before trial reduces the odds of being convicted at trial. While all this new and strong evidence of this causal link matters for policy, it speaks to this review mainly by bolstering the credibility of all the studies. If they failed to reach consensus, that would raise questions about the reliability of their results, or at least signal that impacts vary too much to allow generalization.

More to the point for us are Dobbie, Goldin, and Yang's estimates of the impacts of pre-trial detention on subsequent arrest. Pre-trial detention cut the share of people rearrested in the two years following the bail hearing by 4.1 percentage points (January 2017 version, Table 4, col. 7; se = 5.3%), against a base of 40.4% (January 2017 version, Table 1). The standard error here is large: we cannot easily rule out net impacts of several points up or down. Since the two years in general contain periods of detention and periods of freedom, these findings resemble those of Green and Winik, among others, in whose work incapacitation and aftereffects combine for a net effect indistinguishable from zero.

In a bid to distinguish incapacitation from aftereffects, Dobbie, Goldin, and Yang rerun the regressions after splitting the follow-up period into two parts: before and after a person's case is decided. In the "before" part, which averages a bit over 200 days (January 2017 version, Table A6), pre-trial detention lowered the fraction arrested by 13.4 percentage points (January 2017 version, Table 4, col. 7; se = 4.4%). That looks like incapacitation. But in the reminder of the time—from case disposition to the two-year anniversary of the bail hearing—having experienced pre-trial detention lifted the probability of any rearrest by 15.0 points (se = 4.5%), a roughly 50% reduction from baseline. That suggests that pre-trial detention

[90] Heaton, Mayson, and Stevenson use two designs: ordinary least squares with many controls and a natural experiment exploiting the fact that people who bail hearings are later in the week are more likely to make bail. The latter design assumes there are factors that cause weekly cyclicity in the ability to make bail, such as when paychecks come, but not that cause weekly cyclicity in the sorts of people who are charged with crimes. This assumption seems relatively strong for a natural experiment design, so I find the results less credible than those from judge randomization designs.

[91] Release is defined as exiting jail within three days of a bail hearing.

[92] Because certain crimes might occur more at certain points in the day, week, or month, and because judges' periods of duty might do the same, the leniency measure is then demeaned for each combination of court, year, and day of week; each combination of court, month, and day of week; and, in Philadelphia, each combination of day of week and bail shift, with three shifts per day.

[93] Crystal Yang sent me a revised version of the paper, dated January 2017, which is not publicly posted at this writing.

caused harmful crime aftereffects.[94]

Gupta, Hansman, and Frenchman (Table 10, col. 1) come to similar findings for Philadelphia. Having money bail imposed, which often led to pre-trial detention, increased recidivism by 0.7 percentage points per year (se = 0.8%), where recidivism is defined as having a new charge filed. The authors then add data from Pittsburgh (Gupta, Hansman, and Frenchman, Table 10, col. 2), which leaves the point estimate unchanged but halves the standard error, bringing statistical significance as conventionally defined. However, this gain only comes about by mixing in the "nonrandom judicial assignment" (p. 491) data from Pittsburgh, so I leave this result aside. (I believe the paper's abstract and introduction obscure this trade-off and create the false impression that high-quality evidence shows that imposing money bail raises recidivism.)

Overall, these studies tend to corroborate Green and Winik and Nagin and Snodgrass, both reviewed earlier, The combined effect of incarceration on recidivism via incapacitation and aftereffects is indistinguishable from zero. Incarceration certainly reduces crime outside prison as long as it lasts, but appears to cause more crime later.

## 9.11. Bhuller et al. (2016), "Incarceration, Recidivism and Employment," working paper

To this point in the review, the judge randomization studies have worked at the level of the city—or in the case of Nagin and Snodgrass, the county. Bhuller et al. (2016) brings judge randomization to a *nation*. Admittedly, this is not as impressive as it may seem, since the nation is Norway, which barely surpasses Harris and Cook counties in population, and whose court system presumably sees fewer defendants.[95] At any rate, the study is compelling in its execution. And in contrast with nearly all the US-based judge randomization studies, it documents that in the Norway during 2005–14, being incarcerated *reduced* recidivism. The effect appears confined to people who were not working before they went to prison. Norway's in-prison rehabilitation programs appear to help them gain work after release and leave crime behind.

The study's sample consists of criminal cases processed in Norway during 2005–09 in which the defendant did not confess, i.e., did not plead guilty, sending the case to trial. These instances numbered 33,509.[96] Bhuller et al. confirm the validity of their instrument (p = 0.917 on overall balance test, Table 1) and its strength with respect to the treatment variable, which is whether a defendant was incarcerated after trial (F = 42.04, Table A11).

The authors then present their main results in a graph much like mine in the reanalysis of Green and Winik (Figure 27 and Figure 28, above). They plot the impact of being sentenced to incarceration on the cumulative likelihood of recidivism, by month, over the five years following trial (see Figure 31, derived from Figure 4a in the original; recidivism means facing a new charge; dotted lines show 90% confidence intervals). As expected, the impact starts near zero just after the trial, and then descends, to about –25 percentage points by two years. Incapacitation presumably explains part of the descent. However, unlike in my Green and Winik graphs, the cumulative measure of recidivism does not then reverse course and trend back toward zero. It seems that unlike in Washington, DC, incarceration did not increase post-release criminality in a way that could offset incapacitation.

But Bhuller et al. dig further and find an even greater contrast with Green and Winik, as well as most other judge randomization studies. They argue that despite appearances in Figure 31, incarceration's aftereffects were better than zero: being in prison cut crime after. To reach this conclusion, they first observe that in their data, incapacitation should fade within 24 months. Norway "tries to place prisoners close to home so that they can maintain links with the families" (p. 15), which leads (remarkably, for an American reader) to waiting lists for *entering* prison. Since people sentenced to incarceration waited an average 115 days, then served an average 175 days (Bhuller et al., Table 4, cols 2 & 3), most gained freedom within a year, and nearly all did within two (see Figure 32, derived from Figure 4b in the original).[97,98] Bhuller et al. then exploit the fading of incapacitation by 24 months to focus more sharply on aftereffects. They re-run the graph shown here as Figure 31, but this time only counting new charges after the 24-month mark (Figure 33, derived from Figure 4c in the original). Where before, the impact on recidivism appears to hold steady after 24 months, now it dips anew.

---

[94] One complication in interpreting these estimates is that the dividing point between the two subperiods, the moment when a person's case was decided, could also depend on whether the person was detained pre-trial. In fact, those detained waited 42 fewer days for their trial, 200 instead of 242, which means that within the first two years following their bail hearings, they had 42 more days of freedom in which to be arrested. If each person's arrest probability was constant over time, then this would mechanically create the impression of pre-trial detention leading to more arrests after case disposition. However, the lengthening of this exposure period—from $2 \times 365 - 242$ to $2 \times 365 - 200$—is only 8.6%, which looks inadequate to explain the near-doubling in the any-arrest rate (an increase of 15 points to an average 32% for those detained; Dobbie, Golding, and Yang, January 2017 version, Table 4).

[95] This study is at lower risk of attrition bias, assuming people cross national boundaries less than they cross county boundaries.

[96] The sample is shrunk from 76,609 to 33,509 by a half-dozen restrictions meant to increase reliability (Bhuller et al., Table A1).

The simplest way to explain this counterintuitive combination of findings runs this way. Everyone who goes to prison reforms, and never recidivates. Among those who don't go, one of two post-trial patterns sets in and holds steady: some are never arrested again, other are arrested repeatedly. In this world, the unincarcerated, as a group, would pull ahead of the incarcerated in the first 24 months on the fraction ever recidivating, because all those incarcerated are first incapacitated and then reformed; that would explain Figure 31. After 24 months, the gap in the recidivism probability holds steady because only those who had *already* been charged before 24 months would face further charges. But if, as in Figure 33, we only count new charges after 24 months, then a new gap would open up as the unincarcerated regular recidivists faced their first charges in this delayed time window, while the rest still avoided entanglement with the police.

Reality is of course not so simple. But the implication stands. At least among people who are subject to the natural experiment at the heart of this study—the marginal defendants who would be imprisoned by some judges and not others—time in Norwegian prisons caused something like reform among some of them. It seems to have altered life courses by moving some from a future of repeat offense to one of almost no offense.

One testable implication of this theory is that even if the impact on *whether* people recidivate does not expand beyond 24 months (Figure 31), the impact on *how many* times they do should. Bhuller et al. (Figure 5a) confirms that this is the case, albeit with low statistical significance.

Further analysis yields insight into mechanism. Bhuller et al. split their sample by whether, in the five years leading up to the crimes for which they were tried, defendants had been employed. Variants of Figure 31 restricted to those who had been suggest that incarceration had little impact on recidivism during and after the time in prison (Bhuller et al., Figures 6a and 7a). But the benefits of incarceration were dramatic for those who were *not* working before the crime for which they were charged. A sentence of incarceration slashed their chance of any new charges in the five years after trial from 96% to 50%. And it cut the number of new charges by 22 (see Figure 34, derived from Figure 7b in the original), in a sample in which new charges averaged 9.9 (Bhuller et al., Table 5).

Why did incarceration help those who did not have work beforehand? Evidently by helping them get it after. Among those who *were* previously employed, being sentenced to incarceration immediately cut employment by 25 points (Bhuller et al., Figure 8a); and the rate never recovered even after five years. But among those previously without work, incarceration resulted in a 34-point rise in participation in job-training programs (which are offered in prison; Bhuller et al., Table 8) as well as a 40-point employment rise over five years (Figure 8b). Moreover, among those who were not previously employed, incarceration did not appreciably change the share who recidivated *and* found work within five years. But the fraction who recidivated yet were *never* employed fell dramatically: by 53.5 percentage points, against a baseline of 83% (Table A7, col. 3). In other words, the study's entire recidivism decline went hand in hand with an incarceration-reduced fall, among people who had no work beforehand, in the fraction who had no work after. A final subdivision of the sample, by whether people obtained job training after trial, generates a further refinement: the recidivism decline occurred among people who had not worked before trial, but then got job training and work.

[97] The figure only factors in incarceration resulting from the original charge, not any subsequent ones.

[98] The average incarceration *sentence* was 238 days (Bhuller et al., Table 4, col. 1) "as Norway allows individuals to be released on parole after serving about two-thirds of their prison sentence for good behavior" (p. 25). The presence of parole raises the question of whether parole bias might be at work in some form. If being on parole raises the likelihood of being charged with a new crime, and if parole time is proportional to incarceration time, then this would make the Bhuller et al. more-time-less-crime finding conservative.

In sum, the evidence drives one to the conclusion that in Norway, many people who had not been working, and who were incarcerated for a crime, got job training in prison, which helped them find work after release and thereby avoid a return to crime.

The question arises: what does Norway's experience teach us about the US? And the natural answer is: not much, because Norway is different. Indeed, a 2015 *New York Times* story about a radically humane new prison in Norway recounts this history:

> Norwegian prisons operated much like their American counterparts until 1998. That was the year Norway's Ministry of Justice reassessed the Correctional Service's goals and methods, putting the explicit focus on rehabilitating prisoners through education, job training and therapy. A second wave of change in 2007 made a priority of reintegration, with a special emphasis on helping inmates find housing and work with a steady income before they are even released. (Benko 2015)

But in US prisons since the late 1970s, retribution has dominated rehabilitation as the operating philosophy prisons (Bushway and Paternoster 2009, p. 121–24).

Certainly, then, one should doubt whether the results from Norway carry over to the US. Yet these generalizations about national differences are as easy to make as they are free, to my knowledge, of specific evidence. If the Norwegian system is said to "[focus] on rehabilitation, preparing inmates for life on the outside" (Bhuller et al., p. 4), the granular reality in the country is surely more complex and imperfect than that description implies. And while the US system deprecates rehabilitation, America is a large and diverse country, and many of its prisons still offer training programs, as discussed in the next study in this review.

So while we can doubt, we cannot rule out the possibility that the excellent Bhuller et al. study tells us something about the benefits of incarceration in the US. With more certainty, it tells us what the US *could* achieve if it ran its prisons more like Norway does.

**Figure 31. Estimated impact of being sentenced to incarceration on cumulative any-arrest probability as function of follow-up length, from Bhuller et al. (2016)**

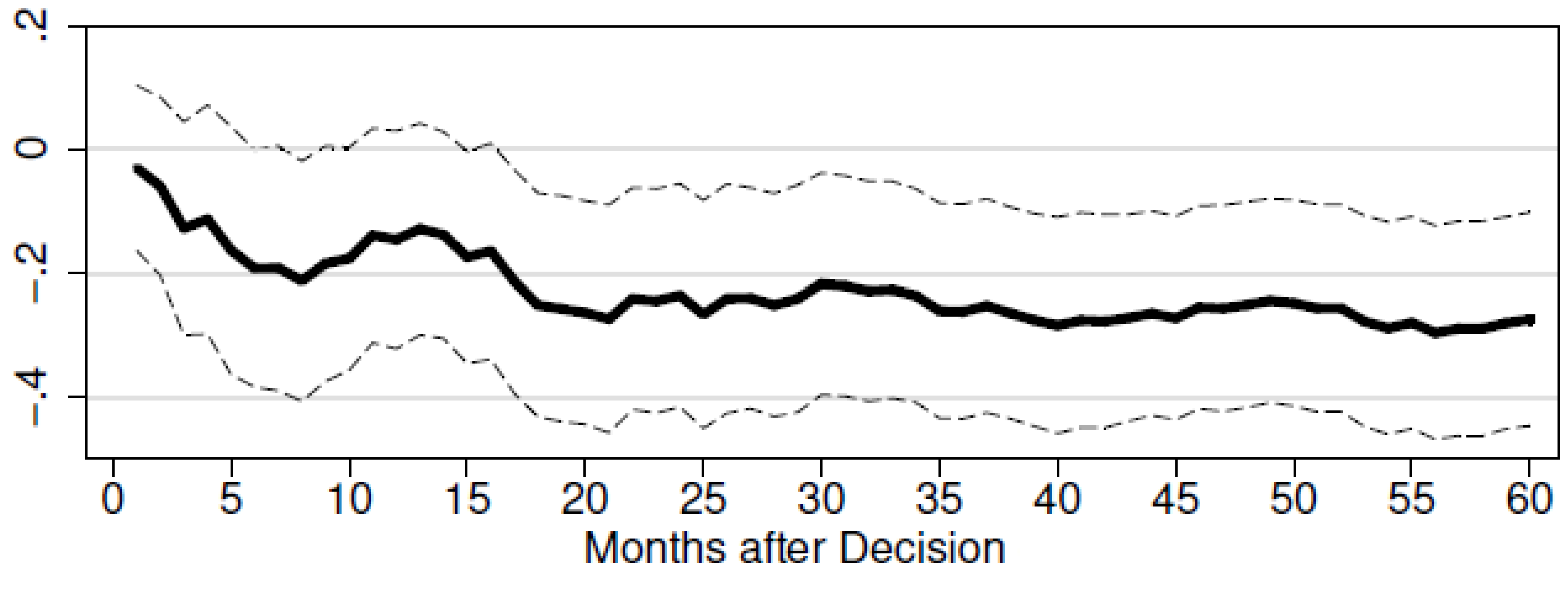

**Figure 32. Estimated impact of being sentenced to incarceration on probability of (still) being incarcerated for that charge as function of follow-up length, from Bhuller et al. (2016)**

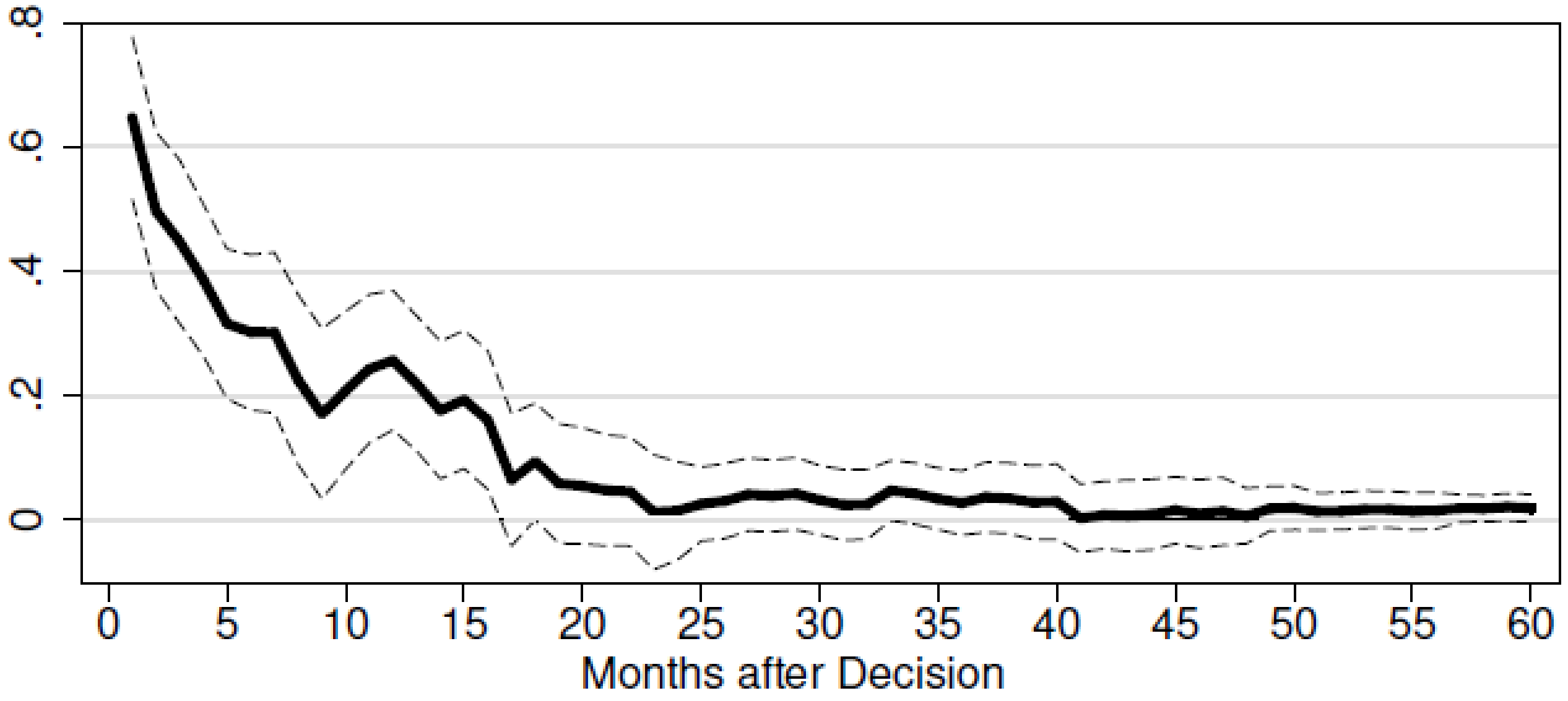

**Figure 33. Estimated impact of being sentenced to incarceration on cumulative post–24 month any-arrest probability as function of follow-up length, from Bhuller et al. (2016)**

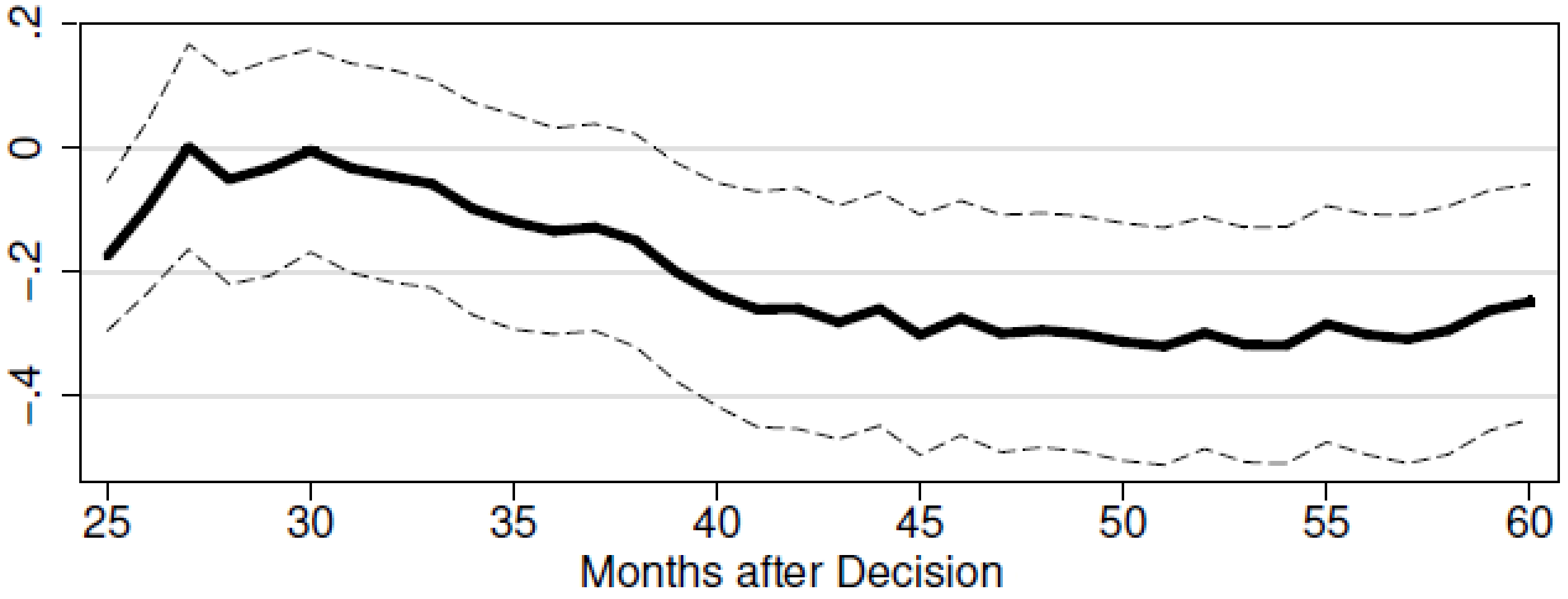

**Figure 34. Estimated impact of being sentenced to incarceration on cumulative number of rearrests as function of follow-up length, among those not previously employed, from Bhuller et al. (2016)**

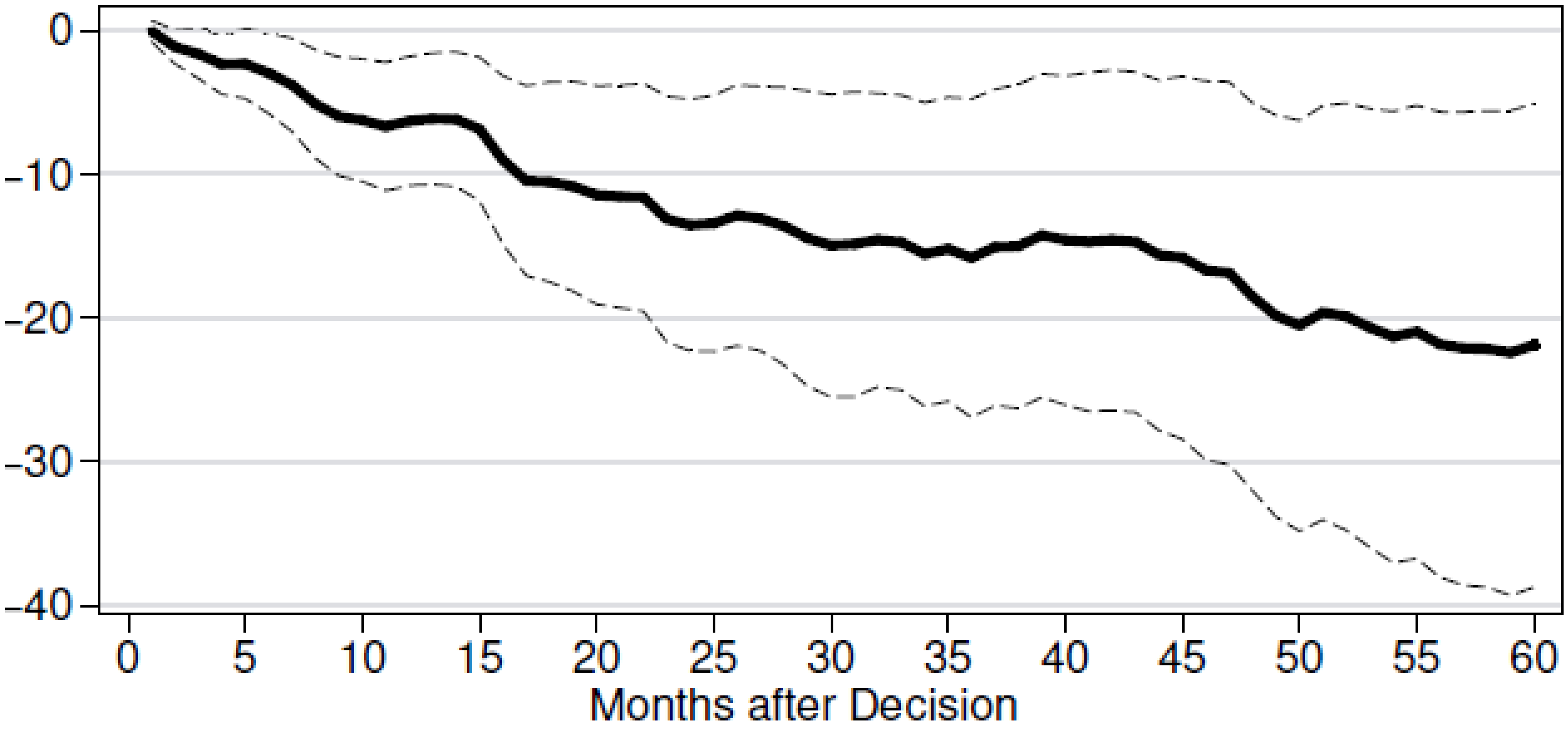

## 9.12. Kuziemko (2013), "How should inmates be released from prison? An assessment of parole versus fixed-sentence regimes," *Quarterly Journal of Economics*

This study is a remarkable four-in-one. Working with individual-level data from Georgia's prison system covering nearly 30 years, Kuziemko discerns and exploits four quasi-experiments. Each involves a different econometric approach. The study concludes that in Georgia:

- Spending more time in prison dramatically reduced recidivism, with each additional month served cutting the probability of return to prison within three years by about 1.3–3.4 percentage points.
- Giving parole boards discretion over individuals' time served is doubly constructive:
    - Parole boards predict which inmates are most likely to recidivate and adjust by retaining them longer. They improve the allocation of the scarce resource of prison space to maximize the crime reduction from incapacitation and aftereffects. Parole boards' foresight is not perfect, but improves on random guesswork.
    - Parole board discretion gives inmates an incentive for good behavior and self-improvement while incarcerated, by holding out the hope of early release. Better conduct in prison leads to lower criminality after release.

The first conclusion, on aftereffects, matters most here. Three of the four quasi-experimental analyses examine the correlation between time served and subsequent recidivism. But having obtained the author's data and code and inspected them closely, I have come to question this conclusion.

### 9.12.1. Discontinuity in parole board length of stay guidance

Since 1979, the Georgia parole board has taken guidance in setting inmate release dates from a table of recommendations called the "grid" (GSBPP 1979, p. 1). The grid has been revised several times (GSBPP 1993, p. 24; 2008, p. 18.). Its rows correspond to the severity of a prisoner's crime, with seven (later eight) categories (j.mp/1MMX69Z). The columns correspond to a "parole success score" that reflects age, prior offense record, and other pre-incarceration traits thought to predict recidivism (j.mp/1MMXdSO). The parole success score too is bracketed into categories. As of April 1, 1993, they were: poor (0–8 points), average (9–13), and excellent (14–20). The cells of the grid contain recommended lengths of prison stay, in months. For example, the 1981 grid recommended that someone convicted of a severity-level-one crime and earning a poor parole success score serve 36 months. (See upper-left corner of Table 16, which shows three editions of the grid.)

Like Chen and Shapiro (above) and Hjalmarsson (2009b, below), Kuziemko takes advantage of the discontinuities between adjacent cells to frame a quasi-experiment. Within a row of the grid, do people who end up on the high side of a threshold, and so serve less time in prison, recidivate more or less than those on the low side? As in those studies, the major caveat is that however sharp the threshold conceptually, the one-point shift to cross it is not infinitesimal, and may be associated with hidden third factors that also affect recidivism.

This concern is one reason that a standard preliminary in such analysis is to check visually for a discontinuity in the treatment, in this case time served in prison. If the treatment rate—in this case, time served—jumps a lot, we can more realistically hope that it will overshadow changes in any hidden causal factors. In the Georgia numbers, the hoped-for discontinuity does manifest, but less so than is suggested in Kuziemko (section IV.B; Figure II). Figure 35 plots average and median months served as a function of parole success score in the relevant sample. Overall, as the parole success score rises, time in prison does fall, presumably because higher-scoring people have been convicted of milder crimes. And, reflecting the grid's jumpiness, the declines accelerate at the two thresholds. However, this acceleration appears more clearly in medians, which are depicted in Kuziemko (Figure II), than in the averages—which matter more econometrically.[99] The fall of the green line (for averages) seems to speed up only slightly as it crosses the thresholds, which could point to instrument weakness and endogeneity bias.

Kuziemko focuses on the 1993 grid's left threshold, and on its first four rows. The paper finds powerful, negative aftereffects. Each additional month served because of being on the "poor" side of the poor-average cut-off led to 1.3 percentage points less chance of return to prison within three years (Kuziemko Table II, col. 3, se=0.291%). Multiplying that by 12 roughly implies that each year of extra time cuts return-to-prison by 15.6 percentage points, nearly half the average rate of 34 percentage points. Though this impact seems very large, it is comparable to that found in Norway. There, being sent to prison led to a stay of 175 days (just under six months) and reduced the probability of rearrest between 24 and 60 months after trial by nearly half (24.8 percentage points against a base of 58%; Bhuller et al. 2016, Table 4, col.3, and Table 5, col. 2).

Access to the Kuziemko data and code allowed me to exactly replicate and then reanalyze these results, which raised four econometric concerns. In correspondence, Ilyana Kuziemko has agreed fully with the first two concerns. The third and fourth are less cut and dried. The four are:

- *Problematic variable construction.* Comparison of the original and replication code revealed a few problems in the definitions of variables, one being important.[100] To determine an inmate's grid row, the original program uses the severity level of the first-listed conviction crime, while the grid referenced the highest severity level of all conviction crimes.[101] (Defendants can be convicted of several crimes at once, and these can differ in severity.) As a result, the original puts some people in the wrong grid row. It does not always hold the grid row fixed while gauging the effect of transiting across columns.
- *A mathematically flawed instrument.* In the analytical set-up, an inmate's grid recommendation, from Table 16, instruments actual time served. The instrumenting equation is:

$$Months\ served_{ips} = \gamma \cdot Grid\ recommendation_{ps} + \lambda_p + \nu_s + \tau + \epsilon_{ips}$$

[99] The paper estimates with Two-stage Least Squares.
[100] The unimportant problems: The code for the control for whether a person was convicted of a property crime contains a bug. And the date on which a sentence began, rather than the date it was handed down, is used to restrict the sample (to 1995–2005). Fixing both hardly changes the results.
[101] The Kuziemko-provided file "ga_2july2012_2.do" defines samples using the variable "pargl_sev1" which is the original data set's "M-PAR-CRIME-SEVERITY-1" rather than "M-HIGHEST-CRIME-SEV-L."

where $i$ indexes individuals, $p$ is parole success points, $s$ is crime severity, $\lambda_p$ and $\nu_s$ are dummy sets, and $\tau$ is additional controls (p. 12). In the section of the grid used in the study—the first four rows of the 1993 grid—the instrument, $Grid\ recommendation$, is an exact linear function of offense severity and parole success score category: each move down a grid row adds 2 months; each move to the right adds 6. See bottom third of Table 16 again. As a result, the instrument is an exact linear combination of the included regressors $\lambda_p$ and $\nu_s$ and is in no useful sense excluded.

Seemingly, the regressions should not run. I think a key reason they do is the previous bullet point: the regressions do not quite define $s$ the way the parole board did. Apparent identification comes from some people being mapped to the wrong grid row.

The revised regressions drop the instruments $\lambda_p$ in favor of $p$, and they replace $Months\ served$ with an above-threshold dummy, all in the spirit of Regression Discontinuity Design.[102]

- *Robustness.* As Table 16 documents, the Georgia's parole board used a different grid before April 1, 1993, with five success categories instead of three. To check robustness, I run the same analysis for inmates whose cases came before the board while it was using a particular edition of this grid from May 1, 1983, to March 31, 1993. I call that the "1983 grid" to distinguish it from the "1993 grid" used in Kuziemko. (An even earlier edition of the grid lasted only two years and the data from that period are poorer, so I do not work with it.) Figure 36 shows that average months served fell substantially in this sample across the poor-fair boundary, more so than at either boundary in Kuziemko's 1993 sample, so it might offer a stronger quasi-experiment. The fall in time served also accelerated discernibly across the fair-average line. Meanwhile, the 1993 grid includes a second threshold, which Kuziemko does not incorporate but I do.
- *Parole bias.* Parole bias could explain the Kuziemko results. In Georgia, a judge would set a maximum sentence and the parole board would decide the split within that time between incarceration and supervised freedom. Thus a (quasi-)experiment in release timing varies two treatments at once: time in prison and time on parole.

  The Georgia data contains enough detail to support several definitions of recidivism. The two main ones are return-to-prison, which Kuziemko uses, and reconviction (meaning conviction of a new, serious crime). The latter excludes returns to prison for technical violations but still often counts revocations triggered by misdemeanor or felony charges even when not fully prosecuted. This means that both variables can be expected to contain parole bias in the direction that would help explain the Kuziemko results, though the second less so. By counting technical violations, the first variable, pure return-to-prison, almost automatically looks worse for those spending less time in prison and more time on parole. The second variable, reconviction, skirts that pitfall, but will still be biased by the asymmetric inclusion of misdemeanors committed by parolees, and possibly by the swifter and more certain conversion of parolees' felony incidents into incarceration.

  Since both the data set's main recidivism variables may contain parole bias, I seek a third way. I focus on the return-to-prison variable but modify it in order to explore the possibility of parole bias, by drawing on other fields in the data set. (Because the data set's unit of observation is the incarceration episode, it generally provides richer information on returns to prison than on reconvictions, which do not always cause return to prison.) My first variant of return-to-prison excludes parole revocations triggered by technical violations or misdemeanor charges, since they can happen to parolees only. The second goes farther by also excluding revocations triggered by felony charges, unless those charges were pursued to conviction.[103]

  Unfortunately, neither of these two modified return-to-prison variables can be presumed unbiased. Indeed, the last one especially may be biased the other way. For it will disproportionately *undercount* new felony charges against parolees to the extent that the government finds it expedient to revoke parole upon evidence of a new crime rather than pursuing full prosecution. For this reason, I use all three versions of return to prison.[104]

[102] In principle, the second change makes no difference because of the collinearity: conditional on $\nu_s$ and $p$, $Months\ served$ is perfectly predicted by this treatment dummy. In practice, the perfect predictions are marred by evident errors in four observations, so the change makes a tiny difference.

[103] Thus, excluded here are cases where the defendant was charged, then reincarcerated after a parole board hearing, or waived his or her right to such a hearing.

**To start the tour of results,**

Table 17 shows Kuziemko's core estimate of the impact of crossing the threshold in the 1993 grid from 8 to 9 points. The first column exactly reproduces the original (Table II, col. 3). The second revises by incorporating the changes described in the first two bullets above. The point estimate in the top row—1.3 points less recidivism per month extra month served—hardly shifts, though the confidence interval widens by about a factor of three.

**Table 18 extends the revised regression to the other grid thresholds. (Its "8 to 9" column, top row, contains the same "revised" estimate as in**

Table 17.) Despite smaller samples, the regressions for the two lower thresholds in the 1983 grid score better on the tests for weak identification than either of the 1993 thresholds. And their coefficient estimates do not suggest significant impact of time served on subsequent return-to-prison (top left). The regressions for the two upper boundaries, on the other hand, perform poorly on tests of weak identification, suggesting that time served fell too little across them to make useful quasi-experiments. Finally, the impact of crossing the upper threshold in the 1993 grid also looks less well identified than that for the lower threshold (the focus of Kuziemko), with borderline results on the weak identification tests and wider standard errors; its point estimate is slightly positive.

Figure 37 graphically unifies those first-row results, using the same weak instrument–robust technique as in the reanalysis of Levitt. For each of the six grid thresholds, it plots the Anderson-Rubin p value as a function of the hypothesized impact rate. Kuziemko's main estimate of −0.0137 (as revised) sets the X coordinate of the leftmost peak. Instrument weakness makes the peak for the other 1993 threshold somewhat wider; yet here the quasi-experiment still clearly favors positive rather than negative impacts of time on subsequent return to prison. The two contours based on reasonably influential 1983 thresholds—5 to 6 and 8 to 9—center near zero or positive values, and with a confidence comparable to Kuziemko's favored negative estimate.

Overall, the original grid-based result, for the three-year return-to-prison rate, appears unrepresentative of the broader patterns in the data. As replicated, its statistical significance is more modest than originally reported, and the other estimates in the top row sit closer to zero or on the other side of it.

Meanwhile, modifying the definition of recidivism by stripping out returns to prison for technical violations and misdemeanors while under supervision—in the second row of Table 18—confirms the impression of lack of impact. So does further removing revocations triggered by felonies not pursued to conviction. The only strongly significant result, in the first column, contradicts the original in sign, possibly because of parole bias in the other direction.

[104] In defense of the choice, the paper also points out that one report estimated that more 80% of parole revocations in California were in fact for new crimes (Kuziemko, note 17).

**Table 16. Parole decision guideline grids, 1981–2007, Georgia (months to serve)**

**~April 1, 1981–April 30, 1983**

| | Parole success score | | | | |
|---|---|---|---|---|---|
| Offense severity | Poor (0–5) | Fair (6–8) | Average (9–10) | Good (11–12) | Excellent (13–20) |
| 1 | 36 | 24 | 15 | 12 | 9 |
| 2 | 42 | 27 | 18 | 15 | 12 |
| 3 | 48 | 30 | 24 | 18 | 15 |
| 4 | 54 | 36 | 30 | 24 | 21 |
| 5 | 60 | 42 | 36 | 30 | 27 |
| 6 | 78 | 66 | 54 | 54 | 42 |
| 7 | 102 | 90 | 84 | 84 | 78 |

**May 1, 1983–March 31, 1993**

| | Poor (0–5) | Fair (6–8) | Average (9–10) | Good (11–12) | Excellent (13–20) |
|---|---|---|---|---|---|
| 1 | 18 | 12 | 8 | 6 | 4 |
| 2 | 21 | 14 | 9 | 8 | 6 |
| 3 | 24 | 15 | 12 | 9 | 8 |
| 4 | 27 | 18 | 15 | 12 | 10 |
| 5 | 52 | 40 | 30 | 25 | 20 |
| 6 | 78 | 60 | 54 | 48 | 36 |
| 7 | 102 | 90 | 78 | 72 | 60 |

**April 1, 1993–December 31, 2007**

| | Poor (0–8) | Average (9–13) | Excellent (14–20) |
|---|---|---|---|
| 1 | 22 | 16 | 10 |
| 2 | 24 | 18 | 12 |
| 3 | 26 | 20 | 14 |
| 4 | 28 | 22 | 16 |
| 5 | 52 | 40 | 34 |
| 6 | 78 | 62 | 52 |
| 7 | 102 | 84 | 72 |

Source: Ganong (2012, Table 1); GSBPP, j.mp/1QNwgT3; GSBPP individual-level data set.

**Figure 35. Mean and median time served by parole success score, grid-based quasi-experimental sample, inmates admitted after 1994 and released before 2006, Georgia, following Kuziemko (2013)**

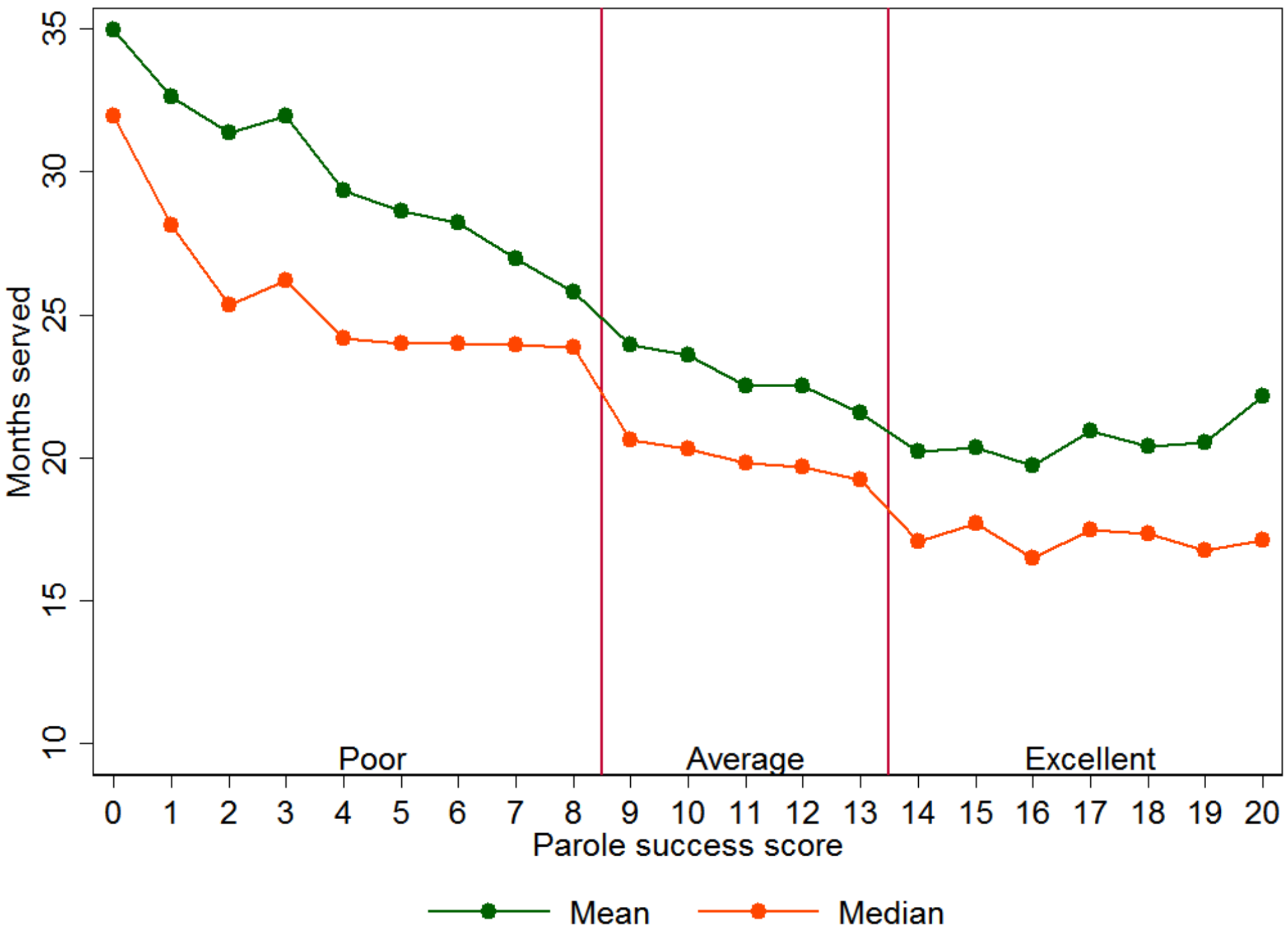

**Figure 36. Mean and median time served by parole success score, grid-based quasi-experimental sample, inmates rated by parole board May 1, 1983–March 31, 1993, Georgia, following Kuziemko (2013)**

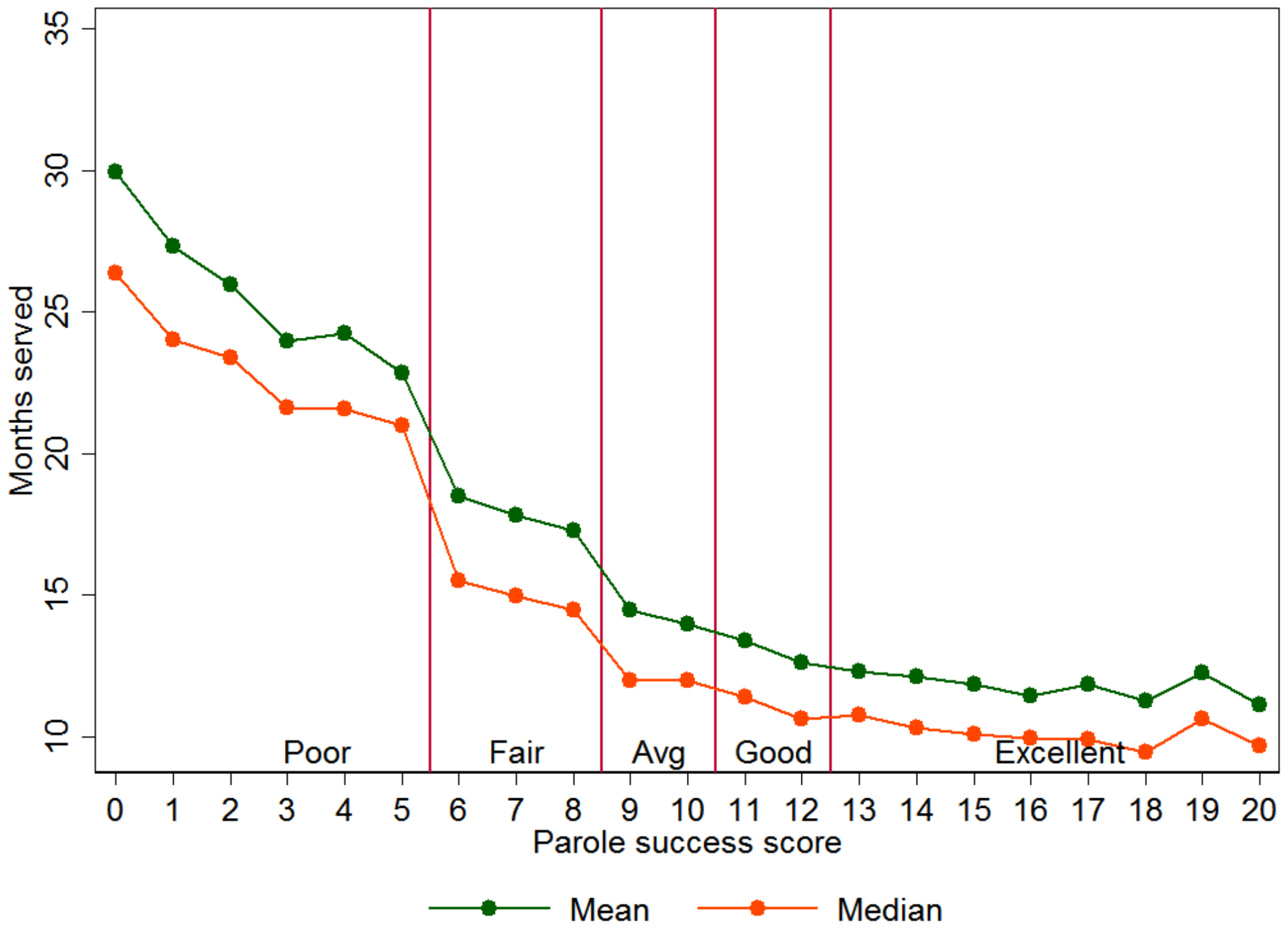

**Table 17. Core Kuziemko (2013) grid-based estimate of three-year return-to-prison rate, original and revised**

| | Original | Revised |
|---|---|---|
| Months in prison | –0.0130*** | –0.0137 |
| | (0.0029) | (0.0082) |
| Black | 0.0316*** | 0.0176** |
| | (0.0073) | (0.0079) |
| Male | 0.0822*** | 0.0880*** |
| | (0.0158) | (0.0277) |
| Age at admission | –0.00712*** | –0.0070*** |
| | (0.0004) | (0.0005) |
| Prior incarcerations | 0.0399*** | 0.0392*** |
| | (0.0029) | (0.0039) |
| Observations | 17,373 | 16,867 |

Controls not shown are dummies for: sentence in years, release year, crime severity score, parole success score (first two columns), and major conviction crime type (violent, property, drug). Regressions restricted by: severity ≤4; 4≤ success score ≤13; parole board made recommendation; age at admission ≥18, sentence between 7 months and 10 years, ≥1 day served; incarceration began with new conviction, not revocation of supervision; conviction crimes not subject to a "90%" or "seven deadly sins" mandatory minimum; beginning of sentence and admission in 1995–2005. Column 1 from Kuziemko (2013, Table II, col. 3). Column 2 introduces these changes: severity code for first-listed conviction offense replaced with severity score used by parole board; sentence begin date replaced with sentencing date; instrument switched from parole board–recommended time served to dummy for parole success score >8; success score entered linearly rather than with a dummy set; bug in definition of property crime fixed. Standard errors clustered by grid cell in parentheses. **p<0.05; ***p<0.01.

**Table 18. Grid-based estimates of effect of time served on recidivism following Kuziemko (2013)**

| Definition of recidivism | 1983 grid 5 to 6 | 1983 grid 8 to 9 | 1983 grid 10 to 11 | 1983 grid 12 to 13 | 1993 grid 8 to 9 | 1993 grid 13 to 14 |
|---|---|---|---|---|---|---|
| Return to prison within 3 years | –0.002 | 0.007 | –1.179 | 0.010 | –0.014 | 0.006 |
| | (0.006) | (0.012) | (5.204) | (0.044) | (0.008) | (0.013) |
| Return to prison within 3 years on felony charge or conviction | 0.000 | –0.002 | 0.259 | –0.015 | –0.009 | –0.002 |
| | (0.004) | (0.007) | (1.188) | (0.034) | (0.006) | (0.009) |
| Return to prison within 3 years on felony conviction | 0.007** | 0.002 | 0.440 | –0.001 | –0.008 | –0.008 |
| | (0.003) | (0.005) | (1.898) | (0.025) | (0.005) | (0.008) |
| Kleibergen-Paap underid. p | 0.00 | 0.01 | 0.81 | 0.08 | 0.00 | 0.04 |
| Kleibergen-Paap rk Wald F | 122.56 | 69.11 | 0.05 | 4.37 | 26.65 | 6.38 |
| Observations | 7,280 | 7,337 | 6,179 | 9,018 | 16,867 | 16,435 |

Sample and control definitions as in

Table 17, col. 2, except that date range for 1983 regression defined by being rated by parole board between May 1, 1983, and March 31, 1993. Samples restricted to 5 points on either side of threshold indicated at tops of columns, but narrowed where necessary to prevent extension across another threshold. Standard errors clustered by grid cell in parentheses. **p<0.05.

**Figure 37. Anderson-Rubin p values for various hypothesized rates of impact on an additional month in prison on probability of return to prison in three years after release, following Kuziemko (2013): all 1983 and 1993 grid thresholds**

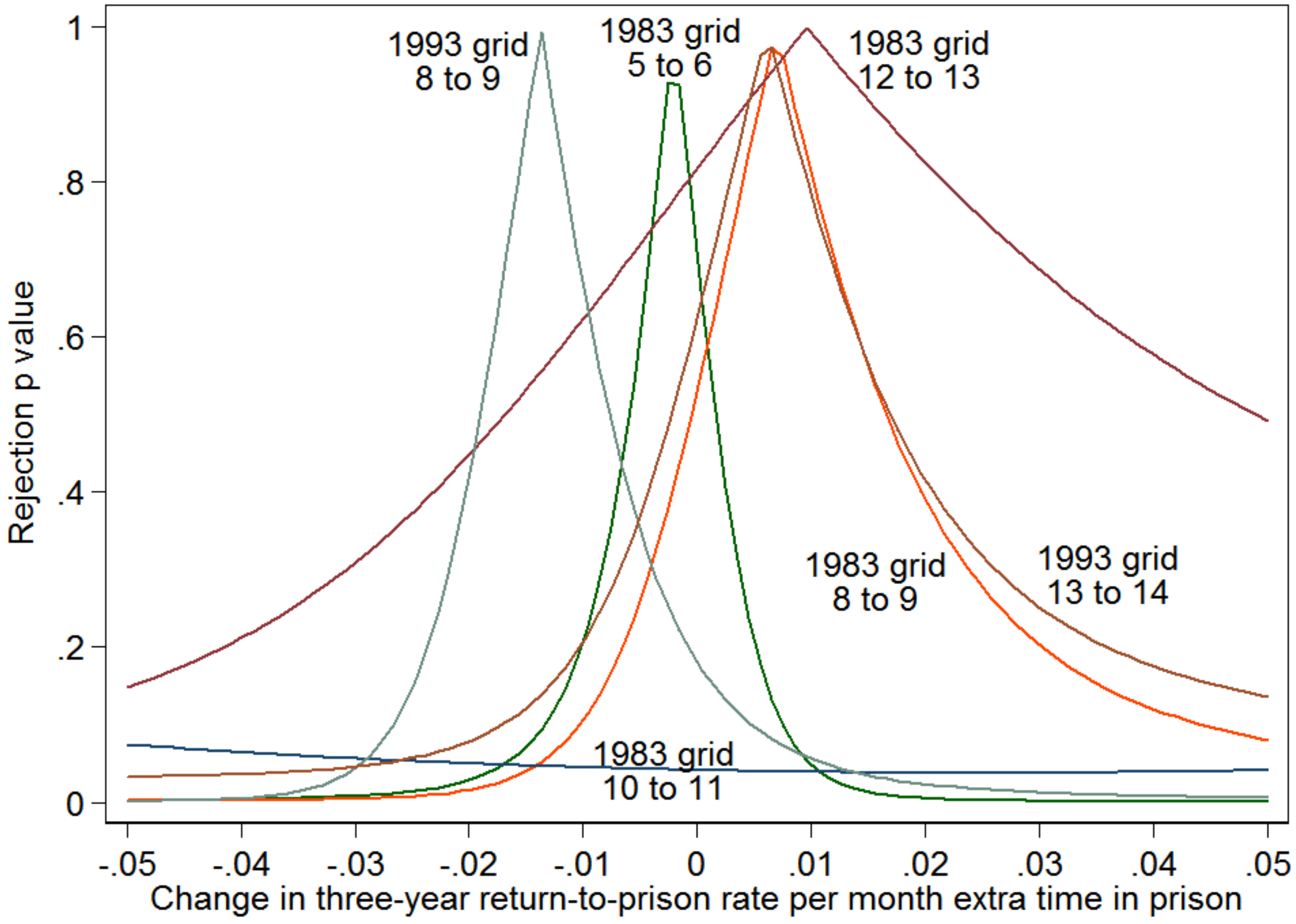

### 9.12.2. Mass prisoner release

Rather like Drago, Galbiati, and Vertova in Italy, Kuziemko examines what happened after Georgia released hundreds of prisoners, on March 18, 1981. The state's *jails* were overcrowded, so to make room for jail inmates, the governor persuaded the parole board to grant early release to non-violent offenders who were closest to release anyway (Kuziemko, p. 14). On average, those released served 13 months instead of the 17 initially recommended by the parole board (pp. 14–15). But some had more time left to serve than others.

In the decoupling of time served from time recommended, Kuziemko (section V.B) spots an opportunity to study the distinct associations of those two variables with recidivism. One could reasonably expect that inmates whom the parole board had recommended to serve longer would recidivate more. The premise of Kuziemko's analysis is that after controlling for this recommendation, taking it as a proxy for criminal propensity, the differences among the inmates in actual time served were arbitrary and formed a good quasi-experiment. Two people who earned the same recommendation from the parole board and both got out that March 18 made good comparators if they happened to have *entered* prison on different days and thus spent different amounts of time there.

The review of this quasi-experiment begins much as last one does. Table 19 starts with the original paper's results for the core regression, which in this case includes no controls. (Kuziemko, Table III, also adds various controls, which does not change results much.) The second column shows the replication, which is not quite exact because the original results come from a snapshot of the Georgia data that predates the one Ilyana Kuziemko shared with me.[105] The two regressions essentially agree (in the second row) that each month of additional time served was associated with 3.4 percentage points less recidivism, again defined as return to prison within three years. That amounts to 9.4% of the sample average of 36% (Kuziemko, Table I, col. 3).

Notice two things about these results. First, the effect is huge: it exceeds that found in the grid-based quasi-experiment (1.3%/month served) by nearly a factor of three, and implies that a season or so in prison is transformational. Aging could not explain it. Nor does parole bias vie as a theory, because the commutations appear to have erased parole time too.[106] Second, the coefficients on the two regressors match in magnitude and standard error but oppose in sign, a pattern that persists strongly through all 15 reported variants (Kuziemko, Tables III and C.II). Kuziemko (section II.A) builds a theoretical model that predicts such symmetry—but only in the coefficients, not the standard errors, and only by assuming implausible perfection in the parole board's knowledge of how each prisoner's recidivism risk depended on time served. The paper emphasizes that this assumption almost certainly does not hold.[107] Indeed, if the parole board operated optimally, it would not need to lean on the grid, with its discontinuities and suddenly instituted revisions, and there would be no basis for the grid-based quasi-experiment above.

But the equal-and-opposite pattern *is* characteristic of regressions in which the two variables are nearly collinear—which Figure 38 shows them to be—*and* in which the strongest explanatory power originates in their *difference*.[108] The difference here is: time recommended − time served = time *commuted*. A theory that tied recidivism directly to time commuted rather than to both time recommended and time served could also explain the results, and would gain credibility from being simpler, in needing to explain one non-zero relationship instead of two, and in removing that otherwise hard-to-explain coincidence in the equal-and-opposite coefficients. The third and fourth columns motivate this perspective shift by adding time commuted as a regressor and dropping time recommended or time served. Since the three variables of interest are collinear, only two can be retained, and all three revised regressions convey exactly the same information, just in different ways. The last, for example, can be read to say that time served is not correlated with return to prison (controlling for time commuted) while receiving a one-month commutation raises the return-to-prison rate by 3.7 percentage points.

---

[105] The edition shared with me has all zeroes in the Old Tentative Release Date variable, apparently because the Georgia data managers viewed it as unreliable (e-mail from Tim Carr, March 8, 2016). So in this subsection I rely on Ganong's snapshot, which is old enough to contain the missing variable but evidently not exactly the same as either of Kuziemko's.

[106] I find that 187 out of 518 releasees returned to prison within three years. Of these, 105 were for new convictions, two were admitted from other custody (perhaps jails), 80 for *probation* violations, and none for parole violations. I thank Ilyana Kuziemko for pointing me to this.

[107] Parole boards "leave some information 'on the table'" (p. 18). "Parole boards appear inclined to make use of heuristics instead of adjusting time served on a truly case-by-case basis" (p. 19).

[108] The Kuziemko results are robust to the exclusion of the two largest outliers in

Figure 38.

If we recast the results as saying that commuting prison time increases the return-to-prison rate, what would explain that? Bushway and Owens offer a theory: having received less punishment in the past than expected makes prospective punishments seem smaller. (See §2.4.4.) The happy surprise blunts deterrence going forward. One appealing aspect of this framing theory is the way it casts study subjects as experiencing large *relative* differences in treatment. If, with Kuziemko, we view the mass release as a quasi-experiment in time served, then we must come to terms with why a single-month increase, against an average base of 14 months, cut recidivism so much. If we instead view the mass release as a quasi-experiment in time commuted, we have much larger relative differences—such as a doubling from one month to two—to tie to the higher recidivism. Possibly the mind imbues framing bias with diminishing returns, so that going from one month to two matters more than going from two to three.[109]

Commenting on an earlier draft of this review, Ilyana Kuziemko suggested and performed a novel test of this theory. We could expect that the commutation especially affected recipients' perceptions of the criminal justice system if they were new to it, whereas people with long experience in prison would revise their perceptions less in response to this single event and demonstrate less change in recidivism. To test this idea, we can split the sample into "first-timers" and "returnees," as in the last two columns of Table 19. The prediction seems to hold: among those in Georgia prison for the first time, each month of time commuted led to 4.8 percentage points more recidivism; among returnees, the rate was 2.4 percentage points. A formal test for equality of the two values returns a one-tail p value of 0.14, rejecting with moderate confidence.

Finally, I test robustness by redefining recidivism—here to reconviction rather than the two narrowed versions of return to prison used in reanalyzing the grid-based quasi-experiment, for lack of the needed detail in the old 1981 data (Table 19, bottom half). For the mass release–based quasi-experiment, switching from return to prison to reconviction produces results that are smaller but match in sign (Table 19, col. 2).

On balance, I view the Kuziemko theory that more time served caused less crime as more strained than the framing theory because the Kuziemko theory needs parole boards to be perfect in order explain the equal-and-opposite coefficients.

[109] At the suggestion of reviewer Steven Raphael, I tried testing for diminishing returns to time commuted by adding a quadratic term to the regression in column 3 of Table 19. The test appears to have little power. The standard error for coefficient on the linear time commuted term, as distinct from the marginal effects at means reported in Table 19, quadruples from 0.034 to 0.138. The coefficient on the quadratic term is 0.018 (se = 0.019). The marginal impact is statistically indistinguishable from zero at p = 0.05 at most values of time commuted.

**Table 19. Replication and extension of core Kuziemko (2013) mass release–based estimate of effect of time served on recidivism, Georgia, 1981**

| | Original | Replication | | | Sample: no prior incarcerations | Sample: at least one prior incarceration |
|---|---|---|---|---|---|---|
| **Recidivism = return to prison within 3 years** | | | | | | |
| Recommended months served | 0.0367*** | 0.0371*** | 0.0028 | | | |
| | (0.0124) | (0.0124) | (0.0024) | | | |
| Actual months served | –0.0339*** | –0.0342*** | | 0.0028 | –0.0009 | 0.0045 |
| | (0.0128) | (0.0128) | | (0.0024) | (0.0031) | (0.0043) |
| Months commuted | | | 0.0342*** | 0.0371*** | 0.0476*** | 0.0240 |
| | | | (0.0128) | (0.0124) | (0.0151) | (0.0214) |
| **Recidivism = felony reconviction within 3 years** | | | | | | |
| Recommended months served | | 0.0142 | 0.0046* | | | |
| | | (0.0120) | (0.0024) | | | |
| Actual months served | | –0.0095 | | 0.0046* | 0.0022 | 0.0069 |
| | | (0.0124) | | (0.0024) | (0.0031) | (0.0042) |
| Months commuted | | | 0.0095 | 0.0142 | 0.0250* | –0.0036 |
| | | | (0.0124) | (0.0120) | (0.0148) | (0.0210) |
| Observations | 519 | 518 | 518 | 518 | 332 | 186 |

All results derive from probit regressions, reported as marginal effects at means. All regressions restricted by: age at admission ≥18, sentence between 7 months and 10 years, incarceration began with new admission rather than probation or parole revocation, and released March 18, 1981, by parole board commutation. All results are average marginal impacts at means, based on probit regressions. Column 1 from Kuziemko (2013, Table III, col. 1). Column 2 applies Kuziemko-provided code to Ganong-provided data. Classical standard errors in parentheses. *$p<0.10$; ***$p<0.01$.

**Figure 38. Months served vs. months recommended, Kuziemko (2013) mass release–based quasi-experiment**

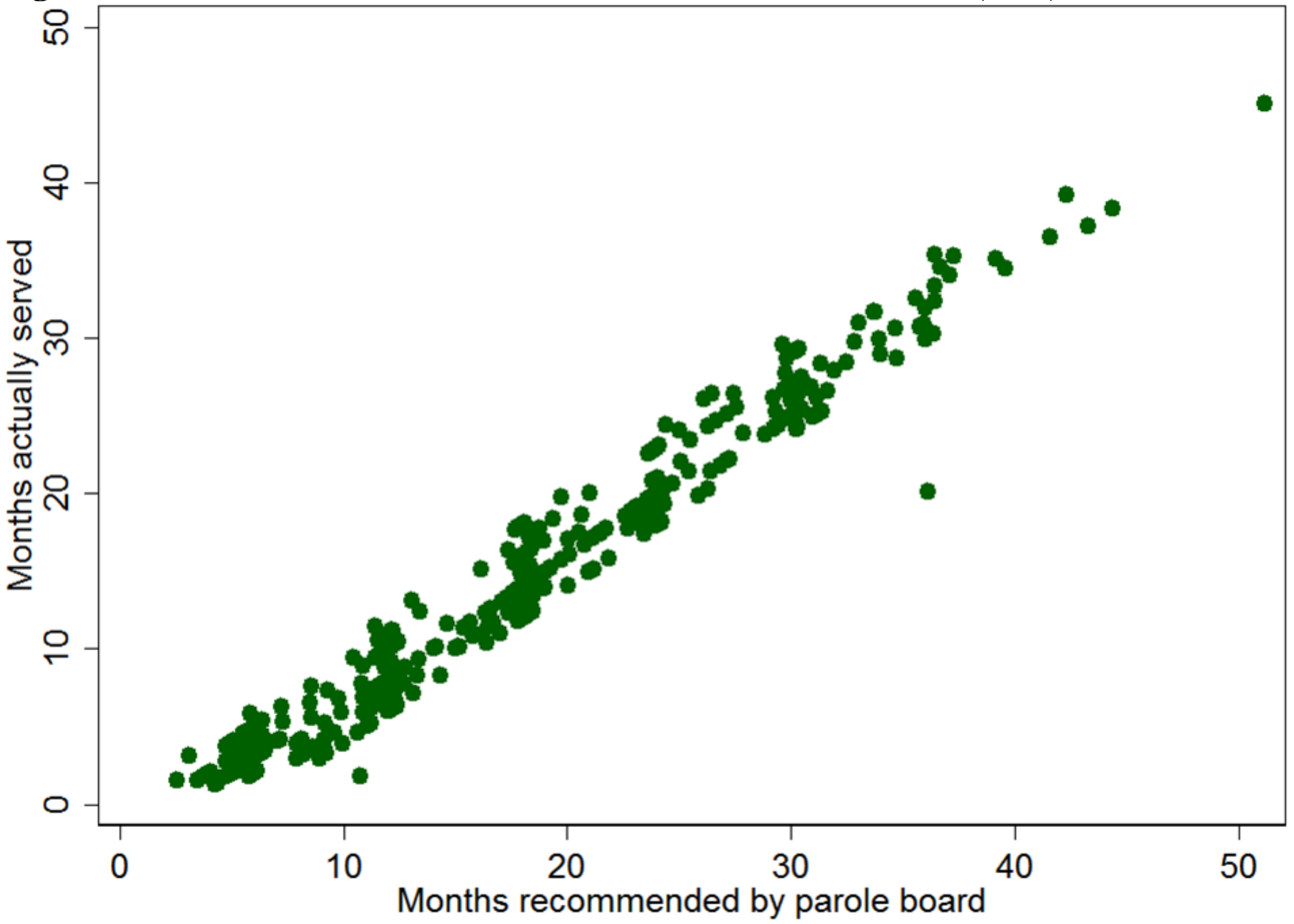

### 9.12.3. Enactment of mandatory minimums for some crimes

With effect January 1, 1998, the Georgia parole board required prisoners convicted of certain serious crimes to serve at least 90% of their original sentences in prison. The board acted voluntarily, formally speaking, but under threat of legislative action (Kuziemko, note 34), and Kuziemko perceives a quasi-experiment in the reduction in parole board discretion.

Since policy change occurred suddenly—it was announced only 23 days before it went into effect—the event invites a Regression Discontinuity Design in the time dimension—a tight examination of whether recidivism jumps up or down as one moves through the data from those sentenced in late 1997 to those sentenced in early 1998.[110] Kuziemko (Figure VII) accepts the invitation, in graphical form, and shows recidivism dropping distinctly at the cut-off date.

**More formally, I find that (fuzzy) RDD puts a coefficient on time served of −1.4 percentage points/month (se = 1.2%), nearly matching the grid-based results in**

Table 17, albeit with a larger standard error. Restricting as in Table 18 to returns to prison for new felony charges or new felony charges pursued to conviction lifts the coefficient to –0.57 and –0.42 (se = 0.46%, 0.42%), consistent with parole bias.[111]

But possibly because such RDD regressions lack the power to emphatically corroborate earlier results—the impact estimate just cited on the three-year return-to-prison rate is insignificant at conventional levels—Kuziemko leaves this approach aside and casts the quasi-experiment differently. The analysis asks not whether increased time in prison affected recidivism, but whether *reduced hope for early release* did. Perhaps the parole board's loss of discretion changed *incentives* for affected prisoners, in a way that affected their lives after returning to freedom. For example, since participating in prison educational programs would hardly accelerate their release, inmates may have participated less, and so have been less equipped to find legal employment once out.

[110] Timing from Jackson v. State Board of Pardons and Paroles, Northern District of Georgia, Case No. 2:01-CV-068-WCO, Order Dated May 29, 2002, j.mp/2bYYkUC.

[111] Regressions cited restrict to people convicted of crimes subject to the 90% policy within a year before or after its adoption. They instrument time served with a post-1997 dummy while controlling for sentencing date and using the same variable definitions, controls, and standard error estimation as in column 4 of Table 20.

Unfortunately, focusing on lost discretion as distinct from time served requires controlling for time served. And since the new policy *did* lift time served a couple of months, controlling for it effectively controls away the discontinuity that is the most plausibly exogenous variation in the sentencing regime. What emerges is a less-compelling difference-in-differences analysis, with the control group being those not convicted of 90% crimes and the before and after periods stretching five years back and four years forward from the policy discontinuity. The regressions do not measure whether future recidivism changed just when the policy changed, but whether it differed between the four years after and the five years before (relative to the control group and conditional on controls).

And while meant to narrow the analytical focus to the incentive channel, controlling for time served still leaves the door open to parole bias. To see why, imagine two subjects, one sentenced before January 1, 1998, and one after, but otherwise matched in time served, crime, and other traits. The one convicted before may have served, say, 50% of her original sentence while the one convicted after served at least 90% of his. Since the two served the same actual time, the second must have received a shorter sentence, and thus have done less parole, which cut his risk of return to prison post-release.

Table 20 exactly replicates four of the five original "ninety percent" regressions (Kuziemko, Table IV) and then revises.[112] The revisions are:

- Using sentencing date instead of sentence-begin date, and highest crime severity rather first-listed-crime severity, as discussed above in connection with the grid-based quasi-experiment.
- Because of the concern about parole bias, again varying the recidivism definition.
- Using the date of sentencing rather than date of crime commission when splitting the sample into "before" and "after" sections. When enacted on December 9, 1997, the 90% policy applied to all people *convicted* after December 31 of eligible crimes—including for crimes committed before enactment. In May 2002, a federal judge found that this structure violated the Constitutional proscription on *ex post facto* laws (j.mp/29i4DkU). In response, in September 2002, the parole board retroactively revised the policy to apply only to eligible crimes *committed* after December 31, 1997 (GSBPP 2002, p. 13). As a result, at the time of Kuziemko's analysis, the official basis of the policy's applicability was indeed the date of crime commission, and that is what Kuziemko uses.

  However, demarcating the boundary between before and after using the date of conviction, as advocated here, better represents what happened in the quasi-experiment. That is how the line was drawn from January 1998 to September 2002, by which time most subjects whose before-after classification would be affected by the 2002 policy revision had served most or all of their 90% terms. (Kuziemko, p. 22, limits the sample to those sentenced to at most five years.) I estimate that only 366 people in the sample of about 30,000 could have benefited from the policy revision, by virtue of still being in prison at the time; only they, in other words, would be cast as subject to the policy if going by conviction dates, yet ultimately have been exempted from the policy.

The first regression in Table 20 is in a sense preliminary, for it does not control for time served. It therefore aims to capture the impact of the 90% rule via both increased time and reduced incentives for good behavior. The rest of the regressions control for time served in the hope of isolating the incentive channel. The second regression appears to be Kuziemko's preferred specification, with an ample control set.[113] The third tries to bolster the reliability of this difference-in-differences estimate—somewhat eroded by the long time span and possibility of secular trends in the recidivism differential for ninety-percenters—by dropping non-ninety-percenters sentenced to less than four years in order to make the control group better resemble

112 For space, Kuziemko's fourth regression, weighting the outcome by crime severity, is omitted.

113 Subsequent regressions "perform robustness checks," suggesting that this is viewed as the base.

the treatment group. The last instead controls for a linear sentencing date trend.

In the first row of Table 20, in the "Revised" columns, the revisions weaken the key coefficients on "90% crime × post-reform," especially in the last variant. Narrowing the definition of recidivism to exclude parole revocations for technical violations (middle section) or felony charges not pursued to conviction as well (last section) further weakens the results.

As noted, this long-span differences-in-difference analysis lacks the surface credibility of most of the quasi-experiments considered in this literature review. Even without the policy change, third factors not controlled for may have influenced the evolution of recidivism among those convicted of 90% crimes. Including a time trend for one of the comparison groups, along with correcting errors in variable construction, greatly weakens the result (upper right of Table 20). Narrowing the outcome to counteract parole bias erases what remains. As a result, I remain unconvinced that imposing the 90% minimums affected return-to-prison rates other than by reducing parole time.

**Table 20. Replication and revision of four Kuziemko (2013) 90%-rule–based estimates of effect of time served on recidivism, Georgia, 1993–2001**

| | Bas regression | | Add controls | | Drop non-90-percenters sentenced <4 years | | Instead, add linear sentence date control | |
|---|---|---|---|---|---|---|---|---|
| | Original | Revised | Original | Revised | Original | Revised | Original | Revised |
| **Recidivism = return to prison within 3 years** | | | | | | | | |
| 90% crime × | 0.0533** | 0.0429 | 0.0685*** | 0.0550* | 0.0626*** | 0.0503* | 0.0722*** | 0.0273 |
| post-reform | (0.0214) | (0.0340) | (0.0156) | (0.0295) | (0.0125) | (0.0289) | (0.0169) | (0.0323) |
| 90% crime | –0.0936*** | –0.0967*** | 0.0388*** | 0.0285* | 0.0353* | 0.0286 | 0.0741 | –0.1600 |
| | (0.0210) | (0.0265) | (0.0140) | (0.0173) | (0.0188) | (0.0198) | (0.2189) | (0.1370) |
| Months served | | | –0.0029*** | –0.0031*** | –0.0031*** | –0.0033*** | –0.0029*** | –0.0031*** |
| | | | (0.0005) | (0.0004) | (0.0005) | (0.0005) | (0.0005) | (0.0004) |
| **Recidivism = return to prison, felony charge or conviction within 3 years** | | | | | | | | |
| 90% crime × | | 0.0283 | | 0.0298 | | 0.0293 | | 0.0010 |
| post-reform | | (0.0269) | | (0.0243) | | (0.0233) | | (0.0221) |
| 90% crime | | –0.0569*** | | –0.0002 | | 0.0013 | | –0.1729*** |
| | | (0.0195) | | (0.0127) | | (0.0112) | | (0.0650) |
| Months served | | | | –0.0002 | | –0.0007*** | | –0.0002 |
| | | | | (0.0003) | | (0.0003) | | (0.0003) |
| **Recidivism = return to prison, felony conviction within 3 years** | | | | | | | | |
| 90% crime × | | 0.0298 | | 0.0208 | | 0.0229 | | –0.0091 |
| post-reform | | (0.0223) | | (0.0190) | | (0.0189) | | (0.0181) |
| 90% crime | | –0.0243 | | 0.0058 | | 0.0075 | | –0.1651*** |
| | | (0.0160) | | (0.0088) | | (0.0073) | | (0.0517) |
| Months served | | | | 0.0029*** | | 0.0020*** | | 0.0029*** |
| | | | | (0.0003) | | (0.0003) | | (0.0003) |
| Observations | 30,481 | 28,359 | 30,480 | 28,358 | 17,437 | 15,288 | 30,480 | 28,358 |

All results from probit regressions, reported as marginal effects at means. Controls not shown are those listed in Table 17, as well as sentence in months and dummies for: Hispanic, release year, crime severity score, sentence year. Regressions restricted by: sentence date in 1993–2001; sentence ≤5 years; crime severity <8; parole board made recommendation; age at admission ≥18; ≥1 day served; incarceration began with new conviction, not revocation of supervision; released before May 24, 2008. First columns in each pair correspond to Kuziemko (2013, Table IV, cols. 1, 2, 3, 5). Second columns apply Kuziemko-provided code to Ganong-provided data. Third columns introduce these changes: severity code for first-listed conviction offense replaced with severity score used by parole board; sentence begin date replaced with sentencing date; treatment period defined relative to sentencing, not crime commission, date. RDD regression restrict to 90-percenters and to sentencing dates within 1 year of policy change; and add controls for sentencing date. Some "original" results deviate slightly from Kuziemko (Table IV) because where Kuziemko uses the now-deprecated Stata command "dprobit" the replication uses the "probit" and "margins" commands. Standard errors clustered by first-listed conviction crime in parentheses. *p<0.1; **p<0.05; ***p<0.01.

### 9.12.4. Kuziemko (2013): Summary

This study appears to make a strong case that more time in prison substantially reduces post-release criminality. But after correcting various technical errors, most of the results appear fragile, or explicable by parole bias or cognitive framing.

## 9.13. Ganong (2012), "Criminal rehabilitation, incapacitation, and aging," *American Law and Economics Review*

Ganong spies yet another quasi-experiment in the Georgia data, also involving the grid. Where Kuziemko exploits administrative discontinues across the rows of the grid, Ganong works off of a jump in the time dimension. As Table 16 shows, the parole board substantially revised the grid with effect April 1, 1993, replacing five parole success categories with three and recommending more time in almost every case. As just discussed in connection with Kuziemko's 90%-rule analysis, this quasi-experiment arguably produces more reliable results because the cleavage in time is so sharp: where a one-point leap from "poor" to "average" can mean having no prior felony convictions instead of one (j.mp/1MMXdSO), the jump from March 31 to April 1 in date of parole board review should ideally signify nothing so substantial.

Peter Ganong publicly posted the paper's code and has shared the data set, making it possible to perfectly match the paper's results. Once more, I have replicated some of the original regressions, then varied them by applying the same research design to a new context—the 1983 revision—and by giving equal space to other definitions of recidivism.

Ganong (p. 3) studies 18,589 people who served between three months and ten years and came up for parole within a year before or after the 1993 rule revision. Once again, a graph nicely shows what happened. Figure 39 plots smoothed moving averages for the two variables of interest while allowing breaks at the two revision dates. It depicts confidence intervals for the return to prison rate since uncertainty in the dependent variable is the mathematical source of uncertainty in regression estimates. Toward its right side, we see that average time served indeed jumped on the revision date, approximately from 27 to 31 months. In tandem fell the three-year return-to-prison rate, which is, as for Kuziemko, Ganong's preferred definition of recidivism. In effect, Ganong's regressions ratify the negative association obvious in the graph: when time served goes up, recidivism goes down. In fact, if one zooms out, one sees that the negative association prevails across the 15 years graphed. The smoothed time served and return-to-prison lines largely mirror each other.[114] In particular, they also moved oppositely at the previous grid revision, on May 1, 1993, though that time the return-to-prison rate changed much less.

One explanation for the strong symmetry is parole bias: when the parole board lengthened (or shortened) average time served, time on parole fell (or rose), and along with it the risk of quick return to prison for technical violations, misdemeanors, and even felonies that might otherwise have gone undetected or unprosecuted.

In reanalyzing the Ganong regressions, I borrow two variations I applied to Kuziemko's grid-based regressions: working with earlier grids—here meaning the switch from the 1981 to 1983 grid as well as from the 1983 to 1993 one (see Table 16)—and narrowing the return-to-prison recidivism variable to explore the role of parole bias.

I also import a change from the reanalysis of Kuziemko's 90% rule quasi-experiment. For as structured, the Ganong paper's core regressions also make a before-after comparison, between the year preceding April 1,

[114] Why the sharp movements in 1989? Under imminent threat of a prison overcrowding lawsuit from Georgia Legal Services, on March 7, 1989, Governor Joe Frank Harris asked the parole board to make an emergency release. "Gov. Harris asked the Parole Board to review cases of misdemeanants plus certain non-violent felony inmates to select those suitable for earlier parole. The offense types he specified were damage to property, habitual traffic violation, forgery, theft, burglary, and revoked parole and revoked probation for technical violations or less serious offenses. Because the resulting inmate pool was not large enough, the Board later had to add low-level drug offenses to find enough acceptable parolees." (GSBPP 1989, p. 1). The data show that accelerated release of less-serious offenders beginning in April 1989 lifted the average time served, and thus age, of the remaining pool coming before the parole board, starting later in 1989.

1993, and the year following. This comparison produces valid impact estimates if little else of relevance changed between periods other than the grid. But the parole board's annual reports (e.g., GSBPP 1979, 1989, 1993, 1994, 2008) reveal an agency in constant flux. The governor asks for more emergency releases to relieve overcrowding. Preliminary decisions on time served come to be made earlier in a prisoner's term. Prisons are built, easing pressures to release people early. And so on. Identification of impacts becomes more reliable the more it focuses on the discontinuity at April 1, 1993 (assuming no other major policy changes on that day). One way to sharpen that focus is to include a time trend in the regressions, causing them then to test whether, relative to the overall trend for the two-year study period, recidivism jumped as inmates came under the guidance of the new grid. In jargon, this replaces difference-in-differences with regression discontinuity design (RDD).[115]

The results of introducing the three changes in all possible combinations—of 1983 vs. 1993 grid revision, difference-in-differences vs. regression discontinuity, three recidivism definitions—appear in Table 21. For each combination, the table reports two regressions side-by-side, which copy Ganong in deploying more or less demanding control sets.[116] In order to match the original, the table expresses all results per year rather than month of time served as in the Kuziemko tables above.

In the first row of Table 21, we find, on the right, exact matches with corresponding regressions in Ganong (Table 4, cols. 3 and 4) and, on the left, comparable results from the earlier grid revision. In the next two rows, we again see how shifting to the more restrictive definitions of return-to-prison recidivism moves the impact estimates in the positive direction. In the second half of the table, the more demanding regression discontinuity framework weakens nearly all the results, if mainly by widening standard errors.

Partly to support the cost-benefit analysis in the conclusion, Table 22 expands on the impact estimates for the 1993 grid revision, by breaking out the impact by cause of return. Copying Ganong (Table 5), it also switches from counting *whether* a person returned to prison within three years to how many times the person returned over 10 years, which matters more for the total crime impact of longer prison spells. As the benign findings in the third and sixth rows of Table 21 suggest, spending more time in prison left no systematic imprint on the subsequent rate of return to prison for murder, assault, drug, or any of the other major crime groups. Impact shines through only on returns to prison by parolees, in the bottom two rows of Table 22. There, I split the outcomes by whether or not the new felony charges against parolees and probationers are pursued to conviction.

That the negatively signed impact in Ganong—more time leading to less crime—is essentially confined to parolees suggests that parole bias is indeed the source of the study's key findings. And while it may seem to overturn Ganong's conclusion that extra time in prison reduces recidivism, it coheres with the paper's benefit analysis. It too breaks out the ten-year impacts on rape, robbery, etc., and attaches dollar values to each. Ganong's (p. 20) preferred estimate is that a year of prison saves society just $50 in crime costs over ten years (se = $1,546)—effectively zero.

Even if one concludes that more time in prison did not reduce criminality in Georgia—that the effect is purely a mirage generated by parole bias—Table 22 hardly argues for the opposite effect: more time did not clearly lead to *more* crime either. Thus within the body of research on aftereffects, the Georgia results still fall on the pessimistic end of the range of credible findings. They are pessimistic, that is, in implying that aftereffects are not so harmful as to cancel out incapacitation. Incarceration may reduce crime, and decarceration may increase it.

---

115 For the same reason, the discontinuity specifications instrument time served with a post-revision dummy rather than grid-recommended time served. Recommended time served changed by different amounts for different grid cells, and this cross-sectional component of variation is not necessarily exogenous. As an instrument, the post-revision dummy removes this variation.

116 As a robustness test, Ganong (Table 4, col. 6) also adds controls for departure status—released with or without parole and with or without probation to follow that. I avoid that specification here for simplicity, and because it controls for some variation in treatment, since serving more time raises the odds "maxing out" and serving no parole after.

**Figure 39. Three-year return-to-prison rate versus time served, smoothed fits allowing breaks for May 1, 1983, and April 1, 1993, grid revisions, Georgia**

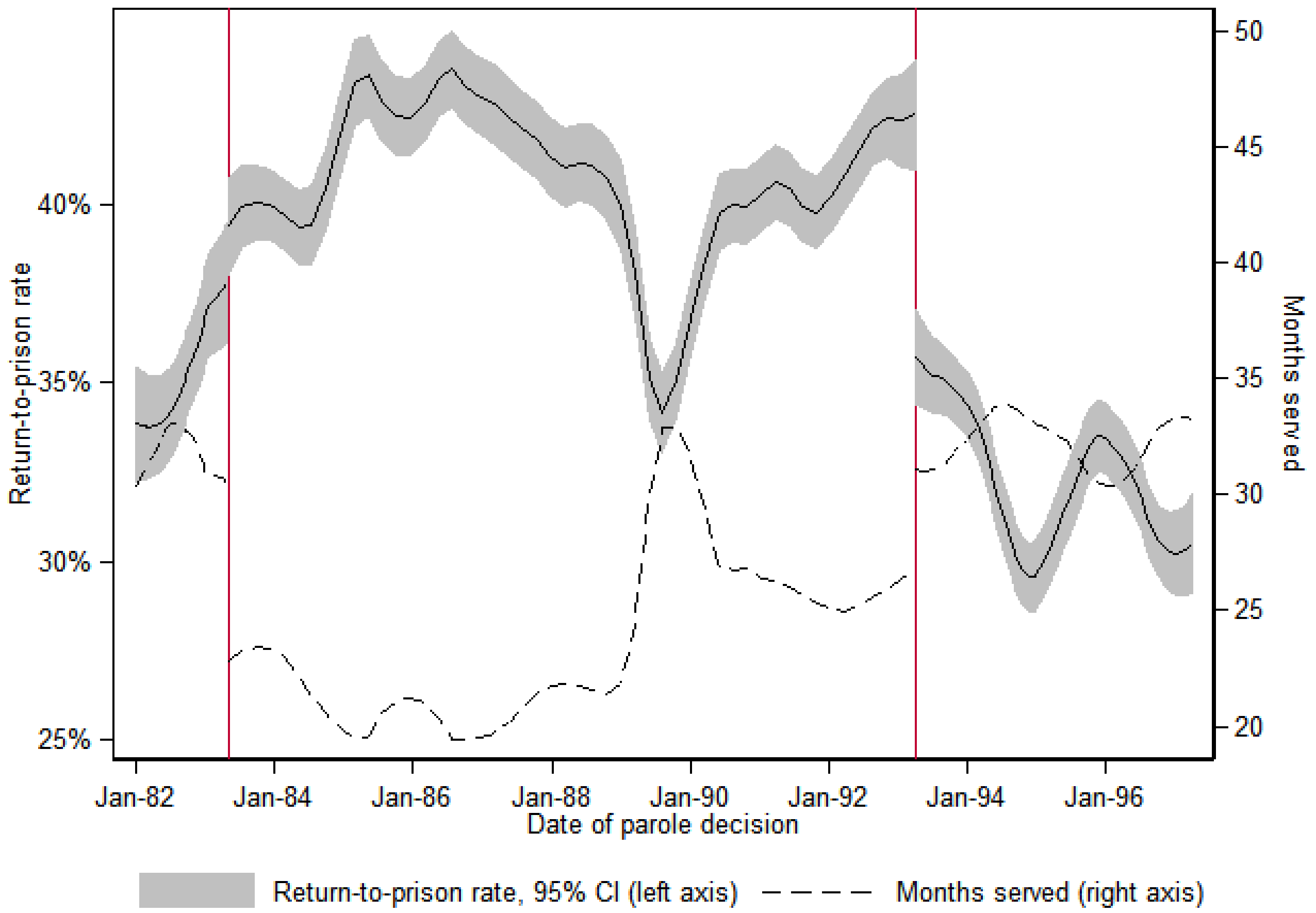

**Table 21. Grid revision–based estimates of effect of time served (in years) on recidivism: Ganong (2012) replication and extensions**

| | May 1, 1983, grid revision | | April 1, 1993, grid revision | |
|---|---|---|---|---|
| Recidivism definition | Grid cell fixed-effect controls | Add demographic and criminal history controls | Grid cell fixed-effect controls | Add demographic and criminal history controls |
| **Difference-in-differences (does not control for date of parole board rating)** | | | | |
| Returned to prison within 3 years | –0.048*** | –0.034 | –0.059*** | –0.080*** |
| | (0.010) | (0.044) | (0.010) | (0.014) |
| Returned to prison within 3 years on felony charge or conviction | 0.010 | 0.124*** | –0.052*** | –0.041** |
| | (0.008) | (0.032) | (0.011) | (0.016) |
| Returned to prison within 3 years on felony conviction | 0.023** | 0.109*** | –0.012 | –0.011 |
| | (0.009) | (0.030) | (0.008) | (0.010) |
| Kleibergen-Paap underid. p | 0.00 | 0.00 | 0.00 | 0.00 |
| Kleibergen-Paap F | 157.90 | 40.49 | 106.83 | 146.81 |
| **Regression discontinuity design (controls for date of parole board rating)** | | | | |
| Returned to prison within 3 years | –0.020 | –0.015 | –0.084** | –0.076*** |
| | (0.071) | (0.062) | (0.033) | (0.027) |
| Returned to prison within 3 years on felony charge or conviction | 0.027 | 0.026 | –0.069 | –0.050 |
| | (0.050) | (0.037) | (0.046) | (0.032) |
| Returned to prison within 3 years on felony conviction | 0.042 | 0.034 | –0.013 | –0.011 |
| | (0.047) | (0.036) | (0.036) | (0.027) |
| Kleibergen-Paap underid. p | 0.00 | 0.00 | 0.00 | 0.00 |
| Kleibergen-Paap F | 19.14 | 32.56 | 64.09 | 86.37 |
| Observations | 15,126 | 15,126 | 18,589 | 18,589 |

Following Ganong (2012), all regressions instrument actual months served, control for some or all of a large set of demographic and criminal history factors, and restrict to those who served between 90 days and 10 years and who were rated by the parole board within one year before or after the relevant revision. Regressions in lower panel control linearly for date of parole board rating, the discontinuity forcing variable, and instrument with a post-revision dummy rather than grid-recommended time served. The upper-right pair exactly matches results in Ganong (Table 4). "Felony charge or conviction" includes revocations of probation or parole triggered by felony charges not pursued to conviction. Standard errors clustered by grid cell in parentheses. **p<0.05; ***p<0.01.

**Table 22. Estimates of aftereffects of incarcerating prisoners one additional year on returns-to-prison per releasee in following 10 years, Georgia, by felony type**

| | Before-after comparison | | | | Regression discontinuity design | | | |
|---|---|---|---|---|---|---|---|---|
| | 1983 grid revision | | 1993 grid revision | | 1983 grid revision | | 1993 grid revision | |
| Crime | Grid FE only | Added controls | Grid FE only | Added controls | Grid FE only | Added controls | Grid FE only | Added controls |
| Homicide | 0.002* | –0.001 | –0.001 | 0.001 | –0.005 | –0.008 | 0.001 | 0.005 |
| | (0.001) | (0.006) | (0.002) | (0.002) | (0.008) | (0.007) | (0.003) | (0.003) |
| Rape | 0.005 | 0.001 | –0.001 | –0.002 | 0.001 | –0.009 | –0.009* | –0.009* |
| | (0.003) | (0.013) | (0.001) | (0.002) | (0.014) | (0.013) | (0.005) | (0.005) |
| Aggravated assault | 0.001 | –0.006 | –0.001 | –0.001 | –0.002 | –0.003 | –0.002 | –0.002 |
| | (0.001) | (0.008) | (0.002) | (0.002) | (0.010) | (0.009) | (0.005) | (0.005) |
| Simple assault | –0.000 | –0.000 | 0.003** | 0.004*** | –0.004 | –0.003 | 0.001 | 0.003 |
| | (0.001) | (0.005) | (0.001) | (0.001) | (0.005) | (0.005) | (0.005) | (0.005) |
| Robbery | 0.006 | 0.024 | 0.002 | 0.005 | 0.005 | –0.004 | 0.017* | 0.016 |
| | (0.005) | (0.015) | (0.002) | (0.006) | (0.023) | (0.021) | (0.010) | (0.011) |
| Burglary | 0.004 | –0.025 | –0.001 | –0.007 | 0.036 | 0.003 | 0.008 | 0.005 |
| | (0.011) | (0.043) | (0.003) | (0.006) | (0.039) | (0.029) | (0.019) | (0.018) |
| Larceny/theft | 0.009 | 0.022 | 0.003 | –0.007 | –0.016 | –0.017 | 0.042* | 0.028* |
| | (0.005) | (0.022) | (0.005) | (0.010) | (0.025) | (0.023) | (0.024) | (0.016) |
| Motor vehicle theft | –0.000 | 0.007 | –0.001 | 0.002 | –0.001 | –0.000 | 0.000 | 0.004 |
| | (0.001) | (0.005) | (0.001) | (0.003) | (0.005) | (0.004) | (0.004) | (0.005) |
| Arson | 0.001 | –0.000 | 0.001 | 0.000 | 0.000 | –0.001 | 0.001 | 0.000 |
| | (0.001) | (0.005) | (0.000) | (0.001) | (0.005) | (0.004) | (0.002) | (0.002) |
| Vandalism | –0.000 | –0.002 | –0.000 | –0.001 | –0.000 | –0.000 | –0.003 | –0.003 |
| | (0.001) | (0.004) | (0.000) | (0.000) | (0.004) | (0.003) | (0.002) | (0.002) |
| Fraud | 0.009 | 0.034 | 0.004* | 0.004 | 0.006 | 0.010 | 0.012 | 0.012 |
| | (0.006) | (0.022) | (0.003) | (0.004) | (0.017) | (0.014) | (0.011) | (0.010) |
| Drug | –0.000 | 0.020 | –0.004 | –0.005 | –0.003 | 0.001 | –0.008 | 0.002 |
| | (0.003) | (0.015) | (0.005) | (0.008) | (0.025) | (0.020) | (0.020) | (0.018) |
| Other | 0.001 | –0.015 | 0.003 | 0.006 | –0.002 | –0.001 | –0.017 | –0.016 |
| | (0.005) | (0.020) | (0.003) | (0.006) | (0.024) | (0.019) | (0.016) | (0.014) |
| Parole/probation revocation for | –0.020* | 0.041 | –0.041*** | –0.033* | –0.037 | –0.045 | –0.100* | –0.073* |
| felony, not pursued to conviction | (0.011) | (0.054) | (0.011) | (0.020) | (0.065) | (0.056) | (0.053) | (0.041) |
| Parole/probation revocation for | –0.032*** | –0.071** | –0.052*** | –0.035** | 0.042 | 0.053 | –0.079* | –0.055 |
| felony, pursued to conviction | (0.008) | (0.033) | (0.009) | (0.016) | (0.041) | (0.038) | (0.042) | (0.036) |

Each cell holds results from a different regression. All regressions parallel those in rows 4 of both panels of Table 21, but with the dependent variables being the number of returns to prison over 10 years for felonies in a given category. Standard errors clustered by grid cell in parentheses. *significant at p<.1. **significant at p<.05. ***significant at p<.01.

## 9.14. Summary: Aftereffects

The preponderance of the evidence says that incarceration in the US increases crime post-release, and enough over the long run to offset incapacitation. A quartet of judge randomization studies (Green and Winik in Washington, DC; Loeffler in Chicago; Nagin and Snodgrass in Pennsylvania; Dobbie, Goldin, and Yang in Philadelphia and Miami) put the net of incapacitation and incarceration aftereffects at about zero. In parallel, Chen and Shapiro find that harsher prison conditions—making for incarceration that is harsher in quality rather than quantity—also increases recidivism. Gaes and Camp concur, though less convincingly because in their study harsher incarceration quality went hand in hand with lower incarceration quantity. Mueller-Smith sides with all these studies and goes farther, finding modest incapacitation and powerful, harmful aftereffects in Houston; but modest hints of randomization failure accompany those results.

Some studies dissent from the majority view that incarceration is criminogenic. Roach and Schanzenbach find beneficial aftereffects in Seattle—a result that is also subject to some doubt about the quality of randomization. Bhuller et al. make a more compelling case that incarceration reduces crime after—in Norway. Berecochea and Jaman, one of the few truly randomized studies in this literature, also looks more likely right than wrong, and is also somewhat distant in its setting, early-1970s California. And there are the two Georgia studies (Kuziemko and Ganong), which upon reanalysis no longer point to beneficial aftereffects, but still do not demonstrate harmful ones either.

Aftereffects must vary by place, time, and person. But the first-order generalization that best fits the credible evidence is that at the margin in the US today, aftereffects offset in the long run what incapacitation does in the short run.

## 10. Juveniles

I have separated the studies of young people from those of adults since incarceration may affect them differently. Here I review four studies. Two touch on deterrence, and one of those also measures incapacitation. The other two look at aftereffects.

### 10.1. Lee and McCrary (2009), "The deterrence effect of prison: Dynamic theory and evidence," working paper

Rather like Helland and Tabarrok, Lee and McCrary compares particular groups of people in order to estimate deterrence—groups whose precise definitions are complicated because of the desire to make statistically compelling comparisons in non-experimental data. Here, the cleavage comes at the 18th birthday, when, in Florida, people attain criminal majority and enter a different punishment regime. Unlike Helland and Tabarrok, this study measures incapacitation in addition to deterrence.

The base sample is defined against a statewide arrest database for 1989–2002. It includes the 64,073 people who were arrested at least once before age 17 (p.11). The logic of the study does not strictly require this restriction. But by focusing on a subset of young people most likely to be arrested around the age of central interest, 18, it may remove noise from the estimates.

Within this group, Lee and McCrary compute the probability *by week of life* that a juvenile's first post-17 arrest occurs, if it has not already occurred. While Lee and McCrary consider pre-17 arrests for any crime in defining their sample, here they only count arrests for serious ("index") crimes. As shown in Figure 40, which is copied from Lee and McCrary (Figure 1A), a smoothed, moving average fit shows that turning 18 brings only the slightest drop in the arrest probability. And it lacks statistical significance (Lee and McCrary, Table 2). In other words, whether for lack of awareness or lack of concern, 18-year-old Floridians appeared undeterred by the additional punishment they risked if accused of a crime. Deterrence in this age group could not be distinguished from zero.

Next, Lee and McCrary ask: If a person is arrested just after the 18th birthday instead of just before, does that increase the time that passes before the *next* arrest? The required comparison is complicated in words: essentially, between those whose first post-17 arrest was at age 17.98 and those whose first post-17 arrest was at age 18.02, in whether rearrested within a certain amount of time of the first post-17 arrest.

Figure 41 (Lee and McCrary, Figure 5A) shows that here the 18th birthday *does* bring sudden drops. For example, eyeballing the graph, the probability of being rearrested within 30 days of one's first post-17 arrest plunges 9.5 percentage points, from 17.9% to 8.4%, as the date of that first arrest passes the 18th birthday. The rate of rearrest within a year falls about 7.7 points, from 54.7% to 47.0%.

Incapacitation most likely explains these drops. The previous graph largely rules out deterrence. Possibly the jail experience changes radically at the 18th birthday causing youngsters who are released after a few days to behave quite differently. But the most straightforward theory is that right after turning 18, fewer teenagers are rearrested within a given number of weeks because more are still behind bars.

Overall, I find this study highly credible. The statistical evidence leaps off the graphs. That vulnerable 18-year-olds underestimate the rise in punishment at 18 and/or fail to immediately factor it into their behavior is plausible. It also coheres with my findings of little or no deterrence in the reanalyses of Helland and Tabarrok and Abrams.

**Figure 40. Probability that first post–age-17 arrest for serious crime occurs in a given week if it has not already occurred, among those arrested at least once before 17, Florida, 1989-2002, from Lee and McCrary (2009)**

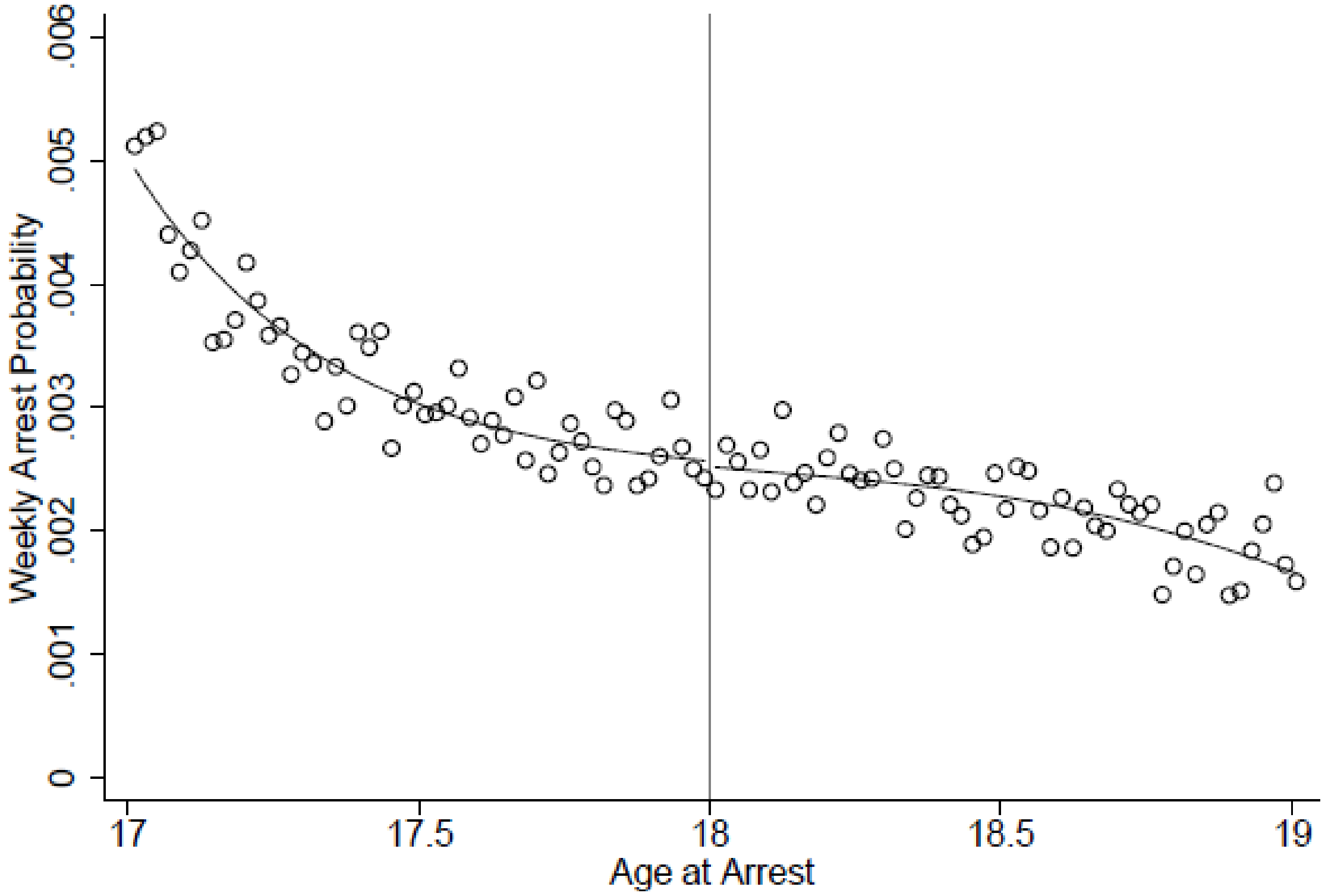

**Figure 41. Probability of second post-17 arrest within 30, 120, or 365 days of first, as function of age of first post-17 arrest, among those arrested at least once before 17, Florida, 1989-2002, from Lee and McCrary (2009)**

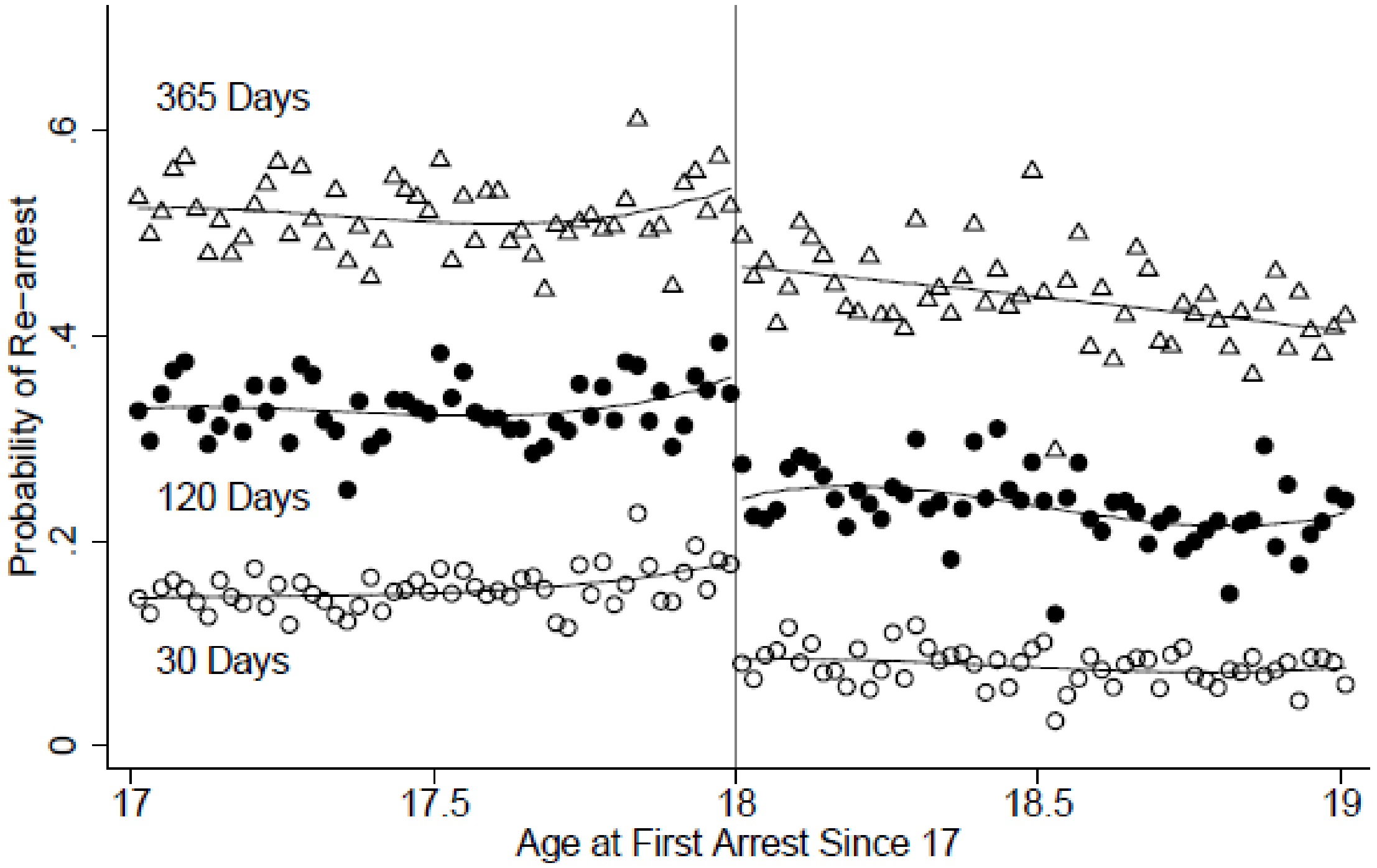

## 10.2. Hjalmarsson (2009a), "Crime and expected punishment: Changes in perceptions at the age of criminal majority," *American Law and Economics Review*

This study distinguishes itself by measuring criminality through self-reports rather than official records. Since 1997, the National Longitudinal Survey of Youth 1997 (NLSY97) has been annually interviewing a fixed set of some 9,000 people who were 12–16 years old as of December 31, 1996 (j.mp/2aAXAEv). The survey has included questions about whether respondents have committed common crimes such as stealing a car or selling drugs (j.mp/1nRyHYT).

Somewhat like Lee and McCrary, Hjalmarsson checks whether the self-reported crime rate of boys falls distinctly as they reach the local age of criminal majority, which varies by state between 16 and 18 (p. 219). Crime self-reports may be biased by reluctance to confess transgressions even in a self-administered questionnaire backed by promises of anonymity. But it seems unlikely that such bias would itself change suddenly at the age of criminal majority, which is what is needed for clean before-after comparison. Meanwhile, the official measures of crime used in other studies—of arrest or trial or conviction or incarceration—harbor measurement problems too, not least because much criminality escapes the net of the criminal justice system.

Figure 42, extracted from Hjalmarsson (Figures 3–7) shows some patterns in the data. Each part plots the self-reported rate of a crime against the number of months until or since the local age of criminal majority. By design, these plots allow sharp breaks only at that age, and indeed breaks appear in all five cases, four of them downward. But these graphs do not display confidence intervals. Separate regressions (Hjalmarsson, Table 7, row 1) find little statistical significance in the jumps, with two-tailed p values I compute as .55, .11, .48, .49, and .22 respectively for auto theft, theft of less than $50, theft of more than $50, drug selling, and assault. Moreover, where the overall trend runs downward with age, we should expect small downward jumps where breaks are allowed—which is largely what we see in Figure 42—purely because of the mathematics behind the plots, even when there is no real break. The points just to the left of the cut-offs are weighted averages only of (higher) data to left, and likewise for the right.

With a distinct data set and analytical set-up, Hjalmarsson thus corroborates Lee and McCrary—and my reanalyses of the studies of adults—on the lack of deterrence for young adults.

**Figure 42. Rate of self-reported criminality, nationwide survey, as function of months until or after local age of criminal majority, from Hjalmarsson (2009a)**

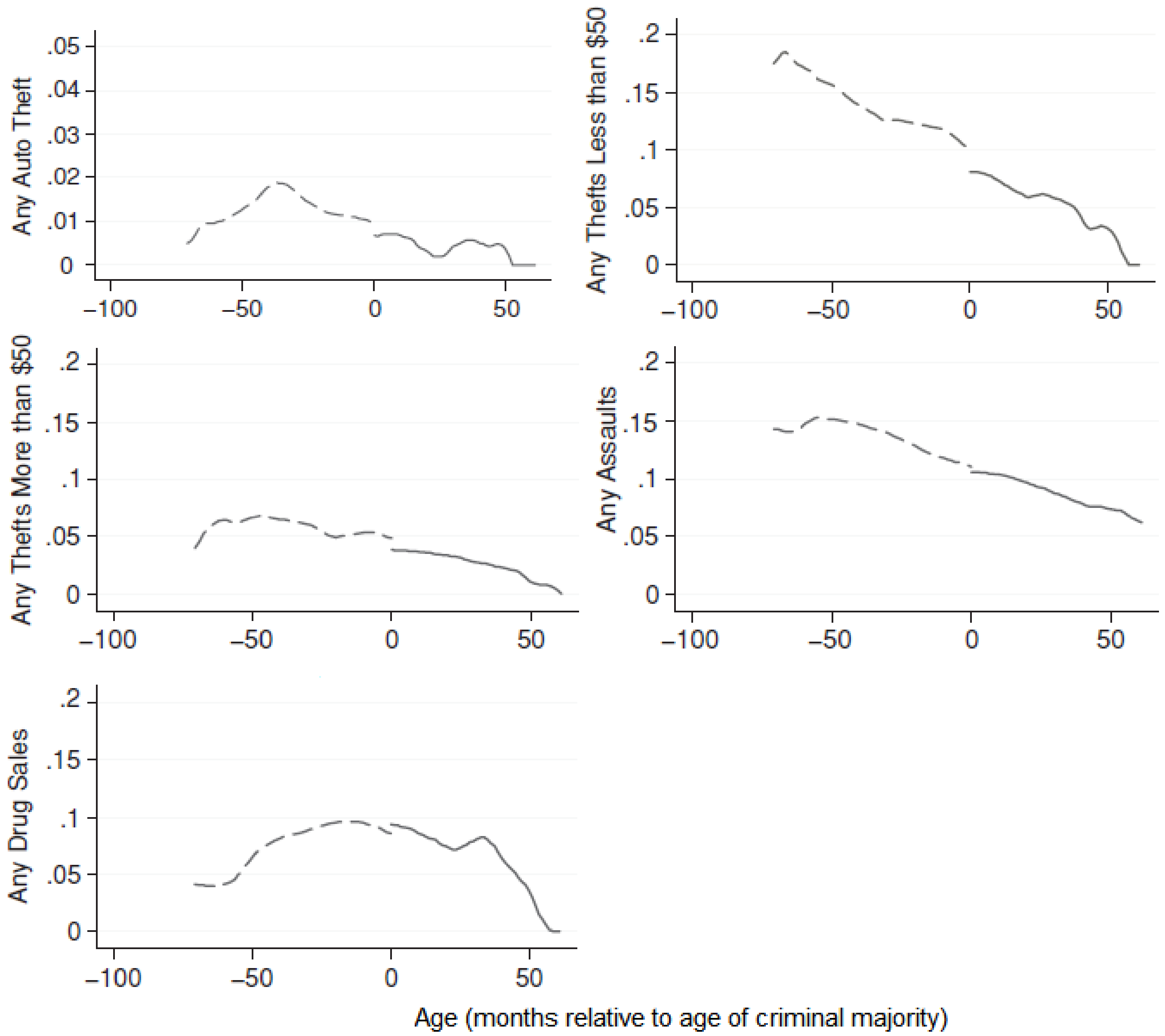

## 10.3. Hjalmarsson (2009b), "Juvenile jails: A path to the straight and narrow or to hardened criminality?", *Journal of Law and Economics*

This paper follows in the footsteps of Chen and Shapiro, as well as Kuziemko, in exploiting a discontinuity in sentencing guidelines. And as its title implies, its interest is also in the aftereffects of incarceration.

The study setting is Washington state between July 1998 and December 2000 (p.785). Hjalmarsson's subjects are juvenile offenders who passed through a critical juncture at sentencing. Some were sentenced to incarceration in a state facility for at least 15 weeks, and some to "local sanctions" such as community service, fines, or up to a month's local detention. Hjalmarsson samples 20,542 youths, of whom 1,147 were sent to state facilities. At the time of offense, youths sent to state facilities averaged 15.6 years old. (Hjalmarsson, Table 1.)

As in Georgia, the Washington sentencing guidelines are embodied in a two-dimensional grid. The vertical dimension links to the severity of the current offense and the horizontal to the seriousness of prior offenses. The boundary between local sanctions and longer-term incarceration zigzags diagonally through the grid. (See Table 23, from Caseload Forecast Council 2014, p. 10.) Hjalmarsson regresses recidivism—whether a youth reappears in court before age 18—on an indicator for whether a youth was placed above or below the boundary, while controlling for dummies for each table row. This causes the regressions to measure the average impact of moving across the boundary *within* a row, which can only happen in rows B, C+, and C. Most subjects in these rows get at most 36 weeks.

Like those in Helland Tabarrok, as well as Gaes and Camp, Hjalmarsson's regressions incorporate information about the timing of recidivism (defined as reappearing in court), not just whether recidivism occurs within some period. They suggest that being to the right of the dividing line—being incarcerated for at least 15 weeks, rather than facing a milder local sanction—cut by a substantial 36% the per-day chance of recidivism among those who had not already recidivated (Hjalmarsson, p. 800; Table 5, col. 4). Tougher punishment time led to less crime.

Two choices in Hjalmarsson raise methodological concerns by pulling the study away from the experimental ideal in ways that could explain the results. However, an early version of the paper includes details that substantially rebut the concerns.

The first problem is that the regressions drop youths who were sentenced in contravention to the guidelines, at Hjalmarsson interprets and applies them to the data at her disposal. In the control group, 424 of 19,395 youths are excluded for having been assigned to the incarceration, while from the treatment group, a substantial 314 out of 1,147 are deleted (p. 793) for not having been assigned to incarceration. Filtering the samples based on events that occur *after* quasi-randomization, such as a judge overriding the default sentence as too harsh or lenient, biases results to the extent those events predict recidivism. This censoring would most likely invite the classic reverse-causation bias in studies of crime and punishment, that people receive harsher sentences *because* they tend to break the law more. That bias would be conservative in this context, since it would tend to filter out the least troubled youths and raise average recidivism in the incarceration group—yet Hjalmarsson finds that group to recidivate less.

Second, the follow-up period begins at the day of sentencing plus the *minimum* possible time of service (p. 785). So if a youth landed in one of the 15–36-week cells in Table 23 and served 36 weeks, the econometrics would still assume release took place after 15. In effect, the study would monitor weeks 16–36 for recidivism, and find none. Incapacitation would be treated as an incarceration aftereffect. The bias would operate much less in the control subjects since they were incarcerated at most five weeks. And this *could* explain why recidivism is measured as lower among incarcerated youths.

An early version of the paper checks both issues by modifying the key regression. Pintoff (2004, Table 4, col. 6) restores the misassigned youths and starts follow-up after the *maximum* sentence. These changes shave the recidivism reduction from incarceration from 36% to 26% while preserving great statistical significance.

In combination with the next study, Hjalmarsson (2009b) poses a puzzle. The next one concludes that putting juveniles in jail increases their criminality, at least over the longer term. I am not sure how to reconcile them, because both look credible.

**Table 23. Washington state juvenile sentencing grid**

| CURRENT OFFENSE CATAGORY | | | | | |
|---|---|---|---|---|---|
| A+ | 180 weeks to age 21 for all category A+ offenses | | | | |
| A | 103-129 weeks for all category A offenses | | | | |
| A- | 15 - 36 WEEKS EXCEPT 30 - 40 WEEKS FOR 15 TO 17 YEAR OLDS | 52 – 65 WEEKS | 80 – 100 WEEKS | 103 – 129 WEEKS | 103 – 129 WEEKS |
| B+ | 15 - 36 WEEKS | 15 - 36 WEEKS | 52 - 65 WEEKS | 80 - 100 WEEKS | 103 - 129 WEEKS |
| B | LS | LS | 15 – 36 WEEKS | 15 – 36 WEEKS | 52 – 65 WEEKS |
| C+ | LS | LS | LS | 15 – 36 WEEKS | 15 – 36 WEEKS |
| C | LS | LS | LS | LS | 15 – 36 WEEKS |
| D+ | LS | LS | LS | LS | LS |
| D | LS | LS | LS | LS | LS |
| E | LS | LS | LS | LS | LS |
| PRIOR ADJUDICATIONS | 0 | 1 | 2 | 3 | 4 or more |

NOTE: References in the grid to days or weeks mean periods of confinement. "LS" means "local sanctions" as defined in RCW 13.40.020.

## 10.4. Aizer and Doyle (2015), "Juvenile incarceration, human capital, and future crime: Evidence from randomly assigned judges," *Quarterly Journal of Economics*

Aizer and Doyle bring judge randomization to the study of juvenile delinquency. Their setting, as in Loeffler, is Chicago. And like Loeffler, as well as Mueller-Smith, the paper links juvenile court data to other data sets in order to study outcomes beyond return to the juvenile courts, including subsequent school attendance, graduation, and adult recidivism. Notably, Aizer and Doyle perform one of the longest follow-ups of any study in this review: subjects are tracked from their first appearance in juvenile court, which must come before age 18, out to age 25 (pp.10–11).[117]

In the Cook County juvenile courts, cases are assigned to judicial calendars according to *neighborhood* (p.7). Normally two judges serve each neighborhood, though sometimes a swing judge serves. (One exception: crimes involving a weapon are routed to distinct calendars.) Within a calendar, assignment to a specific judge appears arbitrary:

> …the judge assignment is a function of the sequence with which cases happen to enter into the system and the judge availability that is set in advance. In particular, there does not appear to be scope for influencing the first judge seen. It is at the first court hearing, for example, that juveniles meet their public defenders (who are also assigned based on day of hearing) and learn who the judge will be. Conversations with court administrators confirm that these assignments are effectively random and that there is no way to influence the judge assigned to the case. (p. 7.)

Arbitrary assignment sets the stage for a judge randomization study—this one of some 40,000 youths who appeared in court between 1990 and 2006.

Aizer and Doyle do not instrument with a large set of judge zero-one indicators but rather, as in Roach and

[117] The only one longer is the ten-year horizon in some of Ganong's regressions.

Schanzenbach, with a single instrument that is a judge's average rate of incarceration (p. 15). The average is recomputed for each defendant, leaving out that defendant's information. Using a single instrument avoids potential degeneracies associated with having a large set of (potentially weak) instruments.

Aizer and Doyle (p. 8) calculate the average incarceration spell for incarcerated youths in the study at 42 days. Perhaps because the young people are not held long, the authors are unable to estimate the impact of an additional month of imprisonment with any precision. The main results pertain not to whether rather than how long a child was put in a state facility.

Those impact estimates, for long-term recidivism, are in Table 24. Incarceration as a juvenile is followed by a substantial 23.4 percentage points more recidivism, meaning entering an Illinois prison before age 25. That central estimate is surrounded by a fairly wide margin of error, albeit one that easily excludes zero. The 95% confidence interval is 15–38%. The impact is rather evenly divided between violent, property, and drug crimes, which are not mutually exclusive in this analysis since a person could be imprisoned for more than one of those. The impact on homicide is positive too, but not clearly different from zero, perhaps because its rarity in the data impedes precise estimation. In addition, incarceration reduces the odds of high school graduation by 12.5 percentage points (se = 4.3%; Table IV, col. 7).

Overall, I find Aizer and Doyle compelling. However, it does deviate from the experimental ideal in one unremarked respect. Since quasi-randomization occurs within neighborhoods, there ought to be dummies for each neighborhood, in order to isolate the quasi-random within-neighborhood, cross-judge differences.[118] Instead, Aizer and Doyle introduce dummies for "communities," which are 76 geographic areas they say are defined at lib.uchicago.edu/e/collections/maps/ssrc. At that site, I find essentially no references to the term "community" and various maps of Chicago from the 1920s and 1930 divided into more than 76 tracts. Even if the definition of community were clear, its congruence or conflict with the court system's geographic divisions would not be, since they are not mapped or counted in the paper. In principle, this switch in geographic unit opens the door to endogeneity, for within a given "community," some judges could serve the more-troubled overlapping "neighborhoods," where juveniles are both more likely run into trouble with the law both as teenagers and young adults, without the first causing the second.

However, Aizer and Doyle report several sets of results that assuage this concern. First, they run a variant of their main regression approach with dummies for each census tract rather than "community." This dispels the concern to the extent that the court's "neighborhoods" can be described as sets of census tracts; though not demonstrated, that extent *may* be great since census tracts are small. And the results easily survive this change (Aizer and Doyle, Table AV). Second, Aizer and Doyle (Table II) show that their sample is reasonably balanced across judges. Overall, low p values in these comparisons do not appear much more common than would occur by chance.[119] If unobserved defendant traits predicting adult recidivism did vary systematically with judge severity—which would invalidate the study—then it would be surprising for observed traits such as age, race, and gender *not* to do so.

Aizer and Doyle point to one potential competing explanation for their findings: perhaps juvenile incarceration does not increase adult criminality, but only the probability that, upon later conviction, a judge will again impose incarceration. (Recall that they measure recidivism as later incarceration.) Two arguments contend against this explanation. First, while prior convictions often influence sentencing, it is less obvious that prior punishments would.[120] Second, as Aizer and Doyle observe, the impact appears strong on violent crime too, "for which incarceration is nearly certain, regardless of [past] juvenile incarceration" (p. 31).

---

118 For example, Di Tella and Schargrodsky (p. 47) and Kling (2006, p. 865) include dummies for the exact geographic domains of randomization.

119 The 10 p values for five traits—Male, African American, Special education, U.S. census tract poverty rate, and age at offense—average 0.40. 0.5 is the ideal.

120 A sentencing guideline summary published by the Illinois General Assembly (2005), refers several times to prior convictions, but never to prior sentences.

Thus while the case is not as hermetic as I would like, because of the mixed definitions of neighborhood and community, Aizer and Doyle's findings looks like the best interpretation of their data: juvenile incarceration increased adult recidivism.

**Table 24. Impact of juvenile incarceration in Cook County, 1990–2006, on chance of subsequent entry into Illinois adult prison before age 25, by type of subsequent conviction, from Aizer and Doyle (2015)**

| Crime category | Average in sample | Estimated impact | Impact standard error |
|---|---|---|---|
| All | 0.327 | 0.234*** | (0.076) |
| Weapon offense | 0.0555 | –0.005 | (0.026) |
| Violent† | 0.121 | 0.149*** | (0.041) |
| Homicide† | 0.043 | 0.035 | (0.030) |
| Assault | 0.0243 | 0.068*** | (0.025) |
| Robbery | 0.049 | 0.065* | (0.035) |
| Property† | 0.06 | 0.142*** | (0.044) |
| Burglary | 0.0238 | 0.061*** | (0.029) |
| Motor vehicle theft | 0.0241 | 0.066*** | (0.025) |
| Drug† | 0.176 | 0.097* | (0.052) |

Regressions include dummies for all neighborhood-year-weapon involvement combinations, as well as other controls, and are instrumented with leave-one-out average judge sentencing frequency. Source: Aizer and Doyle, Tables V (cols. 4, 7), VI (cols 3, 6), AIII (cols. 2, 4, 6, 8, 10). *significant at p<.1; ***significant at p<.01. †Average for those not incarcerated by juvenile court judge.

### 10.5. Summary: Juveniles

Reassuringly, the literature on juveniles looks like the adult literature in miniature. Where incarceration sentences appear to deter adults mildly at most, they perturb juveniles even less. But imprisoning juveniles does incapacitate them too. And as for aftereffects, high-credibility studies again conflict. Hjalmarsson (2009b) finds that time in an institution nudges young people toward a "path to the straight and narrow" while Aizer and Doyle find jail time reducing high school graduation and increasing criminality in young adulthood. The parallels to the adult literature do not end there: the study finding a fall in recidivism exploits a grid, like Kuziemko and Ganong, while the one finding an increase exploits judge randomization. I do not know whether that is more than a coincidence.

## 11. Conclusion

### 11.1. Synthesis

Individual study reviews are summarized in the table in section 4.

From all of this searching, filtering, reading, and reanalyzing, I distill one major lesson about the conduct of social science research and several about the impacts of mass incarceration on crime in the US. As for the first:

- *Many studies of the impacts of incarceration on crime contain problems that could (and in some cases do) overturn the authors' conclusions, once subject to replication and reanalysis.* Of the eight studies selected for this review whose data sets I could obtain or reconstruct, reanalysis revealed minor problems in one (Green and Winik) and significant issues of methodology or interpretation in seven (Helland and Tabarrok, Abrams, Levitt, Lofstrom and Raphael, Green and Winik, Kuziemko, Ganong), which led to major reinterpretations of four (Helland and Tabarrok, Abrams, Kuziemko, Ganong). There is no reason to believe that the studies for which data were unavailable are more reliable.

For this reason, the next section's cost-benefit analysis minimizes reliance on studies whose data I could not access.

As for the implications for the impacts of mass incarceration on crime:

- *Swift and certain punishment can deter, when practical, but perhaps works mainly when complemented with positive incentives and appropriate treatment.* Hawaii established the impressive HOPE program to bring discipline to parole. But it has not replicated well. More generally, the HOPE approach is harder to apply when crimes are hard for the government to observe or when the suspects have not lost certain rights to due process.
- *Longer sentences do not clearly deter crime.* The two key studies of deterrence reviewed here find mild deterrence among adults, with an elasticity of –0.1, but turn out to need major caveats. Helland and Tabarrok's conclusion that California's Three Strikes deterred arises from a treatment group that had more prior offenses, including drug offenses, than its control group (Table 6, above), and more drug arrests after release. And to the extent that individuals in the study exited the drug business for fear of a third strike, probably others entered to replace most of them. The Abrams conclusion that gun add-on laws reduced gun robberies does not extend to another outcome, gun assaults, or another policy, mandatory minimum sentencing laws; and it looks fragile.
- *Intensively supervised release, possibly including electronic monitoring, can increase freedom and save money without increasing crime; but the political will needed to sustain and scale it has been scarce.* In Buenos Aires, releasing inmates to electronic monitoring *reduced* crime overall but the government shut down the program after one releasee murdered a family of four (Di Tella and Schargrodsky 2013, p. 63). Regardless of the overall impact of electronic monitoring on crime, or murder in particular, the image of a program deliberately freeing potentially dangerous criminals became toxic. In the US, a suite of randomized trials of intensively supervised release (Deschenes, Turner, and Petersilia) struggled to recruit subjects, partly because of the wariness among local officials. But studies that did reach reasonable size found no crime cost in moving from prison to intensively monitored release.
- *Incapacitation is real: time inside prison reduces crime outside prison.* Estimates of incapacitation vary greatly, from 0.34 subsequent court appearances/year among in Houston, among people with previous offenses mild enough to split judges on whether to incarcerate them (Table 15, col. 2, above, citing Mueller-Smith); to 3.66 crimes per *month* in the Netherlands in a program targeting prolific offenders (Vollaard). A case relevant to current policy debates is that of California after the 2011 "realignment" reform. If we attribute the property crime rise in California starting in 2011 to the 8% incarceration reduction (Figure 18)—concentrated among non-serious, nonviolent, nonsexual offenders—then each year of incarceration averted by realignment caused some 6.7 more property crimes, within which the association with motor vehicle thefts is clearest, at 1.2 (Table 11).
- *Most relevant, credible studies suggest that incarceration aftereffects are harmful, and strong enough to offset the crime benefits of incapacitation. Yet disagreements remain.* Except for one I deemphasize because of econometric issues, all the judge randomization studies find more time followed by more crime—Green and Winik, Loeffler, Nagin and Snodgrass, Mueller-Smith, Aizer and Doyle. Three studies exploiting discontinuities in guidelines for time served conclude oppositely (Hjalmarsson 2009b, Kuziemko, Ganong), as does an experiment in early release in California in 1970 (Berecochea and Jaman). Of these, I have examined Kuziemko and Ganong closely, and their results appear explicable by various combinations of fragility and parole and framing effects. But if my reinterpretation of their data fails to support beneficial crime aftereffects from longer sentences, it also fails to corroborate the *harmful* aftereffects found in the judge randomization studies.

Drawing together the findings from this long journey of scrutiny leads to a surprisingly simple conclusion: the best estimate of the marginal impact of incarceration on crime in the US today is zero. The claims that increasing the severity of incarceration even mildly deters appear weak. Aftereffects appear to cancel out incapacitation in most contexts. But while zero is my central estimate, I do not view it as certain. On the one hand, the Georgia studies, as reanalyzed here, depart from the rest in not finding harmful crime aftereffects from incarceration. On the other, Mueller-Smith's formidable study goes strongly the other way: aftereffects do not merely cancel out incapacitation but easily surpass it in magnitude, and mostly likely deterrence as well, so that incarceration increases crime at the margin. Meanwhile, all of the studies reviewed probably leave out most of the crime increase *in* prison that comes from putting more people there.[121]

The apparent unreliability of studies that have not undergone replication-based scrutiny argues for setting them all aside. If we focus on the eight replicated in this review the conclusion just voiced further crystalizes. Deterrence looks weak or effectively non-existent (reanalyses of Helland and Tabarrok, Abrams); incapacitation is real (Levitt, Buonanno and Raphael, Lofstrom and Raphael) but the one study that measures incapacitation and aftereffects in the same context finds the second to at least cancel the first. Still, the reanalysis of Ganong dissents: aftereffects are about zero, it says, which cannot cancel out incapacitation. That disagreement drives the split between this review's primary interpretation and the devil's advocate view, which is explored more in the next section.

Important caveats pertain. The "marginal impact of incarceration on crime in the US" is an abstraction. In reality, there are thousands of margins—different people, different crimes, different places, different ways to adjust sentencing guidelines and laws, and so on. With some 30 studies of specific sources of variation in specific contexts, we can reach for the first-order generalization but must remember the complexity behind it. In the same spirit, even if the marginal impact broadly approximates zero, that does not mean that we could eliminate incarceration without raising crime. A conclusion of zero marginal benefit suggests that incarceration ought to fall—since there are no benefits to justify the costs—but does not tell us how far it should fall.

## 11.2. Cost-benefit analysis at the current US margin

Since I am not sure that decarceration would not raise crime, this section takes the analysis one step further by attempting to compare the costs and benefits of decarceration in aggregate.

Cost-benefit analysis implies a utilitarian moral frame. In embracing this frame, I do not mean to dismiss deontology—that is, to imply that notions of right and wrong, justice and injustice have no place in society's decisions around crime and punishment. Rather, I suggest only that a confrontation of costs to benefits should have some significance, morally and politically.

Incarceration affects inmates, their families, actual or would-be victims of crime caused or prevented, public agencies, and the general public. In an attempt to aggregate these consequences and perform a maximally evidence-based assessment of the value of incarceration in the US today, I run a cost-benefit analysis. Inevitably, this entails crude assumptions and vexing choices. Some factors, such as the impact of mass incarceration on the communities most affected by it, go unquantified. Whether a dollar of lost income is more harmful for the typical inmate than the typical taxpayer—if the inmate is poorer to start with—is not considered. (On deep issues in cost-benefit analysis in crime policy, see Dominquez and Raphael 2015.) To partially compensate for these inherent flaws, an accompanying spreadsheet allows the reader to explore the consequences of modifying many parameters in the analysis.

As mentioned, the cost-benefit starts from two interpretations of the evidence. The primary case is simple. It assumes no deterrence benefit, as the reanalyses of Helland and Tabarrok and Abrams suggest, and assumes that incapacitation is exactly cancelled by harmful aftereffects, as conservatively suggested by Green and Winik (along with the unreplicated Loeffler and Nagin and Snodgrass). With zero crime impact, this scenario sees only benefits to decarceration, not costs.

The devil's advocate takes deterrence at the value Helland and Tabarrok and Abrams converge to, an elasticity of –0.1—my skeptical reanalyses notwithstanding. And it estimates incapacitation from the reanalysis of the impacts of realignment in California (Table 11, last row, above). Aftereffects come from the reanalysis of Ganong's study of the long-term impacts of the grid revision in Georgia, in particular the RDD-based estimates by crime for 1993 (Table 22, penultimate column). In this scenario, aftereffects do not offset incapacitation because they approximate zero outside the ambiguous category of returns to prison for felony charges during probation and parole.[122]

[121] Probably murders in prison are reasonably well captured in official statistics while other crimes are not.

The starting point for both scenarios is the national setting in a recent year, as defined by totals for incarceration and crime. About 2.2 million people were incarceration at the end of 2015—1.53 million prisoners and 0.73 million jail inmates (BJS 2016a, Table 1). And in 2015, people reported about 9.2 million "index" crimes to the police, meaning those in the categories that constitute the FBI's official crime rate.[123] (See Table 25, col. 1.) These crimes probably loom largest when people think about public safety. But the list, like most of the studies reviewed here, also omits much: arson, drug crimes, driving under the influence and other traffic violations, white collar crimes including identity theft and online fraud, buying or selling sexual services, and misdemeanors. Leaving out crimes of commerce—implicitly treating them as costing society nothing at the margin—looks reasonably accurate, if only because of the evidently strong replacement effect in the illicit drug industry (§2.4.1). The omission of white collar crimes, which could easily dwarf traditional robbery, burglary, and theft in total cost, is unavoidable for lack of data and impact studies.

These numbers only capture *reported* crimes. A cost-benefit analysis ought to embrace unreported crimes too. The National Crime Victimization Survey lets us estimate the second from the first, by regularly asking a nationally representative sample of households how many crimes they have recently experienced and how many they reported to the police. Column 3 of Table 25 performs the requisite math, raising the estimated index crime total to 26 million in 2014.

In the devil's-advocate scenario, the impacts on crime are expressed in dollars. Researchers have estimated the money cost of crime in several ways, all inevitably problematic (Heaton 2010, pp. 2–4). *Bottom-up* analyses tally concrete expenditures on prevention (e.g., on burglar alarms) and treatment (e.g., for post-assault medical care) and even add in intangible harms inferred from jury awards to crime victims. This approach has generally failed to include intangible costs of crime for communities, as distinct from the victims and perhaps their families. Almost everyone in a neighborhood feels some harm, for example, when assaults and robberies become more common. *Willingness-to-pay* or *contingent valuation* studies attempt to cover all bases by directly asking people how much they would be willing to pay to prevent crimes—or think their community ought to be willing to pay. But this approach has its own limitation: the hypothetical nature of the questions. Expressed and actual willingness to pay may differ. Finally, *hedonic* studies infer actual willingness to pay from prices, such as for property in higher-crime neighborhoods. But they suffer from major endogeneity concerns—does crime depress property values or do low-income neighborhoods just suffer more crime for other reasons? Hedonic studies also are not good at unpacking impacts by crime type.

I follow Mueller-Smith in working with two sets of cost estimates rooted in highly cited studies: the bottom-up accounting of Miller, Cohen, and Wiersema (1996), as updated by McCollister, French, and Fang (2010); and the contingent valuation estimates of Cohen et al. (2004, Table 2). From the first, I include accounting-based "victim costs" and "pain and suffering costs" inferred from jury awards. I leave out that source's estimates of productivity costs (lost earnings while in prison) and estimate them separately. I also exclude criminal justice system costs, since those are dominated by incarceration costs, which are also separately handled. (See second- and fourth-to-last columns of Table 25.)[124]

Impacts on murder turn out to be a wildcard in the results. Because murder is rare, the estimated effects of incarceration upon it are statistically imprecise in nearly all studies. Yet the outsized societal cost of murder magnifies this uncertainty enough to sometimes dominate the overall cost figures. For example, the estimate from Georgia that incapacitation reduced murders by 0.0023 per prisoner-year is quite indistinguishable from zero, with a standard error almost four times as large (0.0090, from Table 22, penultimate column). Even though that impact rate is reasonably interpreted as zero, taking it at face value and multiplying it by $9 million per murder still yields $20,700, sufficient to offset a substantial two-thirds of the cost of imprisoning someone, as discussed below. And the corresponding 95% confidence interval reaches to $60,000. This is why Mueller-Smith (Table 11) drops murder from his cost-benefit analysis while Ganong (p. 20) trims the cost of murder to twice that of rape.

---

[122] This helps explain why Ganong's primary estimate of the crime cost of a year of extra time served is just $50 (Ganong, p. 20; Table 5, col. 5).
[123] In fact, these national magnitudes matter only for deterrence, because the deterrence estimates used here is expressed an elasticity. The incapacitation and aftereffects estimates as used here arrive in per-prisoner-year units, so they do not link mathematically to crime or incarcerated population totals.
[124] Cohen et al. estimate the cost of armed robbery only, as distinct from robbery in general. To recast their figure, $232,000, as being for robbery generally, I follow Heaton (2010, Table 1) in dividing it by $29,000 / $12,000 = 2.42, which is the ratio of corresponding bottom-up estimates in Cohen and Piquero (2009, Table 5), yielding $66,200.

Updating the methodology of Cohen (1988, p. 548) and Miller, Cohen, and Wiersema (1996), McCollister, French, and Fang (2010, §4.1), offers an interesting alternative for taming this wild uncertainty. Using FBI data (e.g., FBI 2015, Expanded Homicide Data Table 12), they calculate the national-level probability that a crime such as rape or aggravated assault leads to death. They then allocate the high costs of homicide to other crime groups in proportion to such probabilities. Using these augmented costs for non-homicide crimes, and discarding homicide per se, removes from the cost-benefit tallies the uncertainty associated with the impact estimates on homicide. In effect, the impacts of incarceration on homicide are taken as directly proportional to those on rape, assault, and robbery, which are themselves more precisely estimated. I prefer this approach for the added stability. By inferring the relevant ratios from McCollister, French, and Fang's bottom-up valuations, I apply the same adjustment to Cohen et al.'s willingness-to-pay valuations as well (see third-to-last and last columns of Table 25, below).

In the devil's-advocate case, the crime increase from decarceration is translated into dollars by each of the two methods in turn, and constitutes the cost side of the decarceration ledger. In both the primary-interpretation and devil's advocate cases, the factors on the *benefits* side run as follows:

- *Reduced prison operation.* The Vera Institute of Justice (Henrichson and Delaney 2012, Figure 4) estimates the all-in annual cost of running prisons at an average $31,286/prisoner (in 2010 dollars) for 40 states with adequate data. This value includes capital and variable costs, both of which are included here as relevant to the long-term implications of policy change. It is weighted by states' prison population counts.
- *Adjustment for food, housing, etc., provided by prison.* Some prison costs are best seen as transfers, since inmates are fed, housed, and perhaps given job training or drug treatment. I crudely estimate their value at $5,000/inmate/year, and count them as a negative benefit. As it happens, Donohue (2009), Table 9.10, uses $2,990 per prisoner for medical care, $1,088 for food, $905 for utilities.
- *Gained liberty.* Despite the free services, most people would rather not live behind bars. To value gained liberty, I start with an estimate of the value of a year of life—more formally, a Quality-Adjusted Life Year (QALY). A traditional valuation in the US context is $50,000/QALY; however, this number has obscure origins (Grosse 2008), and hasn't been adjusted for inflation over decades of use. An analysis by Braithwaite et al. (2008) of the costs and benefits of health care and health insurance suggests that American society is willing to pay $100,000–200,000 to save a year of life. One variant of their analysis excludes the dollar costs and health benefits of healthcare spending for children, making it more relevant to our interest in adult incarceration, and produces a figure of $95,000 (Braithwaite et al., Table 4, Panel A). I use $100,000, taking it to be in 2010 dollars.

  This figure represents the value of a year of life. The question then is how much to discount it in order to value a year of *gained liberty*. Many discounts might be defended; I use 0.5. For comparison, the World Health Organization (2013, p. 77) has used 0.494 for "Amputation of both legs: long term, without treatment."

  Thus, the analysis values a year of lost liberty at $50,000.
- *Prevented earnings/productivity loss during incarceration.* No study replicated here estimates impacts on earnings. But in regressions not discussed in my review above, Mueller-Smith (Table 7, Panel C) estimates that each quarter served in prison for a felony charge cost a Harris County inmate of $1,632 in

earnings.[125]

- *Prevented earnings/productivity loss following incarceration.* Similarly, Mueller-Smith (Table 7, last row) estimates that having been incarcerated cut one's quarterly wages in the first five years post-release by $683.50, plus $246.50 per year served (in 2010 dollars). The need to extrapolate from these two numbers, representing extensive and intensive margins of incarceration, forces me to confront an ambiguity in the scenario I am analyzing: Does an additional prisoner-year of incarceration arise from more incarceration spells or longer ones, or some mix of both? The balance between these two affects the relative weight that should be put on Mueller-Smith's per-incarceration and per-year-of-incarceration figures.

  Through simulations of prison population dynamics, Raphael and Stoll (2013, p. 78) conclude that a rising entrance probability, far more than longer prison spells, caused the state prison populations boom between 1984 and 2004, which in turn dominated over federal prisons in the national trend. Thus for simplicity, I assume that the prison reduction here simulated reverses that history, causing fewer rather than shorter prison spells. Assuming an average length of stay of three years (Pew Center on the States 2012, Table 1, calculates 2.9), Mueller-Smith's estimates point to five-year, undiscounted wage losses of 5 years post-release × 4 quarters/year × ($683.50 wage loss/quarter post-release/incarceration episode + $246.50 wage loss/quarter post-release/year served × 3 years served) = $28,460 per additional *prisoner* and thus a third of that per *prisoner-year* of time served, or $9,487.

  One shortcoming of this estimate is that identification in Mueller-Smith flows mainly from the experiences of people imprisoned less than a year (Mueller-Smith, Figure 4); so by using three years, this calculation extrapolates outside the study sample in length of stay. But this concern is secondary since the total cost here proves modest next to others considered, at about $7,000/prisoner-year (see below).

  Separately, using a judge randomization design with Florida data, Kling (2006, fig. 1) finds that the post-release earnings impacts of incarceration fade within two years, which suggests that going beyond Mueller-Smith's five-year horizon would hardly increase the apparent earnings loss.
- *Other impacts on prisoners, families, and communities.* These are not counted, because of the challenges of valuing them. The President's Council of Economic Advisers (2016, pp. 48–51) lists additional consequences of incarceration for inmates, families, and their communities. While a well-developed literature has attempted to put dollar values on crime, these other considerations have garnered little attempt at valuation. Crime is committed and suffered *in* prison as well as beyond. Families are sundered, with more children missing parents and higher rates of divorce (though some families also benefit from a dangerous loved one being locked up). For released convicts, a felony record can cut off access to public housing and other benefits, and the right to obtain a driver's license. More generally, mass incarceration can engender deep distrust of government, especially in poor and minority communities.

  One option for monetizing the benefits of reducing these harms is to view each averted incarceration as a prevented aggravated assault (meaning one involving a weapon or serious injury). Literature invoked above to monetize crime impacts values an aggravated assault at $22,000–89,000 (in 2010 dollars; see Table 25). A figure in this range might be discounted to the extent that convictions are viewed as more just than criminal assaults, and might partly or fully replace the valuation just put on loss of liberty. I do not pursue this option here, but the reader easily can.

  While the monetized impacts of incarceration on families and communities may be underestimated here, the same may go for the impacts of *crime*. As noted, researchers have not managed to quantify the fear that comes from living in a higher-crime area. And that too disproportionately affects poor and minority communities (Forman 2012, §IV).

The cost-benefit results come together in Table 26, which is denominated in dollars of 2010. The societal cost of an inmate-year of incarceration is put at $92,000, with loss of liberty the largest item ($50,000) and the cost of incarceration next ($26,000 after netting out $5,000 in service transfers). In the primary-interpretation scenario, the offsetting costs are zero. In the devil's-advocate case, they amount to $27,000 or $92,000, depending on the crime valuation methodology, as shown in the bottom-right of the table. This suggests that decarceration is, in the worst case encompassed by the evidence reviewed here, break even for society.

[125] Some people do work while in prison, e.g., under organized prison labor programs. Whether or not they get to keep much of the earnings, this constitutes an economic contribution, which is not counted here, for lack of data.

Vast, informal, invisible confidence intervals surround these figures. The estimates of deterrence, incapacitation, and aftereffects come with standard errors, as do the ratios used to infer total crime from reported crime. Deeper uncertainties pertain to the value of liberty and the consequences of crime and incarceration for families and communities.

Leaving aside imponderables, two big swing variables emerge within the ambit of the analysis. One is which scenario to favor, especially whether to assume incarceration aftereffects are zero, or harm public safety enough to offset incapacitation. The evidence for the latter—the better case for decarceration—looks stronger in my view because it comes from a single study (Green and Winik, in DC) that measures incapacitation and aftereffects in the same context and with the same method and outcome variable, is corroborated by similar (if unreplicated) studies in two other contexts, and is cast as conservative by two more (Mueller-Smith and Aizer and Doyle). In contrast, the devil's advocate takes evidence on incapacitation from post-2011 California, the dependent variable being state-level reported crime, and aftereffects from mid-1990s Georgia, the dependent variable being the individual-level return-to-prison rate. Some other studies do find beneficial aftereffects, but not as many (Berecochea and Jaman, Hjalmarsson 2009b).

The other big swing variable pertains only to the devil's-advocate case: which crime valuations to use. More specifically, *burglary* turns out to dominate the practical disagreement between them, with the low per-burglary cost at $1,700 (in dollars of 2010) and the high one at $32,000 ($25,000 in the source's year-2000 dollars). Both are multiplied by the estimated 1.5 burglaries/prisoner-year caused by realignment in California (last row of Table 11). The high number comes from the willingness-to-pay surveys of Cohen et al. It derives from the facts that at the time of the survey the US had about 100 million households; that they suffered some 4 million burglaries/year; and that respondents on average supported a hypothetical effort costing $100/household if it cut burglaries by 10%, meaning 0.004 fewer burglaries/household/year. (Cohen et al. 2004, Table 2). And $100 / 0.004 = $25,000.

Readers can make their own calls. To me, the willingness-to-pay results look unreliable, for several reasons. Stated willingness to pay is not demonstrated willingness to pay. And the survey respondents may have implicitly overestimated the local burglary rate. If they did—if they overestimated, however tacitly, the denominator in the above fraction—then the method of Cohen et al. would overestimate their willingness to pay to prevent burglary. Several observations give cause to doubt Americans' assessments of crime rates. In almost every year since 1993, a majority of Americans have believed that crime was rising, according to Gallup (McCarthy 2015), even as it almost never was (BJS 2015, Appendix Table 1). And as Dominguez and Raphael (2015, pp. 616–19) point out, a four-state survey found similar willingness to pay for a 30% cut in juvenile offending, across states whose actual offending rates varied by a factor of two (Piquero and Steinberg 2010). Louisiana was at the high end and Washington at the low. Possibly the implication—that where crime is twice as high people are half as willing to pay for absolute reductions—was true and a mere coincidence. The simpler explanation is that across states, respondents' valuations were unmoored from actual local offense rates, so they put about the same values on relative reductions, unaware of the implied per-crime valuations.

Because of the many empirical and philosophical uncertainties, cost-benefit analysis of incarceration cannot be conclusive, only suggestive. The analysis performed here suggests that it is hard to argue from high-credibility evidence that at typical margins in the US today, decarceration would harm society.

**Table 25. Cost-benefit analysis inputs by crime type**

| | Index crimes (thousand), 2015 | | | Incapacitation: California, 2012–14 (committed/ year) | Aftereffects: Georgia, 1993–2004 (returns to prison/ year) | Crime valuations | | | |
|---|---|---|---|---|---|---|---|---|---|
| Crime | A. Reported | B. Reporting rate | Committed (A ÷ B) | | | Bottom-up (2008 $1,000) | | Willingness-to-pay (2000 $1,000) | |
| Murder | 16 | *100.0%* | 16 | –0.002 | 0.0012 | 9,180 | | 9,700 | |
| Rape | 124 | 32.5% | 382 | 0.011 | –0.0093 | 204 | 205 | 239 | 239 |
| Aggravated assault | 764 | 61.9% | 1,235 | 0.013 | –0.0023 | 22.1 | 103.7 | 156 | 156 |
| Robbery | 327 | 61.9% | 529 | –0.034 | 0.0171 | 8.3 | 25.9 | 85 | 85 |
| Burglary | 1,579 | 50.8% | 3,109 | –1.535 | 0.0078 | 1.4 | 1.7 | 25 | 25 |
| Larceny/theft | 5,706 | 28.6% | 19,952 | –3.956 | 0.0416 | 0.5 | 0.5 | 2 | 2 |
| Motor vehicle theft | 708 | 69.0% | 1,026 | –1.197 | 0.0002 | 6.1 | 6.4 | 8 | 8 |
| *Parole/probation revocation* | | | | | –0.1792 | | 3 | 3 | 3 |

Sources: Crime counts from FBI (2016), Table 1; reporting rates from BJS (2016b, Table 4); impacts from Table 11, last row, and Table 22, penultimate column, above; bottom-up values from McCollister, French, and Fang (2010, Table 3, col. 1; Table 4); Cohen et al. (2004, Table 2); valuation of parole/probation revocation from Ganong (2012, p. 32).

**Table 26. Estimated costs and benefits of a person-year of decarceration in the US**

| | Number of crimes caused | | | | Costs & benefits (2010 $1,000) | |
|---|---|---|---|---|---|---|
| | Deterrence | Incapacitation | Aftereffects | Total | Low | High |
| **Benefits (primary & devil's-advocate cases)** | | | | | **92** | |
| Reduced prison operation | | | | | 31 | |
| Less: value of food, housing, etc. | | | | | –5 | |
| Gained liberty | | | | | 50 | |
| Prevented earnings loss during | | | | | 7 | |
| Prevented earnings after, 5 years | | | | | 9 | |
| **Costs (devil's-advocate case)** | **0.90** | **6.70** | **–0.70** | **6.90** | **27** | **92** |
| Murder | 0.0005 | 0.0023 | –0.0015 | 0.0014 | 0 | 0 |
| Rape | 0.01 | –0.01 | 0.05 | 0.05 | 11 | 16 |
| Aggravated assault | 0.04 | –0.01 | 0.01 | 0.04 | 4 | 7 |
| Robbery | 0.02 | 0.03 | –0.10 | –0.04 | –1 | –5 |
| Burglary | 0.11 | 1.53 | –0.11 | 1.53 | 3 | 49 |
| Larceny/theft | 0.69 | 3.96 | –0.73 | 3.91 | 2 | 11 |
| Motor vehicle theft | 0.04 | 1.20 | –0.00 | 1.23 | 8 | 12 |
| Probation/parole revocation for felony charge | | | 0.18 | 0.18 | 1 | 1 |

Sources and methods described in text. In primary-interpretation case, benefits are exactly zero. "Low" crime benefits based on accounting exercise in McCollister, French, and Fang (2010). "High" estimates based on willingness-to-pay survey in Cohen et al. (2004).

## Sources

Abadie, Alberto, Alexis Diamond, and Jens Hainmueller. 2010. "Synthetic Control Methods for Comparative Case Studies: Estimating the Effect of California's Tobacco Control Program." *Journal of the American Statistical Association* 105 (490): 493–505. DOI: 10.1198/jasa.2009.ap08746.

Abrams, David S. 2012. "Estimating the Deterrent Effect of Incarceration Using Sentencing Enhancements." *American Economic Journal: Applied Economics* 4 (4): 32–56. DOI: 10.1257/app.4.4.32.

Ackerberg, Daniel A., and Paul J. Devereux. 2009. "Improved JIVE Estimators for Overidentified Linear Models with and without Heteroskedasticity." *Review of Economics and Statistics* 91 (2): 351–62. DOI: 10.1162/rest.91.2.351.

Aizer, Anna, and Joseph J. Doyle. 2015. "Juvenile Incarceration, Human Capital, and Future Crime: Evidence from Randomly Assigned Judges." *Quarterly Journal of Economics* 130 (2): 759–803. DOI: 10.1093/qje/qjv003.

Alarid, Leanne Fiftal. 2016. *Community Based Corrections*. Cengage Learning.

Alexander, Michelle. 2012. *The New Jim Crow: Mass Incarceration in the Age of Colorblindness*. The New Press.

Allison, Paul D. 2001. Missing Data. SAGE Publications.

Alm, Steven S. 2016. "HOPE Probation: Fair Sanctions, Evidence-Based Principles, and Therapeutic Alliances." *Criminology & Public Policy* 15 (4): 1195–1214. DOI: 10.1111/1745-9133.12261.

Anderson, T. W., Naoto Kunitomo, and Takamitsu Sawa. 1982. "Evaluation of the Distribution Function of the Limited Information Maximum Likelihood Estimator." *Econometrica* 50 (4): 1009–27. DOI: 10.2307/1912774.

Angrist, Joshua D., and Guido W. Imbens. 1995. "Two-Stage Least Squares Estimation of Average Causal Effects in Models with Variable Treatment Intensity." *Journal of the American Statistical Association* 90 (430): 431–42. DOI: 10.1080/01621459.1995.10476535.

Angrist, Joshua D., and Jörn-Steffen Pischke. 2008. *Mostly Harmless Econometrics: An Empiricist's Companion*. Princeton University Press.

Angrist, Joshua D., and Jörn-Steffen Pischke. 2010. "The Credibility Revolution in Empirical Economics: How Better Research Design Is Taking the Con out of Econometrics." *Journal of Economic Perspectives* 24 (2): 3–30. DOI: 10.1257/jep.24.2.3.

Barbarino, Alessandro, and Giovanni Mastrobuoni. 2014. "The Incapacitation Effect of Incarceration: Evidence from Several Italian Collective Pardons." *American Economic Journal: Economic Policy* 6 (1): 1–37. DOI: 10.1257/pol.6.1.1.

Belloni, Alexandre, and Victor Chernozhukov. 2013. "Least Squares after Model Selection in High-Dimensional Sparse Models." *Bernoulli* 19(2): 521–47. DOI: 10.3150/11-BEJ410.

Baum, Christopher F., Mark E. Schaffer, and Steven Stillman. 2007. "Enhanced Routines for Instrumental Variables/Generalized Method of Moments Estimation and Testing." *The Stata Journal* 7 (4): 465–506. DOI: 10.1177/1536867X0800700402.

Beccaria, Cesare. 1819. *An Essay on Crimes and Punishments*. Edward D. Ingraham, trans. Philip H. Nicklin. google.com/books/edition/An_Essay_on_Crimes_and_Punishments/FRDtZqosmnEC.

Berecochea, John B., Dorothy R. Jaman, and Welton A. Jones. 1973. "Time Served in Prison and Parole Outcome: An Experimental Study: Report No. 1." Research Report 49. California Department of Corrections. ncjrs.gov/pdffiles1/Digitization/11444NCJRS.pdf.

Berecochea, John B., and Dorothy R. Jaman. 1981. "Time Served in Prison and Parole Outcome: An Experimental Study: Report No. 2." Research Report 62. California Department of Corrections. ncjrs.gov/pdffiles1/Digitization/82800NCJRS.pdf.

Benson, Bruce L. 2011. *The Enterprise of Law: Justice without the State*. Independent Institute.

Bentham, Jeremy. 1838. *The Works of Jeremy Bentham*. Part II. William Tait. play.google.com/store/books/details?id=uJHRAAAAMAAJ.

Benko, Jessica. 2015. "The Radical Humaneness of Norway's Halden Prison." *New York Times*, March 26. nytimes.com/2015/03/29/magazine/the-radical-humaneness-of-norways-halden-prison.html.

Bhuller, Manudeep, Gordon B. Dahl, Katrin V. Løken, and Magne Mogstad. 2016. "Incarceration, Recidivism and Employment." September 9. web.archive.org/20170119173550/http:/econweb.ucsd.edu/~gdahl/papers/incarceration-recidivism-employment.pdf.

Bureau of Justice Statistics (BJS). 1992. *Correctional Populations in the United States, 1990.* bjs.gov/content/pub/pdf/cpus90.pdf.

Bureau of Justice Statistics (BJS). 2011a. *Correctional Populations in the United States, 2010.* bjs.gov/content/pub/pdf/cpus10.pdf.

Bureau of Justice Statistics (BJS). 2011b. *Criminal Victimization, 2010.* bjs.gov/content/pub/pdf/cv10.pdf.

Bureau of Justice Statistics (BJS). 2013. *Correctional Populations in the United States, 2012.* bjs.gov/content/pub/pdf/cpus12.pdf.

Bureau of Justice Statistics (BJS). 2014. *Correctional Populations in the United States, 2013.* bjs.gov/content/pub/pdf/cpus13.pdf.

Bureau of Justice Statistics (BJS). 2015. *Criminal Victimization, 2014.* bjs.gov/content/pub/pdf/cv14.pdf.

Bureau of Justice Statistics (BJS). 2016a. *Correctional Populations in the United States, 2015.* bjs.gov/content/pub/pdf/cpus15.pdf.

Bureau of Justice Statistics (BJS). 2016b. *Criminal Victimization, 2015.* bjs.gov/content/pub/pdf/cv15.pdf.

Blumstein, Alfred, and Allen J. Beck. 2005. "Reentry as a Transient State between Liberty and Recommitment." In Jeremy Travis and Christy Visher, eds. *Prisoner Reentry and Crime in America.* Cambridge University Press.

Blumstein, Alfred, Jacqueline Cohen, and Paul Hsieh. 1982. "The Duration of Adult Criminal Careers." National Institute of Justice. ncjrs.gov/pdffiles1/Digitization/89569NCJRS.pdf.

Bond, Stephen. 2002. "Dynamic Panel Data Models: A Guide to Micro Data Methods and Practice." Working Paper CWP09/02. Centre for Microdata Methods and Practice. cemmap.ac.uk/wps/cwp0209.pdf.

Braithwaite, R. Scott, David O. Meltzer, Joseph T. King Jr, Douglas Leslie, and Mark S. Roberts. 2008. "What Does the Value of Modern Medicine Say about the $50,000 per Quality-Adjusted Life-Year Decision Rule?" *Medical Care* 46(4): 349–56. DOI: 10.1097/MLR.0b013e31815c31a7.

Buonanno, Paolo, and Steven Raphael. 2013. "Incarceration and Incapacitation: Evidence from the 2006 Italian Collective Pardon." *American Economic Review* 103(6): 2437–65. DOI: 10.1257/aer.103.6.2437.

Bushway, Shawn D., and Emily G. Owens. 2013. "Framing Punishment: Incarceration, Recommended Sentences, and Recidivism." *Journal of Law and Economics* 56(2): 301–31. DOI: 10.1086/669715.

Butterfield, Fox. 1996. "Tough Law on Sentences Is Criticized." *New York Times*. March 8. nytimes.com/1996/03/08/us/tough-law-on-sentences-is-criticized.html.

Caseload Forecast Council. 2014. "2013 Washington State Juvenile Disposition Guidelines Manual." State of Washington. cfc.wa.gov/PublicationSentencing/SentencingManual/Juvenile_Disposition_Manual_2013.pdf.

Census Bureau. 1973. *Statistical Abstract of the United States.* www2.census.gov/library/publications/1973/compendia/statab/94ed/1973-03.pdf.

Census Bureau. 2010. Annual Surveys of State and Local Government Finances. Table 1. www2.census.gov/govs/local/10slsstab1a.xls.

Chalfin, Aaron, and Justin McCrary. 2014. "Criminal Deterrence: A Review of the Literature." eml.berkeley.edu/~jmccrary/chalfin_mccrary2014.pdf.

Chen, M. Keith, and Jesse M. Shapiro. 2007. "Do Harsher Prison Conditions Reduce Recidivism? A Discontinuity-Based Approach." *American Law and Economics Review* 9(1): 1–29. DOI: 10.1093/aler/ahm006.

Clemens, Michael A. 2017. "The Meaning of Failed Replications: A Review and Proposal." *Journal of Economic Surveys* 31 (1): 326–42. DOI: 10.1111/joes.12139.

Cohen, Mark A. 1988. "Pain, Suffering, and Jury Awards: A Study of the Cost of Crime to Victims." *Law & Society*

*Review* 22(3): 537–55. DOI: 10.2307/3053629.

Cohen, Mark A., and Alex R. Piquero. 2009. "New Evidence on the Monetary Value of Saving a High Risk Youth." *Journal of Quantitative Criminology* 25(1): 25–49. DOI: 10.1007/s10940-008-9057-3.

Cohen, Mark A., Roland T. Rust, Sara Steen, and Simon T. Tidd. 2004. "Willingness-to-pay for Crime Control Programs." *Criminology* 42(1): 89–110. DOI: 10.1111/j.1745-9125.2004.tb00514.x.

Council of Economic Advisers. 2016. *Economic Perspectives on Incarceration and the Criminal Justice System*. Executive Office of the President of the United States. obamawhitehouse.archives.gov/sites/whitehouse.gov/files/documents/CEA%2BCriminal%2BJustice%2BReport.pdf.

Cullen, Francis T., Cheryl Lero Jonson, and Daniel S. Nagin. 2011. "Prisons Do Not Reduce Recidivism: The High Cost of Ignoring Science." *Prison Journal* 91 (3_suppl): 48S–65S. DOI: 10.1177/0032885511415224.

Davidson, Russell, and James G. MacKinnon. 2010. "Wild Bootstrap Tests for IV Regression." *Journal of Business & Economic Statistics* 28(1): 128–44. DOI: 10.1198/jbes.2009.07221.

Department of Corrections and Rehabilitation (DCR). 2010. "Second and Third Striker Felons in the Adult Institution Population." State of California. web.archive.org/20121027210830/http:/www.cdcr.ca.gov/Reports_Research/Offender_Information_Services_Branch/Quarterly/Strike1/STRIKE1d1006.pdf.

Department of Corrections and Rehabilitation (DCR). 2013. "Second and Third Striker Felons in the Adult Institution Population." State of California. web.archive.org/20140707004308/http:/www.cdcr.ca.gov/Reports_Research/Offender_Information_Services_Branch/Quarterly/Strike1/STRIKE1d1306.pdf.

Department of Justice (DOJ) and Crime and Justice Institute (CJI). 2004. "Implementing Evidence-Based Practice in Community Corrections: The Principles of Effective Intervention." s3.amazonaws.com/static.nicic.gov/Library/019342.pdf.

Deschenes, Elizabeth Piper, Susan Turner, and Joan Petersilia. 1995. "A Dual Experiment in Intensive Community Supervision: Minnesota's Prison Diversion and Enhanced Supervised Release Programs." *Prison Journal* 75(3): 330–56. DOI: 10.1177/0032855595075003005.

Di Tella, Rafael, and Ernesto Schargrodsky. 2013. "Criminal Recidivism after Prison and Electronic Monitoring." *Journal of Political Economy* 121(1): 28–73. DOI: 10.1086/669786.

Dills, Angela K., Jeffrey A. Miron, and Garrett Summers. 2008. "What Do Economists Know About Crime?" Working Paper 13759. National Bureau of Economic Research. DOI: 10.3386/w13759.

Dobbie, Will, Jacob Goldin, and Crystal Yang. 2016. "The Effects of Pre-Trial Detention on Conviction, Future Crime, and Employment: Evidence from Randomly Assigned Judges." scholar.harvard.edu/files/cyang/files/dgy_bail_august2016.pdf.

Domínguez, Patricio, and Steven Raphael. 2015. "The Role of the Cost-of-Crime Literature in Bridging the Gap Between Social Science Research and Policy Making: Potentials and Limitations." *Criminology & Public Policy* 14(4): 589–632. DOI: 10.1111/1745-9133.12148.

Donohue III, John J. 2009. "Assessing the Relative Benefits of Incarceration: Overall Changes and the Benefit on the Margin." In Steven Raphael and Michael A. Stoll, eds. *Do Prisons Make Us Safer? The Benefits and Costs of the Prison Boom*. Russell Sage Foundation.

Drago, Francesco, Roberto Galbiati, and Pietro Vertova. 2009. "The Deterrent Effects of Prison: Evidence from a Natural Experiment." *Journal of Political Economy* 117(2): 257–80. DOI: 10.1086/599286.

Ehrlich, Isaac. 1981. "On the Usefulness of Controlling Individuals: An Economic Analysis of Rehabilitation, Incapacitation and Deterrence." *American Economic Review* 71(3): 307–22. jstor.org/stable/1802781.

Erwin, Billie S., and Lawrence A. Bennett. 1987. "New Dimensions in Probation: Georgia's Experience with Intensive Probation Services." National Institute of Justice. *Research in Brief*. January. ncjrs.gov/pdffiles1/Digitization/102848NCJRS.pdf.

Farrington, David P. 1986. "Age and Crime." *Crime and Justice* 7: 189–250. jstor.org/stable/1147518.

Farrington, David P., Alex R. Piquero, and Wesley G. Jennings. 2013. *Offending from Childhood to Late Middle Age: Recent Results from the Cambridge Study in Delinquent Development*. Springer Science & Business Media.

Federal Bureau of Investigation (FBI). 2011. *Crime in the United States 2010*. ucr.fbi.gov/crime-in-the-u.s/2010/crime-in-the-u.s.-2010/tables/10tbl05.xls.

Federal Bureau of Investigation (FBI). 2015. *Crime in the United States 2014*. ucr.fbi.gov/crime-in-the-u.s/2014/crime-in-the-u.s.-2014/offenses-known-to-law-enforcement.

Finlay, Keith, Leandro Magnusson, and Mark E. Schaffer. 2013. "weakiv: Weak-instrument-robust Tests and Confidence Intervals for Instrumental-variable (IV) Estimation of Linear, Probit and Tobit models." Statistical Software Components S457684, Boston College Department of Economics.

Fischer, Ryan G. 2005. "Are California's Recidivism Rates Really the Highest in the Nation? It Depends on What Measure of Recidivism You Use." *The Bulletin* 1(1). Center for Evidence-Based Corrections. University of California, Irvine. ucicorrections.seweb.uci.edu/files/2013/06/bulletin_2005_vol-1_is-1.pdf.

Forman, James Jr. 2012. "Racial Critiques of Mass Incarceration: Beyond the New Jim Crow." Faculty Scholarship Series. Paper 3599. Yale Law School. digitalcommons.law.yale.edu/fss_papers/3599.

Gaes, Gerald G., and Scott D. Camp. 2009. "Unintended Consequences: Experimental Evidence for the Criminogenic Effect of Prison Security Level Placement on Post-Release Recidivism." *Journal of Experimental Criminology* 5(2): 139–62. DOI: 10.1007/s11292-009-9070-z.

Ganong, Peter N. 2012. "Criminal Rehabilitation, Incapacitation, and Aging." *American Law and Economics Review* 14(2): 391–424. DOI: 10.1093/aler/ahs010.

Georgia State Bureau of Pardons and Parole (GSBPP). 1979. *Annual Report*. ncjrs.gov/pdffiles1/Digitization/65974NCJRS.pdf.

Georgia State Bureau of Pardons and Parole (GSBPP). 1983. *Annual Report*. archive.org/details/GAStateBoardOfPardonsAndParolesAnnualReportFiscalYear1983P.2.

Georgia State Bureau of Pardons and Parole (GSBPP). 1989. *Annual Report*. ncjrs.gov/pdffiles1/Digitization/121464NCJRS.pdf.

Georgia State Bureau of Pardons and Parole (GSBPP). 1993. *Annual Report*. ncjrs.gov/pdffiles1/Digitization/148288NCJRS.pdf.

Georgia State Bureau of Pardons and Parole (GSBPP). 1994. *Annual Report*. ncjrs.gov/pdffiles1/Digitization/153701NCJRS.pdf.

Georgia State Bureau of Pardons and Parole (GSBPP). 2002. *Annual Report*. pap.georgia.gov/sites/pap.georgia.gov/files/Annual_Reports/2002_Annual_Report0001.pdf.

Georgia State Bureau of Pardons and Parole (GSBPP). 2008. *Annual Report*. pap.georgia.gov/sites/pap.georgia.gov/files/Annual_Reports/08_Annual_Report.pdf.

Green, Donald P., and Daniel Winik. 2010. "Using Random Judge Assignments to Estimate the Effects of Incarceration and Probation on Recidivism among Drug Offenders." *Criminology* 48(2): 357–87. DOI: 10.1111/j.1745-9125.2010.00189.x.

Greene, William. 2003. *Econometric Analysis*. 5th edition. Prentice Hall. web.archive.org/20150226013851/http:/stat.smmu.edu.cn/DOWNLOAD/ebook/econometric.pdf.

Grosse, Scott D. 2008. "Assessing Cost-Effectiveness in Healthcare: History of the $50,000 per QALY Threshold." *Expert Review of Pharmacoeconomics & Outcomes Research* 8(2): 165–78. DOI: 10.1586/14737167.8.2.165.

Gupta, Arpit, Christopher Hansman, and Ethan Frenchman. 2016. "The Heavy Costs of High Bail: Evidence from Judge Randomization." *Journal of Legal Studies* 45(2): 471–505. DOI: 10.1086/688907.

Hansen, Lars Peter, John Heaton, and Amir Yaron. 1996. "Finite-Sample Properties of Some Alternative GMM Estimators." *Journal of Business & Economic Statistics* 14(3): 262–80. DOI: 10.1080/07350015.1996.10524656.

Hawken, Angela, and Mark Kleiman. 2009. "Managing Drug Involved Probationers with Swift and Certain Sanctions: Evaluating Hawaii's HOPE." Department of Justice. ncjrs.gov/pdffiles1/nij/grants/229023.pdf.

Hawken, Angela, Jonathan Kulick, Kelly Smith, Jie Mei, Yiwen Zhang, Sara Jarman, Travis Yu, Chris Carson, and Tifanie Vial. 2016. "Managing Drug Involved Probationers with Swift and Certain Sanctions: Evaluating Hawaii's HOPE." Department of Justice. ncjrs.gov/pdffiles1/nij/grants/249912.pdf.

Heaton, Paul, Sandra Mayson, and Megan Stevenson. 2016. "The Downstream Consequences of Misdemeanor Pretrial Detention." law.upenn.edu/live/files/5693-harriscountybail.

Helland, Eric, and Alexander Tabarrok. 2007. "Does Three Strikes Deter? A Nonparametric Estimation." *Journal of Human Resources* XLII(2): 309–30. DOI: 10.3368/jhr.XLII.2.309.

Henrichson, Christian, and Ruth Delaney. 2012. "The Price of Prisons: What Incarceration Costs Taxpayers." Vera Institute of Justice. vera.org/downloads/Publications/price-of-prisons-what-incarceration-costs-taxpayers/legacy_downloads/price-of-prisons-updated-version-021914.pdf.

Hirschi, Travis, and Michael Gottfredson. 1983. "Age and the Explanation of Crime." *American Journal of Sociology* 89(3): 552–84. jstor.org/stable/2779005.

Hjalmarsson, Randi 2009a. "Crime and Expected Punishment: Changes in Perceptions at the Age of Criminal Majority." *American Law and Economics Review* 11(1): 209–48. DOI: 10.1093/aler/ahn016.

Hjalmarsson, Randi. 2009b. "Juvenile Jails: A Path to the Straight and Narrow or to Hardened Criminality?" *Journal of Law & Economics* 52 (4): 779–809. DOI: 10.1086/596039.

Honaker, James, and Gary King. 2010. "What to Do about Missing Values in Time-Series Cross-Section Data." *American Journal of Political Science* 54(2): 561–81. DOI: 10.1111/j.1540-5907.2010.00447.x.

Institute for Behavior and Health, Inc. (IBH). 2015. "State of the Art of HOPE Probation." courts.state.hi.us/docs/news_and_reports_docs/State_of_%20the_Art_of_HOPE_Probation.pdf.

Illinois General Assembly. 2005. "Penalties for Crimes in Illinois. Legislative Research Unit. ilga.gov/commission/lru/2005PFC.pdf.

Inter-university Consortium for Political and Social Research (ICPSR). National Prisoner Statistics, 1978–2014. Codebook. pcms.icpsr.umich.edu/pcms/performDownload/191b349f-89d5-4d2f-916a-17113f0f2d70.

Iyengar, Radha. 2008. "I'd Rather Be Hanged for a Sheep than a Lamb: The Unintended Consequences of 'Three-Strikes' Laws." Working Paper 13784. National Bureau of Economic Research. DOI: 10.3386/w13784.

Katz, Lawrence, Steven D. Levitt, and Ellen Shustorovich. 2003. "Prison Conditions, Capital Punishment, and Deterrence." *American Law and Economics Review* 5(2): 318–43. jstor.org/stable/42705434.

Killias, Martin, Marcelo Aebi, and Denis Ribeaud. 2000. "Does Community Service Rehabilitate Better than Short-term Imprisonment?: Results of a Controlled Experiment." *Howard Journal of Criminal Justice* 39(1): 40–57. DOI: 10.1111/1468-2311.00152.

Kleiman, Mark. 2009. *When Brute Force Fails: How to Have Less Crime and Less Punishment*. Princeton University Press.

Kleiman, Mark. 2016. "Swift-Certain-Fair: What Do We Know Now, and What Do We Need to Know?" *Criminology & Public Policy* 15(4): 1185–93. DOI: 10.1111/1745-9133.12258.

Klick, Jonathan, and Alexander Tabarrok. 2010. "Police, Prisons, and Punishment: The Empirical Evidence on Crime Deterrence." In *Handbook on the Economics of Crime*. Edward Elgar Publishing. DOI: 10.4337/9781849806206.00014.

Kling, Jeffrey R. 2006. "Incarceration Length, Employment, and Earnings." *American Economic Review* 96 (3): 863–76. DOI: 10.1257/aer.96.3.863.

Koeter, M.J.W. and Bakker, M. (2007). "Effectevaluatie van de Strafrechtelijke Opvang Verslaafden (SOV)." Report 269. Department of Justice. wodc.nl/onderzoeksdatabase/98.071c-effectevaluatie-strafrechtelijke-opvang-verslaafden-sov.aspx.

Kuziemko, I. 2013. "How Should Inmates Be Released from Prison? An Assessment of Parole versus Fixed-Sentence Regimes." *Quarterly Journal of Economics* 128(1): 371–424. DOI: 10.1093/qje/qjs052.

Kuziemko, Ilyana, and Steven D. Levitt. 2004. "An Empirical Analysis of Imprisoning Drug Offenders." *Journal of Public Economics* 88(9): 2043–66. DOI: 10.1016/S0047-2727(03)00020-3.

Lattimore, Pamela K., Doris Layton MacKenzie, Gary Zajac, Debbie Dawes, Elaine Arsenault, and Stephen Tueller. 2016. "Outcome Findings from the HOPE Demonstration Field Experiment: Is Swift, Certain, and Fair an Effective Supervision Strategy?" *Criminology & Public Policy* 15 (4): 1103–41. DOI: 10.1111/1745-9133.12248.

Lee, David S., and Justin McCrary. 2009. "The Deterrence Effect of Prison: Dynamic Theory and Evidence." eml.berkeley.edu/~jmccrary/lee_and_mccrary2009.pdf.

Leslie, Emily, and Nolan G. Pope. 2017. "The Unintended Impact of Pretrial Detention on Case Outcomes: Evidence from New York City Arraignments." *Journal of Law and Economics* 60(3): 529–57. DOI: 10.1086/695285.

Levitt, Steven D. 1996. "The Effect of Prison Population Size on Crime Rates: Evidence from Prison Overcrowding Litigation." *Quarterly Journal of Economics* 111(2): 319–51. DOI: 10.2307/2946681.

Levitt, Steven D. 2004. "Understanding Why Crime Fell in the 1990s: Four Factors That Explain the Decline and Six That Do Not." *Journal of Economic Perspectives* 18(1): 163–90. DOI: 10.1257/089533004773563485.

Loeffler, Charles E. 2013. "Does Imprisonment Alter the Life Course? Evidence on Crime and Employment from a Natural Experiment." *Criminology* 51(1): 137–66. DOI: 10.1111/1745-9125.12000.

Lofstrom, Magnus, and Steven Raphael. 2016. "Incarceration and Crime: Evidence from California's Public Safety Realignment Reform." *Annals of the American Academy of Political and Social Science* 664(1): 196–220. DOI: 10.1177/0002716215599732.

Lofstrom, Magnus, Steven Raphael, and Ryken Grattet. 2014. "Is Public Safety Realignment Reducing Recidivism in California?" Public Policy Institute of California. ppic.org/content/pubs/report/R_614MLR.pdf.

Lott, John R., and John Whitley. 2003. "Measurement Error in County-Level UCR Data." *Journal of Quantitative Criminology* 19(2): 185–98. DOI: 10.1023/A:1023054204615.

Lynch, James P., and John P. Jarvis. 2008. "Missing Data and Imputation in the Uniform Crime Reports and the Effects on National Estimates." *Journal of Contemporary Criminal Justice*. February. DOI: 10.1177/1043986207313028.

Maddala, G. S. 1983. *Limited-Dependent and Qualitative Variables in Econometrics*. Cambridge University Press.

Maltz, Michael D. "Analysis of Missingness in UCR Crime Data." ncjrs.gov/pdffiles1/nij/grants/215343.pdf.

Maltz, Michael D., and Joseph Targonski. 2002. "A Note on the Use of County-Level UCR Data." *Journal of Quantitative Criminology* 18(3): 297–318. DOI: 10.1023/A:1016060020848.

Martin, S. E., S. Annan, and B. Forst. 1993. "The Special Deterrent Effects of a Jail Sanction on First-Time Drunk Drivers: A Quasi-Experimental Study." *Accident Analysis and Prevention* 25(5): 561–68. DOI: 10.1016/0001-4575(93)90008-k.

Martin, Brandon, and Ryken Grattet. 2015. "Alternatives to Incarceration in California." Public Policy Institute of California. ppic.org/content/pubs/report/R_415BMR.pdf.

Marvell, Thomas B., and Carlisle E. Moody. 1994. "Prison Population Growth and Crime Reduction." *Journal of Quantitative Criminology* 10 (2): 109–40. DOI: 10.1007/BF02221155.

Maryland State Commission on Criminal Sentencing Policy (MSCCSP). 1999. *Annual Report*. msccsp.org/Files/Reports/ar1999.pdf.

Maryland State Commission on Criminal Sentencing Policy (MSCCSP). 2001. *Maryland Sentencing Guidelines Manual*. msccsp.org/Files/Guidelines/MSGM/Version_1.0.pdf.

McCarthy, Justin. 2015. "More Americans Say Crime Is Rising in U.S." Gallup. gallup.com/poll/186308/americans-say-crime-rising.aspx.

McCollister, Kathryn E., Michael T. French, and Hai Fang. 2010. "The Cost of Crime to Society: New Crime-Specific Estimates for Policy and Program Evaluation." *Drug and Alcohol Dependence* 108(1-2): 98–109. DOI: 10.1016/j.drugalcdep.2009.12.002.

Miller, Ted R., Mark A. Cohen, and Brian Wiersama. 1996. "Victim Costs and Consequences: A New Look." National

Institute of Justice. ncjrs.gov/pdffiles/victcost.pdf.

Minton, Todd D., and Zhen Zeng. 2015. "Jail Inmates at Midyear 2014." Bureau of Justice Statistics. bjs.gov/content/pub/pdf/jim14.pdf.

Miron, J. A. 1999. "Violence and the U.S. Prohibitions of Drugs and Alcohol." *American Law and Economics Review* 1(1): 78–114. DOI: 10.1093/aler/1.1.78.

Missouri Working Group on Sentencing and Corrections (MWGSC). 2011. Consensus Report. senate.mo.gov/12info/comm/special/MWSC-Report.pdf.

Mueller-Smith, Michael. 2015. "The Criminal and Labor market Impacts of Incarceration." sites.lsa.umich.edu/mgms/wp-content/uploads/sites/283/2015/09/incar.pdf.

Murray, Michael P. 2006. "Avoiding Invalid Instruments and Coping with Weak Instruments." *Journal of Economic Perspectives* 20(4): 111–32. DOI: 10.1257/jep.20.4.111.

Nagin, Daniel S. 2013. "Deterrence: A Review of the Evidence by a Criminologist for Economists." *Annual Review of Economics* 5(1): 83–105. DOI: 10.1146/annurev-economics-072412-131310.

Nagin, Daniel s., Francis T. Cullen, and Cheryl Lero Jonson. 2009. "Imprisonment and Reoffending." *Crime and Justice* 38 (1): 115–200. DOI: 10.1086/599202.

Nagin, Daniel S., and G. Matthew Snodgrass. 2013. "The Effect of Incarceration on Re-Offending: Evidence from a Natural Experiment in Pennsylvania." *Journal of Quantitative Criminology* 29 (4): 601–42. DOI: 10.1007/s10940-012-9191-9.

National Archive of Criminal Justice Data (NACJD). 2007. "Law Enforcement Agency Identifiers Crosswalk [United States], 2005." Interuniversity Consortium for Political and Social Research (ICPSR). DOI: 10.3886/ICPSR04634.V1.

National Highway and Traffic Safety Administration (NHTSA). 2008. *Statistical Analysis of Alcohol-related Driving Trends, 1982–2005*. www-nrd.nhtsa.dot.gov/Pubs/810942.PDF.

National Research Council (NRC). 1986. *Criminal Careers and "Career Criminals.* Vol. I. Alfred Blumstein, Jacqueline Cohen, Jeffrey A. Roth, and Christy A. Visher, eds. National Academies Press. DOI: 10.17226/922.

O'Connell, Daniel J., John J. Brent, and Christy A. Visher. 2016. "Decide Your Time." *Criminology & Public Policy* 15(4): 1073–1102. DOI: 10.1111/1745-9133.12246.

Owens, Emily G. 2009. "More Time, Less Crime? Estimating the Incapacitative Effect of Sentence Enhancements." *Journal of Law & Economics* 52(3): 551–79. DOI: 10.1086/593141.

Owens, Emily Greene. 2011. "Are Underground Markets Really More Violent? Evidence from Early 20th Century America." *American Law and Economics Review* 13(1): 1–44. DOI: 10.1093/aler/ahq017.

Petersilia, Joan, and Susan Turner. 1993. "Intensive Probation and Parole." *Crime and Justice* 17: 281–335. DOI: 10.1086/449215.

Pew Center on the States. 2012. "Time Served: The High Cost, Low Return of Longer Prison Terms." pewtrusts.org/~/media/legacy/uploadedfiles/wwwpewtrustsorg/reports/sentencing_and_corrections/prisontimeservedpdf.pdf.

Pintoff, Randi. 2004. "The Impact of Incarceration on Juvenile Crime: A Regression Discontinuity Approach." economics.yale.edu/sites/default/files/files/Workshops-Seminars/Industrial-Organization/pintoff-041012.pdf.

Piquero, Alex R., David P. Farrington, and Alfred Blumstein. 2003. "The Criminal Career Paradigm." *Crime and Justice* 30: 359–506. DOI: 10.1086/652234.

Piquero, Alex R., and Laurence Steinberg. 2010. "Public Preferences for Rehabilitation versus Incarceration of Juvenile Offenders." *Journal of Criminal Justice* 38(1): 1–6. DOI: 10.1016/j.jcrimjus.2009.11.001.

Quetelet, Adolphe. 1833. *Recherches sur le penchant au crime aux différens âges*. Hayez. play.google.com/store/books/details?id=ZNEiAAAAMAAJ.

Raphael, Steven, and Michael A. Stoll. 2013. *Why Are So Many Americans in Prison?* Russell Sage Foundation.

Rhodes, William, Gerald Gaes, Jeremy Luallen, Ryan Kling, Tom Rich, and Michael Shively. 2016. "Following Incarceration, Most Released Offenders Never Return to Prison." *Crime & Delinquency* 62(8): 1003–25. DOI: 10.1177/0011128714549655.

Roach, Michael A and Schanzenbach, Max Matthew, "The Effect of Prison Sentence Length on Recidivism: Evidence from Random Judicial Assignment" Northwestern Law & Econ Research Paper No. 16-08. DOI: 10.2139/ssrn.2701549.

Roeder, Oliver, Lauren-Brooke Eisen, and Julia Bowling. 2015. "What Caused the Crime Decline?" Brennan Center for Justice. brennancenter.org/sites/default/files/publications/What_Caused_The_Crime_Decline.pdf.

Roodman, David. 2009. "A Note on the Theme of Too Many Instruments." *Oxford Bulletin of Economics and Statistics* 71(1): 135–58. DOI: 10.1111/j.1468-0084.2008.00542.x.

Ross, Hugh Laurence. 1973. "Law, Science, and Accidents: The British Road Safety Act of 1967." *The Journal of Legal Studies* 2(1): 1–78. DOI: 10.1086/467491.

Ross, Hugh Laurence. 1984. *Deterring the Drinking Driver: Legal Policy and Social Control*. Lexington Books.

Schiraldi, Vincent, Jason Colburn, and Eric Lotke 2004. "Three Strikes and You're Out: An Examination of the Impact of 3-Strikes Laws 10 Years after their Enactment." Justice Policy Institute. justicepolicy.org/uploads/justicepolicy/documents/04-09_rep_threestrikesnatl_ac.pdf.pdf.

Simon, Julian L. 1966. "The Price Elasticity of Liquor in the U.S. and a Simple Method of Determination." *Econometrica* 34(1): 193–205. DOI: 10.2307/1909863.

Sourcebook of Criminal Justice Statistics. 2012. University at Albany, Hindelang Criminal Justice Research Center. albany.edu/sourcebook/pdf/t31062012.pdf.

Snyder, Howard N. 2012. "Arrest in the United States, 1990–2010." Bureau of Justice Statistics. bjs.gov/content/pub/pdf/aus9010.pdf.

Stevenson, Megan. 2016. "Distortion of Justice: How the Inability to Pay Bail Affects Case Outcomes." prisonpolicy.org/scans/Distortion-of-Justice-April-2016.pdf.

Strang, Heather, Lawrence W. Sherman, Evan Mayo-Wilson, Daniel Woods, and Barak Ariel. 2013. "Restorative Justice Conferencing (RJC) Using Face-to-Face Meetings of Offenders and Victims: Effects on Offender Recidivism and Victim Satisfaction. A Systematic Review." *Campbell Systematic Reviews* 9(1): 1–59. DOI: 10.4073/csr.2013.12.

Sundt, Jody, Emily J. Salisbury, and Mark G. Harmon. 2016. "Is Downsizing Prisons Dangerous?" *Criminology & Public Policy* 15(2): 315–41. DOI: 10.1111/1745-9133.12199.

Targonski, Joseph. 2012. Missing Data in the Uniform Crime Reports (UCR), 1977-2000 [United States]. Inter-university Consortium for Political and Social Research. DOI: 10.3886/ICPSR32061.v1.

Taxman, Fayes S., Eric S. Shepardson, and James M. Byrne. 2004. *Tools of the Trade: A Guide to Incorporating Science into Practice*. National Institute of Corrections and Maryland Department of Public Safety and Correctional Services. s3.amazonaws.com/static.nicic.gov/Library/020095.pdf.

Toda, Hiro Y., and Taku Yamamoto. 1995. "Statistical Inference in Vector Autoregressions with Possibly Integrated Processes." *Journal of Econometrics* 6 (1): 225–50. DOI: 10.1016/0304-4076(94)01616-8.

Tonry, Michael. 2014. "Why Crime Rates Are Falling throughout the Western World." *Crime and Justice* 43(1): 1–63. DOI: 10.1086/678181.

Ulmer, Jeffrey T., and Darrell Steffensmeier. 2014. "The Age and Crime Relationship: Social Cariation, Social Explanations." In Kevin M. Beaver, J.C. Barnes, and Brian B. Boutwell, eds. *The Nurture versus Biosocial Debate in*

*Criminology: On the Origins of Criminal Behavior and Criminality*. DOI: 10.4135/9781483349114.n24.

Vollaard, Ben. 2013. "Preventing Crime through Selective Incapacitation." *Economic Journal* 123(567): 262–84. DOI: 10.1111/j.1468-0297.2012.02522.x/full.

Washington University Law Review. 1979. "Determinate Sentencing in California and Illinois: Its Effect on Sentence Disparity and Prisoner Rehabilitation." openscholarship.wustl.edu/law_lawreview/vol1979/iss2/10.

Weber Jr., George N. 1996. Memo on Revision of Appendix A of the *Maryland Sentencing Guidelines Manual*. September 23. msccsp.org/Files/Guidelines/MSGM/October_1996.pdf.

Weisburd, David, Tomer Einat, and Matt Kowalski. 2008. "The Miracle of the Cells: An Experimental Study of Interventions to Increase Payment of Court-Ordered Financial Obligations." *Criminology & Public Policy* 7(1): 9–36. DOI: 10.1111/j.1745-9133.2008.00487.x.

Wooldridge, Jeffrey M. 2010. *Econometric Analysis of Cross Section and Panel Data*. MIT Press.

World Health Organization. 2013. *WHO Methods and Data Sources for Global Burden of Disease Estimates 2000–2011*. who.int/healthinfo/statistics/GlobalDALYmethods_2000_2011.pdf.

Zimring, Franklin E., "Populism, Democratic Government, and the Decline of Expert Authority: Some Reflections on Three Strikes in California." *Pacific Law Journal* 28(1): 243–56. scholarlycommons.pacific.edu/mlr/vol28/iss1/9.

Zimring, Franklin E., Gordon Hawkins, and Sam Kamin. 2001. *Punishment and Democracy: Three Strikes and You're out in California*. Oxford University Press.